\documentclass{elsart}

\usepackage{amsmath,amssymb,epsfig,bm}

\numberwithin{equation}{subsection}

\journal{Physics Reports}

\begin{document}



\newcommand{\eV}{\text{eV}}
\newcommand{\GeV}{\text{GeV}}
\newcommand{\arcmin}{\text{arcmin}}
\newcommand{\Mpc}{\text{Mpc}}
\newcommand{\Hunit}{$\text{km\,s}^{-1}\,\text{Mpc}^{-1}$}
\newcommand{\muK}{$\mu\text{K}$}


\newcommand{\D}{\text{D}}
\newcommand{\ud}{\text{d}}
\newcommand{\curl}{\,\text{curl}\,}


\newcommand{\alt}{\lesssim}
\newcommand{\agt}{\gtrsim}


\newcommand{\cla}{\mathcal{A}}
\newcommand{\clb}{\mathcal{B}}
\newcommand{\clc}{\mathcal{C}}
\newcommand{\cld}{\mathcal{D}}
\newcommand{\cle}{\mathcal{E}}
\newcommand{\clf}{\mathcal{F}}
\newcommand{\clg}{\mathcal{G}}
\newcommand{\clh}{\mathcal{H}}
\newcommand{\cli}{\mathcal{I}}
\newcommand{\clj}{\mathcal{J}}
\newcommand{\clk}{\mathcal{K}}
\newcommand{\cll}{\mathcal{L}}
\newcommand{\clm}{\mathcal{M}}
\newcommand{\cln}{\mathcal{N}}
\newcommand{\clo}{\mathcal{O}}
\newcommand{\clp}{\mathcal{P}}
\newcommand{\clq}{\mathcal{Q}}
\newcommand{\clr}{\mathcal{R}}
\newcommand{\cls}{\mathcal{S}}
\newcommand{\clt}{\mathcal{T}}
\newcommand{\clu}{\mathcal{U}}
\newcommand{\clv}{\mathcal{V}}
\newcommand{\clw}{\mathcal{W}}
\newcommand{\clx}{\mathcal{X}}
\newcommand{\cly}{\mathcal{Y}}
\newcommand{\clz}{\mathcal{Z}}


\newcommand{\ve}{\mathbf{e}}
\newcommand{\vehat}{\hat{\mathbf{e}}}
\newcommand{\vn}{\mathbf{n}}
\newcommand{\vnhat}{\hat{\mathbf{n}}}
\newcommand{\vv}{\mathbf{v}}
\newcommand{\vx}{\mathbf{x}}
\newcommand{\vp}{\mathbf{p}}
\newcommand{\vk}{\mathbf{k}}


\newcommand{\Omtot}{\Omega_{\mathrm{tot}}}
\newcommand{\Omb}{\Omega_{\mathrm{b}}}
\newcommand{\Omc}{\Omega_{\mathrm{c}}}
\newcommand{\Omm}{\Omega_{\mathrm{m}}}
\newcommand{\omb}{\omega_{\mathrm{b}}}
\newcommand{\omc}{\omega_{\mathrm{c}}}
\newcommand{\omm}{\omega_{\mathrm{m}}}
\newcommand{\omnu}{\omega_{\nu}}
\newcommand{\Omnu}{\Omega_{\nu}}
\newcommand{\Oml}{\Omega_\Lambda}
\newcommand{\OmK}{\Omega_K}




\newcommand{\Al}{{A_l}}
\newcommand{\TT}{\text{TT}}


\newcommand{\sigt}{\sigma_{\mbox{\scriptsize T}}}


\newcommand{\annphys}{\rm Ann.~Phys.~}
\newcommand{\araa}{\rm Ann.~Rev.~Astron.~\&~Astrophys.~}
\newcommand{\aap}{\rm Astron.~\&~Astrophys.~}
\newcommand{\apj}{\rm Astrophys.~J.~}
\newcommand{\apjl}{\rm Astrophys.~J.~Lett.~}
\newcommand{\apjs}{\rm Astrophys.~J.~Supp.~}
\newcommand{\apss}{\rm Astrophys.~Space~Sci.~}
\newcommand{\cqg}{\rm Class.~Quant.~Grav.~}
\newcommand{\grg}{\rm Gen.~Rel.~Grav.~}
\newcommand{\ijmpd}{\rm Int.~J.~Mod.~Phys.~D~}
\newcommand{\jcap}{\rm JCAP~}
\newcommand{\jetpl}{\rm J.~Exp.~Theor.~Phys.~Lett.~}
\newcommand{\jmp}{\rm J.~Math.~Phys.~}
\newcommand{\mnras}{\rm Mon.~Not.~R.~Astron.~Soc.~}
\newcommand{\nat}{\rm Nature~}
\newcommand{\prd}{\rm Phys. Rev.~D~}
\newcommand{\prl}{\rm Phys.~Rev.~Lett.~}
\newcommand{\plb}{\rm Phys.~Lett.~B~}
\newcommand{\physrep}{\rm Phys.~Rep.~}
\newcommand{\progthp}{\rm Prog.~Theor.~Phys.~}
\newcommand{\rmp}{\rm Rev.~Mod.~Phys.~}

\let\Oldsection\section
\renewcommand{\section}[1]{\Oldsection{\bf #1}}

\newcommand{\be} {\begin{equation}}
\newcommand{\ee} {\end{equation}}
\newcommand{\bea} {\begin{eqnarray}}
\newcommand{\eea} {\end{eqnarray}}


\begin{frontmatter}


\title{Relativistic cosmology and large-scale structure}


\author{Christos G. Tsagas}
\address{Section of Astrophysics, Astronomy and Mechanics, Department of Physics, Aristotle University of Thessaloniki, Thessaloniki 54124, Greece.}

\author{Anthony Challinor}
\address{Institute of Astronomy, Madingley Road, Cambridge, CB3 0HA, UK.} \address{DAMTP, Centre for Mathematical Sciences, Wilberforce Road,\\ Cambridge CB3 0WA, UK.}

\author{Roy Maartens}
\address{Institute of Cosmology \& Gravitation, University of Portsmouth,\\ Portsmouth P01 2EG, UK.}\vspace{1cm}

\tableofcontents

\newpage

\begin{abstract}
General relativity marked the beginning of modern cosmology and it has since been at the centre of many of the key developments in this field. In the present review, we discuss the general-relativistic dynamics and perturbations of the standard cosmological model, the Friedmann-Lemaitre universe, and how these can explain and predict the properties of the observable universe. Our aim is to provide an overview of the progress made in several major research areas, such as linear and non-linear cosmological perturbations, large-scale structure formation and the physics of the cosmic microwave background radiation, in view of current and upcoming observations. We do this by using a single formalism throughout the review, the $1+3$ covariant approach to cosmology, which allows for a uniform and balanced presentation of technical information and physical insight.
\end{abstract}

\begin{keyword}
Cosmology, Large-scale Structure.
\end{keyword}

\end{frontmatter}

\newpage


\section{Relativistic cosmology}\label{sRC}
Cosmology is the study of the dynamics and make-up of the Universe
as a whole, or at least the maximally observable region of the
Universe. Less than 100 years ago, the prevailing view (shared by
Einstein) was that the Universe was static, and the existence of
galaxies beyond our own remained unknown. A revolution was initiated via observations by Leavitt, Hubble and others which showed that the Universe was in fact expanding and contained many distant galaxies.
Friedmann, Lemaitre and other theorists showed how the expansion
could be explained by a spatially homogeneous and isotropic model
obeying the field equations of General Relativity. The expansion
pointed to an extremely hot origin of the Universe, the Big Bang,
and Gamow and others showed how this should leave a thermal relic
radiation, and also how nucleosynthesis of the light elements would
take place in the hot early universe. However, it took many decades
for observations to catch up and confirm this, and to lay the basis
for further developments.

\subsection{Cosmology at the dawn of the 21st century}\label{ssIC}
Cosmology has come of age as an observationally based physical
science in the last few decades, driven by the tremendous growth
in data from increasingly high-precision experiments. Key
milestones since 1990 include:

\begin{itemize}

\item
the Cosmic Background Explorer (COBE), that detected the large-angle anisotropies in the Cosmic Microwave Background (CMB) temperature, and its successors, especially the Wilkinson Microwave Anisotropy Probe (WMAP), that measured the anisotropies at small angles and detected the acoustic peaks;\\

\item
the 2-degree Field (2dF) Galaxy Redshift Survey, that measured the matter power spectrum based on over 200k galaxies, and its
successor, the Sloan Digital Sky Survey (SDSS);\\

\item
the Supernova Cosmology Project (SCP), that measured the
magnitude-redshift data for more than 40 supernovae (SNe), and its successors, including the Supernova Legacy Survey (SNLS).
\end{itemize}

Collectively, these experiments and others, including measurements
of weak lensing, underpin our current understanding of the evolutionary history and contents of the Universe. A broad range
of new and upcoming experiments will aim to refine and extend this
understanding.

The theoretical efforts to interpret the observational data and to
make further predictions that can be tested against observations,
have involved an important interplay between general relativity,
astrophysics, particle physics and computation. The current model
of large-scale structure formation in the Universe is based on the
following:

\begin{itemize}

\item
A spatially homogeneous and isotropic Friedmann-Lemaitre-Robertson-Walker (FLRW) background spacetime,
\begin{equation}
{\rm d}s^2=-{\rm d}t^2+ a^2(t)\left[{\rm d}r^2+f_K^2(r)({\rm d} \theta^2+\sin^2\theta{\rm d}\phi^2)\right]\,,  
\end{equation}
where the form of $f_K(r)$ depends on the model's spatial curvature -- see \S~\ref{sssFLRWM} below, represents the average dynamics on large scales.\\

\item
The expansion rate, $H$, is governed by the Friedmann equation,
\begin{equation}
H^2\equiv \left({\dot{a}\over a}\right)^2=
{1\over3}\,\left(\rho_{(r)}+\rho_{(c)}+\rho_{(b)} +\rho_{(de)}\right)- {K\over a^2}\,,
\end{equation}
with $K=0,\pm1$. Thus, $H$ is determined by the radiation, cold dark matter, baryonic matter and dark energy content of the Universe once $K$, the spatial curvature index, is chosen. The energy densities redshift with expansion according to the conservation law,
\begin{equation}
\dot{\rho}_{(i)}+3H(1+w_{(i)})\rho_{(i)}=0\,,
\end{equation}
where $w_{(i)}={p_{(i)}/\rho_{(i)}}$ with $w_{(r)}=1/3$, $w_{(c)}=0= w_{(b)}$, $w_{(de)}<-1/3$. The primordial radiation-dominated era is preceded by a brief burst of inflationary expansion, driven by a scalar field (or fields). Radiation decouples from baryonic matter soon after the total matter begins to dominate. At relatively recent times, matter begins to give way to a negative-pressure dark energy component which starts to accelerate the expansion again. The simplest model of dark energy has $\rho_{\mathrm{de}}=\Lambda$, the cosmological constant, representing the vacuum energy density, with $w_{\Lambda}=-1$.\\

\item
Large-scale structure emerges as small over-densities begin to grow in the matter-dominated era via gravitational instability. In order to grow the galaxies fast enough, non-baryonic cold dark matter is needed in the standard model based on general relativity. The seeds of these over-densities are provided by the vacuum fluctuations of the inflaton field. The simplest inflation models predict a nearly scale-invariant and Gaussian spectrum of density perturbations, and a sub-dominant component of gravitational wave perturbations. The imprint of these primordial density perturbations is recorded in the CMB anisotropies, and the subsequent evolution is measured via the evolving galaxy distribution. The inflationary model provides not only the seeds for the emergence of observed large-scale structure, but also resolves the critical puzzle within non-inflationary models: that widely separated parts of the CMB sky were never in causal contact yet have the same temperature.

\end{itemize}

\begin{figure*}
\begin{center}
\includegraphics[height=3.0in]{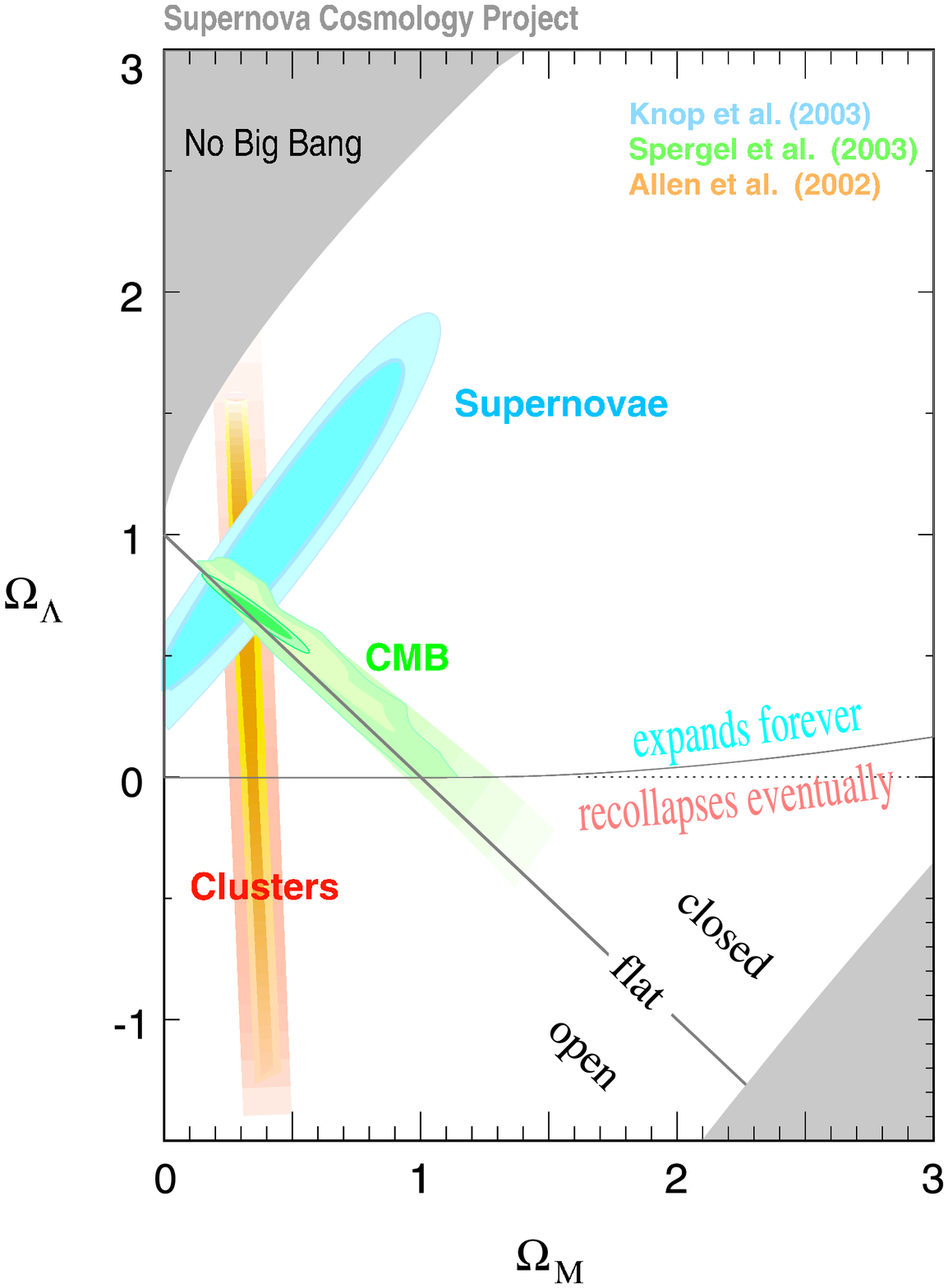}\quad
\includegraphics[height=3.0in]{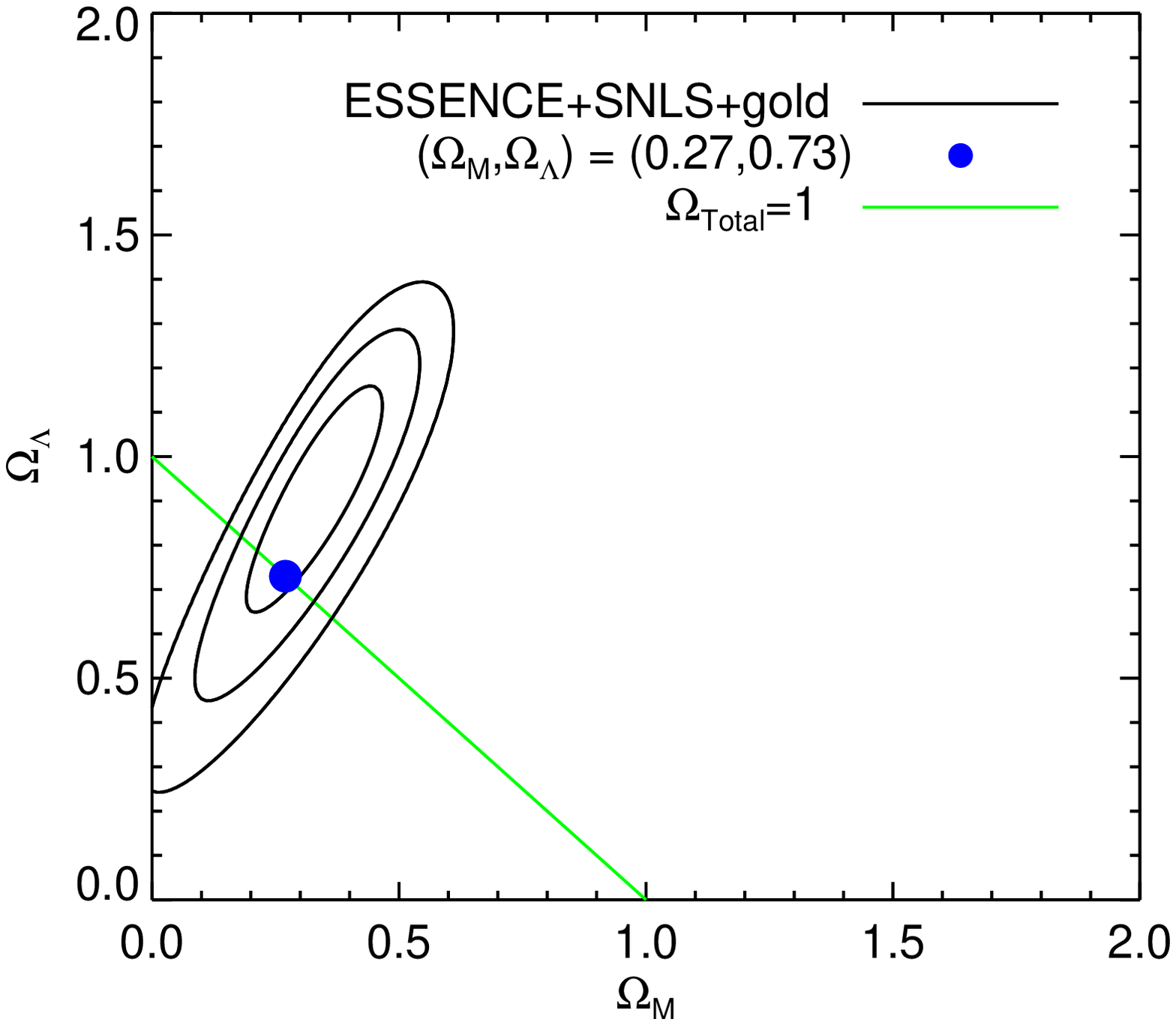}
\end{center}
\caption{Observational constraints in the $(\Omega_{(m)},\Omega_\Lambda)$ plane: joint constraints (left)
(from~\cite{2003ApJ...598..102K}); recent compilation of supernova
constraints (right) (from~\cite{2007ApJ...666..694W}).} \label{sn}
\end{figure*}

\begin{figure*}
\begin{center}
\includegraphics[height=2.25in]{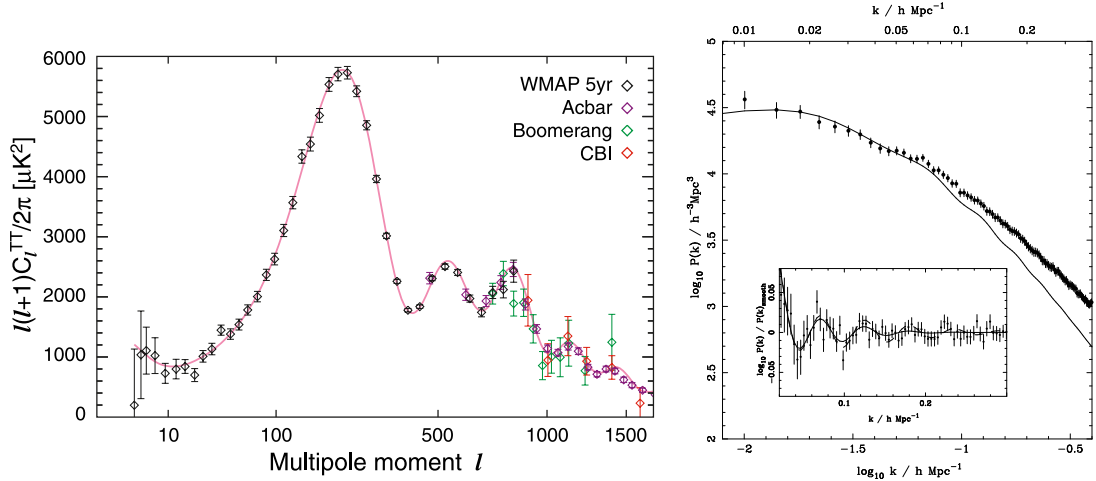}\quad
\includegraphics[height=2.5in]{pk_bigplot}
\end{center}
\caption{Left: Power spectrum of CMB temperature anisotropies,
showing data from WMAP5~\cite{2008arXiv0803.0593N},
the 2003 flight of BOOMERANG~\cite{2006ApJ...647..823J},
CBI~\cite{2004ApJ...609..498R} and the full ACBAR dataset~\cite{2008arXiv0801.1491R}. The red line is the best-fit LCDM model to the data. Right: Matter power spectrum, showing data from the SDSS 2006 data release and the best-fit LCDM curve; the inset shows the imprint (in Fourier space) of the CMB acoustic peaks, known as the baryon acoustic oscillations (from~\cite{2007ApJ...657..645P}).}
\label{cmb}
\end{figure*}

The current ``standard model" of cosmology is the inflationary Cold
Dark Matter (CDM) model with cosmological constant, usually called
LCDM, which is based on general relativity and particle physics
(i.e., the Standard Model, with minimal extensions). The LCDM model
provides an excellent fit to the wealth of high-precision
observational data, on the basis of a remarkably small number of
cosmological parameters (see, e.g.,~\cite{2007ApJS..170..377S,%
2007ApJ...657..645P}). In particular, independent data sets from CMB anisotropies, galaxy surveys and supernova luminosities, lead to a consistent set of best-fit model parameters. This is illustrated in Figs.~\ref{sn} and~\ref{cmb}

The LCDM model is remarkably successful, but we know that its
theoretical foundation, general relativity, breaks down at high
enough energies, $E \gtrsim M_{\text{fundamental}}$, where the
fundamental scale at which new physics kicks in is usually taken to
be the Planck scale, $M_{\text{fundamental}}= M_p \sim
10^{16}\,\mbox{TeV}\,,$ but could possibly be a lower scale, with
particle collider constraints indicating that
$M_{\text{fundamental}}\gtrsim 1\,\mbox{TeV}.$ LCDM can only provide limited insight into the very early universe. Indeed, the crucial
role played by inflation belies the fact that inflation remains an
effective theory without yet a basis in fundamental theory. A
quantum gravity theory will be able to probe higher energies and
earlier times, and should provide a consistent basis for inflation,
or an alternative that replaces inflation within the standard
cosmological model (for recent work, see e.g.
Refs.~\cite{2005LRR.....8...11B,2007PhRvD..75l3507E,%
2006hep.th...10221T,2006PhRvD..74h4003A,2007hep.th....2001B,%
2007hep.th....2059K}).

An even bigger theoretical problem than inflation is that of the
late-time acceleration in the expansion of the
universe~\cite{2006hep.th....3057C,2006AIPC..848..698P,%
2006hep.th....1213N,2006AIPC..861..179P,2006AIPC..861.1013P,%
2006MPLA...21.1083S,2007soch.conf....9B,2007GReGr.tmp....4U}. In
terms of the fundamental energy density parameters,
$\Omega_{(i)}=\rho_{(i)}/3H_0^2$, we can rewrite the Friedmann
equation using the conservation equations,
\begin{equation}
\left({H\over H_0}\right)^2=(\Omega_{(c)}+\Omega_{(b)})(1+z)^3+
\Omega_{(r)}(1+z)^4 +\Omega_\Lambda+ \Omega_K(1+z)^2 \,, \label{H}
\end{equation}
where the redshift is $z=a^{-1}-1$ with $a_0=1$ today. The data
indicates that the present cosmic energy budget is given by
\begin{equation}\label{olom}
\Omega_\Lambda\approx 0.75\,, \hspace{15mm} \Omega_{(m)}\equiv
\Omega_{(c)}+ \Omega_{(b)}\approx 0.25\,, \hspace{15mm}
|\Omega_K|\ll 1\,,
\end{equation}
so that the Universe is currently accelerating, $\ddot {a}_0>0$, and (nearly) spatially flat.

Within the framework of general relativity, the acceleration
typically originates from a dark energy field with negative
pressure. In LCDM, this is the vacuum energy ($w_\Lambda=-1$), but
dynamical dark energy fields have also been considered. For the
simplest option of vacuum energy, the observed value of the
cosmological constant is overwhelmingly smaller than the prediction
of current particle physics. In particular,
\begin{equation}
\rho_{\Lambda,\text{obs}}= \Lambda\sim H_0^2M_p^2\sim
(10^{-33}\,\mbox{eV})^2(10^{19}\,\mbox{GeV})^2\sim
(10^{-3}\,\mbox{eV})^4, \label{v1}
\end{equation}
whereas
\begin{equation}
\rho_{\Lambda,\text{theory}}\sim M_{\text{fundamental}}^4 \gtrsim
1~\mbox{TeV}^4 \gg \rho_{\Lambda,\text{obs}}\,. \label{v2}
\end{equation}
In addition, the $\Lambda$ value needs to be strongly fine-tuned to
be of the same order of magnitude today as the current matter
density, i.e.,
\begin{equation}\label{v3}
\rho_\Lambda \sim \rho_{(m)0}~ \Rightarrow~ \Omega_{\Lambda}\sim
\Omega_{(m)}\,,
\end{equation}
otherwise galaxies and then life could not emerge in the universe.
The question is how this ``coincidence" arises at late times, given
that
\begin{equation}
\rho_\Lambda = \,\mbox{constant}\,, \hspace{5mm} \mbox{while}
\hspace{5mm} \rho_{(m)}\propto (1+z)^3\,.
\end{equation}

No convincing or natural explanation has yet been proposed. String
theory provides a tantalising possibility in the form of the
``landscape" of vacua~\cite{2006hep.th....3249P,%
2006hep.th...10211B}. There appears to be a vast number of vacua
admitted by string theory, with a broad range of energies above and
below zero. The idea is that our observable region of the universe
corresponds to a particular small positive vacuum energy, whereas
other regions with greatly different vacuum energies will look
entirely different. This multitude of regions forms in some sense a
``multiverse". This is an interesting idea, but it is highly
speculative, and it is not clear how much of it will survive the
further development of string theory and cosmology.

A different approach is based on the idea that there is no material
dark energy field, but instead the Universe accelerates due to
gravitational effects. Within general relativity, this has been
proposed via nonlinear effects from structure
formation~\cite{2004JCAP...02..003R,2005PhRvD..71f3537B,%
2005hep.th....3117K,2005PhRvD..72b3517G,2005PhRvL..95o1102C,%
2005gr.qc.....7057N,2006NJPh....8..322K,2005astro.ph.10523M,%
2006astro.ph..1699M,2006AIPC..861..987M,2006CQGra..23.6955P}. As structure forms and the matter density perturbations become nonlinear, there are two questions that are posed: (1)~what is the back-reaction effect of this nonlinear process on the background cosmology?; (2)~how do we perform a covariant and gauge-invariant averaging over the inhomogeneous universe to arrive at the correct FRW background? The simplistic answers to these questions are (1)~the effect is negligible since it occurs on scales too small to be cosmologically significant; (2)~in light of this, the background is independent of structure formation, i.e., it is the same as in the linear regime. A quantitative analysis is needed to fully resolve both issues. However, this is very complicated because it involves the nonlinear features of general relativity in an essential way.

There have been claims that these simplistic answers are wrong, and
that, on the contrary, the effects are large enough to accelerate
the universe. Note the possibility that averaging effects could be significant, even if they do not lead to acceleration. This would indeed be a dramatic and satisfying resolution of the coincidence
problem, without the need for any dark energy field. Of course, this would not solve the problem of the vacuum energy, but would only
re-define the problem as: why does the vacuum not gravitate?
However, the claims for acceleration via nonlinear effects have been widely disputed, and it is fair to say that there is as yet no
convincing demonstration that this is possible.

A more drastic form of a gravitational explanation for late-time
acceleration, is that general relativity breaks down on the largest
scales, and a modified gravity theory takes over on these scales.
Schematically, this means modifying the geometric side of the field
equations,
\begin{equation}
G_{ab}+ G^{({\rm dark})}_{ab}= 8\pi G T_{ab}\,,  \label{mod}
\end{equation}
rather than the matter side,
\begin{equation}
G_{ab}= 8\pi G\left(T_{ab}+ T^{({\rm dark})}_{ab}\right)\,,
\end{equation}
as in the standard general relativity approach. Modified gravity
represents an intriguing possibility for resolving the theoretical
crisis posed by late-time acceleration. However, it turns out to be
extremely difficult to modify general relativity at low energies in
cosmology, without violating the solar system constraints, or without introducing ghosts and other instabilities into the theory, or without altering the expansion rate in the matter-dominated era. Up to now, there is no convincing alternative to the general relativistic dark-energy models. It is indicative of the stir the supernovae observations have caused, that even the Copernican principle itself has been questioned~\cite{2007arXiv0712.3457C}.

In addition to the theoretical problems of inflation and dark
energy, there is also the problem of ``missing mass", i.e., the fact that we cannot account for the observed matter power spectrum, given the observed CMB power spectrum, if we invoke only baryonic matter
and use general relativistic dynamics. The general relativistic
solution to this problem is non-baryonic CDM, dominating over
baryonic matter at roughly 5 to 1. This solution simultaneously
accounts for the rotation curves of spiral galaxies. Extensions of
the Standard Model of particle physics predict various candidate
particles for the cold dark matter, and a range of experiments is
underway or planned to constrain, detect or rule out some of these.

A more radical approach to the missing mass problem is to reject
non-baryonic CDM, and to propose instead a modification to gravity
at low accelerations, similar in spirit to the modified gravity
approach to the dark energy problem. Modified Newtonian dynamics
(MOND) can account for the galactic rotation curves, and there are
covariant relativistic modifications of general relativity that can
reproduce MOND in the Newtonian limit~\cite{2007LNP...720..375S,%
2007PhRvD..75f3508B}. These modified theories typically require both scalar and vector degrees of freedom in the gravitational field, in
addition to the tensor. They lack a simple and natural motivation -- much like most of the modified theories that are alternatives to
dark energy. Future developments may lead to a low-energy
modification of general relativity that does not require dark matter or dark energy, that preserves the successes of general relativity
from solar system to cosmological scales, and that has some
motivation in fundamental theory and a level of internal simplicity
and naturalness.

In this review, we will not further discuss the key theoretical
challenges posed by inflation, by the ``missing mass" problem, and
by the late-time acceleration of the Universe. Instead, we will
adopt the standard view, based on a simple phenomenological model of inflation, on non-baryonic cold dark matter, and on the cosmological constant model of dark energy -- with general relativity applying on all scales from the inflationary energy scale downwards. Our aim is
to study the dynamics of structure formation within this framework
and by means of a single formalism, the 1+3 covariant approach to
cosmology. We begin with a comprehensive presentation of the
covariant formalism, followed by a discussion of the standard
cosmological model, the Friedmann-Lemaitre universe. In
\S~\ref{sIRCs}, we analyse the nonlinear behaviour of a general
cosmological spacetime, containing matter in the form of a single
fluid, a mixture of interacting fluids, a minimally coupled scalar
fields and in the presence of a large-scale magnetic field. We
linearise the nonlinear formulae in \S~\ref{sLCPs} and then use them to discuss the key features and the evolution of perturbed
Friedmannian models in various environments and during different
epochs. Section~\ref{sKINETIC} provides an overview of the covariant kinetic theory, before applying it to the study of the cosmic
microwave and neutrino backgrounds. We conclude this review with a
brief summary of the currents trends in cosmological research and a
look to the future in \S~\ref{sSO}. Finally, in the Appendices, we
provide the reader with all the technical information that is
necessary for the detailed study of this manuscript.

\subsection{The 1+3 covariant description}\label{ss1+3CD}
The covariant approach to general relativity and cosmology dates
back to the work of Heckmann, Sch\"{u}cking, and Raychaudhuri in the 1950s~\cite{1955ZA.....38...95H,1957ZA.....43..161R} and it has
since been employed in numerous applications by many authors
(see~\cite{1971glc..conf....104E,1973clp..conf....1E,%
1997ASSL..211...53E,1999toc..conf....1E} for details). The formalism uses the kinematic quantities, the energy-momentum tensor of the fluid(s) and the gravito-electromagnetic parts of the Weyl tensor, instead of the metric, which in itself does not provide a covariant description. The key equations are the Ricci and Bianchi identities, applied to the fluid 4-velocity vector, while Einstein's equations are incorporated via algebraic relations between the Ricci and the
energy-momentum tensor.

\subsubsection{Local spacetime splitting}\label{sssLStS}
Consider a general spacetime with a Lorentzian metric $g_{ab}$ of
signature ($-,\,+,\,+,\,+$) and introduce a family of observers with worldlines tangent to the timelike 4-velocity vector\footnote{Latin
indices vary between $0$ and $3$ and refer to arbitrary coordinate
or tetrad frames. Greek indices run from $1$ to $3$. We use
geometrised units with $c=1=8\pi G$, which means that all
geometrical variables have physical dimensions that are integer
powers of length.}
\begin{equation}
u^a=\frac{{\rm d}x^a}{{\rm d}\tau}\,,  \label{ua}
\end{equation}
where $\tau$ is the observers' proper time, so that $u_au^a=-1$.
This fundamental velocity field introduces a local 1+3 `threading'
of the spacetime into time and space. The vector $u_a$ determines
the time direction, while the tensor $h_{ab}=g_{ab}+u_au_b$ projects orthogonal to the 4-velocity into the observers' instantaneous rest
space at each event. In the absence of vorticity, the 4-velocity is
hypersurface-orthogonal and $h_{ab}$ is the metric of the
3-dimensional spatial sections orthogonal to $u_a$.

The vector field $u_a$ and its tensor counterpart $h_{ab}$ allow for a unique decomposition of every spacetime quantity into
its irreducible timelike and spacelike parts. These fields are also
used to define the covariant time and spatial derivatives of any
tensor field $S_{ab\cdots}{}^{cd\cdots}$ according to
\begin{equation}
\dot{S}_{ab\cdots}{}^{cd\cdots}=
u^e\nabla_eS_{ab\cdots}{}^{cd\cdots} \hspace{5mm} {\rm and}
\hspace{5mm} {\rm D}_eS_{ab\cdots}{}^{cd\cdots}=
h_e{}^sh_a{}^fh_b{}^ph_q{}^ch_r{}^d\cdots
\nabla_sS_{fp\cdots}{}^{qr\cdots}\,,  \label{deriv}
\end{equation}
respectively.

The effective volume element in the observer's instantaneous rest
space is given by contracting the spacetime volume element
($\eta_{abcd}$) along the time direction,
\begin{equation}
\varepsilon_{abc}=\eta_{abcd}u^d\,.  \label{veps1}
\end{equation}
The totally antisymmetric pseudotensor $\eta_{abcd}$ has
$\eta^{0123}=[-\,{\rm det}(g_{ab})\,]^{-1/2}$, it is covariantly
constant and satisfies the identities $\eta_{abcd}\eta^{efpq}=
-4!\delta_{[a}{}^e\delta_b{}^f\delta_c{}^p\delta_{d]}{}^q$. It
follows that $\varepsilon_{abc}u^a=0$,
\begin{equation}
\eta_{abcd}= 2u_{[a}\varepsilon_{b]cd}- 2\varepsilon_{ab[c}u_{d]}
\hspace{10mm} {\rm and} \hspace{10mm}
\varepsilon_{abc}\varepsilon^{def}= 3!h_{[a}{}^dh_b{}^eh_{c]}{}^f\,.
\label{veps2}
\end{equation}
Note that $\D_ch_{ab}=0=\D_d\varepsilon_{abc}$, while
$\dot{h}_{ab}=2u_{(a}A_{b)}$ and $\dot{\varepsilon}_{abc}=
3u_{[a}\varepsilon_{bc]d}A^d$ (with $A_a=\dot{u}_a$ -- see \S~\ref{sssKs} below).

\subsubsection{The gravitational field}\label{sssGrF}
In the general relativistic geometrical interpretation of gravity,
matter determines the spacetime curvature, while the latter dictates the motion of the matter. This interaction is realised via the
Einstein field equations,
\begin{equation}
G_{ab}\equiv R_{ab}- {1\over2}\,Rg_{ab}= T_{ab}- \Lambda g_{ab}\,,
\label{EFE1}
\end{equation}
where $G_{ab}$ is the Einstein tensor, $R_{ab}=R_{acb}{}^c$ is the
spacetime Ricci tensor (with trace $R$), $T_{ab}$ is the total
energy-momentum tensor of the matter fields and $\Lambda$ is the
cosmological constant. The twice contracted Bianchi identities
guarantee that $\nabla^bT_{ab}=0$ and total energy-momentum
conservation.

The Ricci tensor describes the local gravitational field at each
event due to matter there. The non-local, long-range gravitational
field, mediated via gravitational waves and tidal forces, is encoded in the Weyl conformal curvature tensor $C_{abcd}$. The splitting of
the gravitational field into its local and non-local parts is given
by the decomposition of the Riemann tensor,
\begin{equation}
R_{abcd}= C_{abcd}+ {1\over2}\left(g_{ac}R_{bd}+g_{bd}R_{ac}
-g_{bc}R_{ad}-g_{ad}R_{bc}\right)
-{1\over6}\,R\left(g_{ac}g_{bd}-g_{ad}g_{bc}\right)\,,
\label{Riemann}
\end{equation}
where the Weyl tensor shares all the symmetries of the Riemann
tensor and is also trace-free, $C^c{}_{acb}=0$. Relative to the
fundamental observers, the conformal curvature tensor decomposes
further into its irreducible parts
(e.g.~see~\cite{1973lsss.book.....H,1997PhRvD..55..463M})
\begin{equation}
E_{ab}= C_{acbd}u^cu^d \hspace{15mm} {\rm and} \hspace{15mm}
H_{ab}= {1\over2}\,\varepsilon_a{}^{cd}C_{cdbe}u^e\,.
\label{EabHab}
\end{equation}
Then,
\begin{equation}
C_{abcd}=
\left(g_{abqp}g_{cdsr}-\eta_{abqp}\eta_{cdsr}\right)u^qu^sE^{pr}-
\left(\eta_{abqp}g_{cdsr}+g_{abqp}\eta_{cdsr}\right)u^qu^sH^{pr}\,,
\label{Weyl}
\end{equation}
where $g_{abcd}=g_{ac}g_{bd}-g_{ad}g_{bc}$. Alternatively,
\begin{equation}
C_{ab}{}{}^{cd}=
4\left(u_{[a}u^{[c}+h_{[a}{}^{[c}\right)E_{b]}{}^{d]}+
2\varepsilon_{abe}u^{[c}H^{d]e}+2u_{[a}H_{b]e}\varepsilon^{cde}\,.
\label{rmweyl}
\end{equation}
The spatial, symmetric and trace-free tensors $E_{ab}$ and $H_{ab}$
are known as the electric and magnetic Weyl components. The electric part generalises the tidal tensor of the Newtonian gravitational
potential, but $H_{ab}$ has no Newtonian counterpart. Note that both tensors must be present for a nonzero super-energy flux vector
($P_a=\varepsilon_{abc}E^{bd}H^c{}_d$), which is essential for the
propagation of gravitational waves.

The Weyl tensor represents the part of the curvature that is not
determined locally by matter. However, its dynamics are not
arbitrary because the Riemann tensor satisfies the Bianchi
identities, whose contraction gives~\cite{1973lsss.book.....H}
\begin{equation}
\nabla^dC_{abcd}= \nabla_{[b}R_{a]c}+
{1\over6}\,g_{c[b}\nabla_{a]}R\,,  \label{Bianchi}
\end{equation}
using decomposition (\ref{Riemann}). In a sense the once contracted
Bianchi identities act as the field equations for the Weyl tensor,
determining the part of the spacetime curvature that depends on the
matter distribution at other points. Equation~(\ref{Bianchi}) splits into a set of two propagation and two constraint equations, which
govern the dynamics of the electric and magnetic Weyl components
(see~\S~\ref{sssWC}).

\subsubsection{Matter fields}\label{sssMFs}
With respect to the fundamental observers, the energy-momentum tensor of a general (imperfect) fluid decomposes into its irreducible parts as\footnote{For a multi-component medium, or when allowing for peculiar velocities, one needs to account for the differing 4-velocities of the matter components and the fundamental observers (see~\S~\ref{ssIMfCs}).}
\begin{equation}
T_{ab}= \rho u_au_b+ ph_{ab}+ 2q_{(a}u_{b)}+ \pi_{ab}\,.
\label{Tab1}
\end{equation}
where $\rho=T_{ab}u^au^b$ is the matter energy density, $p=T_{ab}h^{ab}/3$ is the effective isotropic pressure of the fluid, namely the sum between the equilibrium pressure and the associated bulk viscosity, $q_a=-h_a{}^bT_{bc}u^c$ is the total energy-flux vector, and $\pi_{ab}=h_{\langle a}{}^{c} h_{b\rangle}{}^dT_{cd}$ is the symmetric and trace-free anisotropic stress tensor.\footnote{Angled brackets denote the symmetric and trace-free part of spatially projected second-rank tensors and the projected part of vectors according to
\begin{equation}
S_{\langle ab \rangle}= h_{\langle a}{}^{c}h_{b\rangle}{}^dS_{cd}=
h_{(a}{}^{c}h_{b)}{}^dS_{cd}- {1\over3}\,h^{cd}S_{cd}h_{ab} \hspace{10mm} {\rm
and} \hspace{10mm} {V}_{\langle a\rangle}= h_a{}^bV_b\,,
\label{angled}
\end{equation}
respectively (with $S_{\langle ab \rangle}h^{ab}=0$). The reader is
referred to the Appendix (see \S~\ref{AssCD} there) for more details on covariant decomposition.}

The 4-velocity $u_a$ is generally arbitrary and a velocity boost of
the form $u_a\to\tilde{u}_a$ induces changes in the dynamical quantities, given explicitly in Appendix~\ref{AssTU4VB}. When the fluid is perfect, however, there is a unique hydrodynamic 4-velocity, relative to which $q_a$, $\pi_{ab}$ are identically zero and the effective pressure reduces to the equilibrium one. As a result,
\begin{equation}
T_{ab}= \rho u_au_b+ ph_{ab}\,.  \label{pfTab}
\end{equation}
If we additionally assume that $p=0$, we have the simplest case of
pressure-free matter, namely `dust', which includes baryonic matter
(after decoupling) and cold dark matter. Otherwise, we need to
determine $p$ as a function of $\rho$ and potentially of other
thermodynamic variables. In general, the equation of state takes the form $p=p\,(\rho,s)$, where $s$ is the specific entropy. Finally, for a barotropic medium we have $p=p\,(\rho)$ (see \S~\ref{sssETs} below for further discussion).

Expression~(\ref{Tab1}) describes any type of matter, including
electromagnetic fields, scalar fields, etc. (see~\S~\ref{sssEMFs},\,\ref{sssMCSFs}). Since $R=4\Lambda-T$, with
$T=T_a{}^a$, Einstein's equations are recast into
\begin{equation}
R_{ab}= T_{ab}- {1\over2}\,Tg_{ab}+ \Lambda g_{ab}\,.
\label{EFE2}
\end{equation}
The successive contraction of the above, assuming that $T_{ab}$ is
given by Eq.~(\ref{Tab1}), leads to a set of algebraic relations
that will prove useful later
\begin{eqnarray}
R_{ab}u^au^b&=& {1\over2}\,(\rho+3p)- \Lambda\,, \label{EFE3}\\ h_a{}^bR_{bc}u^c&=& -q_a\,, \label{EFE4}\\ h_a{}^ch_b{}^dR_{cd}&=& {1\over2}\,(\rho-p)h_{ab}+ \Lambda h_{ab}+ \pi_{ab}\,.  \label{EFE5}
\end{eqnarray}

\subsection{Covariant relativistic cosmology}\label{ssCRC}
There are various physical choices in cosmology for the fundamental
4-velocity field that defines the $1+3$ splitting of spacetime. Some possibilities include the frame in which the dipole of the CMB
anisotropy vanishes and the local rest-frame of the matter (these
are generally assumed to coincide when averaged on sufficiently
large scales). In specific situations, it may be appropriate to
choose the frame that simplifies the physics (for example, for a
perfect-fluid cosmology it makes sense to adopt the rest-frame of
the fluid), and we shall make several choices for $u_a$ throughout
this review. Once $u_a$ is specified, its integral curves define the worldlines of the fundamental observers introduced in
\S~\ref{sssLStS}.

\subsubsection{Kinematics}\label{sssKs}
The observers' motion is characterised by the irreducible
kinematical quantities of the $u_a$-congruence, which emerge from
the covariant decomposition of the 4-velocity gradient
\begin{equation}
\nabla_bu_a=\sigma_{ab}+ \omega_{ab}+ {1\over3}\,\Theta h_{ab}-
A_au_b\,,  \label{Nbua}
\end{equation}
where $\sigma_{ab}={\rm D}_{\langle b}u_{a\rangle}$,
$\omega_{ab}={\rm D}_{[b}u_{a]}$, $\Theta=\nabla^au_a={\rm D}^au_a$
and $A_a=\dot{u}_a=u^b\nabla_bu_a$ are respectively the shear and the vorticity tensors, the volume expansion (or contraction) scalar, and the 4-acceleration vector. The latter represents non-gravitational forces and vanishes when matter moves under gravity alone. By construction we have $\sigma_{ab}u^a=0= \omega_{ab}u^a=A_au^a$. Also, on using the orthogonally projected alternating tensor $\varepsilon_{abc}$ (with $\dot{\varepsilon}_{abc}=3u_{[a}\varepsilon_{bc]d}A^d$), one defines\footnote{The sign conventions are such that $\vec{\omega}= -\vec{\nabla}\times\vec{v}/2$ in the Newtonian limit and agree with those adopted in the majority of the related articles. Note that in~\cite{1999toc..conf....1E} the vorticity tensor ($\omega_{ab}$) and the orthogonally projected volume element ($\varepsilon_{abc}$) have opposite signs, relative to the ones defined here. The reader should have this in mind when comparing the equations of the two articles.} the vorticity vector $\omega_a=\varepsilon_{abc}\omega^{bc}/2$ (with $\omega_{ab}=\varepsilon_{abc}\omega^c$). We
note that the tensor $v_{ab}={\rm D}_bu_a=\sigma_{ab}+\omega_{ab}+ (\Theta/3)h_{ab}$ describes the relative motion of neighbouring observers. In particular, $v_a=v_{ab}\chi^b$ monitors the relative velocity between the observers' worldlines, with $\chi_a$ representing the relative position vector between the same two flow lines (e.g.~see~\cite{1971glc..conf....104E,1973clp..conf....1E} for details). The volume scalar determines the average separation between two neighbouring observers and is also used to introduce a representative length scale ($a$) by means of the definition $\dot{a}/a=\Theta/3$. The effect of the vorticity is to change the orientation of a given fluid element without modifying its volume or shape. Finally, the shear changes the shape while leaving the volume unaffected.

The non-linear covariant kinematics are determined by a set of
propagation and constraint equations, which are purely geometrical
in origin and essentially independent of the Einstein equations.
Both sets emerge after applying the Ricci identities
\begin{equation}
2\nabla_{[a}\nabla_{b]}u_c= R_{abcd}u^d\,,  \label{Ris}
\end{equation}
to the fundamental 4-velocity vector defined in (\ref{ua}).
Substituting in from (\ref{Nbua}), using decompositions
(\ref{Riemann}), (\ref{Weyl}) and the auxiliary relations
(\ref{EFE3})-(\ref{EFE5}), the timelike and spacelike parts of the
resulting expression lead to a set of three propagation and three
constraint equations. The former contains Raychaudhuri's formula
\begin{equation}
\dot{\Theta}= -{1\over3}\,\Theta^2- {1\over2}\,(\rho+3p)-
2(\sigma^2-\omega^2)+ {\rm D}^aA_a+ A_aA^a+ \Lambda\,, \label{Ray}
\end{equation}
for the time evolution of $\Theta$; the shear propagation equation
\begin{equation}
\dot{\sigma}_{\langle ab\rangle}= -{2\over3}\,\Theta\sigma_{ab}-
\sigma_{c\langle a}\sigma^c{}_{b\rangle}- \omega_{\langle
a}\omega_{b\rangle}+ {\rm D}_{\langle a}A_{b\rangle}+ A_{\langle
a}A_{b\rangle}- E_{ab}+ {1\over2}\,\pi_{ab}\,, \label{sigmadot}
\end{equation}
which describes kinematical anisotropies; and the evolution
equation of the vorticity
\begin{equation}
\dot{\omega}_{\langle a\rangle}= -{2\over3}\,\Theta\omega_a-
{1\over2}\,\curl A_a+ \sigma_{ab}\omega^b\,. \label{omegadot}
\end{equation}
Note that $\sigma^2=\sigma_{ab}\sigma^{ab}/2$ and
$\omega^2=\omega_{ab}\omega^{ab}/2=\omega_a\omega^a$ are
respectively the scalar square magnitudes of the shear and the vorticity, while $E_{ab}$ is the electric component of the Weyl tensor (see \S~\ref{sssGrF}). Also, ${\rm curl}v_a= \varepsilon_{abc}{\rm D}^bv^c$ for any orthogonally projected vector $v_a$, which means that ${\rm D}^b\omega_{ab}={\rm curl}\omega_a$.

The spacelike component of (\ref{Ris}) leads to a set of three
complementary constraints. These are the shear or $(0,\alpha)$
constraint
\begin{equation}
{\rm D}^b\sigma_{ab}= {2\over3}\,{\rm D}_a\Theta+ \curl \omega_a+
2\varepsilon_{abc}A^b\omega^c- q_a\,, \label{shearcon}
\end{equation}
the vorticity-divergence identity
\begin{equation}
{\rm D}^a\omega_a= A_a\omega^a\,,  \label{omegacon}
\end{equation}
and the magnetic Weyl equation
\begin{equation}
H_{ab}= \curl \sigma_{ab}+ {\rm D}_{\langle a}\omega_{b\rangle}+
2A_{\langle a}\omega_{b\rangle}\,.  \label{Hcon}
\end{equation}

Raychaudhuri's formulae (see~\cite{1957ZA.....43..161R} and
also~\cite{2005gr.qc....11123D,2006gr.qc....11123K} for recent
reviews) is the key to the study of gravitational collapse, as it
describes the evolution of the average separation between two
neighbouring observers. For this reason Eq.~(\ref{Ray}) has been
at the core of all the singularity theorems
(see~\cite{1973lsss.book.....H,1984gere.book.....W} and references
therein). Negative terms in the right-hand side of (\ref{Ray})
lead to contraction and positive resist the collapse, which means
that conventional (non-phantom) matter is always attractive unless
$p<-\rho/3$.

\subsubsection{Electromagnetic fields}\label{sssEMFs}
The Maxwell field is determined by the antisymmetric
electromagnetic (Faraday) tensor $F_{ab}$, which relative to a
fundamental observer decomposes into an electric and a magnetic
component as~\cite{1973clp..conf....1E,2005CQGra..22..393T}
\begin{equation}
F_{ab}= 2u_{[a}E_{b]}+ \varepsilon _{abc}B^c\,.  \label{Fab}
\end{equation}
In the above, $E_a=F_{ab}u^b$ and $B_a=\varepsilon_{abc}F^{bc}/2$
are respectively the electric and magnetic fields experienced by the observer (with $E_au^a=0=B_au^a$). The Faraday tensor also
determines the energy-momentum tensor of the Maxwell field according to
\begin{equation}
T_{ab}^{(em)}= -F_{ac}F^c{}_b- {1\over4}\,F_{cd}F^{cd}g_{ab}\,.
\label{Tem1}
\end{equation}
The above expression combines with (\ref{Fab}) to give the
irreducible decomposition for $T_{ab}^{(em)}$, relative to the
$u_a$-frame
\begin{equation}
T_{ab}^{(em)}= {1\over2}\,(E^2+B^2)u_au_b+
{1\over6}\,(E^2+B^2)h_{ab}+ 2\clp_{(a}u_{b)}+ \Pi_{ab}\,,
\label{Tem}
\end{equation}
in the Heaviside-Lorentz units. Here $E^2=E_aE^a$ and $B^2=B_aB^a$ are the square magnitudes of the two fields, $\clp_a=\varepsilon_{abc}E^bB^c$ is the electromagnetic
Poynting vector and $\Pi_{ab}=-E_{\langle a}E_{b\rangle}-B_{\langle
a}B_{b\rangle}$. Expression (\ref{Tem}) allows for a fluid
description of the electromagnetic field and manifests its
generically anisotropic nature. In particular, the Maxwell field
corresponds to an imperfect fluid with energy density $(E^2+B^2)/2$, isotropic pressure $(E^2+B^2)/6$, anisotropic stresses given by
$\Pi_{ab}$ and an energy-flux vector represented by $\clp_a$.
Equation (\ref{Tem}) also ensures that $T_{a}^{(em)\;a}=0$, in
agreement with the trace-free nature of the radiation stress-energy
tensor.

We follow the evolution of the electromagnetic field by means of
Maxwell's equations. In their standard tensor form these read
\begin{equation}
\nabla_{[c}F_{ab]}=0 \hspace{15mm} {\rm and} \hspace{15mm}
\nabla^bF_{ab}=J_a\,,  \label{Max}
\end{equation}
where (\ref{Max}a) reflects the existence of a 4-potential and $J_a$ is the 4-current that sources the electromagnetic field. With
respect to the $u_a$-congruence, the 4-current splits into its
irreducible parts according to
\begin{equation}
J_a= \mu u_a+\clj_a\,,  \label{Ja}
\end{equation}
with $\mu=-J_au^a$ representing the charge density and $\clj_a=h_a{}^bJ_b$ the orthogonally projected current
(i.e.~$\clj_au^a=0$).

Relative to a fundamental observer, each one of Maxwell's equations
decomposes into a timelike and a spacelike component. The timelike
parts of (\ref{Max}a), (\ref{Max}b) lead to a set of two propagation equations
\begin{equation}
\dot{E}_{\langle a\rangle}= -{2\over3}\,\Theta E_a+
\left(\sigma_{ab}+\varepsilon_{abc}\omega^c\right)E^{b}+
\varepsilon_{abc}A^bB^c+ \curl B_a-\clj_a\,, \label{M1}
\end{equation}
\begin{equation}
\dot{B}_{\langle a\rangle}= -{2\over3}\,\Theta B_a+
\left(\sigma_{ab}+\varepsilon_{abc}\omega^c\right)B^b-
\varepsilon_{abc}A^bE^c- \curl E_a\,, \label{M2}
\end{equation}
while their spacelike components provide the constraints
\begin{equation}
{\rm D}^aE_a+ 2\omega^aB_a= \mu \hspace{15mm} {\rm and}
\hspace{15mm} {\rm D}^aB_a- 2\omega^aE_a=0\,. \label{M4}
\end{equation}
Expressions (\ref{M1})-(\ref{M4}) are 1+3 covariant versions of
Amp\`ere's law, Faraday's law, Coulomb's law and Gauss' law
respectively. Therefore, in addition to the usual `curl' and
`divergence' terms, the covariant form of (\ref{M1}) and (\ref{M2})
contains terms generated by the relative motion of the neighbouring
observers. Also, according to (\ref{M4}), the magnetic vector is not solenoidal unless $\omega^aE_a=0$.

\subsubsection{Conservation laws}\label{sssCLs}
The twice contracted Bianchi identities guarantee the conservation
of the total energy momentum tensor, namely that $\nabla^bT_{ab}=0$. This constraint splits into a timelike and a spacelike part, which
respectively lead to the energy and the momentum conservation laws. When dealing with a general imperfect fluid, the former is
\begin{equation}
\dot{\rho}= -\Theta(\rho+p)- {\rm D}^aq_a- 2A^aq_a-
\sigma^{ab}\pi_{ab}\,,  \label{edc1}
\end{equation}
while latter satisfies the expression
\begin{equation}
(\rho+p)A_a= -{\rm D}_ap- \dot{q}_{\langle a\rangle}-
{4\over3}\,\Theta q_a- (\sigma_{ab}+\omega_{ab})q^b- {\rm
D}^b\pi_{ab}- \pi_{ab}A^b\,.  \label{mdc1}
\end{equation}
When the fluid is perfect, the energy-momentum tensor is given by
(\ref{pfTab}) and the above reduce to
\begin{equation}
\dot{\rho}= -\Theta(\rho+p) \hspace{15mm} {\rm and} \hspace{15mm}
(\rho+p)A_a= -{\rm D}_ap\,,  \label{pfcls}
\end{equation}
respectively. It follows, from (\ref{pfcls}b), that the sum $\rho+p$ describes the relativistic total inertial mass of the medium. Then,
when the inertial mass is zero (i.e.~for $\rho+p\rightarrow0$),
consistency demands that the pressure gradients must also vanish.
For a barotropic fluid the latter immediately implies zero density
gradients as well. Also, if the `phantom divide' is
crossed~\cite{2002PhLB..545...23C,2003PhRvL..91g1301C}, the inertial mass becomes negative and the acceleration antiparallel to the force that caused it.\footnote{Phantom cosmologies violate the dominant
energy condition (i.e.~have $\rho+p<0$) and generally lead to future `big rip' singularities (see~\cite{2002PhLB..545...23C,%
2003PhRvL..91g1301C} and also~\cite{2000fufc.conf...71S,%
2003PhLB..562..147N,2003PhRvD..68j3519D,2003PhRvD..68b3522S,%
2003PhRvL..91u1301C,2005PhRvD..71l4036S}). On the other hand,
finite-time (sudden) future singularities can arise even when the
standard energy conditions are observed~\cite{2004CQGra..21L..79B,%
2005CQGra..22.1563B}}.

The energy momentum tensor of the electromagnetic field obeys the
constraint $\nabla^bT_{ab}^{(em)}=-F_{ab}J^b$, with the Faraday
tensor given by (\ref{Fab}) and the quantity in the right-hand side
representing the Lorentz 4-force. Thus, for charged matter the
conservation of the total energy-momentum tensor
$T_{ab}=T_{ab}^{(m)}+T_{am}^{(em)}$ leads to the formulae
\begin{equation}
\dot{\rho}= -\Theta(\rho+p)- {\rm D}^aq_a- 2A^aq_a-
\sigma^{ab}\pi_{ab}+ E_a\clj^a  \label{edc2}
\end{equation}
for the energy density, and
\begin{eqnarray}
(\rho+p)A_a&=& -{\rm D}_ap- \dot{q}_{\langle a\rangle}-
{4\over3}\,\Theta q_a- (\sigma_{ab}+\omega_{ab})q^b- {\rm
D}^b\pi_{ab}- \pi_{ab}A^b \nonumber\\ &&+\mu E_a+
\varepsilon_{abc}\clj^bB^c\,, \label{mdc2}
\end{eqnarray}
for the momentum density. We note the electromagnetic terms in the
right-hand side of the above, the effect of which depends on the
electrical properties of the medium (see \S~\ref{sssIMHDA}). The
last two terms in the right-hand side of (\ref{mdc2}), in
particular, represent the more familiar form of the Lorentz force.

The antisymmetry of the Faraday tensor (see Eq.~(\ref{Fab})) and the second of Maxwell's formulae (see Eq.~(\ref{Max}b)) imply
$\nabla^aJ_a=0$ and the conservation of the 4-current. Then, on
using decomposition (\ref{Ja}), we arrive at the covariant
charge-density conservation law
\begin{equation}
\dot{\mu}= -\Theta\mu- {\rm D}^a\clj_a- A^a\clj_a\,. \label{chcon}
\end{equation}
Hence, in the absence of spatial currents, the evolution of the
charge density depends entirely on the volume expansion (or
contraction) of the fluid element.

\subsubsection{Equilibrium thermodynamics}\label{sssETs}
In relativistic thermodynamics the physical state of a medium is monitored by means of the energy momentum tensor ($T_{ab}$), the particle flux vector ($N_a$) and the entropy flux vector ($S_a$) (e.g.~see~\cite{1976AnPhy.100..310I}). For isolated systems, the former of these three `primary variables' satisfies the conservation law $\nabla^bT_{ab}=0$. The entropy flux, on the other hand, obeys the second law of thermodynamics and, provided the particle number does not change, $N_a$ satisfies the particle number conservation law. Written in covariant terms, these read
\begin{equation}
\nabla^aS_a\geq 0 \hspace{15mm} {\rm and} \hspace{15mm} \nabla^aN_a= 0\,, \label{ThD2lpcl}
\end{equation}
respectively. For a system in equilibrium (or for a perfect fluid) there is no entropy production, which implies that
\begin{equation}
\nabla^aS_a= 0\,.  \label{EqThDs}
\end{equation}
Moreover, all three of the aforementioned  primary variables can be expressed in terms of a unique timelike 4-velocity field, according to
\begin{equation}
T_{ab}= \rho u_au_b+ ph_{ab}\,, \hspace{15mm} S_a= Su_a \hspace{10mm} {\rm and} \hspace{10mm} N_a= nu_a\,,  \label{TSN}
\end{equation}
where $S=-S_au^a$ and $n=-N_au^a$ are the entropy and
particle densities respectively. Note that the last two of the above combine to give
\begin{equation}
S_a= sN_a\,,  \label{Sa-Na}
\end{equation}
with $s=S/n$ representing the specific entropy (the entropy per particle) of the system.

Applying the conservation law $\nabla^bT_{ab}=0$ to the energy-momentum tensor (\ref{TSN}a), leads to the familiar energy and momentum density conservation laws of a perfect fluid (given in \S~\ref{sssCLs} by expressions (\ref{edc1}) and (\ref{mdc1}) respectively). On the other hand, substituting (\ref{TSN}c) into the left-hand side of (\ref{ThD2lpcl}b) provides the conservation equation of the particle number
\begin{equation}
\dot{n}= -\Theta n\,.  \label{dotn}
\end{equation}
Similarly, inserting (\ref{Sa-Na}) into entropy conservation law (\ref{EqThDs}) and then using Eqs.~(\ref{ThD2lpcl}b) and (\ref{TSN}c) we arrive at
\begin{equation}
\dot{s}= 0\,.  \label{dots}
\end{equation}
This ensures that the specific entropy of the system does not change along the fluid motion, which is another way of saying that the flow is adiabatic. When the spatial gradients of the specific entropy also vanish, we have $\nabla_as={\rm D}_as-\dot{s}u_a=0$ and the medium is said to be isentropic. Note that an isentropic fluid has a barotropic equation of state and vise versa~\cite{1996astro.ph..9119M}.

An additional thermodynamic scalar is the temperature ($T$) of the system, which satisfies the Gibbs equation
\begin{equation}
T\,{\rm d}s= {\rm d}\left({\rho\over n}\right)+ p\;{\rm d}\left({1\over n}\right)\,,  \label{Gibbs}
\end{equation}
with ${\rm d}f={\rm d}x^a\nabla_af$ (e.g.~see~\cite{1979AnPhy.118..341I,1996astro.ph..9119M}). Of the five thermodynamic scalars ($\rho$, $p$, $n$, $s$ and $T$), two are needed as independent variables. Selecting the energy density and the specific entropy as our independent quantities, the equation of state of a (perfect) fluid acquires the form $p=p(\rho,s)$. Then,
\begin{equation}
\dot{p}= \left({\partial p\over\partial\rho}\right)_s\dot{\rho}+ \left({\partial p\over\partial s}\right)_{\rho}\dot{s}\,,  \label{dotp}
\end{equation}
which (for $\dot{s}=0$) gives
\begin{equation}
c_s^2\equiv \left({\partial p\over\partial\rho}\right)_s= {\dot{p}\over\dot{\rho}}\,,  \label{cs2}
\end{equation}
namely the square of the adiabatic sound speed. In addition to (\ref{dotp}), the above given equation of state also leads to
\begin{equation}
{\rm D}_ap= \left({\partial p\over\partial\rho}\right)_s{\rm D}_a\rho+ \left({\partial p\over\partial s}\right)_{\rho}{\rm D}_as\,,  \label{Dap}
\end{equation}
thus connecting the spatial gradients of the pressure to perturbations in the energy density and the specific entropy of the system. Consequently, when applied to adiabatic and then to isentropic systems, the relations (\ref{cs2}) and (\ref{Dap}) combine to give
\begin{equation}
{\rm D}_ap= \left({\dot{p}\over\dot{\rho}}\right){\rm D}_a\rho+ \left({\partial p\over\partial s}\right)_{\rho}{\rm D}_as \hspace{10mm} {\rm and} \hspace{10mm} {\rm D}_ap= \left({\dot{p}\over\dot{\rho}}\right){\rm D}_a\rho\,,  \label{Dapi}
\end{equation}
respectively. At this point we note that, according to the standard thermodynamic nomenclature, we distinguish between adiabatic and isentropic perturbations. The former are characterised by $\dot{s}=0$, while the latter demand that the specific entropy is a spacetime invariant. Although the two concepts are distinct, it is not uncommon for cosmologists to say adiabatic and imply isentropic~\cite{1995STIN...9622249B}.

\subsubsection{Spatial curvature}\label{sssSC}
When the fluid flow is irrotational, the rest-space tangent planes
of the fundamental observers mesh together to form spacelike
hypersurfaces orthogonal to their worldlines. These are normal to the $u_a$-congruence and define the hypersurfaces of simultaneity for all the comoving observers. In the presence of vorticity, however, Frobenius' theorem forbids the existence of such integrable hypersurfaces (e.g.~see~\cite{1984gere.book.....W,%
2004rtmb.book.....P}). Then the observers' rest-spaces no longer mesh together smoothly. The projected Riemann tensor is defined by
\begin{equation}
\clr_{abcd}= h_a{}^qh_b{}^sh_c{}^fh_d{}^pR_{qsfp}- v_{ac}v_{bd}+
v_{ad}v_{bc}\,, \label{3Riemann1}
\end{equation}
where $v_{ab}={\rm D}_bu_a$ is the relative flow tensor between two neighbouring observers (see \S~\ref{sssKs}). On using Eqs.~(\ref{EFE1})-(\ref{EFE3}) and decompositions (\ref{Riemann}), (\ref{Weyl}), we find~\cite{2007PhR...449..131B}
\begin{eqnarray}
\clr_{abcd}&=& -\varepsilon_{abq}\varepsilon_{cds}E^{qs}+
{1\over3}\left(\rho-{1\over3}\,\Theta^2+\Lambda\right)(h_{ac}h_{bd}
-h_{ad}h_{bc})\nonumber\\
&{}&+{1\over2}\left(h_{ac}\pi_{bd}+\pi_{ac}h_{bd}
-h_{ad}\pi_{bc}-\pi_{ad}h_{bc}\right)\nonumber\\
&{}&-{1\over3}\,\Theta\left[h_{ac}(\sigma_{bd}+\omega_{bd})
+(\sigma_{ac}+\omega_{ac})h_{bd}-h_{ad}(\sigma_{bc}+\omega_{bc})
-(\sigma_{ad}+\omega_{ad})h_{bc}\right] \nonumber\\
&{}&-(\sigma_{ac}+\omega_{ac})(\sigma_{bd}+\omega_{bd})+(\sigma_{ad}
+\omega_{ad})(\sigma_{bc}+\omega_{bc})\,.  \label{3Riemann2}
\end{eqnarray}
This provides an irreducible decomposition of the projected Riemann
tensor. If $\omega_a=0$, then $\clr_{abcd}$ is the 3-Riemann tensor
of the hypersurfaces of simultaneity orthogonal to $u_a$. In analogy to 4-dimensions, the projected Ricci tensor and Ricci scalar are
respectively defined by
\begin{equation}
\clr_{ab}= h^{cd}\clr_{acbd}= \clr^c{}_{acb} \hspace{15mm} {\rm and}
\hspace{15mm} \clr= h^{ab}\clr_{ab}\,. \label{3Ricci}
\end{equation}

The algebraic symmetries of $\mathcal{R}_{abcd}$ are given by
\begin{equation}
\clr_{abcd}= \clr_{[ab][cd]}  \label{3Rsym1}
\end{equation}
and
\begin{eqnarray}
\clr_{abcd}-\clr_{cdab}&=&
-{2\over3}\,\Theta\left(h_{ac}\omega_{bd}+\omega_{ac}h_{bd}
-h_{ad}\omega_{bc}-\omega_{ad}h_{bc}\right) \nonumber\\
&{}&-2\left(\sigma_{ac}\omega_{bd}+\omega_{ac}\sigma_{bd}
-\sigma_{ad}\omega_{bc}-\omega_{ad}\sigma_{bc}\right)\,.
\label{3Rsym2}
\end{eqnarray}
It follows from the above that $\clr_{abcd}=\clr_{cdab}$ in the
absence of vorticity. In that case, the spatial Riemann tensor
possesses all the symmetries of its 4-dimensional counterpart.

Contracting (\ref{3Riemann2}) on the first and third indices we
arrive at what is usually referred to as the Gauss-Codacci formula
\begin{eqnarray}
\clr_{ab}&=& E_{ab}+ {2\over3}\left(\rho-{1\over3}\Theta^2
+\sigma^2-\omega^2+\Lambda\right)h_{ab}+ {1\over2}\,\pi_{ab}-
{1\over3}\,\Theta(\sigma_{ab}+\omega_{ab})+ \sigma_{c\langle
a}\sigma^c{}_{b\rangle} \nonumber\\
&{}&-\omega_{c\langle a}\omega^c{}_{b\rangle}+
2\sigma_{c[a}\omega^c{}_{b]}\,,  \label{GC}
\end{eqnarray}
while a further contraction leads to the generalised Friedmann
equation
\begin{equation}
\clr= h^{ab}\clr_{ab}= 2\left(\rho-{1\over3}\,\Theta^2+\sigma^2
-\omega^2+\Lambda\right)\,.  \label{3R}
\end{equation}
Finally, one may combine Eqs.~(\ref{GC}) and (\ref{3R}) to obtain
\begin{equation}
\clr_{ab}= {1\over3}\,\clr h_{ab}+ E_{ab}+ {1\over2}\,\pi_{ab}-
{1\over3}\,\Theta(\sigma_{ab}+\omega_{ab})+ \sigma_{c\langle
a}\sigma^c{}_{b\rangle}- \omega_{c\langle a}\omega^c{}_{b\rangle}+
2\sigma_{c[a}\omega^c{}_{b]}\, \label{GC1}
\end{equation}
where all terms on the right, with the exception of the first, are
trace-free.

It should be noted that the matter variables used in this section
represent the total fluid. For example, when dealing with a mixture
of pressure-free dust and isotropic radiation
$\rho=\rho^{(d)}+\rho^{(\gamma)}$ and $\pi_{ab}=0$, in the presence
of an electromagnetic field (see \S~\ref{sssEMFs}) we have
$\rho=\rho^{(m)}+\rho^{(em)}= \rho^{(m)}+(E^2+B^2)/2$,
$\pi_{ab}=\pi_{ab}^{(m)}+\Pi_{ab}$, etc.

\subsubsection{Weyl curvature}\label{sssWC}
The 1+3 splitting of the once contracted Bianchi identities (see
Eq.~(\ref{Bianchi}) in \S~\ref{sssGrF}) leads to a set of two
propagation and two constraint equations that monitor the evolution
of the long range gravitational field, namely tidal forces and
gravity waves. In particular, on using the decomposition
(\ref{Weyl}), the timelike component of (\ref{Bianchi}) leads
to~\cite{1999toc..conf....1E}
\begin{eqnarray}
\dot{E}_{\langle ab\rangle}&=& -\Theta E_{ab}-
{1\over2}\,(\rho+p)\sigma_{ab}+ \curl H_{ab}-
{1\over2}\,\dot{\pi}_{ab}- {1\over6}\,\Theta\pi_{ab}-
{1\over2}\,{\rm D}_{\langle a}q_{b\rangle}-A_{\langle
a}q_{b\rangle} \nonumber\\ &&+3\sigma_{\langle
a}{}^c\left(E_{b\rangle c}-{1\over6}\,\pi_{b\rangle c}\right)+
\varepsilon_{cd\langle
a}\left[2A^cH_{b\rangle}{}^d-\omega^c\left(E_{b\rangle}{}^d+
{1\over2}\,\pi_{b\rangle}{}^d\right)\right]  \label{dotEab}
\end{eqnarray}
and
\begin{eqnarray}
\dot{H}_{\langle ab\rangle}&=& -\Theta H_{ab}- \curl E_{ab}+
{1\over2}\,\curl \pi_{ab}+3\sigma_{\langle a}{}^cH_{b\rangle c}-
{3\over2}\,\omega_{\langle a}q_{b\rangle} \nonumber\\
&&-\varepsilon_{cd\langle a}\left(2A^cE_{b\rangle}{}^d-
{1\over2}\,\sigma^c{}_{b\rangle}q^d+\omega^cH_{b\rangle}{}^d\right)\,.
\label{dotHab}
\end{eqnarray}
Taking the time derivatives of the above, one arrives at a pair of
wavelike equations for the electric and the magnetic parts of the
Weyl tensor, showing how curvature distortions propagate in the form
of gravitational waves like ripples in the spacetime fabric. These
waves are also subjected to a set of constraints, which emerge from
the spacelike component of the decomposed Eq.~(\ref{Bianchi}) and
are given by
\begin{equation}
{\rm D}^bE_{ab}= {1\over3}\,{\rm D}_a\rho- {1\over2}{\rm
D}^b\pi_{ab}- {1\over3}\,\Theta q_a+ {1\over2}\,\sigma_{ab}q^b-
3H_{ab}\omega^b+ \varepsilon_{abc}\left(\sigma^b{}_dH^{cd}
-{3\over2}\,\omega^bq^c\right)  \label{Weylc1}
\end{equation}
and
\begin{equation}
{\rm D}^bH_{ab}= (\rho+p)\omega_a- {1\over2}\,\curl q_a+
3E_{ab}\omega^b- {1\over2}\,\pi_{ab}\omega^b-
\varepsilon_{abc}\sigma^b{}_d\left(E^{cd}
+{1\over2}\,\pi^{cd}\right)\,,  \label{Weylc2}
\end{equation}
respectively~\cite{1999toc..conf....1E}. The above expressions are
remarkably similar to Maxwell's formulae, which explains the names
of $E_{ab}$ and $H_{ab}$. In fact, the Maxwell-like form of the free
gravitational field underlines the rich correspondence between
electromagnetism and general relativity and it has been the subject
of theoretical debate for decades (see~\cite{1958CRAcS..247.1094B,%
1969AnPh..138..59P,1998CQGra..15..705M,2000GReGr..32.1009D} for a
representative list).

\subsection{The Friedmann-Lemaitre universe}\label{ssFLU}
In the previous sections we have considered general inhomogeneous
and anisotropic cosmological spacetimes with imperfect total
energy-momentum tensor. However, the current observational evidence
(principally the CMB) and our theoretical prejudice (the Copernican
principle), strongly support a universe that is homogeneous and
isotropic on cosmological scales, namely a
Friedmann-Lemaitre-Robertson-Walker (FLRW) universe.

\subsubsection{The FLRW metric}\label{sssFLRWM}
The geometry of the simplest non-static, non vacuum solution of the
Einstein field equations is described by the Robertson-Walker line
element. In suitable (comoving) coordinates the latter takes the
form
\begin{equation}
{\rm d}s^2=-{\rm d}t^2+ a^2(t)\left[{\rm d}r^2+f_K^2(r)({\rm
d}\theta^2+\sin^2\theta{\rm d}\phi^2)\right]\,,  \label{FLRWm1}
\end{equation}
where $a$ is the scale factor and $f_K$ depends on the geometry of
the 3-D hypersurfaces. The scale factor defines a characteristic
length scale and leads to the familiar Hubble parameter
$H=3\dot{a}/a$, which determines the rate of the (isotropic)
expansion. The isotropy of the 3-space means that the latter has
curvature equal to $\clr=6K/a^2$, with the curvature index $K$
normalised to $\pm1$ when it is not zero. Then,
\begin{equation}
f_K(r)=\left\{\begin{array}{l} \sin r \hspace{10mm} {\rm
for}~K=+1\,,\\ r \hspace{16mm} {\rm for}~K=0\,,\\ \sinh r
\hspace{8mm} {\rm for}~K=-1\,.\\ \end{array}\right.  \label{FLRWm2}
\end{equation}
When $K=+1$ the 3-space is closed with spherical geometry and finite
total volume. Alternatively, we have flat Euclidean 3-D
hypersurfaces for $K=0$ and open, hyperbolic, ones when $K=-1$. In
either of these two cases the 3-space is unbounded unless nontrivial topologies are employed.

\subsubsection{FLRW cosmologies}\label{sssFLRWCs}
The high symmetry of the Friedmann models means that all kinematical and dynamical variables are functions of time only and any quantity
that represents anisotropy or inhomogeneity vanishes identically.
Thus, in covariant terms an FLRW model has $\Theta=3H(t)\neq0$,
$\sigma_{ab}=0=\omega_a=A_a$, $E_{ab}=0=H_{ab}$, where $H=\dot{a}/a$ is the familiar Hubble parameter. The isotropy of the Friedmann
models also constrains their matter content, which can only have the perfect-fluid form (with $\rho=\rho(t)$ and $p=p(t)$). In addition,
due to the spatial homogeneity, all orthogonally projected gradients (e.g.~${\rm D}_a\rho$, ${\rm D}_ap$, etc) are by definition zero. These mean that the only nontrivial equations are the FLRW version of Raychaudhuri's formula, the equation of continuity and the Friedmann equation. These follow from expressions (\ref{Ray}), (\ref{edc1}) and (\ref{3R}) and are given by
\begin{equation}
\dot{H}= -H^2- {1\over6}\,(\rho+3p)+ {1\over3}\,\Lambda\,,
\hspace{20mm} \dot{\rho}= -3H(\rho+p)  \label{FLRWeqs1}
\end{equation}
and
\begin{equation}
H^2= {1\over3}\,\rho- {K\over a^2}+ {1\over3}\,\Lambda\,,
\label{FLRWeqs2}
\end{equation}
respectively. Note that the isotropy of the FLRW models and
Eq.~(\ref{3Riemann2}) imply that the associated 3-Riemann tensor is
given by $\clr_{abcd}= (K/a^2)(h_{ac}h_{bd}-h_{ad}h_{bc})$.
Introducing the density parameters $\Omega_{\rho}=\rho/3H^2$,
$\Omega_{\Lambda}=\Lambda/3H^2$ and $\Omega_K=-K/(aH)^2$, the
Friedmann equation takes the form
\begin{equation}
1=\Omega_{\rho}+ \Omega_K+ \Omega_{\Lambda}\,.  \label{FLRWeqs3}
\end{equation}
Thus, in the absence of a cosmological constant, the 3-space is flat (i.e.~$K=0$) when the matter density takes the critical value
$\rho=\rho_c=3H^2$ and the Friedmann equation reduces to
$K/a^2=H^2(\Omega_{\rho}-1)$. In that case $\Omega_{\rho}=1$ ensures Euclidean 3-D hypersurfaces, while $\Omega_{\rho}>1$ leads to spherical and $\Omega_{\rho}<1$ to hyperbolic spatial geometry. One
may also combine Eqs.~(\ref{FLRWeqs1}a) and (\ref{FLRWeqs2}) to
obtain an alternative form for the Raychadhuri equation with an
explicit 3-curvature dependence, namely
\begin{equation}
\dot{H}= -{1\over2}\,(\rho+p)+ {K\over a^2}\,.  \label{FLRWeqs4}
\end{equation}
After a little algebra, expression (\ref{FLRWeqs1}a) takes the
alternative form
\begin{equation}
qH^2= {1\over6}\,(\rho+3p)- {1\over3}\,\Lambda\,, \label{FLRWeqs5}
\end{equation}
where $q=-\ddot{a}a/\dot{a}^2=-[1+(\dot{H}/H^2)]$ is the
dimensionless deceleration parameter. When the latter is negative
the universe accelerates, which means that in exact FLRW models with vanishing $\Lambda$ we need to violate the strong energy condition
(i.e.~set $\rho+3p<0$) to achieve accelerated expansion.

The expansion rate also defines a representative length scale, the
Hubble radius,
\begin{equation}
\lambda_H=H^{-1}\,.  \label{Hr}
\end{equation}
In most FLRW models the scale-factor evolution (see
\S~\ref{sssSfEFLRWCs} below) ensures that the Hubble length
effectively coincides with the particle horizon ($d_H\propto t$). In that case, the Hubble radius determines the regions of causal
contact.

The scale factor of an FLRW spacetime with non-Euclidean spatial
geometry also defines the curvature scale ($\lambda_K=a$) of the
model. This is the threshold at which any departures from Euclidean
flatness in the geometry of the spatial hypersurfaces start becoming important (e.g.~see~\cite{1995PhRvD..52.3338L}). Scales smaller than the curvature length are termed subcurvature, while those exceeding
$\lambda_K$ are referred to as supercurvature. The former are essentially immune to the effects of spatial geometry, which become prominent only on supercurvature lengths. However, the
dynamics of fluctuations is sensitive to the spatial geometry on all scales, through its effect on the expansion rate (via the Friedmann
equations). The relation between the curvature scale and the Hubble
radius is determined by Eq.~(\ref{FLRWeqs2}). In the absence of a
cosmological constant, the latter takes the form
\begin{equation}
\left({\lambda_K\over\lambda_H}\right)^2=
-{K\over1-\Omega_{\rho}}\,. \label{K/H}
\end{equation}
Therefore, for $K=-1$ we find that $\lambda_K>\lambda_H$ always,
with $\lambda_K\rightarrow\infty$ as $\Omega_{\rho}\rightarrow1$ and $\lambda_K\rightarrow\lambda_H$ for $\Omega_{\rho}\rightarrow0$. In
practice, this means that supercurvature scales in spatially open
FLRW cosmologies are never causally connected. When dealing with
closed models, on the other hand, expression (\ref{K/H}) shows that
$\lambda_K>\lambda_H$ when $\Omega_{\rho}<2$ and $\lambda_K\leq\lambda_H$ if $\Omega_{\rho}\geq2$. We finally note
that the importance of spatial geometry within a comoving region is
always the same, since the curvature scale simply redshifts with the expansion.

\subsubsection{Luminosity distance}\label{sssLD}
The luminosity distance of an object at redshift $z$ is $D_L=a_0(1+z)r_0$, where $a_0$ and $r_0$ are the current values of the scale factor and of the object's radial distance (e.g.~see~\cite{1983QB981.N3.......}). The latter is determined by integrating the line element ${\rm d}t= (a/\sqrt{1-Kr^2}){\rm d}r$ of a null geodesic. Assuming a spatially flat FLRW model, the result reads
\begin{equation}
r_0= a_0^{-1}\int_0^zH^{-1}{\rm d}x\,, \label{rd}
\end{equation}
and it is easily integrated through the various epochs of the expansion. In view of the recent supernovae observations, however, it helps to express the above in terms of kinematical quantities, and particularly in terms of the deceleration parameter. Following \S~\ref{sssFLRWCs} and recalling that ${\rm d}z=-(1+z)H{\rm d}t$, we have~\cite{2002ApJ...569...18T}
\begin{equation}
\int_H^{H_0}H^{-1}{\rm d}H= \ln\left({H_0\over H}\right)= -\int_0^z(1+q){\rm d}[\ln(1+x)]\,,  \label{int1}
\end{equation}
which substituted into Eq.~(\ref{rd}) leads to
\begin{equation}
a_0r_0= H_0^{-1}\int_0^z{\rm e}^{-\int_0^x(1+q){\rm d} [\ln(1+y)]}{\rm d}x\,. \label{int2}
\end{equation}
Consequently, expressed in terms of the kinematical parameters of a spatially flat Friedmann model, the luminosity distance of an object at redshift $z$ is given by\footnote{For further discussion, extending to FLRW models with $K\neq0$, and for expressions of the luminosity distance in terms of higher order derivatives of the scale factor, the reader is referred to~\cite{2004CQGra..21.2603V}.}
\begin{equation}
D_L= (1+z)H_0^{-1}\int_0^z{\rm e}^{-\int_0^x(1+q){\rm d}[\ln(1+y)]}{\rm d}x\,.  \label{DL2}
\end{equation}
When compared with the measured luminosity distance from remote type Ia supernovae, the above expression indicated that our universe has recently entered a phase of accelerating expansion~\cite{2002ApJ...569...18T,2004ApJ...607..665R}.

\subsubsection{Scale-factor evolution in FLRW
cosmologies}\label{sssSfEFLRWCs}
\underline{\textit{The $K=0$ case:}} To close the (\ref{FLRWeqs1}) system one needs to introduce an equation of state for the matter component. Before doing so, we will first briefly refer to the de Sitter universe. This is an exponentially expanding $K=0$ model, containing no matter and having a positive cosmological constant. Applied to the de Sitter space, expression (\ref{FLRWeqs2}) reduces to $H=\sqrt{\Lambda/3}=\,$constant. The latter integrates immediately, giving $a\propto{\rm e}^{\sqrt{\Lambda/3}t}$ and thus guaranteeing the exponential nature of the expansion.

In what follows we will consider barotropic perfect fluids, mainly in the form of non-relativistic dust or isotropic radiation (with $p=0$ and $p=\rho/3$ respectively). When $w=p/\rho$ is the constant barotropic index of the cosmic medium, the continuity equation (see (\ref{FLRWeqs1}b)) gives
\begin{equation}
\rho= \rho_0\left({a_0\over a}\right)^{3(1+w)}\,.  \label{FLRWrho}
\end{equation}
Substituting this result into the Friedmann equation, assuming
Euclidean spatial sections (i.e.~$K=0$), $w\neq-1$ and setting the cosmological constant to zero, we arrive at
the following expression for the scale factor
\begin{equation}
a= a_0\left({t\over t_0}\right)^{2/3(1+w)}\,,  \label{sfFLRWa}
\end{equation}
having normalised our solution so that $a(t=0)=0$. When dealing with non-relativistic matter with $w=0$ (e.g.~baryonic `dust' or
non-baryonic cold dark matter), we have what is known as the
Einstein-de Sitter universe with $a\propto t^{2/3}$.
Alternatively, we obtain $a\propto t^{1/2}$ in the case of
relativistic species (e.g.~isotropic radiation) and $a\propto
t^{1/3}$ for a stiff medium with $w=1$. An additional special case
is that of matter with zero gravitational mass, which corresponds to $w=-1/3$ and leads to `coasting' expansion with $a\propto t$.
Solution (\ref{sfFLRWa}) does not apply to a medium with $w=-1$ (and therefore with zero inertial mass - see Eq.~(\ref{pfcls}b)). In that case, (\ref{FLRWrho}) guarantees that $\rho=\rho_0=$~constant, which when substituted into (\ref{FLRWeqs2}) leads to $H=H_0=$~constant
and subsequently to exponential expansion (inflation) with
$a\propto{\rm e}^{H_0(t-t_0)}$. Note that during a phase of
exponential (de Sitter-type) expansion, the Hubble radius remains
constant, while the particle horizon increases in the usual manner
(see \S~\ref{sssFLRWCs}).

\underline{\textit{The $K=+1$ case:}} The equation of continuity does not depend on the curvature of the 3-space, which means that expression (\ref{FLRWrho}) monitors the evolution of the matter density irrespective of the model's spatial curvature. When the FLRW spacetime has non-Euclidean spatial geometry it helps to parametrise the scale-factor evolution in terms of the conformal time ($\eta$ -- defined by $\dot{\eta}=1/a$). Then, for $K=+1$, $\Lambda=0$ and $w\neq-1/3$ relations (\ref{FLRWeqs1}), (\ref{FLRWeqs2}) combine to give
\begin{equation}
a= a_0\left\{{\sin[(1+3w)\eta/2+\clc]\over
\sin[(1+3w)\eta_0/2+\clc]}\right\}^{2/(1+3w)}\,,  \label{scFLRWa}
\end{equation}
where $(1+3w)\eta/2+\clc\in(0,\pi)$. Normalising so that
$a(\eta\rightarrow0)\rightarrow0$, the $\eta=\pi/(1+3w)$ threshold
corresponds to the moment of maximum expansion when
$a=a_{max}=a_0\{\sin[(1+3w)\eta_0/2]\}^{-2/(1+3w)}$. For
non-relativistic matter $w=0$ and the above solution reduces to
$a\propto\sin^2(\eta/2)$~\cite{1983QB981.N3.......}, while we obtain $a\propto\sin\eta$ if radiation dominates. When $w=-1/3$ one can no longer use solution (\ref{scFLRWa}). Instead, Eq.~(\ref{FLRWeqs1}a) leads immediately to the familiar coasting-expansion phase with $a\propto t$. Expressions (\ref{FLRWeqs1}b), (\ref{FLRWeqs2}) also provide the relation between scale factor and proper time in the $w=-1$ case. Just like in spatially flat models, Eq.~(\ref{FLRWeqs1}b) ensures that $\rho=\rho_0=$~constant and then (\ref{FLRWeqs2}) leads to $a(1+\sqrt{3/\rho_0}H)\propto {\rm e}^{\sqrt{(\rho_0/3)}\,t}$.

The Einstein universe corresponds to a static $K=+1$ model with positive cosmological constant. In such an environment the density of the matter component is also constant, while Eqs.~(\ref{FLRWeqs1}a), (\ref{FLRWeqs2}) and (\ref{FLRWeqs4}) reduce to the constraints
\begin{equation}
\rho+3p= 2\Lambda\,, \hspace{15mm} {1\over a^2}= {1\over3}\,(\rho+\Lambda) \hspace{10mm} {\rm and} \hspace{10mm} {1\over a^2}= {1\over2}\,(\rho+p)\,,  \label{Eeqs1}
\end{equation}
respectively. The Einstein universe has long been known to be unstable under homogeneous perturbations, though its stability to inhomogeneous distortions is less straightforward (see \S~\ref{sssESU} below).

\underline{\textit{The $K=-1$ case:}} Applied to FLRW cosmologies with hyperbolic spatial geometry, zero cosmological constant and $w\neq-1/3$, the analysis described above leads to the following (ever expanding) evolution law for the scale factor
\begin{equation}
a= a_0\left\{{\sinh[(1+3w)\eta/2+\clc]\over
\sinh[(1+3w)\eta_0/2+\clc]}\right\}^{2/(1+3w)}\,,  \label{soFLRWa}
\end{equation}
where now $(1+3w)\eta/2+\clc>0$. Not surprisingly, the above can be also obtained from (\ref{scFLRWa}), after the trigonometric functions are replaced with their hyperbolic counterparts. Assuming pressure-free `dust' and normalising as before, we find $a\propto\sinh^2(\eta/2)$~\cite{1983QB981.N3.......}. On the other hand, solution (\ref{soFLRWa}) implies $a\propto\sinh\eta$ for a open FLRW universe dominated by relativistic species. We finally note that, similarly to the $K=+1$ case, the system (\ref{FLRWeqs1}), (\ref{FLRWeqs2}) ensures that $a\propto t$ when $w=-1/3$ and $a(1+\sqrt{3/\rho_0}H)\propto {\rm e}^{\sqrt{(\rho_0/3)}\,t}$ for $w=-1$.

A special model with open spatial geometry is the vacuum Milne universe. Similarly to the $w=-1/3$ case, the absence of matter means that (\ref{FLRWeqs1}a) integrates to give a coasting scale factor of the general form $a\propto t$. Here, however, the Friedmann equation -- see expression (\ref{FLRWeqs2}) -- guarantees that $a=t$.

We finally point out that, in the absence of a cosmological constant and after introducing the transformation $y=a^{(1+3w)/2}$, with $w\neq-1/3$, Raychaudhuri's equation (see expression (\ref{FLRWeqs1}a) in \S~\ref{sssFLRWCs}) reduces to a simple harmonic-oscillator of the form
\begin{equation}
y^{\prime\prime}= -K\left({1+3w\over2}\right)^2y\,,  \label{Barrow}
\end{equation}
with primes indicating conformal time derivatives~\cite{1993Obs...113..210B}. This expression is particularly useful when addressing spatially closed or open FLRW models, with $\Lambda=0$ and $w\neq-1/3$. For instance, it is straightforward to verify that for $K=\pm1$ the above given equation leads immediately to solutions (\ref{scFLRWa}) and (\ref{soFLRWa}) respectively.

\subsection{The Bianchi universes}\label{ssBUs}
Despite the success of the Friedmann-Lemaitre models, the structure that we observe today, means that our universe is neither homogeneous nor isotropic, at least on certain scales and to a certain extent. To follow the (late time) evolution of the universe on these scales one needs models with more degrees of freedom than the FLRW ones. The spatially homogeneous and anisotropic Bianchi models have long been used to understand the observed level of isotropy in our universe and also to probe the nature of the initial singularity. Here, we will briefly consider members of the Bianchi family that contain the FLRW models as special cases, referring the reader to review articles and monographs for further details~\cite{1969CMaPh..12..108E,1973CLP.....6...61M,%
1979PhR....56...65C,1986PhR...139....1B,1997icm..book.....K,%
1997dsc..book.....W,2003ASSL..291.....C,2006igrc.book.....P,%
2006GReGr..38.1003E}.

\subsubsection{Classification of Bianchi
cosmologies}\label{sssCBCs}
Time is essentially the only dynamical coordinate in the Bianchi spacetimes, the spatial homogeneity of which has `removed' all the inhomogeneous degrees of freedom and has reduced Einstein's equations to a set of ordinary differential equations. Despite this, the Bianchi family provides a rich set of models where one can study the fully nonlinear theory. We generally distinguish between two differents kinds of Bianchi cosmologies. The non-tilted (or orthogonal) models, with the flow-lines of the fluid normal to the hypersurfaces of homogeneity and the tilted models where this is no longer true. In the latter case the `peculiar' velocity of the matter enters the equations as an additional dynamical variable (see \S~\ref{ssNPKs} below).

The literature contains three basic ways of classifying the orthogonal Bianchi models, all based on the commutation laws of the associated tetrad basis vectors.\footnote{This means classifying the Lie algebras of the Killing vector fields and therefore the associated group of the $G_3$-isometries (see~\cite{1969CMaPh..12..108E,1997dsc..book.....W} for details).} Thus, one may use the tetrad $\{{\bf e}_a,\,a=0,1,2,3\}$, so that ${\bf e}_0$ is the normal vector to the hypersurfaces of homogeneity and
\begin{equation}
\left[{\bf e}_a,{\bf e}_b\right]= \gamma^c{}_{ab}{\bf e}_c\,, \label{tbcl}
\end{equation}
with the commutation functions $\gamma^a{}_{bc}=\gamma^a{}_{bc}(t)$ treated as dynamical variables themselves~\cite{1969CMaPh..12..108E}. The spatial commutators $\gamma^{\alpha}{}_{\beta\gamma}$, with $\alpha,\beta,\gamma=1,2,3$, are then decomposed into the time-dependent pair $n_{\alpha\beta}$ and $a_{\alpha}$ that satisfy the condition
\begin{equation}
n_{\alpha\beta}a^{\beta}= 0\,.  \label{Ji}
\end{equation}
Choosing the tetrad so that $n_{\alpha\beta}$ is a diagonalisable matrix (i.e.~$n_{\alpha\beta}={\rm diag}(n_1,n_2,n_3)$) and $a_{\alpha}=(a,0,0)$, the above reduces to $n_1a=0$. Consequently, one can immediately define two major classes of Bianchi spacetimes. Those with $a=0$ are known as class A models and those with $a\neq0$ are termed class B. Further classification is achieved by the signs of the eigenvalues of $n_{\alpha\beta}$. When dealing with the class B models one may also introduce the scalar $h$, which satisfies the constraint
\begin{equation}
a^2= hn_2n_3\,.  \label{h-p}
\end{equation}
This means that the $h$-parameter is well defined only in class B models with $n_2n_3\neq0$. The general Bianchi classification is given in Tab.~\ref{tab1}, showing that $h<0$ in type $VI_h$ and $h>0$ in $VII_h$. We finally note that some Bianchi groups allow for subspaces of higher symmetry, like isotropic or locally rotationally symmetric models. As a result, the FLRW universes may sometimes appear as special cases in certain Bianchi cosmologies.

\begin{table}
\caption{The non-tilted Bianchi spacetimes classified into two group classes and ten group types (see~\cite{1997dsc..book.....W}).}
\vspace{0.5truecm}
\begin{center}\begin{tabular}{cccccc}
\hline \hline Group class \hspace{5mm} & Group type \hspace{5mm} & $n_1$ & $n_2$ & $n_3$ & \hspace{5mm} FLRW as special case \hspace{10mm} \\ \hline A$\,(a=0)$ & $\left\{\begin{array}{l} I \\ II \\ VI_0 \\ VII_0 \\ VIII \\ IX \end{array}\right.$ & $\begin{array}{c} 0 \\ + \\ 0 \\ 0 \\ - \\ + \end{array}$ & $\begin{array}{c} 0 \\ 0 \\ + \\ + \\ + \\ + \end{array}$ & $\begin{array}{c} 0 \\ 0 \\ - \\ + \\ + \\ + \end{array}$ & $\begin{array}{c} K=0 \\ - \\ - \\ K=0 \\ - \\ K=+1 \end{array}$ \\[1.5truemm]\hline B$\,(a\neq0)$ & $\left\{\begin{array}{l} V \\ IV \\ VI_h \\ VII_h \end{array}\right.$ & $\begin{array}{c} 0 \\ 0 \\ 0 \\ 0 \end{array}$ & $\begin{array}{c} 0 \\ 0 \\ + \\ + \end{array}$ & $\begin{array}{c} 0 \\ + \\ - \\ + \end{array}$ & $\begin{array}{c} K=-1 \\ - \\ - \\ K=-1 \end{array}$ \\ [1.5truemm] \hline \hline
\end{tabular}\end{center}\label{tab1}\vspace{0.5truecm}
\end{table}

\subsubsection{Bianchi~I cosmologies}\label{sssB1Cs}
The simplest anisotropically expanding cosmologies, which are also the simplest generalisation of the spatially flat FLRW universe, are the non-tilted Bianchi~I models. These are class $A$ spacetimes (see Tabble~\ref{tab1}), with Euclidean 3-geometry and line elements of the form
\begin{equation}
{\rm d}s^2= -{\rm d}t^2+ X^2(t){\rm d}x^2+ Y^2(t){\rm d}y^2+ Z^2(t){\rm d}z^2\,,  \label{B1}
\end{equation}
in comoving coordinates.\footnote{For a discussion on the classification of the Bianchi models in two major classes ($A$ and $B$) on the basis of their structures constants the reader is referred to~\cite{1969CMaPh..12..108E,1997dsc..book.....W}.} The above allows for different expansion rates along the three spatial directions, with the average scale factor and the mean Hubble parameter given by $a={}^3\sqrt{XYZ}$ and $H=\Theta/3=\dot{a}/a$ respectively. The spatial homogeneity of the Bianchi~I spacetimes ensures that all invariants depend on time only. The flow lines of the fundamental observers are irrotational geodesics and all spatial gradients vanish identically. Therefore, in covariant terms, all Bianchi~I cosmologies are characterised by
\begin{equation}
\omega_a=0=A_a=H_{ab}=\mathcal{R}_{ab}\,,  \label{B1con}
\end{equation}
which means that the only nonzero quantities are the volume scalar, the shear tensor and the electric part of the Weyl field. Also, because of their generic anisotropy, the Bianchi~I spacetimes can support imperfect fluids with non-vanishing anisotropic pressure (i.e.~$\pi_{ab}\neq0$ though $q_a=0$). For example, the type-I models are natural hosts of large-scale magnetic fields~\cite{1970ApJ...160..147H}.

As with the FLRW case, the 1+3 covariant formulae monitoring the evolution of the Bianchi~I cosmologies are obtained from the general expressions given in \S~\ref{ssCRC}. Thus, applied to a type-I environment and in the absence of a cosmological constant, Eqs.~(\ref{Ray}), (\ref{edc1}) and (\ref{3R}) reduce to
\begin{equation}
\dot{H}= -H^2- {1\over6}\,(\rho+3p)- {2\over3}\,\sigma^2\,, \hspace{10mm} \dot{\rho}= -3H(\rho+p)- \sigma_{ab}\pi^{ab}  \label{B1set1}
\end{equation}
and
\begin{equation}
H^2= {1\over3}\left(\rho+\sigma^2\right)\,.   \label{B1GC1}
\end{equation}
where the latter can be seen as the Bianchi~I analogue of the Friedmann equation. Also note that, on using the density parameter $\Omega_{\rho}$ defined in \S~\ref{sssFLRWCs}, relation (\ref{B1GC1}) takes the form
\begin{equation}
1=\Omega_{\rho}+ \Sigma\,,  \label{B1CG2}
\end{equation}
with $\Sigma=\sigma^2/3H^2$ providing a measure of the model's shear anisotropy. Similarly, in a Bianch~I spacetime, expressions (\ref{sigmadot}), (\ref{GC1}) and (\ref{dotEab}) reduce to
\begin{equation}
\dot{\sigma}_{ab}= -2H\sigma_{ab}- \sigma_{c\langle a}\sigma^c{}_{b\rangle}- E_{ab}+ {1\over2}\,\pi_{ab}\,, \hspace{10mm} E_{ab}= H\sigma_{ab}- \sigma_{c\langle a}\sigma^c{}_{b\rangle}- {1\over2}\,\pi_{ab}  \label{B1set2}
\end{equation}
and
\begin{equation}
\dot{E}_{ab}= -3HE_{ab}- {1\over2}\,(\rho+p)\sigma_{ab}- {1\over2}\,\left(\dot{\pi}_{ab}+H\pi_{ab}\right)+ 3\sigma_{\langle a}{}^c\left(E_{b\rangle c}- {1\over6}\,\pi_{b\rangle c}\right)\,, \label{B1EWl}
\end{equation}
respectively. We also note that Eq.~(\ref{B1set2}b) recasts (\ref{B1set2}a) into
\begin{equation}
\dot{\sigma}_{ab}= -3H\sigma_{ab}+ \pi_{ab}\,.  \label{B1dotsigma}
\end{equation}
The latter ensures that, in the absence of anisotropic pressures, the shear depletes as $a^{-3}$, where $a$ is the average (over the three spatial directions) scale factor.

Once an equation of state for the matter is introduced, the set (\ref{B1set1})-(\ref{B1dotsigma}) governs the dynamics of a Bianchi~I spacetime fully. In the special case of matter in the perfect-fluid form with $p=w\rho$, expression (\ref{B1set1}b) integrates to $\rho\propto a^{-3(1+w)}$. Following (\ref{B1GC1}), this means that the shear will dominate the early expansion, no matter how small the anisotropy may be today (unless the matter component has $w=1$ -- stiff fluid). These shear dominated early stages correspond to the vacuum Kasner regime, in which case Eq.~(\ref{B1GC1}) leads to $a\propto t^{1/3}$. The line element of the Kasner solution has the form
\begin{equation}
{\rm d}s^2= -{\rm d}t^2+ t^{2p_1}{\rm d}x^2+ t^{2p_2}{\rm d}y^2+ t^{2p_3}{\rm d}z^2\,,  \label{Kasner}
\end{equation}
where $p_1+p_2+p_3=1=p_1^2+p_2^2+p_3^2$ (e.g.~see~\cite{1970AdPhy..19..525B,1975ctf..book.....L,%
2003esef.book.....S}). Together, these conditions guarantee that either exactly one of the three exponents is negative or two of them are zero. In the former case the spacetime expands in two directions and contracts along the third, with a cigar-like initial singularity. In the latter case we have motion (expansion) in one direction only, which corresponds to a pancake-type singularity.\footnote{With the exception of the $(p_1,p_2,p_3)=(0,0,1)$ triplet, which corresponds to a flat spacetime, the initial singularity cannot be eliminated by any coordinate transormation~\cite{1975ctf..book.....L}.} Note that the total volume always increases with time, since the average scale factor grows as (recall that $a\propto t^{1/3}$). Given that kinematics dictate the Kasner phase, it is not surprising that the exponents of (\ref{Kasner}) are determined by the model's shear anisotropy according to~\cite{1997dsc..book.....W}
\begin{equation}
p_1= {1\over3}\,(1-2\Sigma_+) \hspace{10mm} {\rm and} \hspace{10mm} p_{2,3}= {1\over3}\,(1+\Sigma_+\pm\sqrt{3}\Sigma_-)\,,  \label{p123}
\end{equation}
with $\Sigma_{\pm}=\sigma_{\pm}/H$. Note that $\sigma_+=(\sigma_2+\sigma_3)/2$ and $\sigma_-=(\sigma_2-\sigma_3)/2\sqrt{3}$, where $\sigma_2$ and $\sigma_3$ are the two independent components of the shear tensor (recall that $\sigma_{\alpha\beta}={\rm diag} (\sigma_1,\sigma_2,\sigma_3)$ and $\sigma_1+\sigma_2+\sigma_3=0$ -- see also~\cite{1997dsc..book.....W} for more technical details and further references).

\subsubsection{Bianchi~$VII_h$ cosmologies}\label{sssB7hCs}
The most general Bianchi spacetimes that contain the open FLRW universe as a special case are the type-$VII_h$ cosmologies (see Table~\ref{tab1}). The late-time attractors of this class B family is, for a broad range of initial data and matter properties, the Lukash plane-wave solution. These vacuum spacetimes are self-similar equilibrium points and have a line element of the form\footnote{Self-similar Bianchi solutions, vacuum or with a non-tilted perfect fluid, have been studied in~\cite{1986CQGra...3.1105H}.}
\begin{equation}
{\rm d}s^2= -{\rm d}t^2+ t^2{\rm d}x^2+ t^{2r}{\rm e}^{2rx} \left[(A{\rm d}y+B{\rm d}z)^2+(C{\rm d}y+A{\rm d}z)^2\right]\,.  \label{Luk1}
\end{equation}
Note that $r$ is a constant parameter in the range $0<r<1$, $A=\cos v$, $B=f^{-1}\sin v$, $C=-f\sin v$ and $v=k(x+\ln t)$. Also, $f$ and $k$ are constants related to $r$ by
\begin{equation}
{k^2(1-f^2)^2\over f^2}= 4r(1-r) \hspace{10mm} {\rm and} \hspace{10mm} r^2=hk^2\,,  \label{Luk2}
\end{equation}
where $h$ is the associated group parameter. Note that $r$ determines the amount and the nature of the model's anisotropy. When $r\rightarrow1$, in particular, the anisotropy vanishes and (\ref{Luk1}) reduces to the metric of the empty Milne universe. At the $r\rightarrow0$ limit, on the other hand, the anisotropy is maximised~\cite{1986CQGra...3.1105H,2005CQGra..22..825B}.

Due to the absence of matter and given its irrotational nature, the Lukash spacetime is covariantly chracterised by the irreducible sets
\begin{equation}
\rho=0=p=q_a=\pi_{ab} \hspace{10mm} {\rm and} \hspace{10mm} A_a=0=\omega_a\,,  \label{Lukcon}
\end{equation}
which imply that the only nonzero quantities are $\Theta$, $\sigma_{ab}$, $E_{ab}$ and $H_{ab}$. Note that the Weyl components have equal magnitudes and are orthogonal to each other (i.e.~$E^2=H^2$ and $E_{ab}H^{ab}=0$ respectively), in line with the Petrov-type N nature of the solution. The absence of matter means that the Lukash universe is Ricci flat, although the curvature of the 3-space (i.e.~the 3-Ricci tensor) is nonzero. Setting the cosmological constant to zero, the kinematics of the model is governed by the propagation formulae~\cite{2005CQGra..22..825B}
\begin{equation}
\dot{H}= -H^2- {2\over3}\,\sigma^2\,, \hspace{15mm} \dot{\sigma}_{ab}= -3H\sigma_{ab}- \clr_{\langle ab\rangle}\,,   \label{Luks1}
\end{equation}
which are supplemented by the constraints
\begin{equation}
H^2= {1\over3}\left(\sigma^2-{1\over2}\,\clr\right) \hspace{15mm} {\rm and} \hspace{15mm} H_{ab}= {\rm curl}\sigma_{ab}\,.  \label{Luks2}
\end{equation}
The evolution of the Weyl field, on the other hand, is monitored by the set
\begin{equation}
\dot{E}_{ab}= -3HE_{ab}+ {\rm curl}H_{ab}+ 3\sigma_{c\langle a}E^c{}_{b\rangle}\,, \hspace{8mm} \dot{H}_{ab}= -3HH_{ab}- {\rm curl}H_{ab}+ 3\sigma_{c\langle a}H^c{}_{b\rangle}\,.  \label{Luks3}
\end{equation}

The average kinematic anisotropy is measured by means of the shear parameter defined by the dimensionless, expansion normalised parameter $\Sigma=\sigma^2/3H^2= (1-r)/(1+2r)$~\cite{2005CQGra..22..825B}. This means that $\Sigma$ remains constant during the model's evolution and lies between zero and unity. Also, minimum anisotropy corresponds to $\Sigma\rightarrow0$ and maximum to $\Sigma\rightarrow1$, ensuring that the shear of the Lukash model is always bounded. On using the $\Sigma$-parameter, Eqs.~(\ref{Luks1}a) and (\ref{Luks2}a) are recast into
\begin{equation}
\dot{H}= -3H^2(1+2\Sigma) \hspace{15mm} {\rm and} \hspace{15mm} \clr= 6H^2(\Sigma-1)\,,  \label{LRay}
\end{equation}
respectively. The former leads to the power-law evolution $a\propto t^{1/(1+2\Sigma)}$ of the average scale factor and the latter ensures that the spatial sections are open. Thus, when the shear is at its minimum, we approach $a\propto t$ and the Milne universe. At the $\Sigma\rightarrow1$ limit, on the other hand, we find the familiar Kasner solution (i.e.~$a\propto t^{1/3}$ -- see \S~\ref{sssB1Cs}). Note that for maximum shear anisotropy the 3-curvature vanishes, whereas the Weyl field tends to zero at both limits~\cite{1986CQGra...3.1105H,2005CQGra..22..825B}. Finally, the deceleration parameter of the Lukash universe is $q=2\Sigma$, with $0<q<2$.

\subsubsection{Bianchi~IX cosmologies}\label{sssB9Cs}
The type-IX model is a class A spacetime and the only Bianci cosmology that contains the closed FLRW universe as a special case (see Table~\ref{tab1}). The model is well known for its oscillatory behaviour with chaotic characteristics and with the matter becoming dynamically negligible as it approaches the initial singularity~\cite{1969PhRvL..22.1071M}. The spacetime metric has the form
\begin{eqnarray}
{\rm d}s^2= -{\rm d}t^2&+& X^2(t)(cos\psi{\rm d}\theta+\sin\psi\sin\theta{\rm d}\phi)^2+ Y^2(t)(\sin\psi{\rm d}\theta-\cos\psi\sin\theta{\rm d}\phi)^2 \nonumber\\ &+&Z^2(t)({\rm d}\psi+\cos\theta{\rm d}\phi)^2\,,  \label{BIX}
\end{eqnarray}
parametrised by the Euler angles with $0\leq\psi\leq4\pi$, $0\leq\theta\leq\pi$ and $0\leq\phi\leq2\pi$. Allowing $H$ to assume all real values and therefore permitting a contracting epoch, the generalised Friedmann equation of the model reads
\begin{equation}
H^2= {1\over3}\left(\rho+\sigma^2-{1\over2}\,\clr\right)\,.  \label{IXFried}
\end{equation}
A key feature of the type-IX cosmologies is that their spatial curvature changes sign during the model's evolution. In particular, the associated 3-Ricci scalar is given by~\cite{1997dsc..book.....W}
\begin{equation}
\clr= -{1\over2}\left[n_1^2+n_2^2+n_3^2 -2(n_1n_2+n_2n_3+n_3n_1)\right]\,,  \label{IXcR1}
\end{equation}
where $n_{1,2,3}>0$ according to Table~\ref{tab1}. Alternatively, one may express $\clr$ in terms of the individual scale factors as~\cite{1982PhR....85....1B}
\begin{equation}
\clr={2(X^2Y^2+X^2Z^2+Y^2Z^2)-(X^4+Y^4+Z^4)\over2(XYZ)^2}\,.  \label{IXcR2}
\end{equation}
The right-hand side in both of the above expressions can take either sign. In fact, $\clr$ is predominantly negative and becomes positive only when the model approaches isotropy (i.e.~for $n_1=n_2=n_3$ or $X=Y=Z$). Since an expansion maximum occurs when the curvature is positive (see Eq.~(\ref{IXFried})), one may argue that the model cannot recollapse while still anisotropic~\cite{1988NuPhB.296..697B}.

The past attractor of the Bianchi-IX spacetimes is the so-called {\em Mixmaster} oscillatory singularity. A qualitative analysis shows that the model approaches the initial singularity through a sequence oscillatory eras. Each era consists of alternating Kasner phases, with a metric given by (\ref{Kasner}) and the negative exponent shifting between two of the directions~\cite{1970AdPhy..19..525B,1975ctf..book.....L}. This means that distances along the associated two axes oscillate, while those in the third decrease. As the model passes through the different eras, the `decreasing axis' bounces from one direction to the next, with the process asymptotically acquiring a random character~\cite{1997PhRvL..78..998C}. The chaotic behaviour of the type-IX model was  suggested as a way of achieving sufficient `mixing' between the three spatial directions that could remove the horizon problem~\cite{1969PhRvL..22.1071M}. As it eventually turned out, however, this mechanism does not work.

\subsubsection{Isotropisation of Bianchi
cosmologies}\label{sssIBCs}
Bianchi cosmologies have been traditionally studied qualitatively, primarily by means of dynamical system methods~\cite{1997dsc..book.....W,2003ASSL..291.....C}. These techniques have revealed an interesting property of many Bianchi models, namely their ``intermediate isotropisation''. The latter occurs because a number of Bianchi-type spacetimes have phase planes where the FLRW solutions are acting as saddle points. This means that these models can isotropise and therefore look very much like a Friemdann universe, over an extended period of their evolution, despite the fact that they start off and end up quite unlike the FLRW spacetimes.

Another issue is whether the Bianchi models show any tendency to isotropise, either at early or at late times. Following \cite{1973ApJ...180..317C}, the set of Bianchi models, with conventional matter, that isotropise asymptotically to the future is of zero measure. It has been shown, however, that Bianchi cosmologies tend towards isotropy at late times when a cosmological constant is present. Taken at face value, this implies that inflation should smooth the anisotropy of these models out. Nevertheless, the existing results apply primarily to non-tilted Bianchi types and also seem to depend on the amount of the initial anisotropy.

\subsubsection{Kantowski-Sachs cosmologies}\label{sssK-SCs}
These are spatially homogeneous spacetimes that do not belong to the Bianchi family~\cite{1965SFJETP...20..1303K,1966JMP.....7..443K}. The Kantowski-Sachs class of models have local rotational symmetry, with metrics which in comoving coordinates read~\cite{1999toc..conf....1E,%
2006igrc.book.....P}
\begin{equation}
{\rm d}s^2= -{\rm d}t^2+ A^2(t){\rm d}r^2+ B^2(t)\left[{\rm d}\theta^2+f^2(\theta)\,{\rm d}\phi^2\right]\,,  \label{K-S}
\end{equation}
where $f(\theta)=\sin\theta$. Note that in general the function $f(\theta)$ obeys an expression of the form (\ref{FLRWm2}) -- see~\S~\ref{sssFLRWM} -- in which case one refers to Kantowski-Sachs-like metrics~\cite{1997PhRvD..55..630B}. However, only the $K=+1$ model falls outside the Bianchi family. Those with zero and negative spatial curvature reduce to axisymmetric type-I and type-III cosmologies respectively~\cite{1997icm..book.....K}. Note that, in the Bianchi classification of \S~\ref{sssCBCs}, the missing type-III spacetime corresponds to the VI$_{-1}$ model~\cite{1997dsc..book.....W}.

\section{Inhomogeneous relativistic cosmologies}\label{sIRCs}
The simplest inhomogeneous cosmologies are spherically symmetric, like the Lemaitre-Tolman-Bondi (LTB) model. The latter has closed spatial sections and matter in the form of irrotational dust. The LTB universe possesses a centre of symmetry, in fact it can allow for up to two such centres, but it is not isotropic about an arbitrary observer; an inevitable consequence of the model's spatial inhomogneity~\cite{1997icm..book.....K}. There also exist inhomogenoeus solutions of the EFE without symmetries, with the Szekeres quasi-spherical model probably being the most celebrated (see~\cite{1997icm..book.....K} for an extended discussion and references). An additional class of cosmologies without (global) symmetry are the so called Swiss-Cheese models, obtained by cutting and pasting segments of spherically symmetric spacetimes (see~\cite{1999toc..conf....1E}). The universe we live in is also believed to be free of symmetries.

\subsection{The gauge problem in cosmology}\label{ssGPC}
It has long been known that the study of cosmological perturbations
is plagued by what is known as the { gauge problem}, reflecting the
fact that in perturbation theory we deal with two spacetime
manifolds~\cite{1946JPhy..10..116L,1963AdPhy..12..185L,%
1967ApJ...147...73S,1980PhRvD..22.1882B,1989PhRvD..40.1804E}. The
first is the physical spacetime, ${\clw}$, that corresponds to the
real universe and the second, denoted here by $\overline{\clw}$, is
a fictitious idealised mathematical model.

\subsubsection{Gauge freedom}\label{sssGF}
In most cosmological studies the idealised, background, spacetime is represented by the homogeneous and isotropic FLRW models. To proceed one needs to establish a one-to-one correspondence, namely a gauge $\phi:\overline{\clw}\rightarrow{\clw}$, between the two
spacetimes. Such a point-identification map is generally arbitrary,
although particular ones may be more suitable for specific cases.
When a coordinate system is introduced in $\overline{\clw}$, the
gauge carries it to ${\clw}$ and vice versa. As a result, a smooth
spacetime is defined into the real universe. Any change in
$\phi:\overline{\clw}\rightarrow{\clw}$, keeping the background
coordinates fixed, is known as a { gauge transformation}. This
introduces a coordinate transformation in the physical spacetime but also changes the event in ${\clw}$ which is associated with a given
event of $\overline{\clw}$. Gauge transformations are therefore
different from coordinate transformations which merely relabel
events. The gauge problem stems from our inherent freedom to make
gauge transformations. Although, the { gauge freedom} is usually
expressed as a freedom of coordinate choice in $\clw$, it should be
understood that it generally changes the point-indentification
between the two spacetimes.

In the study of cosmological perturbations we consider the realistic universe and define perturbations by specifying the map
$\phi:\overline{\clw}\rightarrow{\clw}$ between the $\clw$ and its
fictitious counterpart. However, although we can always perturb away form a given background spacetime, recovering the smooth metric from a given perturbed one is not a uniquely defined process. This is a
problem because it is always possible to choose an alternative
background and therefore arrive at different perturbation values
(see also~\cite{1989PhRvD..40.1804E}). Selecting an unperturbed
spacetime from a given lumpy one corresponds to a gauge choice.
Determining the best gauge is known as the { fitting problem} in
cosmology and there is no unique answer to
it~\cite{1987CQGra...4.1697E}.

\subsubsection{Gauge dependence}\label{sssGD}
By definition, the perturbation of any quantity is the difference
between its value at some event in the real spacetime and its value
at the corresponding, through the gauge, event in the background.
Then, even scalar quantities that have nonzero and position-dependent background values, will lead to gauge-dependent perturbations. Following~\cite{1974RSPSA.341...49S,%
1990CQGra...7.1169S}, we consider a one parameter family of 4-manifolds $\clw_{\epsilon}=\clw(\epsilon)$ embedded in a 5-manifold $\clm$. Each one of these 4-manifolds represents a realistic spacetime, perturbed relative to the background manifold $\bar{\clw}$. We define a point-identification map between $\bar{\clw}$ and $\clw_{\epsilon}$, by introducing in $\clm$ a vector field $X_A$ (with $A=0,\dots,4$), which is everywhere transverse to the embeddings $\clw_{\epsilon}$. Points lying along the same integral curves of $X_A$, which are parametrised by $\epsilon$ for convenience, will be regarded as the `same'. Thus, selecting a specific vector field $X_A$ corresponds to a choice of gauge. If $Q_{\epsilon}$ is some geometrical quantity defined on
$\clw_{\epsilon}$, for small $\epsilon$ we have
\begin{equation}
\bar{h}_{\epsilon}(Q_{\epsilon})= \bar{Q}+ \epsilon
\bar{\cll}_XQ_{\epsilon}+ \clo(\epsilon^2)\,, \label{Qe}
\end{equation}
where an overbar refers ro quantities evaluated in $\bar{\clw}$,
${\cll}_X$ is the Lie derivative along $X_A$ and
$\bar{h}_{\epsilon}$ is the pullback of $\clw_{\epsilon}$ to $\bar{\clw}$. The quantity $\delta Q= \bar{h}_{\epsilon}(Q_{\epsilon})-\bar{Q}=
\epsilon\bar{\cll}_X Q_{\epsilon}$ is what we usually call linear
perturbation of $\bar{Q}$ and it clearly depends on our gauge
choice~\cite{1974RSPSA.341...49S,1990CQGra...7.1169S}.

According to (\ref{Qe}), even quantities that behave like scalars
under coordinate changes will not remain invariant under gauge
transformations. To see this from a different, less technical, point of view we will follow~\cite{1989PhRvD..40.1804E}. Consider the
familiar density perturbation $\delta\rho=\rho-\overline{\rho}$,
where $\rho$ is the matter density. In the right-hand side of the
above we have the difference in the value of the matter density
between two corresponding points in the background and the real
spacetime. A gauge transformation will generally change this
correspondence and therefore the perturbation value. This means that the value of $\delta\rho$ is entirely gauge-dependent and therefore
arbitrary. For instance, one can select the gauge so that the
surfaces of constant background density are the surfaces of constant real density, thus setting $\delta\rho$ to
zero~\cite{1989PhRvD..40.1804E}.\footnote{With this gauge choice the fluid flow lines are not orthogonal to the surfaces of constat
density and comoving observers will still measure a nonzero density
variation.}

\subsection{Covariant and gauge-invariant perturbations}
\label{ssCGiPs}
One way of addressing the gauge problem is by completely fixing the
point-identification map between the background and the real
spacetimes. However, determining the best gauge for a given physical problem is not a trivial task and it might lead to spurious,
gauge-dependent, results. Alternatively, we may only partially fix
the gauge (leaving some residual gauge freedom and always keeping
track of its consequences) or employ gauge-invariant
variables~\cite{1980PhRvD..22.1882B,1989PhRvD..40.1804E}.

\subsubsection{Criteria for gauge invariance}\label{sss(CGI}
Gauge-independent quantities must remain invariant under gauge
transformations between the idealised and the realistic spacetimes.
According to the Stewart and Walker lemma, the simplest cases are
scalars that are constant in the background universe or tensors that vanish there~\cite{1974RSPSA.341...49S}. In both cases the mapped
quantity is also constant and gauge changes are irrelevant because
they all define the same perturbation. The only other possibility
are tensors that can be written as linear combinations of products
of the Kronecker deltas with constant coefficients. The same general criteria also apply to second order perturbations, but this time the Stewart and Walker requirements must be satisfied by the first-order variables~\cite{1997CQGra..14.2585B}.

Most cosmological applications deal with FLRW models. One would
therefore like to know which quantities satisfy this criterion on
Friedmannian backgrounds. Since the only invariantly defined
constant is the cosmological constant and because constant products
of the Kronecker deltas do not occur naturally, the only remaining
option is to look for quantities that vanish in FLRW environments.
Given the symmetries of the Friedmann models, any variable that
describes spatial inhomogeneity or anisotropy must vanish there and
therefore its linear perturbation should remain invariant under
gauge transformations.

\subsubsection{Gauge-invariant inhomogeneities}\label{sssGIIs}
Covariantly, spatial inhomogeneities in the distribution of any
physical quantity are described by the orthogonally projected
gradient of the quantity in question. For the purposes of structure
formation the key variable is the comoving fractional gradient in
the energy density of the matter given by~\cite{1989PhRvD..40.1804E}
\begin{equation}
\Delta_a= {a\over\rho}\,{\rm D}_a\rho\,.  \label{Deltaa}
\end{equation}
The above, which monitors density variations as measured by a pair
of neighbouring fundamental observers (see~\cite{1989PhRvD..40.1804E} and also \S~\ref{sssIIVs} here), is
identically zero in spacetimes with homogeneous spatial sections.
Indeed, by definition we have
\begin{equation}
{\rm D}_a\rho= h_a{}^b\nabla_b\rho= h_a{}^0\nabla_0\rho+
h_a{}^{\alpha}\nabla_{\alpha}\rho=0\,,  \label{gi}
\end{equation}
since $h_a{}^0=0$ in a comoving frame and $\nabla_{\alpha}\rho=0$
because $\rho=\rho(t)$. The vanishing of ${\rm D}_a\rho$ in
spatially-homogeneous models, like the Friedmann universes,
guarantees that $\Delta_a$ satisfies the Stewart and Walker
lemma~\cite{1974RSPSA.341...49S}. Consequently, $\Delta_a$ describes density inhomogeneities within perturbed almost-FLRW spacetimes in a gauge independent way.

The density gradient can be supplemented by a number of auxiliary
variables that describe spatial inhomogeneities in other physical
quantities. Here, for the sake of economy, we will only introduce a
variable for the volume-expansion gradients and for those in the
magneic energy density. Following~\cite{1989PhRvD..40.1804E,%
1997CQGra..14.2539T}, these are defined as
\begin{equation}
{\clz}_a=a{\rm D}_a\Theta \hspace{15mm} {\rm and} \hspace{15mm}
\clb_a={a\over B^2}\,{\rm D}_aB^2\,,  \label{cZa-cBa}
\end{equation}
respectively. Both vanish in spatially homogeneous spacetimes and
therefore they also comply with the Stewart and Walker criterion for gauge invariance. For the rest of this section, we will analyse the
behaviour of these quantities in different cosmological environments and then, in \S~\ref{sLCPs}, we will use them to study the evolution of perturbed almost-FLRW universes.

\subsection{Inhomogeneous single-fluid cosmologies}\label{ssISfCs}
It is broadly accepted that the present large-scale structure of the universe is the result of Jeans-type instabilities, where small
inhomogeneities in the initial density distribution of the cosmic
medium grow gravitationally to form the galaxies and the voids seen
in the universe today. Here, we will present the main equations
governing the nonlinear evolution of density inhomogeneities, within single-fluid cosmologies, in a covariant and gauge-invariant manner.

\subsubsection{Imperfect fluids}\label{sssIFs}
Consider a general spacetime filled with a single imperfect fluid.
Spatial inhomogeneities in the matter density, as measured by a pair of neighbouring observers, are monitored by the orthogonally
projected dimensionless comoving gradient $\Delta_a=(a/\rho){\rm
D}_a\rho$. Taking the covariant derivative of the above and using the energy and momentum conservation laws, respectively given by
(\ref{edc1}) and (\ref{mdc1}), we obtain
\begin{eqnarray}
\dot\Delta_{\langle a\rangle}&=&{p\over\rho}\,\Theta\Delta_a-
\left(1+{p\over\rho}\right){\clz}_a +
{a\Theta\over\rho}\left(\dot{q}_{\langle
a\rangle}+{4\over3}\,\Theta q_a\right)- {a\over\rho}\,{\rm
D}_a{\rm D}^bq_b+
{a\Theta\over\rho}\,{\rm D}^b\pi_{ab} \nonumber\\
&&-\left(\sigma^b{}_a+\omega^b{}_a\right)\Delta_b-
{a\over\rho}\,{\rm D}_a\left(2A^bq_b+\sigma^{bc}\pi_{bc}\right)+
{a\Theta\over\rho}\left(\sigma_{ab}+\omega_{ab}\right)q^b+
{a\Theta\over\rho}\,\pi_{ab}A^b \nonumber\\
&&+{1\over\rho}\left({\rm D}^bq_b
+2A^bq_b+\sigma^{bc}\pi_{bc}\right)\left(\Delta_a -aA_a\right)\,.
\label{indotDeltaa}
\end{eqnarray}
In the right-hand side of the above we notice a number of agents
which act as sources of density perturbations. Hence, even if
$\Delta_a$ is initially zero, it will not generally remain so. One
of the key sources of density inhomogeneities is $\clz_a$, the
volume expansion gradient, the nonlinear evolution of which is
obtained by taking its time derivative and then using the associated expression of the Raychaudhuri equation. The result is
\begin{eqnarray}
\dot{\clz}_{\langle a\rangle}&=& -{2\over3}\,\Theta{\clz}_a-
{1\over2}\,\rho\Delta_a- {3\over2}\,a{\rm D}_ap-
a\left[{1\over3}\,\Theta^2+{1\over2}\,(\rho+3p)
-\Lambda\right]A_a+ a{\rm D}_a{\rm D}^bA_b \nonumber\\&&
-\left(\sigma^b{}_a+\omega^b{}_a\right){\clz}_b- 2a{\rm
D}_a\left(\sigma^2-\omega^2\right)+ 2aA^b{\rm D}_aA_b
\nonumber\\&& -a\left[2\left(\sigma^2-\omega^2\right)-{\rm
D}^bA_b-A^bA_b\right]A_a\,. \label{indotcZa}
\end{eqnarray}

\subsubsection{Perfect fluids}\label{sssPFs}
When dealing with a perfect fluid, a choice of frame can be
made in which the fluid appears isotropic, i.e.\
there is no energy-flux or anisotropic pressure and the associated variables vanish identically (i.e.~$q_a=0=\pi_{ab}$). This choice considerably simplifies Eq.~(\ref{indotDeltaa}) meaning that perfect-fluid density inhomogeneities evolve as
\begin{equation}
\dot\Delta_{\langle a\rangle}= {p\over\rho}\,\Theta\Delta_a-
\left(1+\frac{p}{\rho}\right){\clz}_a-
\left(\sigma^b{}_a+\omega^b{}_a\right)\Delta_b\,.
\label{pndotDeltaa}
\end{equation}
There is no change in the propagation equation of the expansion
gradients, which maintains the algebraic form of (\ref{indotcZa}).
The only difference the perfect fluid makes is that $\Delta_a$ is
now monitored by (\ref{pndotDeltaa}) and the 4-acceleration is given by (\ref{pfcls}b) instead of (\ref{mdc1}). If the medium is also
barotropic (i.e.~for $p=p(\rho)$), the pressure gradients are
directly related to those in the density by $\text{D}_ap=c_{\rm
s}^2\text{D}_a\rho$, where $c_{\rm s}^2=\dot{p}/\dot{\rho}$ is the
adiabatic sound speed (see \S~\ref{sssETs}).

\subsubsection{Covariant conserved quantities}\label{sssCCQs}
In the metric-based perturbative formalism, the curvature
perturbation on uniform density hypersurfaces is conserved for the
adiabatic growing mode on super-Hubble scales, and the geometric
interpretation of this can be understood via the perturbation of the expansion e-folds, $N=\ln a$, using the so-called ``separate
universe" picture~\cite{1985JETPL..42..152S}. A covariant version of this result is based on defining an appropriate spatial-gradient quantity, and leads to a simple geometric nonlinear conserved
quantity for a perfect fluid~\cite{2007JCAP...02...17L}.

Along each worldline of the perfect fluid, we define the generalised, covariant local e-fold function
\begin{equation}
\alpha= {1\over3}\int\Theta\,{\rm d}\tau\,,  \label{alpha}
\end{equation}
where $\tau$ is proper time. Applying the commutation law (\ref{A4}) to the above scalar, as well as the density of the fluid,
gives\footnote{Using the Lie derivative along $u_a$, simplifies the
identity (\ref{A4}) to $\D_a\dot{f}={\cll}_u(\D_af)-\dot{f}A_a$.}
\begin{equation}
{1\over3}\,\D_a\Theta= {\cll}_u\D_a\alpha -\dot\alpha A_a\,,
\label{DaTheta}
\end{equation}
where ${\cll}_u$ is the Lie derivative along $u_a$. Then, the
projected gradient of the energy-density conservation law (see
Eq.~(\ref{pfcls}a)) leads to
\begin{equation}
{\cll}_u(\D_a\rho)+ 3(\rho+p){\cll}_u(\D_a \alpha)+
\Theta\D_a(\rho+p)= 0\,.  \label{ludr}
\end{equation}

Defining the auxiliary projected vector $\zeta_a=\D_a\alpha-(\dot\alpha/\dot\rho)\,\D_a\rho$, we can
simplify expression (\ref{ludr}) into
\begin{equation}
{\cll}_u\zeta_a= -{\Theta\over3(\rho+p)}\,\Gamma_a\,,  \label{Luz}
\end{equation}
with $\Gamma_a=\D_ap-(\dot{p}/\dot{\rho})\D_a\rho.$ For isentropic perturbations, $\Gamma_a$ vanishes identically (see expression (\ref{Dapi}) in \S~\ref{sssETs}) and the above guarantees that
\begin{equation}
{\cll}_u\zeta_a =0\,.  \label{cz}
\end{equation}
In other words, $\zeta_a$ is a conserved quantity in the isentropic/barotropic case on all scales and at all perturbative orders.

\subsection{Inhomogeneous multi-fluid cosmologies}\label{ssIMfCs}
During its evolution the universe goes through epochs where the
matter is better described by a mixture of several fluids, rather
than a single component. This brings about the need for studies of
inhomogeneous multi-component systems
(see~\cite{1980PhRvD..22.807L,1984PThPS..78....1K} for non-covariant treatments). When studying the effects of inhomogeneities on the
CMB, for example, one usually considers a mixture of radiation,
baryonic matter and neutrinos (e.g.~see~\cite{1999ApJ...513....1C}
and \S~\ref{sKINETIC} here). Studies of nonlinear gravitational
collapse also require the use of a multi-fluid description, in order to incorporate the effects of peculiar velocities
(see \S~\ref{ssNPKs} for more details).

\subsubsection{4-velocity fields}\label{sss4vFs}
Consider spacetime filled with a mixture of different fluids.
Suppose that $u_a$ is the 4-velocity of the fundamental observers
and $u_a^{(i)}$ that of the $i$-th fluid component
(i.e.~$u_a^{(i)}u_{(i)}^a=-1$). The tensors projecting orthogonal to $u_a$ and $u_a^{(i)}$ are
\begin{equation}
h_{ab}=g_{ab}+ u_au_b \hspace{15mm} \mathrm{and} \hspace{15mm}
h_{ab}^{(i)}=g_{ab}+ u_a^{(i)}u_b^{(i)}\,,  \label{h-hi}
\end{equation}
respectively. The relation between $u_a$ and $u_a^{(i)}$ is determined by the Lorentz boost
\begin{equation}
u_a^{(i)}=\gamma^{(i)}\left(u_a+v_a^{(i)}\right)\,, \label{Lboost}
\end{equation}
where $v_a^{(i)}u^a=0$. Here, $\gamma_{(i)}=(1-v_{(i)}^2)^{-1/2}$ is the Lorentz-boost factor and $v_a^{(i)}$ is the peculiar velocity of the $i$-th component relative to $u_a$. For non-relativistic
peculiar motions $\gamma_{(i)}\simeq1$.

\begin{figure*}
\begin{center}
\includegraphics[height=3in,width=5in,angle=0]{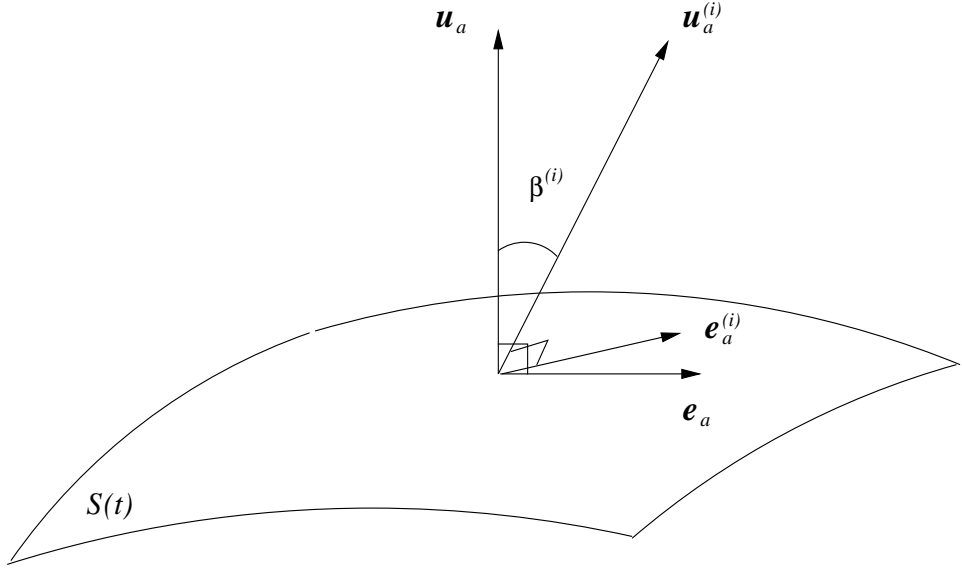}\quad
\end{center}
\caption{In a multi-component system, the 4-velocity $u_a^{(i)}$ of the $i$-th fluid makes a hyperbolic angle $\beta^{(i)}$ with the fundamental 4-velocity field $u_a$, normal to the hypersurfaces of homogeneity $S(t)$. The unit vectors $e_a$ and $e_a^{(i)}$ are orthogonal to $u_a$ and $u_a^{(i)}$ respectively. Following definition (\ref{Lboost}), the peculiar velocity of the $i$-th species is $v_a^{(i)}=v^{(i)}e_a$, with $v_{(i)}^2= v_a^{(i)}v_{(i)}^a$.}  \label{mfluid}
\end{figure*}

The boost relation can also be recast in terms of the hyperbolic,
tilt, angle $\beta_{(i)}$ between the two 4-velocity vectors (see Fig.~\ref{mfluid}). Noting that $\cosh\beta^{(i)}=-u_a^{(i)}u^a=\gamma^{(i)}$ and
$\sinh\beta^{(i)}e_a=\gamma^{(i)}v_a^{(i)}=h_a{}^bu_b^{(i)}$, where
$v_a^{(i)}=v^{(i)}e_a$, expression (\ref{Lboost})
reads~\cite{1973CMaPh..31..209K}
\begin{equation}
u_a^{(i)}= \cosh\beta^{(i)}u_a+ \sinh\beta^{(i)}e_a\,.  \label{ui}
\end{equation}
In addition, $v_{(i)}=\tanh\beta_{(i)}$, which means that when the tilt angle is small (i.e.~for $\beta_{(i)}\ll1$) we have
$v_{(i)}\simeq\beta_{(i)}$ and non-relativistic peculiar velocities.

\subsubsection{Multi-component perfect fluids}\label{ssMcPFs}
The non-equilibrium state of a material medium is described by its
energy momentum tensor $T_{ab}$, its particle flux vector $N_a$ and
by the entropy flux vector $S_a$. The former two are conserved
($\nabla_bT^{ab}=0=\nabla_aN^a$) and the last obeys the second
law of thermodynamics ($\nabla_aS^a\geq0$ -- see~\cite{1976AnPhy.100..310I} and also \S~\ref{sssETs} here). When the strong energy condition holds, the energy-momentum tensor of a fluid has a unique timelike eigenvector $u_a^E$, normalised so that
$u_a^Eu_E^a=-1$. One may also define a unitary timelike vector parallel to $N_a$ by $u_a^N=N_a/\sqrt{-N_aN^a}$. Provided that the fluid is perfect (or
in equilibrium) all three vectors $u_a^E$, $u_a^N$ and $S_a$ are
parallel and define a unique hydrodynamic 4-velocity vector (the
rest-frame of the fluid flow). This is the only frame the energy-momentum tensor of the matter assumes the perfect-fluid form (see expression (\ref{pfTab})).

When dealing with an imperfect fluid, however, there is no a
uniquely defined hydrodynamic 4-velocity. Then, the energy momentum
tensor and the particle flux vector decompose as
\begin{equation}
T_{ab}=\rho u_au_b+ ph_{ab}+ 2q_{(a}u_{b)}+ \pi_{ab} \hspace{15mm}
{\rm and} \hspace{15mm} N_a= nu_a+ \cln_a\,,  \label{imTabNa}
\end{equation}
respectively. Here $\cln_a=h_a{}^bN_b$ is the particle drift and recall that $n=-N_au^a$ is the particle number density. Also, the total energy-flux vector in Eq.~(\ref{imTabNa}a) is given by
\begin{equation}
q_a= Q_a+ {1\over n}\,(\rho+p)\cln_a\,,  \label{qa}
\end{equation}
with $Q_a$ representing the associated heat flux. In the case of an imperfect medium two frames are of special status. The first, which has $u_a\equiv u_a^E$ and zero total energy flux, is known as the energy -- or Landau -- frame~\cite{1975ctf..book.....L}. There, following (\ref{qa}), the comoving observers see a nonzero particle drift equal to $\cln_a=-[n/(\rho+p)]Q_a$. The alternative option is the Eckart -- or particle -- frame with $u_a\equiv u_a^N$~\cite{1940PhRv...58..919E}. Here, the observers see no particle drift and therefore $q_a=Q_a$.

Consider a mixture of perfect fluids, where the $i$-th component has energy density $\rho_{(i)}$, isotropic pressure $p_{(i)}$ and moves
along the timelike 4-velocity field $u^{(i)}_a$. Relative to this
frame, the energy-momentum tensor and the particle flux of the
individual species respectively read
\begin{equation}
T_{ab}^{(i)}= \rho^{(i)}u_a^{(i)}u_b^{(i)}+ p^{(i)}h_{ab}^{(i)}
\hspace{15mm} {\rm and} \hspace{15mm} N_a^{(i)}= n^{(i)}u_a^{(i)}\,,
\label{TabiNai1}
\end{equation}
with $h_{ab}^{(i)}$ given by (\ref{h-hi}b) and $n_{(i)}$
representing the number density of each component in their own rest
frame. With respect to the $u_a$-frame, however, the above become
\begin{eqnarray}
T_{ab}^{(i)}= \hat{\rho}^{(i)}u_au_b+ \hat{p}^{(i)}h_{ab}+
2u_{(a}\hat{q}_{b)}^{(i)}+ \hat{\pi}_{ab}^{(i)} \hspace{10mm} {\rm
and} \hspace{10mm}N_a^{(i)}= \hat{n}^{(i)}u_a+
\hat{\cln}_a^{(i)}\,, \label{TabiNai2}
\end{eqnarray}
respectively. The former corresponds to the energy-momentum tensor
of an imperfect fluid with
\begin{equation}
\hat{\rho}^{(i)}= \gamma_{(i)}^2\left(\rho^{(i)}+p^{(i)}\right)-
p^{(i)}\,, \hspace{15mm} \hat{p}^{(i)}= p^{(i)}+
{1\over3}\left(\gamma_{(i)}^2-1\right)
\left(\rho^{(i)}+p^{(i)}\right)\,, \label{hat1}
\end{equation}
\begin{equation}
\hat{q}_a^{(i)}=
\gamma_{(i)}^2\left(\rho^{(i)}+p^{(i)}\right)v_a^{(i)}\,,
\hspace{10mm} {\rm and} \hspace{10mm} \hat{\pi}_{ab}^{(i)}=
\gamma_{(i)}^2\left(\rho^{(i)}+p^{(i)}\right)\left(v_a^{(i)}v_b^{(i)}-
{1\over3}\,v_{(i)}^2h_{ab}\right)\,.  \label{hat2}
\end{equation}
Similarly, expression (\ref{TabiNai2}b) is that of an imperfect
fluid with particle number density and particle drift given by
\begin{equation}
\hat n^{(i)}=\gamma^{(i)}n^{(i)} \hspace{10mm} {\rm and}
\hspace{10mm} \hat{\cln}_a{}^{(i)}=\hat{n}^{(i)}v_a^{(i)}\,,
\label{hat3}
\end{equation}
respectively. Clearly, when dealing with non-relativistic peculiar
velocities, we may ignore terms quadratic in $v_{(i)}$ and therefore set $\gamma_{(i)}$ to unity. Then, expressions (\ref{hat1}) and
(\ref{hat2}) reduce to $\hat{\rho}_{(i)}=\rho_{(i)}$,
$\hat{p}_{(i)}=p_{(i)}$, $\hat{q}_a^{(i)}=
(\rho_{(i)}+p_{(i)})v_a^{(i)}$ and $\hat{\pi}_{ab}^{(i)}=
0$~\cite{1992ApJ...395...54D}. Also, (\ref{hat3}) simplify to $\hat{n}^{(i)}=n^{(i)}$ and $\hat{\cln}^{(i)}_a=n^{(i)}v_a^{(i)}$. From now on, unless otherwise stated, we will always assume that $v_{(i)}\ll1$.

\subsubsection{Conservation laws of the total fluid}\label{sssCLTF}
Measured relative to the $u_a$-frame the total fluid has an
effective energy density $\rho=\Sigma_i\rho^{(i)}$, pressure
$p=\Sigma_ip^{(i)}$, energy flux $q_a=\Sigma_i\hat{q}_a^{(i)}$ and
\begin{equation}
T_{ab}= \rho u_au_b+ ph_{ab}+ 2u_{(a}q_{b)}\,.  \label{tTab}
\end{equation}
The latter is conserved which means that $\nabla^bT_{ab}=0$.
Assuming an effective equation of state of the form $p=p(\rho,s)$,
where $s$ is the associated specific entropy, we have
\begin{equation}
{\rm D}_ap= {c_s^2\rho\over a}\,\Delta_a+ {p\over a}\,\cle_a\,,
\label{tDap}
\end{equation}
where $c_s^2=(\partial p/\partial\rho)_s$ and $\cle_a=(a/p)(\partial p/\partial s)_{\rho}{\rm D}_as$ are the square of the effective sound speed and the effective entropy perturbation respectively. Therefore, the momentum-density conservation law of the total fluid
reads
\begin{equation}
(\rho+p)A_a= -{c_s^2\rho\over a}\,\Delta_a- {p\over
a}\,\cle_a- \dot{q}_{\langle a\rangle}- {4\over3}\,\Theta q_a-
\left(\sigma_a{}^b+\omega_a{}^b\right)q_b\,,  \label{tmdc}
\end{equation}
while the effective total energy satisfies (\ref{edc1}). We note
that the expansion dynamics is determined by the total fluid and
spatial inhomogeneities in the volume expansion are monitored via
Eq.~(\ref{indotcZa}) with ${\rm D}_ap$ and $A_a$ given by
(\ref{tDap}) and (\ref{tmdc}) respectively.

\subsubsection{Conservation laws of the i-th fluid}\label{sssCLiF}
Assuming a mixture of interacting and non-comoving perfect fluids,
the energy-momentum of the $i$-th species reads (see
\S~\ref{ssMcPFs})
\begin{equation}
T_{ab}^{(i)}= \rho^{(i)}u_au_b+ p^{(i)}h_{ab}+
2u_{(a}q_{b)}^{(i)}\,,  \label{mpfTab}
\end{equation}
with $q_a^{(i)}=(\rho^{(i)}+p^{(i)})v_a^{(i)}$ and $v_a^{(i)}$
representing the peculiar velocity of the component. The above
satisfies the conservation law
\begin{equation}
\nabla^bT_{ab}^{(i)}= I_a^{(i)}\,,  \label{mpfcTab}
\end{equation}
where the interaction term has $\sum_iI_a^{(i)}=0$ due to the
conservation of $T_{ab}=\sum_iT_{ab}^{(i)}$. Setting
$\cli^{(i)}=-I_a^{(i)}u^a$ and $\cli_a^{(i)}=h_a{}^bI_a^{(i)}$, the
timelike and spacelike parts of (\ref{mpfcTab}) give
\begin{equation}
\dot{\rho}^{(i)}=-\Theta\left(\rho^{(i)}+p^{(i)}\right)- {\rm
D}^aq_a^{(i)}- 2A^aq_a^{(i)}+ \cli^{(i)}  \label{mpfedc}
\end{equation}
and
\begin{equation}
\left(\rho^{(i)}+p^{(i)}\right)A_a= -{c_s^{2(i)}\rho^{(i)}\over
a}\,\Delta_a^{(i)}- {p^{(i)}\over a}\,\cle_a^{(i)}- \dot{q}_{\langle
a\rangle}^{(i)}- {4\over3}\,\Theta q_a^{(i)}-
\left(\sigma_a{}^b+\omega_a{}^b\right)q_b^{(i)}+ \cli_a^{(i)}\,,
\label{mpfmdc}
\end{equation}
where $c_s^{2(i)}=(\partial p^{(i)}/\partial\rho^{(i)})_{s^{(i)}}$
and $\cle_a^{(i)}=(a/p^{(i)})(\partial p^{(i)}/\partial
s^{(i)})_{\rho^{(i)}}{\rm D}_as^{(i)}$ are respectively the square
of the sound speed and the entropy perturbation of the species in question. Also, following definition (\ref{Deltaa}),
$\Delta_a^{(i)}=(a/\rho^{(i)}){\rm D}_a\rho^{(i)}$ describes
inhomogeneities in the density distribution of the $i$-th fluid
component, relative to the $u_a$-frame. Finally we note that
$\sum_i\cli^{(i)}=0=\sum_i\cli_a^{(i)}$.

\subsubsection{Nonlinear density perturbations in the i-th
species}\label{sssNDPiSs}
Taking the time derivative of $\Delta_a^{(i)}$ and using the
conservation laws (\ref{mpfedc}), (\ref{mpfmdc}) we arrive at the
following nonlinear expression
\begin{eqnarray}
\dot\Delta_{\langle a\rangle}^{(i)}&=&
{p^{(i)}\over\rho^{(i)}}\,\Theta\Delta_a^{(i)}-
\left(1+\frac{p^{(i)}}{\rho^{(i)}}\right){\clz}_a +
\frac{a\Theta}{\rho^{(i)}}\left(\dot{q}_{\langle
a\rangle}^{(i)}+{4\over3}\,\Theta q_a^{(i)}\right)-
{a\over\rho^{(i)}}\,{\rm D}_a\left({\rm
D}^bq_b^{(i)}-\cli^{(i)}\right)\nonumber\\
&&-\left(\sigma^b{}_a+\omega^b{}_a\right)\Delta_b^{(i)}-
{2a\over\rho^{(i)}}\,{\rm D}_a\left(A^bq_b^{(i)}\right)+
\frac{a\Theta}{\rho^{(i)}}\left(\sigma_a{}^b
+\omega_a{}^b\right)q^{(i)}_b\nonumber\\
&&+\frac{1}{\rho^{(i)}}\left({\rm
D}^bq_b^{(i)}+2A^bq_b^{(i)}-\cli^{(i)}\right)\left(\Delta_a^{(i)}
-aA_a\right)- {a\Theta\over\rho^{(i)}}\,\cli_a^{(i)}\,.
\label{mpfdotDeltaa}
\end{eqnarray}
When the interaction term is specified, the above describes the
propagation of spatial inhomogeneities in the density distribution
of the $i$-th species. Recall that
$q_a^{(i)}=(\rho^{(i)}+p^{(i)})v_a^{(i)}$ and that the nonlinear
evolution of $\clz_a$ is governed by Eq.~(\ref{indotcZa}).

\subsection{Inhomogeneous magnetised cosmologies}\label{ssIMCs}
From the Earth and the nearby stars, to galaxies, galaxy clusters
and remote high-redshift protogalaxies, magnetic fields have been
repeatedly observed~\cite{1994RPPh...57..325K,%
1997ApJ...480..481K,2002ChJAA...2..293H}. Although the origin of
cosmic magnetism is still a mystery, it appears that we live in a
magnetised universe~\cite{2004IJMPD..13..391G}.

\subsubsection{The ideal MHD approximation}\label{sssIMHDA}
With the exception of any period of inflation and early reheating,
the universe has been a good conductor throughout its lifetime. As a result, $B$-fields of cosmological origin have remained frozen into
the expanding cosmic fluid during most of their evolution. This
allows us to study the magnetic effects on structure formation
within the limits of the ideal magnetohydrodynamics (MHD)
approximation. The latter is described by means of Ohm's law, which
has the covariant form~\cite{1971ApJ...164..589G,1975clel.book.....J}
\begin{equation}
J_a= \mu u_{a}+ \varsigma E_a\,,  \label{Ohm}
\end{equation}
with $\varsigma$ representing the scalar conductivity of the medium. Here, quantities on the right are in the rest-frame of the plasma.
Equation~(\ref{Ohm}) splits the 4-current into a timelike part due to the charge density and a spacelike part from conduction,
\begin{equation}
\clj_a= \varsigma E_a\,.  \label{Ohm1}
\end{equation}
This form of Ohm's law covariantly describes the resistive
magnetohydrodynamic (MHD) approximation in the single-fluid
approach.\footnote{Ohm's law relates the induced 3-current with the electric field and it is generally given in the form of a propagation equation of the electric current. For the fully nonlinear 1+3 covariant version of the generalised Ohm's law in the case of a hot multi-component plasma the reader is referred to~\cite{2008MNRAS.tmp..150K}.} Note the absence of the induced electric field from the above, reflecting the fact that the covariant form of Maxwell's formulae (see expressions (\ref{M1})-(\ref{M4})) already incorporates the effects of relative motion. According to (\ref{Ohm1}), zero electrical conductivity implies that the spatial currents vanish, even when the electric field is non-zero. On the other hand, non-zero spatial currents are compatible with a vanishing electric field as long as the conductivity of the medium is infinite (i.e.~for $\varsigma\rightarrow\infty$). Thus, at the ideal MHD limit, the electric field vanishes in the frame of the fluid and the energy-momentum tensor of the residual magnetic field simplifies to~\cite{1997CQGra..14.2539T}
\begin{equation}
T_{ab}^{(B)}= {1\over2}\,B^2u_au_b+ {1\over6}\,B^2h_{ab}+
\Pi_{ab}\,, \label{TB}
\end{equation}
with $\Pi_{ab}=-B_{\langle a}B_{b\rangle}$. Accordingly, the
$B$-field corresponds to an imperfect fluid with energy density
$\rho_{B}=B^2/2$, isotropic pressure $p_{B}=B^2/6$ and anisotropic
stresses represented by $\Pi_{ab}$. Similarly, Maxwell's equations
(see \S~\ref{sssEMFs}) reduce to a single propagation formula,
\begin{equation}
\dot{B}_{\langle a\rangle}=
\left(\sigma_{ab}+\varepsilon_{abc}\omega^c -{2\over3}\,\Theta
h_{ab}\right)B^b\,,  \label{Bdot}
\end{equation}
and to the following three constraints
\begin{equation}
\curl B_{a}+ \varepsilon_{abc}A^bB^c=\clj_a\,,  \label{C1}
\end{equation}
\begin{equation}
2\omega^aB_a= \mu\,, \hspace{25mm} {\rm D}^aB_a=0\,. \label{C2}
\end{equation}
The right-hand side of (\ref{Bdot}) is due to the relative motion of the neighbouring observers and guarantees that the magnetic
forcelines always connect the same matter particles, namely that the field remains frozen-in with the highly conducting fluid. Expression (\ref{C1}) provides a direct relation between the spatial currents,
which are responsible for keeping the field lines frozen-in with the matter, and the magnetic field itself (e.g.~see~\cite{1983mfa..book.....Z}).

Contracting the magnetic induction equation (\ref{Bdot}) along $B_a$ leads to the evolution law for the energy density of the field,
which takes the nonlinear form
\begin{equation}
\left(B^2\right)^{\cdot}= -{4\over3}\,\Theta B^2-
2\sigma_{ab}\Pi^{ab}\,.  \label{B2dot}
\end{equation}
This shows that in a highly conducting cosmic medium we have
$B^2\propto a^{-4}$ always unless there is substantial anisotropy,
in which case the $B$-field behaves as a dissipative radiative
fluid. Note that in a spatially homogeneous, radiation-dominated
universe with weak overall anisotropy, the shear term in the
right-hand side of (\ref{B2dot}) means that the ratio
$B^2/\rho^{(\gamma)}$ is no longer constant but displays a slow
`quasi static' logarithmic decay (see~\cite{1970SvA....13..608Z} and also~\cite{1997PhRvD..55.7451B,2000CQGra..17.2215T}).

\subsubsection{Conservation laws}\label{sssCLsm}
Covariant studies of cosmic electromagnetic fields date back to the
work of Elhers and Ellis (see also~\cite{1982MNRAS.198..339T} for an analogous approach), while the Newtonian version of the relativistic approach was recently given in~\cite{2008arXiv0804.1702S}. Following~\cite{1973clp..conf....1E}, the energy momentum tensor corresponding to a magnetised single perfect
fluid of infinite conductivity is
\begin{equation}
T_{ab}= \left(\rho+{1\over2}\,B^2\right)u_au_b+
\left(p+{1\over6}\,B^2\right)h_{ab}+ \Pi_{ab}\,, \label{MHDTab}
\end{equation}
and the medium behaves as an imperfect fluid with effective density
$\rho+B^2/2$, isotropic pressure $p+B^2/6$ and solely magnetic
anisotropic stresses represented by $\Pi_{ab}$. Applied to the
above, and using the MHD form of Maxwell's equations, the
conservation law $\nabla^bT_{ab}=0$ splits into the following
expressions that respectively describe energy and momentum density
conservation\footnote{One can also obtain the conservation laws
(\ref{MHDedc}) and (\ref{MHDmdc}) by taking the MHD limit of
expressions (\ref{edc2}), (\ref{mdc2}) of \S~\ref{sssCLs} and then
substituting in the 3-current from constraint (\ref{C1}).}
\begin{equation}
\dot{\rho}= -(\rho+p)\Theta\,,  \label{MHDedc}
\end{equation}
\begin{equation}
\left(\rho+p+{2\over3}\,B^2\right)A_a= -{\rm D}_ap-
\varepsilon_{abc}B^b\curl B^c- \Pi_{ab}A^b\,.  \label{MHDmdc}
\end{equation}
Note the absence of magnetic terms in Eq.~(\ref{MHDedc}), since only the electric field contributes to (\ref{edc1}). The magnetic energy is separately conserved, as guaranteed by the magnetic induction equation (\ref{Bdot}) and reflects the fact that the magnetic energy density is separately conserved. The second last term in (\ref{MHDmdc}), which is often
referred to as the magnetic Lorentz force, is always normal to the
$B$-field lines and decomposes as
\begin{equation}
\varepsilon_{abc}B^b\curl B^c= {1\over2}\,{\rm D}_aB^2- B^b{\rm
D}_bB_a\,.  \label{Lorentz}
\end{equation}
The last term in the above is the result of the magnetic tension. In so far as this tension stress is not balanced by the pressure
gradients, the field lines are out of equilibrium and there is a
non-zero Lorentz force acting on the particles of the magnetised
fluid.

\subsubsection{Nonlinear density perturbations of the magnetised
fluid}\label{sssNlDPMF}
In the presence of magnetic fields, the nonlinear evolution of
spatial inhomogeneities in the density distribution of a single,
highly conducting perfect fluid is described by the expression
\begin{eqnarray}
\dot{\Delta}_{\langle a\rangle}&=& \frac{p}{\rho}\,\Theta\Delta_a-
\left(1+\frac{p}{\rho}\right)\clz_a+
\frac{a\Theta}{\rho}\,\varepsilon_{abc}B^b\curl B^c+
{2a\Theta B^2\over3\rho}\,A_a-
\left(\sigma_{ba}+\omega_{ba}\right)\Delta^b \nonumber\\
&{}&+\frac{a\Theta}{\rho}\,\Pi_{ab}A^b\,.  \label{mdotcDa}
\end{eqnarray}
This is obtained after taking the projected time-derivative of
definition (\ref{Deltaa}) and then using relations (\ref{MHDedc}),
(\ref{MHDmdc}). Similarly, starting form (\ref{cZa-cBa}a) and using
(\ref{MHDedc}) we arrive at the nonlinear evolution equation for the expansion gradients
\begin{eqnarray}
\dot{\clz}_{\langle a\rangle}&=& -{{2\over3}}\Theta\clz_a-
{{1\over2}}\rho\Delta_a- {{1\over2}}B^2\clb_a+
{{3\over2}}a\varepsilon_{abc}B^b\curl B^c+ a{\rm D}_a{\rm D}^bA_b+ 2aA^b{\rm D}_aA_b \nonumber\\ &{}&+\left[{{1\over2}}\clr-
3\left(\sigma^2-\omega^2\right)+{\rm D}^bA_b +A_bA^b\right]aA_a-
\left(\sigma_{ba}+\omega_{ba}\right)\clz^b+ {3\over2}\,a\Pi_{ab}A^b
\nonumber\\ &{}&-2a{\rm D}_a\left(\sigma^2-\omega^2\right)\,,
\label{mdotcZa}
\end{eqnarray}
where $\clr$ is the Ricci scalar of the magnetised 3-D space
orthogonal to the fluid motion. Finally, the non-linear propagation
formula monitoring spatial inhomogeneities in the magnetic energy
density (see definition (\ref{cZa-cBa}b)) is
\begin{eqnarray}
\dot{\clb}_{\langle a\rangle}&=&
\frac{4\rho}{3(\rho+p)}\,\dot{\Delta}_{\langle a\rangle}-
\frac{4p\Theta}{3(\rho+p)}\,\Delta_a-
\frac{4a\Theta}{3(\rho+p)}\,\varepsilon_{abc}B^b\curl B^c-
{{4\over3}}a\Theta\left[1+\frac{2B^2}{3(\rho+p)}\right]A_a \nonumber\\
&{}&-\left(\sigma_{ba}+\omega_{ba}\right)\clb^b+
\frac{4\rho}{3(\rho+p)}\left(\sigma_{ba}+\omega_{ba}\right)\Delta^b-
\frac{4a\Theta}{3(\rho+p)}\,\Pi_{ab}A^b-
\frac{2a}{B^2}\,\Pi^{bc}{\rm D}_a\sigma_{bc} \nonumber\\
&{}&-\frac{2a}{B^2}\,\sigma^{bc}{\rm D}_a\Pi_{bc}+
\frac{2}{B^2}\,\sigma_{bc}\Pi^{bc}\clb_a-
\frac{2a}{B^2}\,\sigma_{bc}\Pi^{bc}A_a\,.  \label{dotcBa}
\end{eqnarray}
The above results from the time-derivative of (\ref{cZa-cBa}b) by
means of (\ref{B2dot}), (\ref{mdotcZa}) and cannot be used when the
magnetised medium has $p=-\rho$.

\subsection{Inhomogeneous scalar-field cosmologies}\label{ssISCs}
Scalar-field dominated universes have come into prominence primarily through the inflationary scenarios. Covariantly, scalar-field
cosmologies have been discussed in a series of papers by Madsen and
Ellis~\cite{1985Ap&SS.113..205M,1988CQGra..5.627M,1991CQGra...8..667E}
and more recently by Vernizzi and
Langlois~\cite{2007JCAP...02...17L,2005PhRvD..71f1301V}

\subsubsection{Minimally coupled scalar fields}\label{sssMCSFs}
Consider a general, pseudo-Riemannian spacetime filled with a single scalar field ($\varphi$), which is minimally coupled to gravity. The associated Lagrangian density is
\begin{equation}
\cll_{\varphi}=
-\sqrt{-g}\left[{1\over2}\;\nabla_a\varphi\nabla^a\varphi
+V(\varphi)\right]\,, \label{cL}
\end{equation}
where $g$ is the determinant of the spacetime metric and
$V(\varphi)$ is the effective potential that describes the
self-interaction of the scalar field. The stress-energy tensor
associated with (\ref{cL}) has the form
\begin{equation}
T_{ab}^{(\varphi)}= \nabla_a\varphi\nabla_b\varphi-
\left[{1\over2}\;\nabla_c\varphi\nabla^c\varphi
+V(\varphi)\right]g_{ab}\,.  \label{sfTab1}
\end{equation}
Applying the twice contracted Bianchi identities, namely the
conservation law $\nabla^bT_{ab}=0$, to the above and assuming that
$\nabla_a\varphi\neq0$ we arrive at the familiar Klein-Gordon
equation
\begin{equation}
\nabla^a\nabla_a\varphi- V^{\prime}(\varphi)= 0\,,  \label{KG1}
\end{equation}
where in this case the prime indicates differentiation with respect
to $\varphi$. We note that when $\nabla_a\varphi=0$ expression
(\ref{sfTab1}) reduces to $T_{ab}^{(\varphi)}=-V(\varphi)g_{ab}$.
The latter ensures that $\nabla_aV(\varphi)=0$, since
$\nabla^bT_{ab}=0$, and $\varphi$ behaves as an effective
cosmological constant rather than a dynamical scalar field.

\subsubsection{Scalar-field kinematics}\label{SfKs}
In order to achieve an 1+3 covariant fluid-description of scalar
fields, one first needs to assign a 4-velocity vector to the
$\varphi$-field itself. Suppose that the 4-vector $\nabla_a\varphi$
is timelike, namely that $\nabla_a\varphi\nabla^a\varphi<0$ over an
open spacetime region. In this case $\nabla_a\varphi$ defines the
normals to the spacelike hypersurfaces $\varphi(x^a)=$~constant and
we define our 4-velocity field
by~\cite{1980PhRvD..22.807L,1988CQGra..5.627M}
\begin{equation}
u_a= -{1\over\dot{\varphi}}\,\nabla_a\varphi\,,  \label{sfua}
\end{equation}
with $\dot{\varphi}=u^a\nabla_a\varphi\neq0$. This means that
$\dot{\varphi}^2=-\nabla_a\varphi\nabla^a\varphi>0$ and $u_au^a=-1$ as required. Also, the flow vector $u_a$ defines our time direction and introduces a unique threading of the spacetime into time and space. The metric of the 3-space orthogonal to $u_a$ is represented by the projection tensor
\begin{equation}
h_{ab}=g_{ab}+
{1\over\dot{\varphi}^2}\,\nabla_a\varphi\nabla_b\varphi\,,
\label{sfhab}
\end{equation}
which satisfies the standard requirements $h_{ab}=h_{(ab)}$,
$h_{ab}u^b=0$, $h_a{}^a=3$ and $h_{ab}h^b{}_c=h_{ac}$. In addition,
$h_{ab}$ defines the covariant derivative operator ${\rm
D}_a=h_a{}^b\nabla_b$ orthogonal to $u_a$ and therefore guarantees
that
\begin{equation}
{\rm D}_a\varphi=0\,,  \label{Davphi}
\end{equation}
always. This result is a key feature of covariant scalar-field
cosmologies and, as we will see in the following sections, it
`dominates' the fluid description of the $\varphi$-field and
essentially dictates all aspects of its evolution.

The irreducible variables describing the kinematics of
$u_a=-\nabla_a\varphi/\dot{\varphi}$ in a covariant manner are
obtained in the usual way by means of the decomposition (\ref{Nbua}). More specifically, starting from definition (\ref{sfua}), using the Klein-Gordon equation and the fact ${\rm D}_a\varphi=0$, we arrive at~\cite{1992CQGra...9..921B}
\begin{equation}
\Theta= -{1\over\dot{\varphi}}\left[\ddot{\varphi}
+V^{\prime}(\varphi)\right]\,, \hspace{5mm} \omega_{ab}=0\,,
\hspace{5mm} \sigma_{ab}=
-{1\over\dot{\varphi}}h_a{}^ch_b{}^d\nabla_c\nabla_d\varphi+
{1\over3\dot{\varphi}}
\left[\ddot{\varphi}+V^{\prime}(\varphi)\right] h_{ab} \label{sfkin1}
\end{equation}
and
\begin{equation}
A_a= -{1\over\dot{\varphi}}\,{\rm D}_a\dot{\varphi}\,.
\label{sfAa1}
\end{equation}
Hence, the 4-velocity field (\ref{sfua}) is irrotational and
$\dot{\varphi}$ acts as an acceleration potential for the fluid
flow. These expressions provide the irreducible kinematical
variables associated with a minimally coupled scalar field. The
reader is referred to~\cite{1988CQGra..5.627M} for a more general
description of scalar-field kinematics, of which relations
(\ref{sfkin1}) and (\ref{sfAa1}) emerge as a special case.

\subsubsection{Scalar fields as perfect fluids}\label{sssSFsPFs}
The introduction of the timelike velocity field (\ref{sfua}) also
allows for a dynamically convenient fluid-description of scalar
fields. In particular, by means of (\ref{sfua}) the energy-momentum
tensor (\ref{sfTab1}) has perfect-fluid form\footnote{A non-minimally coupled scalar field corresponds to an imperfect
medium~\cite{1988CQGra..5.627M}.}
\begin{equation}
T_{ab}^{(\varphi)}= \rho^{(\varphi)}u_au_b+ p^{(\varphi)}h_{ab}\,,
\label{sfTab2}
\end{equation}
with
\begin{equation}
\rho^{(\varphi)}={1\over2}\,\dot{\varphi}^2+ V(\varphi)
\hspace{10mm} {\rm and} \hspace{10mm}
p^{(\varphi)}={1\over2}\,\dot{\varphi}^2- V(\varphi)\,.
\label{sfrho-p}
\end{equation}
Clearly, if we demand a positive definite energy density for the
field, then $\dot{\varphi}^2+2V(\varphi)>0$.

The two simplest cases correspond to a free scalar field with purely kinetic energy (i.e.~$V(\varphi)=0$), which has $p^{(\varphi)}=\rho^{(\varphi)}$ and behaves like a stiff-matter
component. If the field's energy is purely potential, on the other
hand, $p^{(\varphi)}=-\rho^{(\varphi)}$. Such an (effective)
equation of state is approximately achieved during the slow-rolling
regime of the standard inflationary scenarios. Then, the potential
energy dominates over the kinetic, which means that
$\dot{\varphi}^2\ll V(\varphi)$.

In general, either
\begin{equation}
p^{(\varphi)}=\rho^{(\varphi)}-2V(\varphi)\,,  \label{sfeos}
\end{equation}
or $p^{(\varphi)}=\dot{\varphi}^2-\rho^{(\varphi)}$. Therefore,
minimally coupled scalar fields do not generally behave like
barotropic media. Instead, the $\varphi$-field may be regarded as
bulk viscous fluid with the potential playing the role of bulk
viscosity~\cite{1985Ap&SS.113..205M}.

\subsubsection{Conservation laws}\label{SFCLs}
When applied to the energy-momentum tensor of a minimally coupled
scalar field, the twice-contracted Bianchi identities lead to the
familiar Klein-Gordon equation (see expression (\ref{KG1}) in
\S~\ref{sssMCSFs}). Once the timelike 4-velocity vector (\ref{sfua}) is introduced, we may use condition ${\rm D}_a\varphi=0$ to write
$\nabla_a\varphi=-\dot{\varphi}u_a$. Then, the Klein-Gordon equation acquires its 1+3 covariant form
\begin{equation}
\ddot{\varphi}+ \Theta\dot{\varphi}+ V^{\prime}(\varphi)=0\,.
\label{KG2}
\end{equation}
The above can also be seen as the energy-density conservation law
associated with the perfect fluid description of a minimally coupled scalar field. Indeed, after substituting expressions (\ref{sfrho-p}) into the standard energy conservation equation we immediately recover (\ref{KG2}). On the other hand, the conservation of the momentum density is given by expression (\ref{sfAa1}). This can be verified by inserting relations (\ref{sfrho-p}) into the familiar
momentum-density conservation law of a single perfect fluid (see
Eqs.~(\ref{pfcls}b) in \S~\ref{sssCLs}).

\subsubsection{Nonlinear scalar-field perturbations}\label{NlSfPs}
Taking the orthogonally projected gradients of Eq.~(\ref{sfeos}),
which can be seen as the equation of state of a minimally coupled
scalar field, we obtain
\begin{equation}
{\rm D}_a p^{(\varphi)}= {\rm D}_a\rho^{(\varphi)}=
\dot{\varphi}{\rm D}_a\dot{\varphi}\,.  \label{sfDap1}
\end{equation}
This relation is guaranteed by our 4-velocity choice, which ensures
that ${\rm D}_a\varphi=0$ always (see Eq.~(\ref{Davphi})) and
therefore that ${\rm D}_aV(\varphi)=0$ as well. Introducing the variable $\Delta_a^{(\varphi)}= (a/\rho^{(\varphi)}){\rm D}_a\rho^{(\varphi)}$ that describes inhomogeneities in the (effective) energy density of the $\varphi$-field, expression (\ref{sfDap1}) leads to
\begin{equation}
a{\rm D}_a p^{(\varphi)}= \rho^{(\varphi)}\Delta_a^{(\varphi)}\,,  \label{sfDap2}
\end{equation}
or, equivalently, $\Delta_a^{(\varphi)}=
(a\dot{\varphi}/\rho^{(\varphi)}){\rm D}_a\dot{\varphi}$. From (\ref{sfDap1}) follows that, despite its non-barotropic equation of state, in hydrodynamic terms the scalar field behaves as an effective stiff fluid. The non-barotropic nature of the $\varphi$-field emerges in the associated $\Gamma_a$-parameter (see \S~\ref{sssCCQs}), which does not vanish. Indeed, for a minimally coupled scalar field, we find that
\begin{equation}
\Gamma_a= {\rm D}_a p^{(\varphi)}- \left({\dot{p}^{(\varphi)}\over\dot{\rho}^{(\varphi)}}\right) {\rm D}_a\rho^{(\varphi)}= -{2V^{\prime}(\varphi)\over\Theta\dot{\varphi}}\,{\rm D}_a\rho^{(\varphi)}\neq 0\,,  \label{sfGa}
\end{equation}
unless of course $V^{\prime}(\varphi)=0$. The aforementioned `duality' in the nature of scalar fields represents a major departure from conventional perfect-fluid behaviour and is reflected in the statement that scalar-field perturbations are `non-adiabatic'.

Setting $\Delta_a^{(\varphi)}=(a/\rho^{(\varphi)}){\rm
D}_a\rho^{(\varphi)}$, $\Lambda=0$ and using expression (\ref{sfDap2}), we may adapt the system (\ref{indotDeltaa}), (\ref{indotcZa}) to a minimally coupled scalar field. To be precise, we obtain the propagation equations
\begin{equation}
\dot\Delta_{\langle a\rangle}^{(\varphi)}= {p^{(\varphi)}\over\rho^{(\varphi)}}\,\Theta\Delta_a^{(\varphi)}-
\left(1+\frac{p^{(\varphi)}}{\rho^{(\varphi)}}\right){\clz}_a-
\left(\sigma^b{}_a+\omega^b{}_a\right)\Delta_b^{(\varphi)}\,,
\label{sfdotDeltaa}
\end{equation}
and
\begin{eqnarray}
\dot{\clz}_{\langle a\rangle}&=& -{2\over3}\,\Theta{\clz}_a-
2\rho^{(\varphi)}\Delta_a^{(\varphi)}- a\left[{1\over3}\,\Theta^2
+{1\over2}\,(\rho^{(\varphi)}+3p^{(\varphi)})\right]A_a+ a{\rm
D}_a{\rm D}^bA_b- \left(\sigma^b{}_a+\omega^b{}_a\right){\clz}_b
\nonumber\\&& -2a{\rm D}_a\left(\sigma^2-\omega^2\right)+
2aA^b{\rm D}_aA_b- a\left[2\left(\sigma^2-\omega^2\right)-{\rm
D}^bA_b-A^bA_b\right]A_a\,,  \label{sfdotcZa}
\end{eqnarray}
respectively. Finally, combining (\ref{sfAa1}) and (\ref{sfDap2}) we arrive at the following expression for the 4-acceleration of the
$u_a$-congruence defined in \S~\ref{SfKs} (see Eq.~(\ref{sfua})
there)
\begin{equation}
a(\rho^{(\varphi)}+p^{(\varphi)})A_a=
-\rho^{(\varphi)}\Delta_a^{(\varphi)}\,. \label{sfAa2}
\end{equation}
The above, together with Eqs.~(\ref{sfdotDeltaa}) and
(\ref{sfdotcZa}), monitors the evolution of nonlinear density
perturbations in scalar-field dominated cosmologies.

\section{Linear cosmological perturbations}\label{sLCPs}
\subsection{Linearisation}\label{ssL}
Before linearising the full equations of the previous section, one
needs to select the exact solution that will provide the unperturbed background about which the nonlinear formulae will be linearised. Terms with nonzero unperturbed value are assigned zero perturbative order, while those that vanish in the background are treated as first order perturbations~\cite{1989PhRvD..40.1804E,%
1989PhRvD..40.1819E}. This guarantees that linear variables satisfy the criterion for gauge-invariance~\cite{1974RSPSA.341...49S}. By definition, all the first-order quantities are assumed weak relative to the background ones and of perturbative order $\clo(\epsilon)$ in a smallness parameter $\epsilon$ (e.g.~see~\cite{1997dsc..book.....W}). When linearising, products of $\clo(\epsilon)$ variables are neglected.

Cosmological applications use the homogeneous and isotropic FLRW
spacetime as the unperturbed zero-order model. When dealing with a
Friedmann-Lemaitre background, the only zero-order quantities are the matter energy density $\rho$, isotropic pressure $p$ and the volume expansion $\Theta=3H$. If the geometry of spatial sections is not Euclidean, these are supplemented by the 3-Ricci scalar $\clr$.

\subsubsection{Interpreting the inhomogeneity
variables}\label{sssIIVs}
To linear order, the inhomogeneity variables of \S~\ref{sssGIIs}
describe measurable spatial differences in the distribution of the
physical quantities in question. To show this, consider a connecting vector between two worldlines of $u_a$ (so that its Lie derivative
along $u_a$ vanishes). The projection of this vector perpendicular
to $u_a$ defines a spatial vector $\chi_a$, which connects events on neighbouring worldlines that are simultaneous (as seen by $u_a$ -- see~\S~\ref{sssKs}). Then, $\chi_au^a=0$ and
\begin{equation}
\dot{\chi}_{\langle a \rangle}= \frac{1}{3} \Theta \chi_a+
(\sigma_{ab}+\omega_{ab})\chi^b\,,
\label{deltaxdot}
\end{equation}
which means that in a FLRW universe $\chi_a=a\chi_a^0$ (with
$\chi_a^0=$ constant)~\cite{1999toc..conf....1E}. If we Taylor
expand the density about some worldline, then the density in the
local rest space of $u_a$ at displacement $\chi_a$ has
\begin{equation}
\tilde{\rho}-\rho=\chi^a{\rm D}_a\rho\,,  \label{deltarho}
\end{equation}
to leading order. Using the definition (\ref{Deltaa}) and the fact that $\chi_a\propto a$ in the FLRW background, the above translates into
\begin{equation}
\delta\rho\propto\chi_0^a\Delta_a\,,  \label{Deltaa2}
\end{equation}
where $\delta\rho=(\tilde{\rho}-\rho)/\rho$. In other words, the
comoving fractional gradient $\Delta_a$ describes the measurable
local density variation between two neighbouring fundamental
observers. Moreover, $\Delta_a$ closely corresponds to the familiar
energy-density contrast of the non-covariant studies.

\subsection{Single-fluid perturbations}\label{ssSfPs}
\subsubsection{Linear evolution equations}\label{sssLEEs}
Consider an FLRW universe filled with a single barotropic perfect
fluid (i.e.~assume that $p=p(\rho)$). On this background linear
inhomogeneities in the energy density of the medium evolve according to
\begin{equation}
\dot\Delta_a=3wH\Delta_a- (1+w){\clz}_a\,,  \label{lbdotDeltaa}
\end{equation}
The above is obtained from (\ref{pndotDeltaa}) after dropping its
nonlinear terms and assuming a barotropic equation of state for the
cosmic fluid. Also the parameter $w=p/\rho$ is the barotropic index
of the fluid with
\begin{equation}
\dot{w}=-3H(1+w)(c_s^2-w)\,.  \label{dotw}
\end{equation}
This means that $w=$~constant when $c_s^2=w$ provided $H\neq0$ and
$w\neq-1$ (as happens during the radiation and dust eras, for example). Within the same almost-FLRW environment expression (\ref{indotcZa}) linearises to
\begin{equation}
\dot{\clz}_a= -2H{\clz}_a- {1\over2}\,\rho\Delta_a-
{3\over2}\,a{\rm D}_ap- a\left[3H^2
+{1\over2}\,\rho(1+3w)-\Lambda\right]A_a+ a{\rm D}_a{\rm
D}^bA_b\,, \label{lbdotcZa}
\end{equation}
where $\rho(1+w)A_a=-{\rm D}_ap$ to linear order (see (\ref{pfcls}b)).
Given the barotropic nature of the medium, we may define $c_{\rm
s}^2={\rm d}p/{\rm d}\rho$ as the square of the adiabatic sound
speed. Then, the linearised conservation law for the momentum
density reads
\begin{equation}
a(1+w)A_a=-c_{\rm s}^2\Delta_a\,.  \label{lbmdc}
\end{equation}
Solving this for $A_a$, substituting the result into the right-hand
side of (\ref{lbdotcZa}) and keeping terms of up to linear order, we obtain
\begin{equation}
\dot{\clz}_a= -2H{\clz}_a- {1\over2}\,\rho\Delta_a- \frac{c_{\rm
s}^2}{1+w}\left({\rm D}^2\Delta_a+{K\over a^2}\,\Delta_a\right)-
6ac_{\rm s}^2H\curl \omega_a\,, \label{lbdotcZa1}
\end{equation}
where $K=0,\pm1$ is the 3-curvature index of the FLRW background and ${\rm D}^2={\rm D}^a{\rm D}_a$ is the orthogonally projected Laplacian operator. Note that in deriving the above we have also used the linear commutation laws between the orthogonally projected gradients of scalars and spacelike vectors (see Appendix~\ref{AssCCLs}), as well as the zero-order Friedmann equation $\rho-3H^2+\Lambda=3K/a^2$ (see \S~\ref{sssFLRWCs}).

The linear expansion of this model is determined by the associated
expression of Raychaudhuri's formula (\ref{Ray}), written in the
form
\begin{equation}
qH^2= {1\over6}\,\rho(1+3w)+ {c_s^2\over3a^2(1+w)}\,\Delta-
{1\over3}\,\Lambda\,,  \label{lbRay}
\end{equation}
on using (\ref{lbmdc}). Note that $q$ is the deceleration parameter
of the perturbed spacetime (see Eq.~(\ref{FLRWeqs5}) in
\S~\ref{sssFLRWCs}) and $\Delta=a{\rm D}^a\Delta_a$ describes scalar density perturbations (see \S~\ref{sss3TI} below). When positive,
the latter represents overdensities and adds to the gravitational pull of the matter. In the opposite case, $\Delta$ corresponds to an underdense region and tends to accelerate the expansion.

Substituting (\ref{lbmdc}) into the linearised counterpart of
expression (\ref{sigmadot}), we find that, to first order, the shear anisotropy evolves as
\begin{equation}
\dot{\sigma}_{ab}= -2H\sigma_{ab}- E_{ab}- {c_s^2\over
a^2(1+w)}\Delta_{\langle ab\rangle}\,, \label{lbdotsigma}
\end{equation}
where $\Delta_{\langle ab\rangle}$ represents anisotropies in the
distribution of the density gradients (see \S~\ref{sss3TI} below).
Finally, the rotational behaviour of a perturbed FLRW universe,
containing a single barotropic fluid, is monitored by the
propagation equation of the vorticity vector. Starting from
(\ref{omegadot}) and using (\ref{lbmdc}), together with the
linearised commutation law (\ref{A1}), we arrive at
\begin{equation}
\dot{\omega}_a= -2\left(1-{3\over2}\,c_s^2\right)H\omega_a\,.
\label{lbdotomega}
\end{equation}
Thus, kinematic vortices decay with the expansion unless the
barotropic medium has an equation of state `stiffer' than $w=2/3$
(e.g.~see~\cite{1977MNRAS.179P..47B}).

The above propagation equations are supplemented by a set of three
linear constraints (see (\ref{shearcon})-(\ref{Hcon}) for the
nonlinear expressions), namely by
\begin{equation}
{\rm D}^b\sigma_{ab}={2\over3}\,{\rm D}_a\Theta+ \curl \omega_a\,,
\hspace{15mm} {\rm D}^a\omega_a=0  \label{lbkcon1}
\end{equation}
and
\begin{equation}
H_{ab}= \curl \sigma_{ab}+ {\rm D}_{\langle a}\omega_{b\rangle}\,.
\label{lbkcon2}
\end{equation}
When the right-hand side of (\ref{lbkcon1}a) vanishes, the shear is
also transverse and describes pure-tensor perturbations, namely gravitational waves (see \S~\ref{sssIGWs}). Also, following
(\ref{lbkcon1}b) and (\ref{lbkcon2}), $\omega_a$ is a linear
solenoidal vector and in the absence of rotation the magnetic Weyl
component is fully determined by the shear.

Additional constraints between the kinematical and the dynamical
quantities are obtained by the linearised Gauss-Codacci formula (see \S~\ref{sssSC}), which takes the form
\begin{equation}
\clr_{ab}= {1\over3}\,\clr h_{ab}- H(\sigma_{ab}+\omega_{ab})+
E_{ab}\,,  \label{lbGC}
\end{equation}
where
\begin{equation}
\clr= 2\left(\rho-{1\over3}\,\Theta^2+\Lambda\right)\,,  \label{lbFried}
\end{equation}
may be seen as the linear counterpart of Friedmann's equation.

\subsubsection{Three types of inhomogeneity}\label{sss3TI}
The comoving, orthogonally projected gradient $\Delta_a$ contains
collective information about three types of inhomogeneities: density perturbations (a scalar mode); vortices (a vector mode) and shape distortions. Noting that the latter is not a pure tensor mode, but a combination of scalar and vector perturbations, we refer the reader to Appendix~\ref{AssSVTMs} for details on the covariant definition of pure scalar, vector and tensor modes. All the information is encoded in the dimensionless projected gradient $\Delta_{ab}=a{\rm D}_b\Delta_a$ and it is decoded by splitting $\Delta_{ab}$ into its irreducible components as
\begin{equation}
\Delta_{ab}=\Delta_{\langle ab\rangle}+ \Delta_{[ab]}+
{1\over3}\,\Delta h_{ab}\,,  \label{Delta}
\end{equation}
where $\Delta_{\langle ab\rangle}=a{\rm D}_{\langle
b}\Delta_{a\rangle}$, $\Delta_{[ab]}=a{\rm D}_{[b}\Delta_{a]}$ and
$\Delta=a{\rm D}^a\Delta_a$~\cite{1990PhRvD..42.1035E}. The first of these variables describes variations in the anisotropy pattern of
the gradient field (e.g.~pancakes or cigar-like structures). The
second term in the right-hand side of (\ref{Delta}) is related to
magnitude preserving changes of $\Delta_a$ (i.e.~vortex-like
distortions), while $\Delta$ monitors scalar variations in the
spatial distribution of the matter (i.e.~overdensities or
underdensities).

\subsubsection{Density perturbations}\label{sssLPs}
Linearising the orthogonally projected gradients of
(\ref{lbdotDeltaa}), (\ref{lbdotcZa}) and then taking the trace of
the resulting expressions we find that
\begin{equation}
\dot{\Delta}= 3wH\Delta- (1+w){\clz}  \label{lbdotDelta}
\end{equation}
and
\begin{equation}
\dot{\clz}= -2H{\clz}- \left[{1\over2}\,\rho+\frac{3Kc_{\rm
s}^2}{a^2(1+w)}\right]\Delta- \frac{c_{\rm s}^2}{1+w}{\rm
D}^2\Delta\,,  \label{lbdotcZ}
\end{equation}
respectively. This system governs the linear gravitational clumping
of matter in a perturbed almost-FLRW universe filled with a single
perfect fluid. Note that in deriving Eq.~(\ref{lbdotcZ}) we have used the linear result ${\rm D}^a\curl \omega_a=0$. Taking the time derivative of (\ref{lbdotDelta}), recalling expression (\ref{dotw}) and using the background relations (\ref{FLRWeqs2}), (\ref{FLRWeqs4}), we arrive at~\cite{1989PhRvD..40.1819E}
\begin{eqnarray}
\ddot{\Delta}&=&
-2\left(1-3w+{3\over2}\,c_s^2\right)H\dot{\Delta} \nonumber\\
&&{} +\left[\left({1\over2}+4w-3c_s^2-{3\over2}\,w^2\right)\rho
+(5w-3c_s^2)\Lambda-{12(w-c_s^2)K\over a^2}\right]\Delta \nonumber\\
&&{}+c_{\rm s}^2{\rm D}^2\Delta\,.  \label{lbddotDelta1}
\end{eqnarray}
This wave-like equation, with extra terms due to gravity and the
universal expansion, monitors the linear evolution of matter
aggregations in a almost-FLRW universe filled with a single
barotropic perfect fluid.

In deriving expression (\ref{lbddotDelta1}) we have assumed that
$\dot{w}\neq0$. Although the equation of state of the cosmic medium
changes during the lifetime of the universe, it remains essentially constant throughout several cosmological epochs of interest (e.g.~$w=1/3$ during the radiation era and $w=0$ after equality). Setting $w=$~constant means $w=c_{\rm s}^2$ (see (\ref{dotw})) and
simplifies Eq.~(\ref{lbddotDelta1}) to
\begin{equation}
\ddot{\Delta}= -2\left(1-{3\over2}\,w\right)H\dot{\Delta}+
\left[{1\over2}\,\rho(1-w)(1+3w)+2\Lambda w\right]\Delta+ c_{\rm
s}^2{\rm D}^2\Delta\,,  \label{lbddotDelta2}
\end{equation}
which is now independent of the background 3-curvature. On
introducing the standard scalar harmonics ${\clq}^{(k)}$, we may
expand the perturbed variable in a series of harmonics modes as
$\Delta=\Sigma_{k}\Delta^{(k)}{\clq}^{(k)}$, where ${\rm
D}_a\Delta^{(k)}=0$ and $k$ is the comoving wavenumber of the
associated mode (see Appendix~\ref{AssSVTMs} for details). Then,
expression (\ref{lbddotDelta2}) takes the form
\begin{equation}
\ddot{\Delta}^{(k)}=
-2\left(1-{3\over2}\,w\right)H\dot{\Delta}^{(k)}+
\left\{{1\over2}\,\left[\rho(1-w)(1+3w)+4\Lambda w\right]
-\frac{k^2c_{\rm s}^2}{a^2}\right\}\Delta^{(k)}\,.
\label{klbddotDelta}
\end{equation}
The last term in the right-hand side of the above demonstrates the
competing effects of gravitational attraction and pressure support,
with collapse occurring when the quantity within the angled brackets in positive. Noting that $\lambda_k=a/k$ is the physical wavelength
of the perturbed scalar mode, we conclude that gravitational
contraction will take place only on scales larger than the
associated { Jeans length}, $\lambda_k>\lambda_J$, where
\begin{equation}
\lambda_{J}\simeq\frac{c_{\rm s}}{\sqrt{\rho(1-w)(1+3w)+4\Lambda
w}}\,.  \label{dJeans}
\end{equation}
The above means that when the cosmic medium has a `stiff' equation
of state (i.e.~for $w=1$), the linear gravitational aggregation of
the perturbed matter component is only supported by a positive
cosmological constant (see also (\ref{klbddotDelta})). The Jeans
scale also determines an associated { Jeans mass}. In the absence of a cosmological constant, the latter is given by
\begin{equation}
M_J\propto \rho_b\lambda_J^3\simeq
10^{16}\left({\Omega_b\over\Omega}\right) \left(\Omega
h^2\right)^{-1/2}\,{\rm M}_{\odot}\,,  \label{rJmass}
\end{equation}
which is approximately the mass of a supercluster of galaxies. Note
that $\Omega$, $\Omega_b$ are respectively the total and the
baryonic density parameters, while $H=100h~{\rm km}\,{\rm
sec}^{-1}\,{\rm Mpc}^{-1}$ (e.g.~see~\cite{1993sfu..book.....P}).

Setting $K=0=\Lambda$ we will seek analytical solutions to
Eq.~(\ref{klbddotDelta}). During the radiation era (i.e.~when
$w=1/3=c_{\rm s}^2$), $H=1/(2t)$, $\rho=3/(4t^2)$ and in a comoving
frame we have
\begin{equation}
\frac{{\rm d}^2\Delta^{(k)}}{{\rm d}t^2}+ \frac{1}{2t}\frac{{\rm
d}\Delta^{(k)}}{{\rm d}t}- \frac{1}{2t^2}\left[1-\frac{1}{6}
\left(\frac{k}{aH}\right)^2\right]\Delta^{(k)}=0\,,
\label{cklbddotDelta}
\end{equation}
where $k/aH=\lambda_H/\lambda_k$ and $\lambda_H=1/H=2t$. On
super-Hubble scales, where $k/aH\ll1$ and the pressure support is
negligible, the solution is
\begin{equation}
\Delta= \Delta_1t+ \Delta_2t^{-1/2}\,,  \label{rls}
\end{equation}
with $\Delta_{1,2}=$~constant. Therefore, during the radiation era,
large-scale matter aggregations grow as $\Delta\propto a^2$. Well
inside the horizon, $k/aH\gg1$, pressure gradients support against
gravitational collapse and the over-density oscillates according to
\begin{equation}
\Delta^{(k)}= \Delta_1^{(k)} \sin\left[\sqrt{3}\frac{k}{a_0H_0}
\left(\frac{t}{t_0}\right)^{1/2}\right]+ \Delta_2^{(k)}
\cos\left[\sqrt{3}\frac{k}{a_0H_0}
\left(\frac{t}{t_0}\right)^{1/2}\right]\,,  \label{rss}
\end{equation}
where $\Delta_{1,2}^{(k)}=$~constant and the zero suffix corresponds to a given initial time.

After matter-radiation equality $w=0=c_{\rm s}^2$ and (for
$K=0=\Lambda$) we have $H=2/(3t)$ and $\rho=4/(3t^2)$. In these
conditions Eq.~(\ref{klbddotDelta}) leads to the following
scale-independent evolution
\begin{equation}
\Delta= \Delta_1t^{2/3}+ \Delta_2t^{-1}\,,  \label{das}
\end{equation}
for the density contrast. Accordingly, matter aggregations in the
post-recombination universe grow proportionally to the scale factor. Not surprisingly this relativistic result is identical to the one
obtained through a Newtonian treatment (e.g.~see~\cite{1993sfu..book.....P,1980lssu.book.....P,%
1995coec.book.....C,2002LNP...592..223T}).

Beyond decoupling, the photons can no longer provide pressure
support and gravitational attraction is only counterbalanced by
ordinary baryonic gas pressure. At the time of recombination the
latter is $p^{(b)}\simeq n^{(b)}k_BT_{rec}$, while its radiation
counterpart is given by $p^{(\gamma)}\simeq n^{(\gamma)}k_BT_{rec}$
(with $k_B$ representing Boltzmann's constant). Because
$n^{(b)}\simeq10^{-8}n^{(\gamma)}$, the pressure support drops
drastically at decoupling and soon after recombination the Jeans
mass reduces to
\begin{equation}
M_J\simeq 10^{4}\left({\Omega_b\over\Omega}\right) \left(\Omega
h^2\right)^{-1/2}\,{\rm M}_{\odot}\,,  \label{dJmass}
\end{equation}
which is close to that of a star cluster~\cite{1993sfu..book.....P}.

\subsubsection{Isocurvature perturbations}\label{sssIPs}
One can define as isocurvature perturbations those occurring on
hypersurfaces of uniform curvature, namely fluctuations which
maintain ${\rm D}_a\clr=0$ at all times~\cite{1989PhRvD..40.1804E}.
This should be distinguished from the definition typically found in
the literature, where the term isocurvature usually means distortions in multi-component systems with zero perturbation in the total energy-density initially (e.g.~see~\cite{1992PhR...215..203M,2000cils.book.....L}).

Isocurvature fluctuations also require zero vorticity to guarantee
the integrability of the 3-D hypersurfaces orthogonal to $u_a$.
Then, the condition for isocurvature perturbations is obtained by
linearising the orthogonally projected gradient of (\ref{3R}). To be precise, using definitions (\ref{Deltaa}), (\ref{cZa-cBa}a) we
arrive at
\begin{equation}
a{\rm D}_a\clr= 2\rho\Delta_a- 4H\clz_a\,. \label{lDacR}
\end{equation}
When ${\rm D}_a\clr=0,$ the right-hand side of the above ensures
that linear expansion gradients and those in the fluid are connected by a simple algebraic relation. The projected comoving divergence of the latter translates into the following linear constraint between
the associated scalar variables
\begin{equation}
2H\clz= \rho\Delta\,. \label{licon}
\end{equation}
Assuming a spatially flat background, this is a self-maintained condition for pressure-free dust, but holds on large scales only when the fluid has non-zero pressure~\cite{1989PhRvD..40.1804E}. In particular, propagating (\ref{lDacR}) along the observer's worldline and keeping up to linear order terms gives
\begin{equation}
({\rm D}_a\clr)^{\cdot}= -3H{\rm D}_a\clr+ {4c_s^2H\over
a(1+w)}\,{\rm D}^2\Delta_a\,,  \label{dotlicon}
\end{equation}
since the vorticity has already been switched off. This result shows that the linear isocurvature condition is self-maintained when the
fluid is pressure-free or at long-wavelengths, where the Laplacian
terms on the right-hand side of Eq.~(\ref{dotlicon}) can be
neglected.

Imposing the isocurvature condition (\ref{licon}) to the linear
system (\ref{lbdotDelta}), (\ref{lbdotcZ}), allows us to eliminate
the expansion inhomogeneities (i.e.~the variable $\clz$) from
Eq.~(\ref{lbdotDelta}). The latter then acquires the simple form
\begin{equation}
\dot{\Delta}^{({\rm iso})}= -{3\over2}\,(1-w)H\Delta^{({\rm
iso})}\,, \label{ildotDelta}
\end{equation}
which monitors the linear evolution of isocurvature scalar/density
perturbations on a spatially flat FLRW background filled with a single perfect fluid. Thus, within our scheme, isocurvature perturbations can be defined and treated in single as well as in multi-fluid cosmologies. Solving Eq.~(\ref{ildotDelta}) shows that linear isocurvature modes always decay. In particular, we find $\Delta^{({\rm iso})}\propto t^{-1/2}$ during the radiation era and $\Delta^{({\rm iso})}\propto t^{-1}$ after equipartition.

\subsubsection{Density vortices}\label{sssDVs}
Magnitude preserving changes in the distribution of the density
gradient are monitored via the antisymmetric orthogonally projected
tensor $\Delta_{[ab]}$. To first order, the latter is directly
related to the vorticity tensor according to
$\Delta_{[ab]}=-3a^2(1+w)H\omega_{ab}$, given that the orthogonally
projected gradients of scalars do not commute in the presence of
rotation (see Appendix~\ref{AssCCLs}). Similarly we find that
${\clz}_{[ab]}=3a^2\dot{H}\omega_{ab}$, which means that
\begin{equation}
{\clz}_{[ab]}=-\frac{\dot{H}}{(1+w)H}\Delta_{[ab]}\,,
\label{cZ-Delta}
\end{equation}
at the linear perturbative level. Thus, linearising the skew part of the orthogonally projected gradient of (\ref{lbdotDeltaa}), and then using the above and the background relations (\ref{FLRWeqs1}a) and
(\ref{FLRWeqs2}) we obtain
\begin{equation}
\dot{\clw}_a= -{1\over2H}\,\left[(1-w)\rho -\frac{2(1-3w)K}{a^2}
-2\Lambda w\right]{\clw}_a\,, \label{lbdotcWa}
\end{equation}
where ${\clw}_a\equiv \varepsilon_{abc}\Delta^{bc}/2$.
Accordingly, when $w<1/3$, positive background 3-curvature tends to
increase ${\clw}_a$, while a negative curvature index will have the
opposite effect. Also, for $w>0$, the effect of a positive
cosmological constant is to increase vortex-like density
perturbations.

Introducing the vector harmonic functions ${\clq}^{(k)}_a$ (with
$\dot{\clq}_a^{(k)}=0={\rm D}^a{\clq}_a^{(k)}$), we may write
${\clw}_a=\Sigma_{k}{\clw}^{(k)}{\clq}_a^{(k)}$ with ${\rm
D}_a{\clw}^{(k)}=0$. Then, the $k$-th harmonic mode evolves as
\begin{equation}
\dot{\clw}^{(k)}= -{1\over2H}\,\left[(1-w)\rho
-\frac{2(1-3w)K}{a^2} -2\Lambda w\right]{\clw}^{(k)}\,.
\label{klbdotcWa}
\end{equation}
Setting $K=0=\Lambda$ we find that ${\clw}\propto t^{-1/2}$ during
radiation and ${\clw}\propto t^{-1}$ throughout dust. Therefore,
linear vortices in the density distribution of a perturbed FLRW
universe decay with time on all scales.

\subsubsection{Dissipative effects}\label{sssDEs}
So far we have treated the cosmic medium as a perfect fluid, ignored dissipation and have established a physical scale, the Jeans length, that plays an important role in structure formation
scenarios. There are other processes, however, which can modify the
purely gravitational evolution of matter perturbations. For baryons, the most important interaction is their coupling to the pre-recombination photons. As the latter diffuse from high-density
to low-density regions, they drag the baryons along with them,
erasing inhomogeneities in their distribution. This process, which
is known as { Silk damping}, can wipe out small-scale fluctuations in the baryonic component~\cite{1967Natur.215.1155S,%
1968ApJ...151..459S}. In particular, if $\ell_S$ is the scale
associated with this effect, the corresponding { Silk mass} is
given by (e.g.~see~\cite{1993sfu..book.....P})
\begin{equation}
M_S\propto \rho_b\ell_S^3\simeq
6.2\times10^{12}\left({\Omega\over\Omega_b}\right)^{3/2}
\left(\Omega h^2\right)^{-5/4}\,{\rm M}_{\odot}\,.  \label{Smass}
\end{equation}
The effect of photon free-streaming rises steeply as we approach
recombination and then ceases. Thus, before recombination,
subhorizon-sized baryonic perturbations on scales below $\ell_S$ are obliterated by Silk damping, those with masses between $M_S$ and
$M_J$ oscillate like acoustic waves, while modes having $M>M_J$ can
grow. The latter, however, are of little cosmological interest.

In an analogous way, the free geodesic motion of the collisionless
(dark matter) species erases any structure that tries to form in
their small-scale distribution. Therefore, the ideal fluid
description of the dark component is a good approximation on
sufficiently large scales only. The dissipative process is known as
{\em free streaming} (or {\em Landau damping}) and its full study requires integrating the collisionless Boltzmann equation of the species in question (see \S~\ref{sKINETIC}). Nevertheless, one can obtain an estimate of the effect by calculating the maximum distance traveled by a free-streaming particle. Following~\cite{1993sfu..book.....P,2002LNP...592..223T}, hot thermal relics (see \S~\ref{sssCDM-BU} below for a brief discussion on dark matter candidates), have
\begin{equation}
\ell_{FS}\simeq 0.5\left({m_{DM}\over1\,{\rm keV}}\right)^{-4/3}
\left(\Omega_{DM} h^2\right)^{1/3}\,{\rm Mpc}\,,  \label{FSscale}
\end{equation}
where $m_{DM}$ is the mass of the species in units of $1\,$keV. This means that the minimum scale to survive collisionless dissipation
depends crucially on the particles' mass. Applied to neutrinos with
$m_{\nu}\simeq30\,$eV, the above gives $\ell_{FS}\simeq28\,$Mpc and
a corresponding mass-scale of approximately $10^{15}\,{\rm M}_{\odot}$. For much heavier candidate with, say, $m_{DM}\simeq1$~keV, free streaming will wipe out perturbations on
scales below $\ell_{FS}\simeq0.5$~Mpc (or smaller than
$M_{FS}\simeq10^9\,{\rm M}_{\odot}$). Overall, the lighter the dark
matter is, the less power survives on small scales.

Cold thermal relics (CDM - see \S~\ref{sssCDM-BU}) have very small
dispersion velocities and the values of their free-streaming masses
are very low. As a result, perturbations in the dark-matter
component grow unimpeded by damping processes on all scales of
cosmological interest, although they suffer stagnation due to the
Meszaros effect (see \S~\ref{sssR-DU} below) until the time of
matter-radiation equality. After recombination, the potential wells
of the collisionless species serve to boost the growth of baryonic
perturbations (see \S~\ref{sssCDM-BU}).

\subsection{Multi-fluid perturbations}\label{ssMP}
\subsubsection{Linearised evolution equations for the $i$-th
species}\label{ssLEEiS}
In the FLRW background all the members of the multi-component system are perfect fluids sharing the same 4-velocity $u_a$ by construction. As a result, the effective flux terms, which depend on the peculiar velocities of the species vanish to zero order. This ensures that these quantities are gauge invariant first-order perturbations. Then, Eq.~(\ref{mpfdotDeltaa}) linearises to
\begin{eqnarray}
\dot\Delta_a^{(i)}&=& 3w^{(i)}H\Delta_a^{(i)}-
\left(1+w^{(i)}\right){\clz}_a +
\frac{3aH}{\rho^{(i)}}\left(\dot{q}_{\langle
a\rangle}^{(i)}+4Hq_a^{(i)}\right)- {a\over\rho^{(i)}}\,{\rm
D}_a\left({\rm
D}^bq_b^{(i)}-\cli^{(i)}\right)\nonumber\\
&&-{1\over\rho^{(i)}}\,\cli^{(i)}\Delta_a^{(i)}+
{a\over\rho^{(i)}}\,\cli^{(i)}A_a-
{3aH\over\rho^{(i)}}\,\cli_a^{(i)}\,, \label{lmpfdotDeltaa1}
\end{eqnarray}
where $w^{(i)}=p^{(i)}/\rho^{(i)}$ and
$q_a^{(i)}=\rho^{(i)}(1+w^{(i)})v_a^{(i)}$, with $v_a^{(i)}$
representing the peculiar velocity of each fluid relative to the
$u_a$-frame. The associated momentum-density conservation law reads
\begin{equation}
a\rho^{(i)}\left(1+w^{(i)}\right)A_a=
-c_s^{2(i)}\rho^{(i)}\Delta_a^{(i)}- p^{(i)}\cle_a^{(i)}-
a\left(\dot{q}_{\langle a\rangle}^{(i)}+4Hq_a^{(i)}\right)+
a\cli_a^{(i)}\,, \label{lmpfmdc}
\end{equation}
with $c_s^{2{(i)}}=\dot{p}^{(i)}/\dot{\rho}^{(i)}$ to zero order.
Using the above to eliminate the third term in the right-hand side
of (\ref{lmpfdotDeltaa1}) and then employing the linear part of
(\ref{tmdc}) we obtain
\begin{eqnarray}
\dot\Delta_a^{(i)}&=&
3\left(w^{(i)}-c_s^{2{(i)}}\right)H\Delta_a^{(i)}-
3w^{(i)}H\cle_a^{(i)}- \left(1+w^{(i)}\right){\clz}_a \nonumber\\
&&-{a\over\rho^{(i)}}\,{\rm D}_a\left({\rm D}^bq_b^{(i)}
-\cli^{(i)}\right)- {1\over\rho^{(i)}}\,\cli^{(i)}\Delta_a^{(i)}
\nonumber\\ &&+{a\over\rho(1+w)}\left[3(1+w^{(i)})H
-{1\over\rho^{(i)}}\,\cli^{(i)}\right] \left({c_s^2\rho\over a}\,
\Delta_a+{p\over a}\,\cle_a +\dot{q}_a+4Hq_a\right)\,,
\label{lmpfdotDeltaa2}
\end{eqnarray}
where in the FLRW background $w=(1/\rho)\sum_i\rho^{(i)}w^{(i)}$ and $c_s^2=[1/\rho(1+w)]\sum_ic_s^{2(i)}\rho^{(i)}(1+w^{(i)})$. This is
coupled to the linear propagation equation of the expansion
gradient, which depends on the total fluid and obeys
\begin{eqnarray}
\dot{\clz}_a&=& -2H{\clz}_a- {1\over2}\,\rho\Delta_a-
{3\over2}\,c_s^2\rho\Delta_a- {3\over2}\,p\,\cle_a-
a\left[3H^2+{1\over2}\,(\rho+3p) -\Lambda\right]A_a \nonumber\\
&&{}+a{\rm D}_a{\rm D}^bA_b\,,  \label{lmpfdotcZa1}
\end{eqnarray}
obtained from (\ref{indotcZa}) by means of decomposition
(\ref{tDap}). Using the linearised part of Eq.~(\ref{tmdc}), the
linear commutation laws between the projected gradients of scalars
and recalling that $\dot{s}=0$ in an FLRW spacetime, the above
transforms to
\begin{eqnarray}
\dot{\clz}_a&=& -2H{\clz}_a- {1\over2}\,\rho\Delta_a-
\frac{c_s^2}{1+w}\left({\rm D}^2\Delta_a+{K\over
a^2}\,\Delta_a\right)- \frac{w}{1+w}\left({\rm D}^2\cle_a+{K\over
a^2}\,\cle_a\right)\nonumber\\
&&+{3a\over\rho(1+w)}\left[{1\over2}\,\rho(1+w)
-\frac{K}{a^2}\,\right]\left(\dot{q}_a+4Hq_a\right)-
{a\over\rho(1+w)}\,{\rm D}_a{\rm D}^b
\left(\dot{q}_b+4Hq_b\right)\nonumber\\
&&-6ac_s^2H{\rm D}^b\omega_{ab}\,.  \label{lmpfdotcZa2}
\end{eqnarray}
Expressions (\ref{lmpfdotDeltaa2}) and (\ref{lmpfdotcZa2}) are the basic members of a system of equations that monitors the linear evolution of density inhomogeneities in an almost-FLRW universe filled with several interacting and non-comoving perfect fluids.

\subsubsection{Entropy perturbations in a multi-fluid
system}\label{sssEPMfS}
In a fluid mixture, inhomogeneities in the effective total energy
density ($\rho$) are related to those in densities of the individual
members by
\begin{equation}
\Delta_a=\frac{1}{\rho}\sum_{i}\rho^{(i)}\Delta_a^{(i)}\,.
\label{tDeltaa}
\end{equation}
Treating the whole multi-system as an effective single (total)
fluid, we may combine (\ref{lmpfmdc}) with the linear part of
(\ref{tmdc}) and the above to obtain
\begin{equation}
p\,\cle_a=\sum_ip^{(i)}\cle_a^{(i)}+
\sum_ic_s^{2(i)}\rho^{(i)}\Delta_a^{(i)}-
c_s^2\sum_i\rho^{(i)}\Delta_a^{(i)}\,,  \label{ltcEa1}
\end{equation}
which provides the total effective entropy perturbation of the
multi-component system relative to the properties of its members.
Note that in the FLRW background all species share the same
4-velocity and the effective total sound speed satisfies the
relation
\begin{equation}
c_s^2=\frac{1}{\rho(1+w)}\sum_ic_s^{2(i)}
\left[\rho^{(i)}\left(1+w^{(i)}\right)\right]\,,  \label{tcs2}
\end{equation}
to zero order. Using the above we may recast expression
(\ref{ltcEa1}) into the following linear expression
\begin{equation}
\cle_a=\frac{1}{p}\sum_ip^{(i)}\cle_a^{(i)}+
\frac{1}{2ph}\sum_{i,j}h^{(i)}h^{(j)}
\left(c_s^{2(i)}-c_s^{2(j)}\right)\cls_a^{(ij)}\,,  \label{ltcEa2}
\end{equation}
where $h=\rho+p$, $h^{(i)}=\rho^{(i)}+p^{(i)}$ and
\begin{equation}
\cls_a^{(ij)}=\frac{\rho^{(i)}}{h^{(i)}}\Delta_a^{(i)}-
\frac{\rho^{(j)}}{h^{(j)}}\Delta_a^{(j)}\,.  \label{cSa}
\end{equation}
with $\cls_a^{(ij)}=-\cls_a^{(ji)}$. Accordingly, in a
multi-component fluid the total effective entropy perturbation has
one part coming from the intrinsic entropy perturbations of the
individual species and another due to their different dynamical
behaviour. Following (\ref{ltcEa2}), the latter vanishes if the species share the same sound speed or when $\cls_a^{(ij)}=0$.

\subsubsection{Density perturbations in the $i$-th
species}\label{sssLMAiS}
The scalar $\Delta^{(i)}=a{\rm D}^a\Delta_a^{(i)}$ describes
overdensities (or underdensities) in the matter distribution of the
$i$-th fluid. Thus, assuming no interactions and taking the comoving projected divergence of (\ref{lmpfdotDeltaa2}) we obtain
\begin{eqnarray}
\dot\Delta^{(i)}&=&
3\left(w^{(i)}-c_s^{2{(i)}}\right)H\Delta^{(i)}-
3w^{(i)}H\cle^{(i)}- \left(1+w^{(i)}\right)\clz-
{a^2\over\rho^{(i)}}\,{\rm D}^2\left({\rm D}^aq_a^{(i)}\right)
\nonumber\\ &&+\frac{3(1+w^{(i)})H}{\rho(1+w)}
\left(c_s^2\rho\Delta+p\,\cle\right)+
{3a^2(1+w^{(i)})H\over\rho(1+w)}\,{\rm D}^a
\left(\dot{q}_a+4Hq_a\right)\,, \label{lmpfdotDelta1}
\end{eqnarray}
where $\cle^{(i)}=a{\rm D}^a\cle_a^{(i)}$ and $\cle=a{\rm D}^a\cle_a$. In an analogous way Eq.~(\ref{lmpfdotcZa2}) leads to
the linear expression
\begin{eqnarray}
\dot{\clz}&=& -2H{\clz}- {1\over2}\,\rho\Delta-
\frac{c_s^2}{1+w}\left({\rm
D}^2\Delta+\frac{3K}{a^2}\Delta\right)- \frac{w}{1+w}\left({\rm
D}^2\cle+\frac{3K}{a^2}\cle\right) \nonumber\\
&&-\frac{a^2}{\rho(1+w)}\left[{\rm D}^2{\rm
D}^a\left(\dot{q}_a+4Hq_a\right)+ \frac{3K}{a^2}{\rm
D}^a\left(\dot{q}_a+4Hq_a\right)\right] \nonumber\\
&&+{3\over2}\,a^2{\rm D}^a\left(\dot{q}_a+4Hq_a\right)\,,
\label{lmpfdotcZ1}
\end{eqnarray}
since ${\rm D}^a{\rm D}^b\omega_{ab}=0$ to first approximation.
Also, in deriving the above we have used the first-order auxiliary result $a{\rm D}^a{\rm D}^2\Delta_a={\rm D}^2\Delta+(2K/a^2)\Delta$ and an exactly analogous relation between the Laplacians of $\cle_a$ and $\cle$.

Expressions (\ref{lmpfdotDelta1}) and (\ref{lmpfdotcZ1}) govern the
linear evolution of matter aggregations in the density of the $i$-th species provided the total flux vector
$q_a=\sum_iq_a^{(i)}=\sum_i\rho^{(i)}(1+w^{(i)})v_a^{(i)}$ is specified. Using the energy frame (i.e.~setting $q_a=\sum_iq_a^{(i)}=0$) the system (\ref{lmpfdotDelta1}),
(\ref{lmpfdotcZ1}) reduces to
\begin{eqnarray}
\dot\Delta^{(i)}&=&
3\left(w^{(i)}-c_s^{2{(i)}}\right)H\Delta^{(i)}-
3w^{(i)}H\cle^{(i)}- \left(1+w^{(i)}\right)\clz-
a\left(1+w^{(i)}\right){\rm D}^2v^{(i)} \nonumber\\
&&+\frac{3(1+w^{(i)})H}{1+w} \left(c_s^2\Delta+w\cle\right)
\label{lmpfdotDelta2}
\end{eqnarray}
and
\begin{equation}
\dot{\clz}= -2H{\clz}- {1\over2}\,\rho\Delta-
\frac{c_s^2}{1+w}\left({\rm
D}^2\Delta+\frac{3K}{a^2}\Delta\right)- \frac{w}{1+w}\left({\rm
D}^2\cle+\frac{3K}{a^2}\cle\right)\,, \label{lmpfdotcZ2}
\end{equation}
respectively. The peculiar motion of the species, relative to the
$u_a$-frame, is described by the scalar $v^{(i)}=a{\rm
D}^av_a^{(i)}$ (not to be confused with the magnitude of the peculiar peculiar velocity field -- see \S~\ref{sss4vFs}). To linear order, the evolution of the latter is obtained by combining the conservation laws for the momentum density of the $i$-th component (i.e.~Eq.~(\ref{lmpfmdc}) with $\cli_a^{(i)}=0$) and that of the effective total fluid (the linear part of (\ref{tmdc})). In particular, recalling that
$q_a^{(i)}=\rho^{(i)}(1+w^{(i)})v_a^{(i)}$, a straightforward
algebraic calculation leads to
\begin{eqnarray}
\dot{v}^{(i)}&=& -\left(1-3c_s^{2(i)}\right)Hv^{(i)}-
\frac{1}{a(1+w^{(i)})}\left(c_s^{2(i)}\Delta^{(i)}
+w^{(i)}\cle^{(i)}\right)\nonumber\\
&&-\frac{1}{a(1+w)}\left(c_s^2\Delta+w\cle\right).
\label{lmpfdotv}
\end{eqnarray}

\subsubsection{A radiation and dust universe}\label{sssR-DU}
Consider a spatially flat FLRW background containing a mixture of
radiation and dust. This is a good approximate description
of the pre-decoupling universe, if the radiation is identified with
the photons and the neutrinos, $\rho^{(r)}=\rho^{(\gamma)}+\rho^{(\nu)}$, while the pressureless
component accounts for the nonrelativistic species, baryonic and
cold dark matter, $\rho^{(d)}=\rho^{(b)}+\rho^{(c)}$. Neglecting the photon-baryon interaction terms we have
$\rho^{(r)}=\rho_0^{(r)}(a_0/a)^4$ and
$\rho^{(d)}=\rho_0^{(d)}(a_0/a)^3$, with the zero suffix indicating
a chosen initial time. The total energy of the mixture is
$\rho=\rho^{(r)}+\rho^{(d)}$, the total pressure is
$p=\rho^{(r)}/3$, the effective total sound speed is
$c_s^2=4\rho^{(r)}/[3(4\rho^{(r)}+3\rho^{(d)})]$ and the unperturbed background satisfies the condition $3H^2=\rho$.

According to expressions (\ref{ltcEa2}) and (\ref{cSa}) in \S\ref{sssEPMfS}, the total effective entropy perturbation of the above described radiation-dust mixture is
\begin{equation}
\cle_a= -{4\rho^{(d)}\over3\rho^{(d)}+4\rho^{(r)}} \left(\Delta_a^{(d)}-{3\over4}\Delta_a^{(r)}\right)\,.  \label{cEa0}
\end{equation}
Consequently, imposing the condition of zero effective entropy perturbation corresponds to setting $\Delta_a^{(d)}= 3\Delta_a^{(r)}/4$ and vice-versa. An additional special case emerges when neither $\Delta_a^{(d)}$ nor $\Delta_a^{(r)}$ vanishes, but the total energy density is homogeneously distributed (i.e.~$\Delta_a=0$). Then, Eq.~(\ref{tDeltaa}) guarantees that $\rho^{(d)}\Delta_a^{(d)}= -\rho^{(r)}\Delta_a^{(r)}$.

In what follows we will assume that the radiation field has a homogeneous density distribution (i.e.~for $\Delta^{(r)}=0$), which holds, for example, inside the sound horizon after averaging over acoustic oscillations, or on scales that are damped by diffusion. We may then consider aggregations in the dust component only. Following (\ref{lmpfdotDelta2})-(\ref{lmpfdotv}) these are monitored by the system
\begin{equation}
\dot\Delta^{(d)}= -\clz- a{\rm D}^2v^{(d)}\,, \hspace{20mm}
\dot{\clz}= -2H{\clz}- {1\over2}\,\rho^{(d)}\Delta^{(d)} \label{lrddotsDeltacZ}
\end{equation}
and
\begin{equation}
\dot{v}^{(d)}= -Hv^{(d)}, \label{lrdfdotv}
\end{equation}
given that $c_s^2\Delta+w\cle=0$ to first
order~\cite{1992ApJ...395...54D}. The latter is verified by first
applying (\ref{ltcEa2}) to our two-component system, which shows
that $\cle=-(\rho^{(d)}h^{(r)}/3ph)\Delta^{(d)}$. Then, since
$\Delta=(\rho^{(d)}/\rho)\Delta^{(d)}$ and using the earlier given
expression for the effective sound speed of the total fluid, one can easily ensure that $w\cle=-c_s^2\Delta$.

The system (\ref{lrddotsDeltacZ}), (\ref{lrdfdotv}) monitors the
linear evolution of overdensities in the matter distribution of the
dust component within a perturbed spatially flat FLRW universe
filled with a mixture of radiation and dust, where the radiative
component is homogeneously distributed. Taking the time derivative
of (\ref{lrddotsDeltacZ}a), and using the linear commutation law $({\rm D}^2v^{(d)})^{\cdot}={\rm D}^2\dot{v}^{(d)}-2H{\rm D}^2v^{(d)}$, the above system reduces to the scale independent equation~\cite{1974A&A....37..225M}
\begin{equation}
\Delta^{\prime\prime{(d)}}=
-{2+3a\over2a(1+a)}\,\Delta^{\prime{(d)}}+
{3\over2a(1+a)}\,\Delta^{(d)}\,.  \label{hlrdddotDelta}
\end{equation}
Here primes indicate differentiation with respect to the scale
factor and we have normalised the parameters so that $a=1$ at the
time of matter-radiation equality~\cite{1992ApJ...395...54D}. The general solution has the
form~\cite{1975A&A....41..143G}
\begin{equation}
\Delta^{(d)}= \clc_1\left(1+{3\over2}\,a\right)-
\clc_2\left[\left(1+{3\over2}\,a\right) \ln\left({\sqrt{1+a}+1\over\sqrt{1+a}-1}\right)-
3\sqrt{1+a}\;\right]\,,  \label{hlrdDelta}
\end{equation}
which shows that $\Delta^{(d)}$ grows proportionally to the scale
factor at late times in agreement with a single-fluid Einstein-de
Sitter model. Deep into the radiation era on the other hand, $a\ll1$ and the density contrast is effectively constant (it grows only logarithmically). This stagnation, or freezing-in, of matter perturbations prior to equality is generic to models with a period of expansion that is dominated by relativistic particles and it is sometimes referred to as the Meszaros effect~\cite{1974A&A....37..225M}. Finally, we note that the Newtonian analysis also leads to the same results (e.g.~see~\cite{1993sfu..book.....P,1995coec.book.....C,%
2002LNP...592..223T}).

\subsubsection{A CDM and baryon universe}\label{sssCDM-BU}
It has long been known that purely baryonic scenarios cannot explain the structure observed in the universe today. The main reason is the tight coupling between the photons and baryons in the
pre-recombination era, which washes out perturbations in the
baryonic component. Dark matter, on the other hand, interacts only
gravitationally and therefore it is not subjected to the photon
drag. High energy physics provides a long catalogue of dark matter
candidates. Thermal relics, namely those kept in thermal equilibrium with the rest of the universe until the time they decouple,
typically classify as Hot Dark Matter (HDM) and Cold Dark Matter
(CDM) species.\footnote{Non-thermal relics, such as axions, magnetic monopoles and cosmic strings, remain out of equilibrium throughout
their lifetime.} CDM has small dispersion velocities by the time of
decoupling and does not suffer free-streaming dissipation.

Following \S~\ref{sssR-DU} above, perturbations in the CDM component grow between equipartition and recombination by a factor of $a_{\rm
rec}/a_{\rm eq}=T_{\rm eq}/T_{\rm rec}\simeq21\Omega h^2$, where
$h=H/100\,\mathrm{km}\,\mathrm{s}^{-1}\,\mathrm{Mpc}^{-1}$
is the rescaled, dimensionless Hubble parameter. After
decoupling the universe becomes effectively transparent to radiation and perturbations in the ordinary matter can start growing, driven
by the gravitational potential of the collisionless species. Applied to a mixture of CDM and non-relativistic baryons, the system
(\ref{lmpfdotDelta2})-(\ref{lmpfdotv}) leads to
\begin{equation}
\dot\Delta^{(b)}= -\clz- a{\rm D}^2v^{(b)}\,, \hspace{20mm}
\dot{\clz}= -2H{\clz}- {1\over2}\,\rho\Delta
\label{lDMbdotsDeltacZ}
\end{equation}
and
\begin{equation}
\dot{v}^{(b)}= -Hv^{(b)},  \label{lDMbfdotv}
\end{equation}
where now $\rho=\rho^{(c)}+\rho^{(b)}$ and
$\rho \Delta= \rho^{(c)} \Delta^{(c)}+\rho^{(b)}\Delta^{(b)}$.
Proceeding as in the previous section, the time derivative of (\ref{lDMbdotsDeltacZ}a) leads to
\begin{equation}
\ddot{\Delta}^{(b)}+ 2H\dot{\Delta}^{(b)}=
{1\over2}\,\rho^{(c)}\Delta^{(c)}\,.  \label{DMbdddotDelta1}
\end{equation}
Note that the dark component dominates the baryonic one
(i.e.~$\rho^{(b)}\ll\rho^{(c)}$) and, just after recombination,
$\Delta^{(b)}\ll\Delta^{(c)}$. Recalling that $\Delta^{(c)}\propto
a$ after recombination (see \S~\ref{sssR-DU}) and also that
$\rho^{(c)}\propto a^{-3}$, we
find~\cite{1993sfu..book.....P,2002LNP...592..223T}
\begin{equation}
\Delta^{(b)}= \Delta^{(c)}\left(1-{a_{\rm rec}\over a}\right)\,,
\label{lDMbDelta}
\end{equation}
which shows that $\Delta^{(b)}\rightarrow\Delta^{(c)}$ when $a\gg
a_{\rm rec}$. In other words, after decoupling, baryonic
fluctuations fall in the potential wells of the collisionless
species and quickly catch up with perturbations in the dark-matter
component. This result demonstrates how the presence of the
non-baryonic species accelerates the gravitational collapse of
ordinary matter and therefore the onset of structure formation.
Moreover, because the dark-matter perturbation dominates the
baryonic one, we expect to see the baryonic fluctuations manifested
as a small acoustic peaks in the large-scale correlation function of galaxy surveys. Recent observations seem to confirm this, thus
giving further support to dark-matter and to the CDM structure
formation scenarios~\cite{2005ApJ...633..560E,2005MNRAS.362..505C}.

Cold relics have been proposed since the early 1980's in order to
reproduce the small-scale structure of the universe
(see~\cite{1982ApJ...263L...1P,1984Natur.311..517B,%
1992Natur.356..489D,1993PhR...231....1L,1995MNRAS.276L..69W} and
references therein). Although purely CDM models do not seem to agree with observation, CDM and dark-energy, the latter as an effective
cosmological constant, (i.e.~$\Lambda$CDM scenarios) appear in very
good agreement with the current data~\cite{2003ApJS..148..175S}.

\subsection{Magnetised perturbations}\label{ssMPs}
\subsubsection{The Alfv\'en speed}\label{sssAS}
We assume a weakly magnetised, spatially flat FLRW background
containing a sufficiently random magnetic field. This means that
$\langle B_a\rangle=0$, while $\langle B^2\rangle\neq0$ and $\langle B^2\rangle/\rho\ll1$ on all scales of interest.\footnote{Even a
random vector field will introduce a degree of anisotropy to the
FLRW background. Nevertheless, it sounds plausible that a
sufficiently weak $B$-field can be adequately studied within
almost-FLRW models. This has been verified by studies of perturbed
magnetised Bianchi~I universes~\cite{2000CQGra..17.2215T}.}
Therefore, the energy density of the background $B$-field has only a time dependence (i.e.~$\langle B^2\rangle=B^2(t)$) and decays
adiabatically as
\begin{equation}
B^2=-4HB^2 \hspace{5mm} \Rightarrow \hspace{5mm} B^2\propto
a^{-4}\,, \label{zdotB2}
\end{equation}
according to expression (\ref{B2dot}). The relative strength of the
field is measured by the dimensionless ratio $\beta=B^2/\rho$, which is used to define the Alfv\'en speed
\begin{equation}
c_{\rm a}^2=\frac{\beta}{1+w+\beta}\,.  \label{Alfven}
\end{equation}
Provided $w\neq-1$, the above definition satisfies the constraint
$c_{\rm a}^2<1$ always and for a weak magnetic field, with
$\beta\ll1$, reduces to $c_{\rm a}^2=\beta/(1+w)$.

\subsubsection{Magnetised density perturbations}\label{sssMDPs}
Taking the comoving, orthogonally projected divergence of
Eq.~(\ref{mdotcDa}), using expression (\ref{MHDmdc}) and linearising we arrive at the following equation~\cite{2007PhR...449..131B}
\begin{equation}
\dot{\Delta}= 3w\left(1-{2\over3}\,c_{\rm a}^2\right)H\Delta
-(1+w)\clz+ {3\over2}\,c_{\rm a}^2(1+w)H\clb- c_{\rm a}^2
(1+w)H\clk\,,  \label{lmdotDelta}
\end{equation}
for the linear evolution of $\Delta$. Note that $\clk=a^2\clr$ is the rescaled 3-Ricci scalar of the perturbed spacetime. Also, in deriving the above we have kept up to $c_{\rm a}^2$-order terms, given the weakness of the magnetic field. Finally, we have assumed that $\dot{w}=0$ to zero order, which means that $w=c_s^2=$ constant in the background, as it happens during the radiation and dust eras for example. Following (\ref{lmdotDelta}), the field will generally act as a source of density perturbations even when there are no such distortions present initially. Also, the magnetic field's presence
has a direct and an indirect effect on $\Delta$. The former results
from the pressure part of the Lorentz force (see decomposition
(\ref{Lorentz}) in \S~\ref{sssCLsm}) and carries the effects of the
isotropic magnetic pressure. The latter comes from the tension
component of the Lorentz force and is triggered by the magnetic
coupling to the spatial curvature of the perturbed model.
Surprisingly, a positive 3-curvature perturbation causes $\Delta$ to decrease, while a negative $\clk$ has the opposite effect. This
rather counter-intuitive behaviour of the magneto-curvature term in
(\ref{lmdotDelta}) is a direct consequence of the elasticity of the
field lines (see also Eq.~(\ref{lmdotcZ}) below).

Similarly, the linearised, orthogonally-projected, comoving
divergences of (\ref{mdotcZa}) and (\ref{dotcBa}) lead
to~\cite{2007PhR...449..131B}
\begin{eqnarray}
\dot\clz&=& -2\left(1+{2\over3}\,c_{\rm a}^2\right)H\clz-
{1\over2}\,\left(1-{4\over3}\,c_{\rm a}^2\right)\rho\Delta+
{1\over4}\,c_{\rm a}^2(1+w)\rho\clb-
{1\over2}\,c_{\rm a}^2(1+w)\rho\clk \nonumber\\
&{}&-\frac{c_s^2}{1+w}\left(1-{2\over3}\,c_{\rm a}^2\right){\rm
D}^2\Delta- {1\over2}\,c_{\rm a}^2{\rm D}^2\clb  \label{lmdotcZ}
\end{eqnarray}
and
\begin{equation}
\dot\clb=\frac{4}{3(1+w)}\dot{\Delta}\,, \label{lmdotcB}
\end{equation}
respectively. The latter is a key linear result, ensuring that
perturbations in the magnetic energy density evolve in step with
those in the density of the matter~\cite{1998CQGra..15.3523T,%
2000PhRvD..61h3519T}. Finally, starting from the linear propagation
equation of the 3-Ricci scalar we obtain~\cite{2007PhR...449..131B}
\begin{equation}
\dot\clk= -{4\over3}\,c_{\rm a}^2H\clk+
{4c_s^2\over(1+w)}\,\left(1-{2\over3}\,c_{\rm a}^2\right)H\Delta+
2c_{\rm a}^2H\clb\,. \label{lmdotcK}
\end{equation}
The system (\ref{lmdotDelta})-(\ref{lmdotcK}) monitors the linear
evolution of scalar inhomogeneities in the density distribution of
the matter in a weakly magnetised, spatially flat almost-FLRW
universe.

\subsubsection{Radiation era}\label{sssRE}
During the radiation epoch the background dynamics are determined by the parameters $w=1/3=c_s^2$, $H=1/(2t)$ and $\rho=3/(4t^2)$. Also, the weakness of the magnetic field means that $c_{\rm a}^2=3\beta/4$, where $\beta=B^2/\rho=$~constant $\ll1$. At the same time, the expansion scale factor is $a\propto t^{1/2}$, which means that the Hubble radius at comoving proper time $t$ is given by
$\lambda_{H}\equiv1/H=2t$. Then, recalling that physical wavelengths and comoving wavenumbers are related by $\lambda_k=a/k$, the harmonically decomposed system (\ref{lmdotDelta})-(\ref{lmdotcK}) reads
\begin{eqnarray}
\dot{\Delta}^{(k)}&=& {1\over2}\,\left(1
-{1\over2}\,\beta\right)t^{-1}\Delta^{(k)}- {4\over3}\,\clz^{(k)}-
{1\over2}\,\beta t^{-1}\clk^{(k)}+ {3\over4}\,\beta
t^{-1}\clb^{(k)}\,,  \label{mrldotDel}\\ \dot\clz^{(k)}&=&
-\left(1+{1\over2}\,\beta\right)t^{-1}\clz^{(k)}-
{3\over8}\,(1-\beta)t^{-2}\left[1
-{1\over6}\,\left(\frac{k}{aH}\right)^2
\left(1+{1\over2}\,\beta\right)\right]\Delta^{(k)} \nonumber\\
&{}&-{3\over8}\,\beta t^{-2}\clk^{(k)}+ {3\over16}\,\beta
t^{-2}\left[1+{1\over2}\,
\left(\frac{k}{aH}\right)^2\right]\clb^{(k)}\,,
\label{mrldotZeta}\\ \dot\clk^{(k)}&=& -{1\over2}\,\beta
t^{-1}\clk^{(k)}+ {1\over2}\,\left(1
-{1\over2}\,\beta\right)t^{-1}\Delta^{(k)}+
{3\over4}\,\beta t^{-1}\clb^{(k)}\,,  \label{mrldotcK}\\
\dot\clb^{(k)}&=& \dot{\Delta}^{(k)}\,.  \label{mrldotcB}
\end{eqnarray}

On super-Hubble scales, $k\ll aH$, and, since $\beta\ll1$ always, the system (\ref{mrldotDel})-(\ref{mrldotcB}) is essentially scale-independent and has the power-law solution~\cite{2007PhR...449..131B}
\begin{equation}
\Delta= \clc_0+ \clc_1t^{1-{\textstyle{1\over3}}\beta}+
\clc_2t^{-{\textstyle{1\over2}}+ {\textstyle{5\over6}}\beta}+
\clc_3t^{-\beta}\,,  \label{mrlDelta}
\end{equation}
where the $\clc_\imath$s are constants that depend on the initial
conditions. We note that in the absence of the magnetic field
(i.e.~for $\beta=0$), we recover the standard evolution of the
density contrast in perturbed non-magnetised FLRW models (compare to solution (\ref{rls}) in \S~\ref{sssLPs}). Thus, in the weak-field
limit the main magnetic effect is to reduce the growth rate of the
dominant density mode. In addition, the field also decreases the
rate of the standard decay mode and introduces a new `non-adiabatic' decay mode.

Well below the Hubble radius, $k\gg aH$, the scale-dependent terms
inside the brackets of Eq.~(\ref{mrldotZeta}) become important.
Thus, on sub-Hubble lengths the system (\ref{mrldotDel})-(\ref{mrldotcB}) has the oscillatory
solution~\cite{2007PhR...449..131B}
\begin{equation}
\Delta^{(k)}= \clc_0+
\clc_1\sqrt{\frac{2}{\pi\alpha}}\sin\left(\alpha t^{1/2}\right)+
\clc_2\sqrt{\frac{2}{\pi\alpha}} \cos\left(\alpha
t^{1/2}\right)\,, \label{rsDelta3}
\end{equation}
with $\alpha=c_s(k/a_0H_0)t_0^{-1/2}(1+\beta/2)$. Therefore, well
inside the horizon magnetised matter aggregations oscillate like
magneto-sonic waves. The field's presence tends to reduce the
amplitude of the oscillation and increase its
frequency~\cite{1998CQGra..15.3523T,2000PhRvD..61h3519T}. In both
cases the effect of the $B$-field is proportional to to the ratio
$\beta=B^2/\rho$ and its relative strength. As pointed out
in~\cite{1996PhLB..388..253A}, the increased frequency of
$\Delta^{(k)}$ should bring the peaks of short-wavelength
oscillations in the density of the radiation component closer. This
in turn could produce a potentially observable signature in the CMB. An additional magnetic effect arises from the presence of a constant mode in solution (\ref{rsDelta3}). The latter suggests that, unlike
the magnetic-free case (e.g.~see~\cite{1993sfu..book.....P}),
magnetic matter aggregations in the pre-equilibrium universe
oscillate around a generally non-zero average value.

\subsubsection{Dust era}\label{sssDE}
After the end of the radiation era, the unperturbed background is
well approximated by $w=0=c_s^2$, $H=2/(3t)$, $\rho=4/(3t^2)$.
Also, $c_{\rm a}^2=\beta$ and it is no longer constant but decreases with time according to $\beta\propto a^{-1}\propto t^{-2/3}$.
Applying the usual harmonic decomposition to the perturbation
variables, and keeping up to $\beta$-order terms, the system
(\ref{lmdotDelta})-(\ref{lmdotcK}) reads
\begin{eqnarray}
\dot{\Delta}^{(k)}&=& -\clz^{(k)}-{2\over3}\,\beta
t^{-1}\clk^{(k)}+ \beta t^{-1}\clb^{(k)}\,, \label{ldmdotDelta}\\
\dot{\clz}^{(k)}&=& -{4\over3}\,\left(1
+{2\over3}\,\beta\right)t^{-1}\clz^{(k)}- {2\over3}\,\left(1
-{4\over3}\,\beta\right)t^{-2}\Delta^{(k)}-
{2\over3}\,\beta t^{-2}\clk^{(k)} \nonumber\\
&{}&+{1\over3}\,\beta\left[1+{2\over3}\,
\left(\frac{k}{aH}\right)^2\right] t^{-2}\clb^{(k)}\,,
\label{ldmdotZeta}\\ \dot\clk^{(k)}&=& -{8\over9}\,\beta
t^{-1}\clk^{(k)}+ {4\over3}\,\beta t^{-1}\clb^{(k)}\,,
\label{ldmdotcK}\\ \dot\clb^{(k)}&=&
{4\over3}\,\dot{\Delta}^{(k)}\,.  \label{ldmdotcB}
\end{eqnarray}
After equipartition the dimensionless parameter $\beta$ is no longer constant but decays in time. Without the weak and decaying terms,
the above has the following late-time
solution~\cite{2000PhRvD..61h3519T,2007PhR...449..131B}
\begin{equation}
\Delta^{(k)}= \clc_+t^{\alpha_+}+ \clc_-t^{\alpha_-}+
\clc_3t^{-2/3}+ \clc_4\,,  \label{dltDelta}
\end{equation}
with
\begin{equation}
\alpha_{\pm}= -{1\over6}\,\left[1\pm5\sqrt{1-{32\over75}\,\beta_0
\left(\frac{k}{a_0H_0}\right)^2}\right]\,. \label{alphas}
\end{equation}
In the absence of the $B$-field (i.e.~when $\beta_0=0$), we
immediately recover the standard non-magnetic solution with
$\alpha_+=2/3$ and $\alpha_-=-1$ (see solution (\ref{das}) in \S~\ref{sssLPs}). Therefore, the main magnetic effect is to reduce the growth rate of density perturbations by an amount proportional to its relative strength (i.e.~to the ratio $\beta_0=(B^2/\rho)_0$). It should be noted that the inhibiting role of the field was first observed in the Newtonian treatment of~\cite{1971SvA....14..963R} and later in the relativistic studies of~\cite{1998CQGra..15.3523T,2000PhRvD..61h3519T}. According to
(\ref{dltDelta}) and (\ref{alphas}), the magnetic impact is
inversely proportional to the scale in question. Hence, on large
scales the introduction of the $B$-field simply adds the decaying
$t^{-2/3}$ mode to the standard magnetic-free result. Note also that $\Delta$ describes the directionally averaged gravitational clumping of the matter. Generally, the perturbations will grow at different
rates parallel and perpendicular to the magnetic field and so there
will also be non-spherical evolution in the shapes of these
distortions.

After matter-radiation equality the magnetic field is essentially
the sole source of pressure support. The associated magnetic Jeans
length is obtained by means of the following wave-like equation
\begin{equation}
\ddot{\Delta}^{(k)}= -{4\over3}\,\left(1
-{1\over3}\,\beta\right)t^{-1}\dot{\Delta}^{(k)}+
{2\over3}\,\left\{1-{8\over3}\,\beta \left[1+{1\over6}\,
\left(\frac{k}{aH}\right)^2 \right]\right\}t^{-2}\Delta^{(k)}\,,
\label{ldmddotDelta}
\end{equation}
obtained by taking the time derivative of (\ref{ldmdotDelta}),
ignoring the 3-curvature effects and setting
$\clb_0^{(k)}=4\Delta_0^{(k)}/3$. The latter condition guarantees
that $\clb^{(k)}=4\Delta^{(k)}/3$ always (see Eq.~(\ref{ldmdotcB})), while dropping the magneto-curvature terms from (\ref{ldmdotDelta})
means that only the effects of the isotropic magnetic pressure are
accounted for. When the factor in braces in the last term of
Eq.~(\ref{ldmddotDelta}) is positive, gravity prevails and the density contrast grows. This happens on scales larger than the magnetic Jeans length
\begin{equation}
\lambda_{J_B}= \sqrt{\frac{4\beta}{9-24\beta}}\;\lambda_{H}\,.
\label{mJeans}
\end{equation}
The latter is considerably smaller than the corresponding Hubble
radius since $\beta\ll1$, but for $B$-fields with current strengths
of the order of $10^{-7}$~G is intriguingly close to the size of a
galaxy cluster (i.e.~$\lambda_{J_B}\sim1$~Mpc --
see~\cite{2007PhR...449..131B}).

\subsection{Scalar-field perturbations}\label{ssSFPs}
\subsubsection{The effective fluid characteristics}\label{sssEFCs}
In line with the fluid description of scalar fields, we may
introduce the familiar dimensionless $w$-parameter. Thus, based on
Eqs.~(\ref{sfrho-p}) we define
\begin{equation}
w_{\varphi}= {p^{(\varphi)}\over\rho^{(\varphi)}}=
\frac{\dot{\varphi}^2-2V(\varphi)}{\dot{\varphi}^2+2V(\varphi)}\,.
\label{sfw}
\end{equation}
Then, provided that $\rho^{(\varphi)}>0$, we have
$\dot{\varphi}^2/2+V(\varphi)>0$ and the familiar condition $-1\leq
w_{\varphi}\leq1$ follows from $V(\varphi)\geq0$. Taking the time
derivative and using the Klein-Gordon equation, as seen in
expression (\ref{KG2}) we arrive at
\begin{equation}
\dot{w}_{\varphi}= -3H(1+w_{\varphi})
\left({\dot{p}^{(\varphi)}\over\dot{p}^{(\varphi)}}
-w_{\varphi}\right)\,.  \label{sfdotw}
\end{equation}
The above agrees with the evolution law of the $w$-parameter in
conventional perfect-fluid cosmologies (see Eq.~(\ref{dotw}) in
\S~\ref{sssLEEs}), although in the scalar-field case the ratio
\begin{equation}
{\dot{p}^{(\varphi)}\over\dot{\rho}^{(\varphi)}}= 1+
{2V^{\prime}(\varphi)\over3H\dot{\varphi}}\,,  \label{sfcs2}
\end{equation}
no longer represents an associated thermodynamic sound speed.
According to expression (\ref{sfdotw}), for $H\neq0$ and $w_{\varphi}\neq-1$, the $w^{(\varphi)}$-parameter is time-invariant when $\dot{p}^{(\varphi)}/\dot{\rho}^{(\varphi)}=w^{(\varphi)}$. In that case, relations (\ref{sfw}) and (\ref{sfcs2}) combine to give the evolution law $\dot{V}(\varphi)=-3H(1+w^{(\varphi)})V(\varphi)$ for the scalar-field potential. In the latter case it is straightforward to show that $\dot{p}^{(\varphi)}/\dot{\rho}^{(\varphi)}=w^{(\varphi)}=
1-2V(\varphi)/\rho^{(\varphi)}=$~constant. As expected,
$\dot{p}^{(\varphi)}/\dot{\rho}^{(\varphi)}=
w^{(\varphi)}\rightarrow1$ when $V(\varphi)\rightarrow0$ and
$\dot{p}^{(\varphi)}/\dot{\rho}^{(\varphi)}=
w^{(\varphi)}\rightarrow-1$ for $\rho^{(\varphi)}\rightarrow
V(\varphi)$ (i.e.~for~$\dot{\varphi}\rightarrow0$).

\subsubsection{Density perturbations}\label{sssLSfAs}
Aggregations in the effective energy density of a minimally coupled
scalar field are described by the comoving divergence
$\Delta^{(\varphi)}=a{\rm D}^a\Delta_a^{(\varphi)}$, which to linear order is given by $\Delta^{(\varphi)}= (a^2/\rho^{(\varphi)}){\rm
D}^2\rho^{(\varphi)}= (a^2\dot{\varphi}/\rho^{(\varphi)}){\rm
D}^2\dot{\varphi}$. At this perturbative level, the comoving
divergences of (\ref{sfdotDeltaa}), (\ref{sfdotcZa}) read
\begin{equation}
\dot\Delta^{(\varphi)}= 3w^{(\varphi)}H\,\Delta^{(\varphi)}-
\left(1+w^{(\varphi)}\right){\clz}\,,  \label{lsfdotDelta}
\end{equation}
and
\begin{equation}
\dot{\clz}= -2H\clz- \left[{1\over2}\,\rho^{(\varphi)}+{3K\over
a^2(1+w_{\varphi})}\right]\Delta^{(\varphi)}-
{1\over1+w_{\varphi}}\;{\rm D}^2\Delta^{(\varphi)}\,,
\label{lsfdotcZ}
\end{equation}
respectively. Note that in deriving (\ref{lsfdotcZ}) and in order to express the 4-acceleration and its divergence with respect to the
density gradients, we have used expression (\ref{sfAa1}). The later
ensures that ${\rm D}^aA_a=-[1/a^2(1+w_{\varphi})]\Delta$ to first
approximation. According to Eq.~(\ref{lsfdotDelta}), $\Delta$
decouples from $\clz$ when $w^{(\varphi)}\rightarrow-1$ (see
\S~\ref{sssSSrI} below). For $w^{(\varphi)}\neq-1$, on the other
hand, the time derivative of (\ref{lsfdotDelta}) gives the following wavelike equation for the $k$-th harmonic mode~\cite{1992CQGra...9..921B}
\begin{eqnarray}
\ddot{\Delta}_{(k)}^{(\varphi)}&=& -2\left(1-3w_{\varphi}
+{3\dot{p}^{(\varphi)}\over2\dot{\rho}^{(\varphi)}}\right)
H\dot{\Delta}_{(k)}^{(\varphi)}+
{3\over2}\left(1+8w_{\varphi}-3w_{\varphi}^2
-6{\dot{p}^{(\varphi)}\over\dot{\rho}^{(\varphi)}}\right)
H^2\Delta_{(k)}^{(\varphi)}  \nonumber\\
&&+{1\over a^2}
\left[{9\over2}(1-w_{\varphi}^2)K-k^2\right]\Delta_{(k)}^{(\varphi)}\,,
\label{lsfddotDelta}
\end{eqnarray}
where $K=0,\pm1$. The above governs the evolution of linear
inhomogeneities in the density distribution of a minimally coupled
scalar field in a perturbed FLRW universe. It becomes clear that to
a very large extent this evolution is determined by the effective
equation of state of the $\varphi$-field, which in turn depends on
the latter's kinetic energy and potential.

\subsubsection{Standard slow-roll inflation}\label{sssSSrI}
Standard inflation corresponds to approximately exponential de
Sitter expansion, with $H$ and $\rho^{(\varphi)}$ nearly constants.
This is achieved when the scalar field is rolling very slowly down
its potential. For $\dot{\varphi}^2 \ll V(\varphi)$ and
$|\ddot{\varphi}| \ll H |\dot{\varphi}|$, we have
$w^{(\varphi)}=p^{(\varphi)}/\rho^{(\varphi)}\rightarrow-1$ (see
\S~\ref{sssSFsPFs}). This means that as we approach the de Sitter
regime, Eq.~(\ref{lsfddotDelta}) no longer depends on the background spatial curvature and the linear evolution of
$\Delta_{(k)}^{(\varphi)}$ is governed by
\begin{equation}
\ddot{\Delta}_{(k)}^{(\varphi)}=
-5H\dot{\Delta}_{(k)}^{(\varphi)}- 6H^2\left[1+{1\over6}
\left(\frac{k}{aH}\right)^2\right] \Delta_{(k)}^{(\varphi)}\,,
\label{ldSddotDelta}
\end{equation}
where $H\simeq$~constant. Thus, once the mode has crossed the Hubble radius and $k\ll aH$, the solution is (e.g.~see~\cite{1993sfu..book.....P})
\begin{equation}
\Delta^{(\varphi)}= \clc_1\,{\rm e}^{-2Ht}+ \clc_2\,{\rm
e}^{-3Ht}\,.  \label{ldSDelta}
\end{equation}
Following \S~\ref{sssSfEFLRWCs}, this means that during a phase of
exponential, de Sitter-type expansion $\Delta^{(\varphi)}\propto
a^{-2}$ (e.g.~see~\cite{1993sfu..book.....P}). Therefore, any
overdensities (or underdensities) that may exist in the spatial
distribution of the inflaton field will decay exponentially
irrespective of their scale and the background
curvature.\footnote{The covariant variable corresponding to the
coordinate-based canonical variable for quantisation of scalar field fluctuations, is given by~\cite{2007PhRvD..75h7302P}
\begin{equation}
v_a= {a\dot\varphi\over3H}
\left(\int\D_a\Theta\,d\tau-\D_a\int\Theta\,d\tau\right).
\end{equation}
This gradient variable corresponds to the variable $v=aQ$,
where $Q$ is the Mukhanov-Sasaki variable.}

We note that the above analysis applies to a slowly rolling scalar
field, with $w^{(\varphi)}\simeq-1$. When $w^{(\varphi)}=-1$, on the other hand, the $\varphi$-field has zero kinetic energy and
$p^{(\varphi)}=-\rho^{(\varphi)}=-V(\varphi)$. In that case it is
straightforward to show that ${\rm D}_ap^{(\varphi)}=-{\rm
D}_a\rho^{(\varphi)}=-V^{\prime}(\varphi){\rm D}_a\varphi=0$, which
means that there are no inhomogeneities in the effective energy
density and pressure of the field. This result is a direct
consequence of our spacetime slicing, which guarantees that ${\rm
D}_a\varphi=0$.

\subsubsection{Coasting universe}\label{sssCU}
When $w^{(\varphi)}=-1/3$ the effective gravitational mass of the
$\varphi$-field vanishes. This leads to a `coasting' FLRW universe
with $a\propto t$ and $H=1/t$. During this phase of `minimal'
inflation Eq.~(\ref{lsfddotDelta}) reduces to
\begin{equation}
\ddot{\Delta}_{(k)}^{(\varphi)}=
-3H\dot{\Delta}_{(k)}^{(\varphi)}+ {1\over
a^2}\,(4K-k^2)\Delta_{(k)}^{(\varphi)}  \label{lcddotDelta}
\end{equation}
and has the solution
\begin{equation}
\Delta_{(k)}^{(\varphi)}= \clc_+t^{\alpha_+}+
\clc_-t^{\alpha_-}\,, \label{lcDelta}
\end{equation}
with
\begin{equation}
\alpha_{\pm}=-1\pm {1\over a_0}\sqrt{a_0^2+(4K-k^2)t_0^2}
\label{lcalpha1}
\end{equation}
and $a_0=a(t_0)$. Recall that when dealing with a coasting FLRW
universe we have $\Omega=\Omega_0$~constant and
$a_0/t_0=\sqrt{K/(\Omega_0-1)}$ (see \S~\ref{sssSfEFLRWCs}). The
nature of the above given solution is determined by the sign of the
sum $a_0^2+(4K-k^2)t_0^2$, which in turn depends on the background
3-curvature and the scale of the inhomogeneity~\cite{1992CQGra...9..921B}. Assuming a spatially flat
unperturbed model, expression (\ref{lcalpha1}) gives
\begin{equation}
\alpha_{\pm}=-1\pm \sqrt{1-\left(\frac{k}{a_0H_0}\right)}\,.
\label{lcalpha2}
\end{equation}
Following this, we find $\alpha_+\simeq0$ and $\alpha_-\simeq-2$ on
super-Hubble lengths, which implies that superhorizon-sized
inhomogeneities remain constant. Well inside the horizon, on the
other hand, solution (\ref{lcDelta}), (\ref{lcalpha2}) takes the
oscillatory form
\begin{equation}
\Delta_{(k)}^{(\varphi)}= \clc\,t^{-1}
\cos\left[\left(\frac{k}{a_0H_0}\right)\ln t\right]\,,
\label{slcDelta}
\end{equation}
meaning that small-scale perturbations in the density of the
$\varphi$-field fluctuate with decreasing amplitude. Finally, at the $k=aH$ threshold we find that $\Delta_{(k)}^{(\varphi)}$ decays as
$t^{-1}$.

\subsection{Gravitational wave perturbations}\label{ssGWPs}
Gravitational waves are propagating fluctuations in the geometry of
the spacetime fabric, usually described as weak perturbations of the background metric. Alternatively, one can monitor gravity-wave
distortions covariantly by means of the electric and magnetic
components of the Weyl tensor~\cite{1966ApJ...145..544H},which
describe the free gravitational field (see \S~\ref{sssGrF}).

\subsubsection{Isolating tensor modes}\label{sssIGWs}
Gravitational waves are covariantly described by the transverse
degrees of freedom in the electric ($E_{ab}$) and magnetic
($H_{ab}$) parts of the Weyl tensor. The transversality is necessary to ensure that the pure tensor modes of the locally free
gravitational field have been isolated. The same condition is also
imposed on the shear and any other orthogonally projected,
traceless, second-rank tensor that might be present. Thus, when
studying the propagation of gravitational radiation in perturbed
FLRW models with perfect-fluid matter, we demand that (see
\S~\ref{sssGrF} and \S~\ref{sssKs})
\begin{eqnarray}
{\rm D}^bE_{ab}&=& {1\over3}\,{\rm D}_a\rho= 0\,,  \label{ltrans1}\\
{\rm D}^bH_{ab}&=& \rho(1+w)\omega_a= 0\,,  \label{ltrans2}\\ {\rm
D}^b\sigma_{ab}&=& {2\over3}\,{\rm D}_a\Theta+ \curl \omega_a=
0\,,  \label{ltrans3}
\end{eqnarray}
to linear order and at all times~\cite{1997CQGra..14.1215D}. In our
cosmological environment, this is achieved by switching the
vorticity off and by setting ${\rm D}_a\rho=0={\rm D}_ap={\rm
D}_a\Theta$ (for a barotropic medium it suffices to ensure that
${\rm D}_a\rho=0={\rm D}_a\Theta$). These constraints, which are
self-consistent (i.e.~preserved in time) at the linear perturbative
level, guarantee that the 4-acceleration also vanishes to first
approximation. Then, the only nontrivial linear constraints left are
\begin{equation}
H_{ab}= \curl \sigma_{ab} \hspace{15mm} {\rm and} \hspace{15mm}
\clr_{\langle ab\rangle}= -H\sigma_{ab}+ E_{ab} \label{lGWcon}
\end{equation}
where $\curl \sigma_{ab}=\varepsilon_{cd\langle a}{\rm
D}^c\sigma_{b\rangle}{}^d$ (see Eqs.~(\ref{Hcon}) and (\ref{GC}) in
\S~\ref{sssKs} and \S~\ref{sssSC} respectively). Note that,
according to (\ref{lGWcon}b), the linear condition ${\rm D}^bE_{ab}= 0={\rm D}^b\sigma_{ab}$ guarantees that ${\rm D}^b\clr_{\langle ab\rangle}=0$ as well. When allowing for anisotropic pressure, one needs to impose the additional constraint
\begin{equation}
{\rm D}^b\pi_{ab}=0\,.  \label{lmGWcon}
\end{equation}
In the case of a magnetised, highly conductive environment, for
example, we demand that ${\rm D}^b\Pi_{ab}=(1/3){\rm D}_aB^2-
B^b{\rm D}_bB_a=0$ at all times~\cite{2001PhRvD..63l3507M,%
2002CQGra..19.3709T}.

\subsubsection{Covariant description of the gravitational-wave
energy density}\label{sssCDGwED}
In a perturbed FLRW universe, the energy density of gravitational radiation is determined by the
pure tensor part ($\clh_{\alpha\beta}^{TT}$, with $\alpha,\beta=1,2,3$) of the metric perturbation,
according to (e.g.~see~\cite{2002PhRvD..65b3517C})
\begin{equation}
\rho_{GW}={(\clh_{\alpha\beta}^{TT})^{\prime}
(\clh_{TT}^{\alpha\beta})^{\prime}\over2a^2}\,,  \label{GWrho}
\end{equation}
where a prime indicates differentiation with respect to conformal
time (recall that $c=1=8\pi G$ throughout this review). In a
comoving frame, with $u^a=\delta_0^au^0$, we
have~\cite{1992ApJ...395...34B,1989PhRvD..39.2882G}
\begin{equation}
\sigma_{\alpha\beta}= a(\clh_{\alpha\beta}^{TT})^{\prime} \hspace{15mm} {\rm and} \hspace{15mm} \sigma^{\alpha\beta}= a^{-3}(\clh_{TT}^{\alpha\beta})^{\prime}\,,
\label{TTsigma}
\end{equation}
so that~\cite{2002CQGra..19.3709T}
\begin{equation}
\rho_{GW}=\sigma^2\,,  \label{cGWrho}
\end{equation}
which provides a simple covariant expression for the energy density
of gravitational-wave distortions in an almost FLRW universe.

\subsubsection{Evolution of gravitational waves}\label{sssEGWs}
In a FLRW spacetime the Weyl tensor vanishes identically, which
means that $E_{ab}$ and $H_{ab}$ provide a covariant and
gauge-invariant description of perturbations in the
free-gravitational field. Once the pure tensor modes have been
isolated, we can proceed to linearise the propagation equations
(\ref{dotEab}) and (\ref{dotHab}) of \S~\ref{sssWC}. Around a
Friedmannian background filled with a single perfect fluid, the
latter reduce to
\begin{equation}
\dot{E}_{ab}= -3HE_{ab}- {1\over2}\,\rho(1+w)\sigma_{ab}+ \curl
H_{ab}  \label{ldotEab1}
\end{equation}
and
\begin{equation}
\dot{H}_{ab}= -3HH_{ab}- \curl E_{ab}\,, \label{ldotHab}
\end{equation}
respectively. Because the magnetic part of the Weyl tensor satisfies the constraint (\ref{lGWcon}a) the linear evolution of $H_{ab}$ is
determined by that of the shear, which propagates according to (see
expression (\ref{sigmadot}) in \S~\ref{sssKs})
\begin{equation}
\dot{\sigma}_{ab}= -2H\sigma_{ab}- E_{ab}\,. \label{ldotsigma}
\end{equation}
Furthermore, on using the commutation law between the orthogonally
projected gradients of spacelike tensors (see Appendix~\ref{AssCCLs}) and the zero-order expression $\clr_{abcd}=
(K/a^2)(h_{ac}h_{bd}-h_{ad}h_{bc})$ for the 3-Riemann tensor (see
\S~\ref{sssSC}), constraint (\ref{lGWcon}a) leads to the
auxiliary relation $\curl H_{ab}= (3K/a^2)\sigma_{ab}-{\rm
D}^2\sigma_{ab}$. The latter transforms Eq.~(\ref{ldotEab1}) into
\begin{equation}
\dot{E}_{ab}= -3HE_{ab}- {1\over2}\,\rho(1+w)\sigma_{ab}+ {3K\over
a^2}\,\sigma_{ab}- {\rm D}^2\sigma_{ab}\\, \label{ldotEab2}
\end{equation}
which together with (\ref{ldotsigma}) monitors the linear evolution
of gravitational waves in perturbed FLRW universes.

Taking the time derivative of (\ref{ldotsigma}), using
Eq.~(\ref{ldotEab2}), the background Raychaudhuri and Friedmann
formulae (see (\ref{FLRWeqs1}a), (\ref{FLRWeqs2}) respectively) and
keeping terms of up to linear order only, we arrive at the following
wave-like equation for the gravitationally induced shear
\begin{equation}
\ddot{\sigma}_{ab}= -5H\dot{\sigma}_{ab}-
{1\over2}\,\rho(1-3w)\sigma_{ab}+ {K\over a^2}\,\sigma_{ab}+ {\rm
D}^2\sigma_{ab}\,,  \label{lddotsigma}
\end{equation}
with $K=0,\pm1$. The above is no longer coupled to the propagation
equation of the electric Weyl tensor, which means that the shear
wave-equation alone can describe the propagation of gravitational
radiation in perturbed, perfect-fluid FLRW cosmologies. We proceed
by introducing the standard tensor harmonics $\clq_{ab}^{(k)}$, with $\clq_{ab}^{(k)}=\clq_{\langle ab\rangle}^{(k)}$,
$\dot{\clq}_{ab}^{(k)}=0={\rm D}^b\clq_{ab}^{(k)}$ and ${\rm
D}^2\clq_{ab}^{(k)}=-(k/a)^2\clq_{ab}^{(k)}$. Then, setting
$\sigma_{ab}=\sum_{k}\sigma_{(k)}\clq_{ab}^{(k)}$, with ${\rm D}_a\sigma_{(k)}=0$, expression (\ref{lddotsigma}) provides the following wave equation for the $k$-th shear mode.
\begin{equation}
\ddot{\sigma}_{(k)}= -5H\dot{\sigma}_{(k)}-
\left[{1\over2}\,\rho(1-3w)-{1\over
a^2}\,(K-k^2)\right]\sigma_{(k)}\,. \label{lkddotsigma}
\end{equation}
It should be noted here that, in order to account for the different
polarisation states of gravitational radiation, one expands the
pure-tensor perturbations in terms of the electric and the magnetic
parity harmonics (see~\cite{2000CQGra..17..871C} and references
therein). Nevertheless, the coupling between the two states means
that Eq.~(\ref{lkddotsigma}) still holds.

Assuming a spatially flat background and a radiation-dominated
universe, we have $w=1/3$, $H=1/(2t)$ and $\rho=3/(4t^2)$. Then,
(\ref{lkddotsigma}) simplifies to
\begin{equation}
\ddot{\sigma}_{(k)}= -{5\over2t}\,\dot{\sigma}_{(k)}-
\left({k\over a}\right)^2\sigma_{(k)}\,, \label{rlkddotsigma}
\end{equation}
with the last term vanishing on super-Hubble lengths (asymptotically). The above admits the solution
\begin{eqnarray}
\sigma_{(k)}&=& t^{-3/2}\left({k\over aH}\right)
\left[\clc_1\,\sin\left({k\over aH}\right)+
\clc_2\,\cos\left({k\over aH}\right)\right] \nonumber\\
&&+t^{-3/2} \left[\clc_1\,\cos\left({k\over aH}\right)-
\clc_2\,\sin\left({k\over aH}\right)\right]\,, \label{rsigma}
\end{eqnarray}
where $k/aH\propto t^{1/2}$. Therefore, the amplitude of
gravitational waves on small scales in a radiation-dominated and
spatially-flat FLRW universe decays as $a^{-2}$. On super-Hubble
scales, on the other hand, $k/aH\ll1$ and (\ref{rsigma}) is
approximated by the power law
\begin{equation}
\sigma_{(k)}=\clc_1t^{-3/2}\left[1-{1\over3}{\clc_2\over\clc_1}
\left({k\over a_0H_0}\right)^3\left({t\over t_0}\right)^{3/2}\right]= \sigma_0+ {2\over3}\,\dot{\sigma}_0t_0\left[1-\left({t\over t_0}\right)^{3/2}\right]\,.
\label{rlssigma}
\end{equation}
Consequently, when $\dot{\sigma}_0=0$ we have $\sigma_{(k)}=$~constant. On the other hand, for $\dot{\sigma}_0=-3\sigma_0/2t_0$ we find $\sigma_{(k)}\propto t^{-3/2}\propto a^{-3}$.

After equality, we have $w=0$, $H=2/(3t)$ and $\rho=4/(3t^2)$. In
this environment, Eq.~(\ref{lkddotsigma}) takes the form
\begin{equation}
\ddot{\sigma}_{(k)}= -{10\over3t}\,\dot{\sigma}_{ab}- {2\over3t^2}
\left[1+{2\over3}\left({k\over aH}\right)^2\right]\sigma_{(k)}\,.
\label{dlkddotsigma}
\end{equation}
Hence, on superhorizon scales,
\begin{equation}
\sigma_{(k)}= \clc_1\,t^{-1/3}+\clc_2\,t^{-2}\,,  \label{dlsigma}
\end{equation}
implying that after equality large-scale gravitational wave
perturbations decay as $a^{-1/2}$.

\subsection{Perturbed non-Friedmannian cosmologies}\label{ssPNFC}
Current observational data strongly support the homogeneous and
isotropic FLRW spacetimes as the best model for our universe on very large scales. On the other hand, the presence of nonlinear
structures, in the form of galaxies, galaxy clusters and
superclusters, shows that on the relevant scales the universe is neither homogeneous nor isotropic. For this reason and also because we have no real information about the very early stages of our universe, it may be unwise to exclude a priori all the non-FLRW cosmologies from our studies. In what follows we will consider the stability of such unconventional universes against linear perturbations.

\subsubsection{The Bianchi I universe}\label{sssBIU}
Bianchi models have been studied by several authors in an attempt to achieve better understanding of the observed small amount of anisotropy in the universe. The same models have also been used to examine the role of certain anisotropic sources during the formation of the large-scale structure we see in the universe today. Some Bianchi cosmologies, for example, are natural hosts of large-scale magnetic fields and therefore their study can shed light on the implications of cosmic magnetism for galaxy formation. The simplest Bianchi family that contains the flat FLRW universe as a special case are the type-I spacetimes (see Table~\ref{tab1}).

Using covariant techniques, the stability of the Bianchi~I models against linear perturbations was studied in~\cite{1993PhRvD..48.3562D}, for the case of a single perfect fluid, and also in~\cite{2000CQGra..17.2215T}, where a magnetic field was also present. Assuming non-magnetised dust for simplicity, Eqs.~(\ref{pndotDeltaa}) and (\ref{indotcZa}) linearise to
\begin{equation}
\dot\Delta_a= -{\clz}_a- \sigma_{ab}\Delta^b  \label{lB1dotDeltaa}
\end{equation}
and
\begin{equation}
\dot{\clz}_a= -{2\over3}\,\Theta{\clz}_a-
{1\over2}\,\rho\Delta_a- \sigma_{ab}{\clz}^b- 2a{\rm
D}_a\sigma^2\,, \label{lB1dotcZa}
\end{equation}
respectively. The above monitor the linear evolution of density inhomogeneities in the dust component on a Bianchi~I background. The system is obviously not closed and one also needs the propagation formula of the shear and the shear constraint (see expressions (\ref{sigmadot} and (\ref{shearcon}) respectively). To first order and in the absence of fluid pressure, these are respectively given by
\begin{equation}
\dot{\sigma}_{ab}= -{2\over3}\,\Theta\sigma_{ab}-
\sigma_{c\langle a}\sigma^c{}_{b\rangle}- E_{ab} \label{lB1sigmadot}
\end{equation}
and
\begin{equation}
{\rm D}^b\sigma_{ab}= {2\over3a}\,\clz_a+ \curl\omega_a\,.  \label{lB1shearcon}
\end{equation}
The above immediately bring into play the electric component of the Weyl field, the linear evolution of which depends on its magnetic counterpart (see Eq.~(\ref{dotEab}) in \S~\ref{sssWC}), as well as the vorticity. The complexity of the mathematics means that analytical progress can achieved in spacial cases only. In~\cite{1993PhRvD..48.3562D}, for example, the perturbed model is assumed both irrotational and axially symmetric. It was then possible to obtain analytical solutions for the two independent components of the density contrast in the form of power series. Confining to early times, when $t\rightarrow0$, the latter converge to
\begin{equation}
\Delta_1= \clc_1t^{-2/3}+\clc_2t^{-5/3} \hspace{10mm} {\rm and} \hspace{10mm} \Delta_{2,3}= \clc_3t^{10/3}+\clc_4t^{1/3}\,, \label{BIDelta1}
\end{equation}
with the $\clc$s representing the integration constants. This result argues for growth along two of the directions that is faster than the standard Einstein-de Sitter rate (compare to Eq.~(\ref{das}) in \S~\ref{sssLPs}) and agrees qualitatively with the analysis of~\cite{1972PhRvD...6..969P}.

In the magnetised study of~\cite{2000CQGra..17.2215T} the anisotropic pressure of the $B$-field (see expression (\ref{TB}) in \S~\ref{sssIMHDA}) adds further complications to the system of (\ref{lB1dotDeltaa}) and (\ref{lB1dotcZa}). Treating the magnetic field as the sole source of anisotropy, makes $B_a$ a shear eigenvector and considerably simplifies the mathematics. Then, for magnetic strengths compatible with the high isotropy of the CMB spectrum, one can obtain analytical solutions and establish the corrections to the FRW-related results of \S~\ref{ssMPs}. Following~\cite{2000CQGra..17.2215T}, the anisotropy of the background model brings about the tension properties of the $B$-field. This happens through the general relativistic coupling between the field and the spatial curvature of the perturbed model and makes the overall magnetic effect sensitive to the amount of the 3-curvature distortion.

\subsubsection{The Einstein static universe}\label{sssESU}
The possibility that our universe might have started out in an
asymptotically static state, reminiscent of the
Eddington-Lema\^{i}tre cosmology, has been discussed by several
authors (e.g.~see~\cite{2004CQGra..21..223E,2004CQGra..21..233E,%
2005PhRvD..72l3003C,2005PhRvD..71l3512M,2006CQGra..23.6927M} and
references therein). It is therefore useful to investigate the
stability of the family of the Einstein static spacetimes, which has long been known to be unstable against homogeneous and isotropic
perturbations~\cite{1930MNRAS..90..668E}. In fact, the instability
of this model is very well established among the community, despite
later work showing that the issue is not as clear cut as the
Newtonian intuition may suggest~\cite{1987NuPhB.292..784G}.

In an Einstein static background $H=0=\dot{H}$ and $K=+1$. Then,
formulae (\ref{FLRWeqs1}) and (\ref{FLRWeqs2}) in \S~\ref{sssFLRWCs} ensure that  $\rho$ (and therefore $p$, $w$) is time invariant,
$\Lambda=(1+3w)\rho/2$ and $1/a^2=(1+w)\rho/2$. In this environment
expression (\ref{lbddotDelta1}) reduces
to~\cite{2003CQGra..20L.155B}
\begin{equation}
\ddot{\Delta}= {1\over2}\,(1+w)(1+3c_s^2)\rho\Delta+ c_{\rm
s}^2{\rm D}^2\Delta\,,  \label{ESddotDelta1}
\end{equation}
which monitors the linear evolution of scalar/density perturbations
in a perturbed Einstein static universe. By employing the standard
scalar harmonics (see \S~\ref{sssLPs} for more details), the above
decomposes into
\begin{equation}
\ddot{\Delta}^{(k)}=
{1\over2}\,(1+w)\left[1+(3-k^2)c_s^2\right]\rho\Delta^{(k)}\,,
\label{ESddotDelta2}
\end{equation}
and has a (neutrally) stable, oscillatory solution as long as
$(k^2-3)c_s^2>1$. Recalling that in spatially closed models the
eigenvalue of a given mode and its comoving wavenumber are discrete
and related by $k^2=\nu(\nu+2)$, with $\nu=1,2,3,\dots$, we conclude that a perturbed mode is stable against gravitational collapse
if~\cite{2003CQGra..20L.155B}
\begin{equation}
\left[\nu(\nu+2)-3\right]c_s^2>1\,,  \label{lstcon}
\end{equation}
to linear order. The first inhomogeneous mode has $\nu=1$ and
therefore it does not satisfy the stability condition. Nevertheless,this is simply a gauge mode reflecting our freedom to change the fundamental 4-velocity vector. Hence, all physical modes have $\nu\geq2$ and for them linear stability is guaranteed as long as $c_s^2>1/5$.\footnote{The same condition for the linear stability of the Einstein static universe was also obtained in~\cite{1987NuPhB.292..784G} in the restricted case of conformal
metric perturbations. The stability of the radiation-filled model
and the instability of the one containing pressureless dust was
demonstrated in~\cite{1967RvMP...39..862H}.}

The physical explanation for this rather unexpected stability lies
in the Jeans length of the Einstein static
model~\cite{1987NuPhB.292..784G}. Although there are always unstable modes in flat spaces, namely wavelengths larger than the associated
Jeans scale, a closed universe sets an upper limit to the allowed
wavelengths. The above analysis shows that, for a sufficiently large sound speed, all physical wavelengths fall below the Jeans scale.
Noting that the linear stability of the Einstein static model does
not guarantee its overall stability~\cite{2005PhRvD..71d4011L}, we
point out that current data seem to favour a slightly closed
universe~\cite{2003ApJS..148..175S}. Also, spatially closed static
models violate the inflationary singularity theorems
of~\cite{1994PhRvL..72.3305B,1997PhRvD..56..717B,2003PhRvL..90o1301B}
and therefore can avoid the quantum-gravity
era~\cite{1999PhLB..455...90M,2004CQGra..21..223E}.

\subsubsection{The G\"{o}del universe}\label{sssGU}
G\"{o}del's universe is an exact solution of the Einstein field
equations that is both stationary and spatially
homogeneous~\cite{1949RvMP...21..447G}. The G\"{o}del spacetime is
also rotationally symmetric about each point and well known for its
unusual properties. The most intriguing among them is the existence
of closed timelike curves, which violates global causality and makes time travel a theoretical possibility.\footnote{For additional
examples of spacetimes with closed timelike curves the reader is
referred to~\cite{2003GReGr..35.1721S,2007gr.qc.....3139G}.}
Although G\"{o}del's world is not a realistic model of the universe
we live in, it has been widely used to study and illustrate the
effects of global vorticity within the realm of general relativity
(e.g.~see~\cite{1998JMP....39.2148K,2003AmJPh..71..801O,%
2006physics..12253B} and references therein).

Relative to a timelike 4-velocity field, tangent to the world lines
of the fundamental observers, the kinematics of the G\"{o}del
spacetime are covariantly described by~\cite{1971glc..conf....104E,1973lsss.book.....H}
\begin{equation}
\Theta=0=A_a=\sigma_{ab}=H_{ab} \hspace{15mm} {\rm and}
\hspace{15mm} \omega_a,\,E_{ab}\neq0\,,  \label{Gkin}
\end{equation}
where $\nabla_b\omega_a=0=\dot{E}_{ab}={\rm D}_cE_{ab}$ and
$\nabla_cE_{ab}\neq0$. Also, the stationary nature of the model
means that all the propagation equations have been transformed into
constraints, some of which are trivial. In particular, the G\"{o}del analogues of the continuity, the Raychaudhuri and the Friedmann
formulae (see \S~\ref{sssKs}, \S~\ref{sssCLs} and \S~\ref{sssSC}) read
\begin{equation}
\dot{\rho}=0, \hspace{15mm} {1\over2}\,\rho(1+3w)- 2\omega^2-
\Lambda=0 \label{Geqs1}
\end{equation}
and
\begin{equation}
\rho- \omega^2+ \Lambda=0\,,  \label{Geqs2}
\end{equation}
respectively (see also~\cite{2004CQGra..21.1773B} for more technical details). We note that expressions (\ref{Geqs1}b) and (\ref{Geqs2})
combine to give $\rho(1+w)=2\omega^2$ and $p-\rho=2\Lambda$. The
former of these relations shows that the vorticity provides a direct measure of the model's inertial mass. The latter ensures that the
value of the cosmological constant depends on the equation of state
of the fluid that fills the G\"{o}del spacetime. The amount of
rotation also determines the radius,
$R_G=\sqrt{2}\ln(1+\sqrt{2})/\omega$, of the observers causal region.

Using the above given relations, one can linearise the nonlinear
formulae of \S~{\ref{ssISfCs} around a G\"{o}del background. More
specifically, Eqs.~(\ref{indotDeltaa})-(\ref{indotcZa}) reduce to
the system~\cite{2004CQGra..21.1773B}
\begin{equation}
\dot\Delta_a= -(1+w){\clz}_a+ \omega_{ab}\Delta^b\,,
\label{lGdotDeltaa}
\end{equation}
\begin{equation}
\dot{\clz}_a= -{1\over2}\,\rho\Delta_a- {3\over2}\,a{\rm D}_ap+
a{\rm D}_a{\rm D}^bA_b+ 2a{\rm D}_a\omega^2+
\omega_{ab}{\clz}_b\,, \label{lGdotcZa}
\end{equation}
with $\rho(1+w)A_a=-{\rm D}_ap$. Assuming rigid rotation, we may
ignore the second-last term in the right-hand side of
(\ref{lGdotcZa}). Then, the orthogonally projected derivatives of
the above combine to give the following wavelike
equation~\cite{2004CQGra..21.1773B}
\begin{equation}
\ddot{\Delta}^{(k)}=\left[\omega^2(1+3c_s^2) -\left({kc_s\over a
}\right)^2\right]\Delta^{(k)}\,, \label{lGddotDeltaa}
\end{equation}
for the linear evolution of the $k$-th harmonic mode. Thus, for dust there is no pressure support and matter aggregations grow
unimpeded.\footnote{G\"{o}del's solution was originally given for
dust (i.e.~$\rho\neq0$, $p=0$ and $\Lambda\neq0$). Nevertheless, by
introducing the transformation $\rho\rightarrow\rho^{\prime}=\rho+p$ and $\Lambda\rightarrow\Lambda^{\prime}=\Lambda+p$, we can
reinterprete the G\"{o}del spacetime as a perfect-fluid model.}
Otherwise, there is an effective Jeans length, equal to
$\lambda_{J_G}=c_s/\omega\sqrt{1+3c_s^2}$, below which the
inhomogeneities oscillate~\cite{2004CQGra..21.1773B}. The same type
of neutral stability was also claimed for gravitational wave
peturbations, while the evolution of linear rotational distortions
was found to depend on the amount of the shear anisotropy
(see~\cite{2004CQGra..21.1773B} for further discussion).

We point out that the value of the Jeans length is comparable to the radius of the smallest closed timelike curve (see above). This means that the causal regions in G\"{o}del models with nonzero pressure
are stable against linear matter aggregations. Also, expression
(\ref{lGddotDeltaa}) demonstrates how the vorticity of the G\"{o}del universe contributes to the overall gravitational pull. This result
suggests that rotational energy has `weight' and seems to favour the de Felice~\cite{1991MNRAS.252..197D} and the
Barrab\`{e}s~et~al~\cite{1995MNRAS.276..432B} interpretation of the
Abramowicz-Lasota `centrifugal-force reversal'
effect~\cite{1974MNRAS.5..327A}.

The above analysis does not specifically address the stability of
G\"{o}del's closed timelike curves, a issue recently considered
in~\cite{2007gr.qc.....3100R}. The appearance of closed timelike
curves in general relativistic G\"{o}del-type cosmologies, namely in rigidly rotating homogeneous spacetimes, can be avoided by
introducing extra matter sources, higher-order terms in the
Laplacian or string-theory corrections~\cite{1983PhRvD..28.1251R,%
1998PhRvD..58j3502B,1999PhRvD..60j3519K}. This does not seem to be
the case, however, in G\"{o}del-type brane
models~\cite{2004PhRvD..69f4007B}.

\subsubsection{The Lukash Bianchi~$VII_h$ universe}
The most general Bianchi universes that contain the spatially open
FLRW model as a special subcase, are those of type $VII_h$. The
late-time asymptotes of the non-tilted Bianchi~$VII_h$ cosmologies
evolve towards a vacuum plane-wave solution known as the Lukash
universe~\cite{1973SJETP..37..739D,1976NCimB..35..268L}. These
spacetimes describe the most general effects of spatially
homogeneous perturbations on open Friedmann universes and the Lukash metric plays a guiding role in these investigations
(see~\cite{1986PhR...139....1B} and references therein).

In the absence of matter, the plane-wave attractors of the
Bianchi~$VII_h$ models are covariantly characterised by
\begin{equation}
A_a=0=\omega_a \hspace{15mm} {\rm and} \hspace{15mm}
\Theta,\,\sigma_{ab},\,E_{ab},\,H_{ab}\neq0\,,  \label{Lcon}
\end{equation}
with $E_{ab}E^{ab}=H_{ab}H^{ab}$ and
$E_{ab}H^{ab}=0$~\cite{1986CQGra...3.1105H,2005CQGra..22..825B}.
Then, the Lukash analogues of the Raychaudhuri and Friedmann
equations read
\begin{equation}
\dot{\Theta}= -{1\over3}\,\Theta^2- 2\sigma^2 \hspace{15mm} {\rm
and} \hspace{15mm} \clr= -{2\over3}\,\Theta^2+ 2\sigma^2\,,
\label{Leqs1}
\end{equation}
respectively. We note that the 3-curvature scalar $\clr$ is
negative, thus guaranteeing the hyperbolic spatial geometry of the
model (e.g.~see~\cite{1986CQGra...3.1105H,2005CQGra..22..825B}).
Also, despite the absence of matter fields, the Lukash metric is not Ricci flat and the associated Gauss-Codacci formula is
\begin{equation}
\clr_{\langle ab\rangle}= -{1\over3}\,\Theta\sigma_{ab}+
\sigma_{c\langle a}\sigma^c{}_{b\rangle}+ E_{ab}\,.  \label{Leqs2}
\end{equation}
Using the dimensionless parameter $\Sigma=3\sigma^2/\Theta^2$ to
measure the expansion anisotropy, one can also recast the
Raychaudhuri equation into the following alternative
expression~\cite{2005CQGra..22..825B}
\begin{equation}
\dot{\Theta}=-{1\over3}\,\Theta^2(1+2\Sigma)\,,  \label{Leqs3}
\end{equation}
where $0<\Sigma<1$ in accord with $\clr<0$ in (\ref{Leqs1}b). Then,
the average scale factor obeys a simple power-law evolution with
$a\propto t^{1/(1+2\Sigma)}$. In the absence of shear effects, we recover the familiar $a\propto t$ evolution of the Milne universe. As $\Sigma\rightarrow1$, on the other hand, we approach the Kasner vacuum solution (i.e.~$a\propto t^{1/3}$). Note that for maximum shear the spatial curvature of the model vanishes, while only its isotropic part survives at the opposite end.

The nature of the Lukash solution makes it a good testing ground for studying the final stages of ever-expanding FLRW cosmologies with
positive gravitational mass (i.e.~$\rho+3p>0$). Allowing for a
low-density, pressure-free matter component, we may identify our
fundamental 4-velocity field with that of the fluid. Then,
linearising Eq.~(\ref{sigmadot}) around the Lukash background and
using (\ref{GC1}), we find that shear perturbations are monitored by the system~\cite{2005CQGra..22..825B}
\begin{equation}
\dot{\Sigma}= -{4\over3}\,\bar{\Theta}(1-\bar{\Sigma})\Sigma-
3\bar{\Theta}^{-2}\clr_{\langle ab\rangle}\bar{\sigma}^{ab}
\label{lLdotSigma}
\end{equation}
and
\begin{equation}
\dot{S}= {4\over3}\,\bar{\Theta}(1-\bar{S})S+
3\bar{\Theta}^{-2}\clr_{\langle ab\rangle}\bar{\sigma}^{ab}\,,
\label{lLdotS}
\end{equation}
where the overdots indicate zero-order quantities. Here $S=1-\Sigma$ by definition and overbars indicate quantities of
zero perturbative order. Expression (\ref{lLdotSigma}) allows for a
gauge-independent description of linear shear anisotropies at the
$\bar{\Sigma}\rightarrow0$ limit, while Eq.~(\ref{lLdotS}) does the
same as $\bar{\Sigma}\rightarrow1$ (i.e.~when
$\bar{S}\rightarrow0$). In the first instance the last term in the
right-hand side of (\ref{lLdotSigma}) vanishes and therefore
$\Sigma\propto a^{-4}$, meaning that any kinematic anisotropies that may occur will quickly disperse. When $\bar{\Sigma}\rightarrow1$, on the other hand, the shear can increase further and therefore force
the Lukash solution away from the Bianchi~$VII_h$ family. In other
words, the linear stability of the Lukash universe appears to depend on the amount of the background shear anisotropy, something also
seen in the study of linear vortices and gravitational wave
perturbations (see~\cite{2005CQGra..22..825B} and
also~\cite{2005CQGra..22.3391H}).

\section{Kinetic theory and the cosmic neutrino and microwave backgrounds}\label{sKINETIC}

Kinetic theory provides a self-consistent description of a gas of
particles and naturally includes the free-streaming limit, where
collisions are negligible, and the hydrodynamic limit, where
collisions maintain tight-coupling and the gas approaches
fluid-like behaviour. In cosmology, only the free-streaming limit
is appropriate for the neutrino background for all times after
decoupling (at temperature $\sim 10^{11}\,\text{K}$). However, for
the cosmic microwave background (CMB), Compton scattering off
electrons in the electron-baryon plasma prior to recombination (at
temperature $\sim 3000\,\text{K}$) makes a fluid description
adequate on comoving scales larger than $30\, \text{Mpc}$. On
smaller scales, perturbations in the density of the radiation are
damped due to photon diffusion~\cite{1968ApJ...151..459S}. Kinetic
theory provides a seamless description of the associated
transitions from ideal fluid behaviour, through that of an
imperfect fluid to a free-streaming (collisionless) gas.

In this section we review the $1+3$-covariant formulation of
relativistic kinetic theory and its application to the (massive)
neutrino background and the anisotropies and polarization of the
CMB. Our emphasis is on the physics involved, but we do include
brief discussions of current observations where this is helpful.
For textbook discussions on relativistic kinetic theory, see
Refs.~\cite{1971LNP....10....1S,1980rkt.book.....G,1988kteu.book.....B}.

\subsection{Distribution functions and the Liouville equation}\label{ssDIST}

We consider a gas of identical particles each of mass $m$. The
four-momentum of a particle $p^a$ can be decomposed into an energy
and three-momentum with respect to the velocity field $u^a$ as
\begin{equation}
p^a = E u^a + \lambda^a = Eu^a + \lambda e^a \; ,
\end{equation}
where $\lambda = \sqrt{E^2 - m^2}$ is the magnitude of the
three-momentum and the projected vector $e^a$ is the propagation
direction. For massless particles, like photons, $\lambda=E$. In
situations where the polarization (or helicity) is not important,
we can describe the gas by a scalar-valued one-particle
distribution function $f(x^a,p^a)$. The number of particles in a
proper phase space element $\ud^3 \vx \ud^3 \vp$ is then $f \ud^3
\vx \ud^3 \vp$. For a given set of particles, their phase space
volume is both Lorentz invariant (i.e.\ the same for all
observers) and, in the absence of collisions, constant along their
path. Introducing an affine parameter $\tau$, normalised such that
$p^a = \ud x^a / \ud\tau$, the collisionless evolution of the gas
is described by the Liouville equation
\begin{equation}
{\ud f \over \ud \tau }= 0 \; , \label{eq:liouville}
\end{equation}
where the derivative is along the path in phase space. In the
presence of collisions, the right-hand side should be replaced by
the appropriate Lorentz-invariant collision operator $C[f]$, which
is a functional of $f$, to give a Boltzmann equation.

The 4-momentum of a free particle is parallel-transported so that
$p^b \nabla_b p^a = 0$. We can use this to find how the energy
(or $\lambda$) and direction propagate by projecting along and
perpendicular to $u^a$ respectively. For $\lambda$,
\begin{equation}
\frac{\ud \lambda}{\ud\tau} = - E^2 A_a e^a -
E\lambda\left(\sigma_{ab}e^ae^b + \frac{1}{3}\Theta\right).
\label{geodesic}
\end{equation}
In the Robertson-Walker limit, $\ud \lambda / \ud \tau = - E
\lambda H$ where $H = \Theta/3$ is the Hubble parameter. The
momentum thus redshifts as the inverse of the scale factor, $1/a$.
For the direction,
\begin{equation}
\frac{\ud e^{\langle a \rangle}}{\ud \tau} = - \frac{E^2}{\lambda}
s^a{}_b A^b - E \left(\omega^a{}_b e^b + s^{ab} \sigma_{bc} e^c
\right) , \label{direction}
\end{equation}
where $s_{ab} \equiv h_{ab} - e_a e_b$ is the {screen-projection
tensor} which projects into the two-dimensional screen
perpendicular to the propagation direction $e^a$ in the local
rest-space of $u^a$. Note that the derivative on the left of
Eq.~(\ref{direction}) is the covariant derivative along the particle's path.
The normalisation condition $e^a
e_a = 1$ is preserved since $\ud e^{\langle a \rangle} / \ud \tau$
is perpendicular to $e^a$. In the Robertson-Walker limit, $\ud
e^{\langle a \rangle} / \ud \tau=0$, i.e.\ $e^a$ is as constant as the
constraint $e^a u_a = 0$ allows. In the real universe this is no longer so and
Eq.~(\ref{direction}) then describes the action of gravitational
lensing (see e.g.\ Ref.~\cite{2001PhR...340..291B} for a recent
review).

The distribution function depends on spacetime position, energy
and direction. For the direction dependence, it is convenient to
expand in spherical multipoles
as~\cite{1981MNRAS.194..439T,1983AnPhy.150..455E}
\begin{equation}
f(x^a,p^a)=\sum_{l=0}^\infty F_{A_l}(x^a,E) e^{A_l} = F(E) +
F_a(E) e^a + F_{ab}(E) e^a e^b + \dots, \label{eq:multipoles}
\end{equation}
where the tensors $F_{A_l}(E) = F_{\langle a_1\dots
a_l\rangle}(E)$ are projected (orthogonal to $u^a$) symmetric and
trace-free so are irreducible under the action of
three-dimensional rotations. The expansion~(\ref{eq:multipoles})
is equivalent to an expansion in spherical harmonics, but has the
advantage of being fully covariant. We can invert the expansion as
follows:
\begin{equation}
F_{A_l}(E) = \frac{1}{\Delta_l} \int f e_{\langle A_l \rangle} \,
\ud \Omega \quad \text{where} \quad \Delta_l \equiv \frac{4\pi 2^l
(l!)^2}{(2l+1)!} ,
\end{equation}
and we have used
\begin{equation}
\int e_{\langle A_l \rangle} e^{\langle B_{l'} \rangle} \, \ud
\Omega = \Delta_l h_{\langle A_l \rangle}^{\langle B_l \rangle}
\delta_{ll'} = \Delta_l h_{\langle a_1}^{\langle b_1} \dots h_{a_l
\rangle}^{b_l \rangle} \delta_{ll'} .
\end{equation}
For further details on the covariant multipole expansion, see
e.g.\ Ref.~\cite{2000AnPhy.282..285G}.

The propagation equations for the multipole moments follows from
substituting Eq.~(\ref{eq:multipoles}) into the Boltzmann
equation, using Eqs.~(\ref{geodesic}) and~(\ref{direction}), and
extracting the irreducible terms. The result is
\begin{multline}
E \dot{F}_{\langle \Al\rangle} -  \lambda^2 \frac{\Theta}{3}
\frac{\partial F_\Al}{\partial E} + \frac{l+1}{2l+3} \lambda \D^a
F_{a\Al} + \lambda \D_{\langle a_l} F_{A_{l-1}\rangle}+ l E
F_{a\langle
A_{l-1}}\omega_{a_l\rangle}{}^a \\
-\left[ \lambda E \frac{\partial F_{\langle A_{l-1} }}{\partial E}
- (l-1)\frac{E^2}{\lambda} F_{\langle A_{l-1}}\right]
A_{a_l\rangle}
 - \frac{l+1}{2l+3}\left[ (l+2)\frac{E^2}{\lambda} F_{a\Al} + \lambda
E \frac{\partial F_{a\Al}}{\partial E} \right] A^a  \\
-\frac{l}{2l+3} \left[ 3E F_{a \langle A_{l-1}} + 2\lambda^2
\frac{\partial F_{a \langle A_{l-1}}}{\partial E} \right]
\sigma_{a_l\rangle}{}^a - \frac{(l+1)(l+2)}{(2l+3)(2l+5)} \left[
(l+3)E
F_{ab\Al} + \lambda^2\frac{\partial F_{ab\Al}}{\partial E}\right]\sigma^{ab} \\
-\left[\lambda^2\frac{\partial F_{\langle A_{l-2}}}{\partial E} -
(l-2)E F_{\langle A_{l-2}} \right]\sigma_{ a_{l-1}a_l\rangle}=
C_{\Al}[f] , \label{FAl}
\end{multline}
where $C_{\Al}[f]$ are the multipoles of the invariant collision
term. The spacetime derivatives are all taken at fixed $E$ (or
$\lambda$). This equation was first obtained in
Ref.~\cite{1983AnPhy.150..455E} (and in
Ref.~\cite{1981MNRAS.194..439T} for the massless case), but the
form given here benefits from the streamlined notation introduced
in Ref.~\cite{1997PhRvD..55..463M}. The original Boltzmann
equation contains a spacetime derivative along the particle path
which, when split into multipoles, connects the derivative of
$F_{\Al}(E)$ along $u^a$ with the projected derivatives of the
$l-1$ and $l+1$ moments. This describes the generation of
anisotropy from spatial inhomogeneities in the distribution
function. For a fluctuation of characteristic size $a/k$, the
timescale for propagating anisotropy through $\Delta l = 1$ is
$\sim a/(k v)$ where $v = \lambda / E$ is the magnitude of the
particle's 3-velocity. This is simply the time taken for a
particle to traverse the inhomogeneity. The terms in
Eq.~(\ref{FAl}) containing derivatives with respect to $E$ arise
from the redshifting of the particle's energy. We see from
Eq.~(\ref{geodesic}) that the isotropic expansion, acceleration
and shear source anisotropy at multipole $l$ from $l$, $l \pm 1$
and both $l \pm 2$ and $l$ respectively. The remaining terms in
Eq.~(\ref{FAl}) arise from the evolution of the particle's
direction: vorticity, acceleration and shear source anisotropy at
$l$ from $l$, $l\pm 1$ and both $l\pm 2$ and $l$ respectively.

\subsubsection{Bulk properties}\label{sssSTRESS}

The stress-energy tensor of the gas of particles is determined
from the one-particle distribution function by
\begin{equation}
T_{ab} = \int f p_a p_b \, \frac{\ud^3 \vp}{E} .
\end{equation}
Note that $\ud^3 \vp /E$ is the Lorentz-invariant volume element
on the positive-energy mass shell $p^a p_a = -m^2$. Decomposing
$T_{ab}$ into energy and momentum densities, isotropic pressure
and anisotropic stress, as in Eq.~(\ref{Tab1}), we have
\begin{eqnarray}
\rho &=& \Delta_0 \int_0^\infty \ud \lambda \, \lambda^2 E F ,
\label{eq:rhodef} \\
q_a &=& \Delta_1 \int_0^\infty \ud \lambda \, \lambda^2 E
(\lambda/E) F_a ,
\label{eq:qdef} \\
\pi_{ab} &=& \Delta_2 \int_0^\infty \ud \lambda \, \lambda^2 E
(\lambda/E)^2 F_{ab} , \label{eq:pidef} \\
p &=& \frac{\Delta_0}{3} \int _0^\infty \ud \lambda \, \lambda^2 E
(\lambda/E)^2 F ,
\end{eqnarray}
where $F$ is the monopole of the distribution function. The
equations of motion for these quantities follow from integrating
Eq.~(\ref{FAl}) with $\lambda^2 \ud \lambda$ and appropriate
powers of the velocity-weight $\lambda/E$. For the energy and
momentum densities,
\begin{eqnarray}
\dot{\rho} &=& - \Theta(\rho + p) - \D^a q_a - 2 A^a q_a -
\sigma^{ab} \pi_{ab}
+ \Delta_0 \int_0^\infty \ud \lambda \, \lambda^2 C_0[f] \label{eq:rhodot} \\
\dot{q}_{\langle a \rangle} &=& -\frac{4}{3}\Theta q_a + (\rho +
p) A_a
- \D_a p - \D^b \pi_{ab} - (\omega_a{}^b + \sigma_a{}^b)q_b \nonumber \\
&&\mbox{} - A^b \pi_{ab} + \Delta_1 \int_0^\infty \ud \lambda
\frac{\lambda^3}{E} C_a[f] , \label{eq:qdot}
\end{eqnarray}
where $C_0[f]$ is the monopole of the collision term. The gas can
exchange energy and momentum through interactions with external
particles and/or fields, and these processes are described by the
final terms in Eqs.~(\ref{eq:rhodot}) and~(\ref{eq:qdot}). In
their presence, the stress-tensor is not conserved but instead has
divergence
\begin{eqnarray}
\nabla^b T_{ab} &=& u_a \Delta_0 \int_0^\infty \ud \lambda \,
\lambda^2 C_0[f]
+ \Delta_1 \int \ud \lambda \frac{\lambda^3}{E} C_a[f] \\
&=& \int C[f] p_a \, \frac{\ud^3 \vp}{E} ,
\end{eqnarray}
which is manifestly a 4-vector.

The propagation equations for the energy and momentum densities do
not form a closed system even when there are no interactions. For
massive particles, the pressure is not simply related to $\rho$ and
so the equation of state is dynamical. Moreover, a propagation
equation for the anisotropic stress is also needed to close the
system. The required information is, of course, contained in the
original Boltzmann equation which can be recast as a
two-dimensional, infinite closed hierarchy for the moments of $f$
integrated over energy with positive (integer)
velocity-weights~\cite{1983AnPhy.150..455E,2002PhRvD..66b3531L}. The
integrated moments contain all those that appear in the
stress-tensor as a subset. The two-dimensional hierarchy simplifies
in a number of important special cases. For relativistic matter,
$\lambda \sim E$, the hierarchy becomes one-dimensional and this is
appropriate for the cosmological neutrino background at temperatures
$T \gg m$. For tightly-coupled collisional matter, such as the CMB
in the pre-recombination era when Thomson scattering is efficient
(see \S~\ref{ssCMB}), anisotropies at multipole $l$ are suppressed
by $(v_* k t_\mathrm{coll}/a)^l$, where $a/k$ is the scale of
inhomogeneity, $t_\mathrm{coll}$ is the collision time and $v_*$ is
a typical particle speed. Finally, for non-relativistic matter the
hierarchy can be truncated at low velocity weight\footnote{This is
accurate provided that the typical free-streaming distance per
Hubble time is small compared to the size of the inhomogeneity.} in
which case only a small number of moments need be propagated. The
latter truncation scheme is used to study the effect of velocity
dispersion on linear structure formation in
Refs.~\cite{2002PhRvD..66b3531L,1999PhRvD..60j3503M}.

The other important bulk properties of the gas are the particle and
entropy fluxes. We shall only consider the former here; for
relativistic thermodynamics, see e.g.\
Ref.~\cite{1934rtc..book.....T}. The particle flux is given in terms
of the distribution function by
\begin{equation}
N_a = \int f p_a \, \frac{\ud^3 \vp}{E} .
\end{equation}
The number density and particle drift evaluate to
\begin{eqnarray}
n &=& \Delta_0 \int_0^\infty \, \ud \lambda \lambda^2 F ,\\
\cln_a &=& \Delta_1 \int_0^\infty \, \ud \lambda \lambda^2
(\lambda/E) F_a .
\end{eqnarray}
The propagation equation for $n$ follows from integrating the
$l=0$ moment of Eq.~(\ref{FAl}) over $\lambda^2 \ud \lambda$:
\begin{equation}
\dot{n} = - \Theta n - \D^a \cln_a - A^a \cln_a + \Delta_0
\int_0^\infty \ud \lambda \frac{\lambda^2}{E} C_0[f] ,
\label{eq:ndot}
\end{equation}
so that the divergence of the particle flux is
\begin{eqnarray}
\nabla^a N_a &=& \Delta_0 \int_0^\infty \ud \lambda
\frac{\lambda^2}{E} C_0[f] \\
&=& \int C[f] \, \frac{\ud^3 \vp}{E} .
\end{eqnarray}
As for the stress-tensor, Eq.~(\ref{eq:ndot}) is part of a
two-dimensional infinite hierarchy of integrated
moments~\cite{1983AnPhy.150..455E}. With an approximate
truncation, this can be solved to determine the evolution of the
particle flux.

We end by noting that if we chose the frame $u^a$ such that $q_a =
0$ (the energy frame), the particle flux will generally not
vanish. Hence the energy and particle frame generally differ. An
important exception is for linear CMB fluctuations, where the
energy-dependence of the dipole $F_a(E)$ (and higher multipoles)
factorises; see \S~\ref{ssCMB}.

\subsubsection{Linearisation around FLRW cosmologies}\label{sssLIN_KIN}

An important result due to Ehlers, Geren \&
Sachs~\cite{ehlers:1344} follows from the exact multipole
equations~(\ref{FAl}): if there exists a family of free-falling
observers who measure freely-propagating self-gravitating
radiation to be exactly isotropic in some domain of a
dust-dominated universe, the spacetime is \emph{exactly} FLRW in
that region. Of course, the CMB is not exactly isotropic but the
result can be shown to be stable in the sense that if a family of
observers sees almost isotropic radiation, the universe is close
to FLRW~\cite{1995ApJ...443....1S} (but see
Ref.~\cite{2003GReGr..35..969C} for a critique of the technical
assumptions involved). Then, if we accept the (spacetime)
Copernican assumption, the currently observed isotropy of the CMB
implies that the geometry of the universe is well described by an
almost-FLRW model (at least since recombination).

For an FLRW model, isotropy demands that the distribution function
be isotropic and homogeneous, i.e.\ the only non-zero multipole is
the monopole, and this has vanishing projected gradient at fixed
energy. According to the discussion in \S~\ref{sss(CGI}, the $l
\geq 1$ multipoles and the projected derivative of the monopole
are, therefore, gauge-invariant measures of perturbations in the
distribution function about an FLRW model. As for most covariant
and gauge-invariant perturbations, the variables do, however,
depend on the choice of frame $u^a$. This dependence is discussed
further for massless particles in \S~\ref{sssTRANS}. If we
consider small departures from FLRW, the covariant and
gauge-invariant variables will themselves be small and we can
safely ignore products between small quantities.

In an FLRW background, the Liouville equation~(\ref{eq:liouville})
for collisionless matter reduces to
\begin{equation}
\frac{\partial f}{\partial t} - H \lambda \frac{\partial
f}{\partial \lambda} =0 .
\end{equation}
This is solved by $f = f(a\lambda)$ where $a$ is the
Robertson-Walker scale factor, as follows also from noting that
$a\lambda$ is conserved along the particle path. In the perturbed
universe, it is convenient to introduce the comoving momentum $q
\equiv a \lambda$ and energy $\epsilon \equiv a E$. It is $q$ that
is conserved in the background, while $\epsilon^2 = q^2 + a^2 m^2$
exactly. We can then write the distribution function as
$f(x^a,q,e^a)$ or in terms of angular multipoles $F_\Al(x^a,q)$.

Before proceeding, it is worth making the following remarks about
the scale factor $a$ in the perturbed universe. The scale factor is
defined by integrating $\dot{a}/a = \Theta/3$ and, for a given
choice of $u^a$, is only defined up to a hypersurface. The initial
hypersurface on which $a = \mathrm{const.}$ should be chosen
physically so as to ensure that $h_a \equiv \D_a a = 0$ if the model
is FLRW. The exact propagation equation for $h_a$ follows from
commuting the space and time derivatives to find
\begin{equation}
\dot{h}_{\langle a \rangle} = \frac{1}{3} a \left(\Theta A_a +
\D_a \Theta \right) - \sigma_a{}^b h_b + \omega_a{}^b h_b .
\label{eq:hdot}
\end{equation}
It follows that $\dot{h}_{a}$ is well-defined at first-order,
despite the first-order hypersurface ambiguity in both $h_a$ and
$a$.

The multipole form of the Boltzmann equation~(\ref{FAl}) contains
spacetime derivatives taken at fixed $E$ (or $\lambda$). If
instead we take the derivative at fixed $q$, we have
\begin{equation}
\nabla_a F_{A_l} |_\lambda = \nabla_a F_{A_l}|_q +
\left(\frac{1}{a}h_a - \frac{1}{3} \Theta a u_a \right) q
\frac{\partial F_{A_l}}{\partial q} .
\end{equation}
Using this result in Eq.~(\ref{FAl}), and dropping terms that are
second-order, we obtain the linearised multipole
equations~\cite{2002PhRvD..66b3531L}
\begin{eqnarray}
\dot{F}_{\langle \Al \rangle} &+& \frac{q}{\epsilon}
\frac{l+1}{2l+3} \D^b F_{b\Al} + \frac{q}{\epsilon} \D_{\langle
a_l}F_{A_{l-1}\rangle}+ ~\delta_{l1}\left(
\frac{1}{a}\frac{q}{\epsilon} h_{a_1} -\frac{\epsilon}{q}
A_{a_1}\right)q \frac{\partial F}{\partial q} \nonumber\\
&&{}\phantom{xxxxxxxxxxxx}-\delta_{l2}\sigma_{a_1 a_2} q
\frac{\partial F}{\partial q} = \frac{a}{\epsilon}C_{\Al}[f] ,
\label{qpoles}
\end{eqnarray}
where all spacetime derivatives are at fixed $q$. The $l=1$
equation contains the first-order combination
\begin{equation}
\clv_a(q) \equiv a \D_a F + h_a q \frac{\partial F}{\partial q} =
a \D_a F |_\lambda ,
\end{equation}
which removes the hypersurface ambiguity of $\D_a F$ and $h_a$.
The propagation equation for $\clv_a(q)$ follows from
Eq.~(\ref{qpoles}) for $l=0$. Taking the projected derivative and
commuting time and spatial derivatives gives
\begin{equation}
\dot{\clv}_{\langle a \rangle} = - \frac{a}{3}\frac{q}{\epsilon}
\D_a \D^b F_b + \dot{h}_{\langle a \rangle} q \frac{\partial
F}{\partial q} + a \D_a \left.
\left(\frac{C_0[f]}{E}\right)\right|_\lambda + a A_a
\frac{C_0[f]}{E} , \label{propv}
\end{equation}
with all spacetime derivatives at fixed $q$ except where stated
otherwise. In the absence of collisions, this equation, together
with the $l>0$ multipole equations, form a closed system (once
supplemented with the usual kinematic equations) with which one can
propagate the perturbations to the distribution function directly.
The collisionless form of these equations is well known in the
synchronous and Newtonian gauge~\cite{1995ApJ...455....7M} and are
what is usually used to propagate massive neutrino perturbations
numerically~\cite{1995ApJ...455....7M,1996ApJ...469..437S,%
1996ApJ...467...10D}.

The general strategy for solving Eq.~(\ref{qpoles}) is to
decompose the spatial dependence of $F_{\Al}$ into scalar, vector
and tensor parts which evolve independently in linear theory. The
essential idea is that the scalar component of $F_{\Al}$ is
obtained by taking the PSTF part of $l$ projected derivatives of
some scalar field, the vector component by the PSTF part of $l-1$
projected derivatives of a (projected) divergence-free vector
field, and the tensor component from the PSTF part of $l-2$
derivatives of a PSTF, divergence-free rank-2 tensor field. The
decomposition is unique, although the tensor  {potentials} are
generally not, provided the tensor field being expanded satisfies
appropriate boundary conditions (for non-compact
spaces)~\cite{1990CQGra...7.1169S}. In general, for $l>2$, a
rank-$l$ PSTF tensor can have higher-rank tensor contributions.
However, in linear theory there are no gravitational source terms
for the higher-rank contributions and so, starting from an early
epoch when interactions are efficient in maintaining isotropy, the
higher-rank contributions will not be present.

The tensor potentials for the scalar, vector and tensor
contributions can be expanded in terms of sets of appropriate
harmonic functions that are complete in FLRW spaces. Given FLRW
symmetry, a convenient choice is the eigenfunctions of the
(comoving) projected Laplacian $a^2 \D^2$. These can be chosen to be
orthogonal, and this property will be inherited by the tensors
derived from them. In the following paragraphs we briefly summarise
the properties of these harmonic functions, bringing together and
extending results in Refs.~\cite{1992ApJ...395...34B,%
1982PThPh..68..310T,1986ApJ...308..546A}.

\emph{Scalar perturbations}. For these we expand in terms of
scalar-valued eigenfunctions satisfying
\begin{equation}
a^2 \D^2 \clq^{(0)} + k^2 \clq^{(0)} = 0,
\end{equation}
with $\dot{\clq}^{(0)}=0$. These equations hold only at zero-order, i.e.
the harmonic functions are defined on the FLRW background. The
superscript $(0)$ denotes scalar perturbations, and, to avoid
clutter, in this section we suppress the index $(k)$. The allowed
eigenvalues $k^2$ depend on the spatial curvature of the
background model. Defining $\nu^2 = (k^2+K)/|K|$, where $6K/a^2$
is the curvature scalar of the FLRW spatial sections, the regular,
normalisable eigenfunctions have $\nu \geq 0$ for open and flat
models ($K \leq 0$). In Euclidean space, this implies all $k^2
\geq 0$. The $k=0$ solutions are homogeneous and, therefore, do
not appear in the expansion of first-order tensors, for example
$\Delta_a \equiv a \D_a \rho / \rho$. In open models, the modes
with $\nu \geq 0$ form a complete set for expanding
square-integrable functions, but they necessarily have $k \geq
\sqrt{|K|}$ and so cannot describe correlations longer than the
curvature scale~\cite{1995PhRvD..52.3338L}. Super-curvature
solutions (with $-1 < \nu^2 < 0$) can be constructed by analytic
continuation and Ref.~\cite{1995PhRvD..52.3338L} argues that these
should be included in an expansion of a general random field. A
super-curvature mode is generated in some models of open
inflation~\cite{1995PhRvD..52.3314B}. In closed models $\nu$ is
restricted to integer values $\geq
1$~\cite{1982PThPh..68..310T,1986ApJ...308..546A} and there are
$\nu^2$ linearly-independent modes per $\nu$. The mode with
$\nu=1$ cannot be used to construct perturbations (its projected
gradient vanishes globally), while the modes with $\nu=2$ can only
describe perturbations where all perturbed tensors with rank $>1$
vanish~\cite{1980PhRvD..22.1882B}.

For the scalar contribution to a rank-$l$ tensor, such as the
$l$-th multipole of the distribution function $F_\Al$, we expand
in rank-$l$ PSTF tensors $\clq^{(0)}_\Al$ derived from the
$\clq^{(0)}$ via~\cite{2000AnPhy.282..285G,1998PhRvD..58b3001C}
\begin{equation}
\clq^{(0)}_\Al = \left(\frac{-a}{k}\right)^{l}\D_{\langle
a_1}\dots \D_{a_l \rangle} \clq^{(0)}. \label{eq:52}
\end{equation}
The recursion relation for the $\clq^{(0)}_{A_l}$,
\begin{equation}
\clq^{(0)}_{A_l} = -\frac{a}{k} \D_{\langle
a_l}\clq^{(0)}_{A_{l-1}\rangle}, \label{eq:57}
\end{equation}
follows directly. The factor $a^l$ in the definition of the
$\clq^{(0)}_{A_l}$ ensures that $\dot{\clq}^{(0)}_{A_l}=0$ at
zero-order. The multipole equation~(\ref{qpoles}) also involves
the divergence of $\clq^{(0)}_{\Al}$, for which we need the
result~\cite{1999ApJ...513....1C,2000AnPhy.282..285G}
\begin{equation}
\D^{a_l} \clq^{(0)}_{A_l} =
\frac{k}{a}\frac{l}{(2l-1)}\left[1-(l^2-1)
\frac{K}{k^2}\right]\clq^{(0)}_{A_{l-1}} . \label{eq:58}
\end{equation}
When we discuss CMB polarization in \S~\ref{ssCMB}, we shall also
require the result that~\cite{2000PhRvD..62d3004C}
\begin{equation}
\text{curl}\, \clq^{(0)}_{A_l} = 0 , \label{eq:scalcurl}
\end{equation}
where the curl of a general rank-$l$ PSTF tensor is defined by
\begin{equation}
\text{curl}\, S_{\Al} = \varepsilon_{bc\langle a_l}\D^b
S_{A_{l-1}\rangle}{}^c . \label{eq:curl}
\end{equation}
In closed models, the $\clq^{(0)}_{\Al}$ vanish for $l \geq \nu$,
so only modes with $\nu > l$ contribute to rank-$l$ tensors.

The decomposition of the distribution function into angular
multipoles $F_{A_l}$, and the subsequent expansion in the
$\clq^{(0)}_{A_l}$, combine to give a normal mode expansion which
involves the objects $\D_{\langle \Al \rangle} \clq^{(0)}
e^{\Al}$. For $K=0$, with the $\clq^{(0)}$ taken to be Fourier
modes, this is equivalent to the usual Legendre expansion
$P_l(\hat{\vk}\cdot \ve)$ where $\hat{\vk}$ is the Fourier
wavevector (e.g.\ Ref.~\cite{1995ApJ...455....7M}). In non-flat
models, the expansion is equivalent to the Legendre tensor
approach, first introduced by Wilson~\cite{1983ApJ...273....2W}.
The advantage of handling the angular and scalar harmonic
decompositions separately is that the former can be applied quite
generally for an arbitrary cosmological model. Furthermore,
extending the normal-mode expansions to cover polarization and
vector and tensor modes in non-flat models is then rather trivial.

\emph{Vector perturbations}. For these we use the PSTF rank-1
eigenfunctions of the Laplacian,
\begin{equation}
a^2 \D^2 \clq^{(\pm 1)}_{a} + k^2 \clq^{(\pm 1)}_{a} = 0 ,
\label{eq:72}
\end{equation}
that are divergence-free, $\D^a \clq^{(\pm 1)}_{a}=0$, and have
vanishing time derivative, $\dot{\clq}^{(\pm 1)}_{a}=0$. The
superscript $(\pm 1)$ labels the two possible parities (electric
and magnetic) of the vector harmonics, e.g.\
Ref.~\cite{1982PThPh..68..310T}. We can always choose the parity
states so that
\begin{equation}
\text{curl}\, \clq^{(\pm 1)}_{a} = \frac{k}{a} \sqrt{1+
\frac{2K}{k^2}} \clq^{(\mp 1)}_{a} , \label{eq:curl1}
\end{equation}
which ensures that both parities have the same normalisation. For
vector modes we define $\nu^2=(k^2+2K)/|K|$. The regular,
normalisable eigenmodes have $\nu \geq 0$ for flat and open
models, while for closed models $\nu$ is an integer $\geq 2$.

We now differentiate the $\clq^{(\pm 1)}_a$ vectors $l-1$ times to
form PSTF tensors:
\begin{equation}
\clq^{(\pm 1)}_{A_l} \equiv \left(\frac{-a}{k}\right)^{l-1}
\D_{\langle A_{l-1}} \clq^{(\pm 1)}_{a_l\rangle} ,
\end{equation}
which are constant in time. They satisfy the same recursion
relation~(\ref{eq:57}) as the scalar harmonics. For the projected
divergences, we have~\cite{2004PhRvD..70d3518L}
\begin{equation}
\D^{a_l} \clq^{(\pm 1)}_{A_l} =  \frac{k}{a}
\frac{(l^2-1)}{l(2l-1)} \left[1-(l^2-2)\frac{K}{k^2}\right]
\clq^{(\pm 1)}_{A_{l-1}}. \label{eq:78}
\end{equation}
Finally, for polarization we shall require the result
\begin{equation}
\text{curl}\, \clq^{(\pm 1)}_{\Al} = \frac{1}{l}\frac{k}{a}
\sqrt{1+ \frac{2K}{k^2}} \clq^{(\mp 1)}_{\Al} . \label{eq:81}
\end{equation}
As for scalar perturbations, the $\clq^{(\pm 1)}_{\Al}$ vanish for
$l \geq \nu$ in closed models.

\emph{Tensor perturbations}. Here we use the PSTF rank-2
eigenfunctions of the Laplacian,
\begin{equation}
a^2 \D^2 \clq^{(\pm 2)}_{ab} + k^2 \clq^{(\pm 2)}_{ab} = 0 ,
\end{equation}
that are transverse, $\D^b \clq^{(\pm 2)}_{ab}=0$, and have
vanishing time derivative, $\dot{\clq}^{(\pm 2)}_{ab}=0$. The
superscript $(\pm 2)$ labels the two possible parity states for the
tensor harmonics~\cite{2000CQGra..17..871C,1982PThPh..68..310T,%
1980RvMP...52..299T}.For our purposes, the states can be
conveniently chosen so that
\begin{equation}
\text{curl}\, \clq^{(\pm 2)}_{ab} = \frac{k}{a} \sqrt{1+
\frac{3K}{k^2}} \clq^{(\mp 2)}_{ab} . \label{eq:curl2}
\end{equation}
For tensor modes we define $\nu^2=(k^2+3K)/|K|$. The regular,
normalisable eigenmodes have $\nu \geq 0$ for flat and open
models, while for closed models $\nu$ is an integer $\geq 3$.

Following our treatment of scalar perturbations, we form rank-$l$
PSTF tensors $\clq^{(\pm 2)}_{A_l}$ by differentiation:
\begin{equation}
\clq^{(\pm 2)}_{A_l} \equiv \left(\frac{-a}{k}\right)^{l-2}
\D_{\langle A_{l-2}} \clq^{(\pm 2)}_{a_{l-1} a_l\rangle} .
\label{eq:73}
\end{equation}
The $\clq^{(\pm 2)}_{A_l}$ satisfy the same recursion
relation~(\ref{eq:57}) as the scalar harmonics but their projected
divergences are~\cite{2000CQGra..17..871C}
\begin{equation}
\D^{a_l} \clq^{(\pm 2)}_{A_l} =  \frac{k}{a}
\frac{(l^2-4)}{l(2l-1)} \left[1-(l^2-3)\frac{K}{k^2}\right]
\clq^{(\pm 2)}_{A_{l-1}}.
\end{equation}
Finally, for polarization we require
\begin{equation}
\text{curl}\, \clq^{(\pm 2)}_{\Al} = \frac{2}{l}\frac{k}{a}
\sqrt{1+ \frac{3K}{k^2}} \clq^{(\mp 2)}_{\Al} .
\end{equation}
As before, in closed models, the $\clq^{(\pm 2)}_{\Al}$ vanish for
$l \geq \nu$.

Combining the angular and spatial expansions gives a set of
normal-mode functions going like $\D_{\langle A_{l-2}}\clq^{(\pm
2)}_{a_{l-1} a_l \rangle} e^\Al$ for tensor modes. This
generalises Wilson's approach~\cite{1983ApJ...273....2W} for
scalar perturbations to tensor modes. For the special case of a
flat universe, and working with circularly-polarized Fourier
modes,\footnote{Note that the (complex) Fourier modes are not
parity states and the $\text{curl}$ relations~(\ref{eq:curl1})
and~(\ref{eq:curl2}) do not hold directly.} the normal modes
reduce to $m= \pm 2$ spherical harmonics, $Y_{l \pm 2}(\ve)$, when
the Fourier wavevector lies along the $z$-axis. Similarly, for
vector modes $\D_{\langle A_{l-1}}\clq^{(\pm 1)}_{a_l \rangle}
e^\Al$ reduces to $m= \pm 1$ spherical harmonics, $Y_{l \pm
1}(\ve)$. In this limit, the normal-mode expansions are equivalent
to those in the  {total angular momentum} method of Hu \&
White~\cite{1997PhRvD..56..596H}.

Quite generally, the normal-mode functions for constructing
anisotropy from rank-$m$ perturbations are $\clq^{(m)}_\Al
e^{\Al}$. The evolution of these quantities along the line of
sight from some point $R$ is governed by the
recursion~\cite{1983ApJ...273....2W,1998PhRvD..57.3290H}
\begin{equation}
\frac{1}{k}\frac{\ud }{\ud \chi} \clq^{(m)}_\Al e^{\Al}
=\clq^{(m)}_{A_{l+1}} e^{A_{l+1}} -
\frac{(l^2-m^2)}{(2l+1)(2l-1)}\left[1-(l^2 - m - 1)\frac{K}{k^2}
\right] \clq^{(m)}_{A_{l-1}} e^{A_{l-1}} \label{eq:clqrecurs}
\end{equation}
where we have used the recursion relation~(\ref{eq:57}) and the
divergence result~(\ref{eq:58}) suitably generalised to rank-$m$
perturbations. Here, $\chi$ is the comoving radial distance along
the line of sight, so $\chi=0$ at $R$. The coupling of the $l$th
normal mode to $l\pm 1$ induces a similar coupling between the
anisotropy at multipoles $l$ and $l\pm 1$ (c.f.\ the advective
coupling in Eq.~(\ref{FAl})). The solution for the $l=m$ normal
mode along the line of sight can be written in terms of the $l
\geq m$ modes at $R$ as
\begin{equation}
\left. \clq^{(m)}_{A_m}e^{A_m} \right|_\chi = 4\pi
\sum_{l=m}^\infty \frac{1}{\Delta_l} \frac{l!}{(l-m)!} \frac{\nu
(\nu^2 + m + 1)^{(l-m)/2}}{\prod_{n=0}^l \sqrt{\nu^2 + n^2}}
\frac{\Phi_l^\nu(x)}{\sinh^m x} \left(\clq^{(m)}_{\Al}
e^{\Al}\right)_{\chi=0} , \label{eq:clqmproj}
\end{equation}
where $\nu^2 = [k^2 + (m+1)K]/|K|$, $x= \sqrt{|K|} \chi$ and the
$\Phi_l^\nu(x)$ are the ultra-spherical Bessel functions (see
e.g.\ Refs.~\cite{1986ApJ...308..546A}).\footnote{Note that
$\nu x \rightarrow k \chi$ in the limit $K \rightarrow 0$.} They are the
generalisation of spherical Bessel functions to $K \neq 0$ spaces
-- $\Phi^\nu_l(x) \rightarrow j_l(k\chi)$ in the limit of a flat
model -- and we see that $\Phi_l^\nu(x) / \sinh^m x$ give the
radial dependence of the rank-$m$ eigenfunctions
$\clq^{(m)}_{A_m}$.\footnote{ This agrees with the explicit
constructions in
Refs.~\cite{1982PThPh..68..310T,1986ApJ...308..546A} for the $m
\leq 2$ cases.} We have written Eq.~(\ref{eq:clqmproj}) in a form
appropriate for $K < 0$; for closed models $\nu^2 + n$ should be
replaced by $\nu^2 - n$, where $n$ is an integer, and the
hyperbolic functions by their trigonometric counterparts. For $K >
0$ the sum over $l$ truncates at $l = \nu-1$. Note that
Eq.~(\ref{eq:clqmproj}) is independent of the specific
representation of the $\clq^{(m)}_{A_m}$ that we choose.
Physically, it describes the linearised anisotropy pattern
generated at $R$ from a first-order source in the Boltzmann
equation of the form $\clq^{(m)}_{A_m} e^{A_m}$ acting at radial
distance $\chi$. For a given rank-$m$ perturbation, the response
to sources with $l > m$ local angular dependence, i.e.\ of the
form $\clq^{(m)}_\Al e^{\Al}$, follows from differentiating
Eq.~(\ref{eq:clqmproj}) and using the
recursion~(\ref{eq:clqrecurs}). In FLRW models, the sources for
the scalar distribution function are always of this form. We shall
see meet examples of these projections in the integral solutions
for the CMB anisotropies in \S~\ref{sssCMB_S}--\ref{sssCMB_T}.

\subsection{Cosmic neutrino background}\label{ssNEUTRINO}

In this subsection we discuss briefly the perturbations of the
cosmic neutrino background from the perspective of the
1+3-covariant kinetic theory developed above.

Flavour oscillations imply that at least two of the three neutrino
mass eigenstates have non-zero masses. The inferred squared-mass
differences $\Delta m_{21}^2 = 7.9 \times 10^{-5}\, \text{eV}^2$
and $|\Delta m_{31}^2| = 2.2 \times 10^{-3}\, \text{eV}^2$ imply
that the summed masses $\sum_i m_i \geq 0.056 \, \text{eV}$ for
the normal hierarchy ($\Delta m_{31}^2 > 0$) and $\sum_i m_i \geq
0.095 \, \text{eV}$ in the inverted hierarchy. All other things
being equal, the effect of non-zero neutrino masses is to suppress
the matter perturbations on scales below the neutrino Jeans scale
(see e.g.\ Ref.~\cite{2006PhR...429..307L} for a recent review and Refs.~\cite{1980SvAL....6..252D,1983ApJ...274..443B} for early pioneering work). Combining their three-year data with large-scale structure data, the WMAP team find $\sum_i m_i < 0.9 \, \text{eV}$ (95\% confidence) via this route in a flat
universe~\cite{2007ApJS..170..377S}. Cosmology therefore places an
important constraint on the absolute neutrino mass scale. Better
constraints can be obtained by including smaller scale
measurements of the matter power spectrum from the
(Lyman-$\alpha$) absorption spectra of distant
quasars~\cite{2006JCAP...10..014S,2006MNRAS.370L..51V}, although
there are some apparent inconsistencies between the CMB and
Lyman-$\alpha$ data that may invalidate the conclusions on
neutrino masses.

Neutrinos decouple at temperatures $\sim 1 \,\text{MeV}$ when they
are ultra-relativistic. Since they were in thermal equilibrium
before this time, the distribution function in the FLRW limit is
\begin{equation}
f(q) \approx \left[\exp\left(\frac{q}{k_{\text{B}}T_d
a_d}\right)+1 \right]^{-1} , \label{eq:sn1}
\end{equation}
where we have set the chemical potential to zero. Here, $T_d$ is
the temperature at neutrino decoupling when the scale factor is
$a_d$. Note that when expressed in terms of the comoving momentum,
the distribution function is time independent. Once the neutrinos
become non-relativistic (typical momentum $\lambda \ll m$), the
distribution function starts to depart from Fermi-Dirac form since
it is the momentum not energy that redshifts with the expansion of
the universe. Defining the neutrino temperature $T_\nu \equiv a_d
T_d / a$, at redshift zero $T_\nu$ is related to the CMB
temperature by $T_\nu = (4/11)^{1/3} T_{\text{CMB}}=1.96 \,
\text{K}$; the CMB temperature is higher due to photon production
at electron-positron annihilation.

The current neutrino temperature and the inferred mass differences
imply that at least two of the mass eigenstates are
non-relativistic at the present epoch. Noting that the average
momentum
\begin{equation}
\langle \lambda \rangle = \frac{7\pi^4}{180 \zeta(3)} k_{\text{B}}
T_\nu \approx 3.15 k_{\text{B}} T_\nu ,
\end{equation}
the current upper limit of $0.3 \, \text{eV}$ for the mass of any
eigenstate implies that neutrinos were relativistic at the time of
hydrogen recombination. For this reason, light neutrino
\emph{masses} only affect the CMB indirectly through changes to
the angular-diameter distance to last scattering. Their effect on
the growth of small-scale matter perturbations is significant
though~\cite{1980PhRvL..45.1980B} and is the mechanism by which
cosmological observations constrain the (summed) absolute neutrino
mass.

In the following subsections we discuss the dynamics of scalar and
tensor-mode neutrino perturbations. We include the latter since it
has the potentially observable consequence of damping sub-horizon
gravitational waves during radiation
domination~\cite{2004PhRvD..69b3503W}.

\subsubsection{Scalar perturbations}\label{sssNEUTRINO_S}

For scalar perturbations we expand the multipoles $F_\Al(q)$ in
terms of the $\clq^{(0)}_{\Al}$:
\begin{equation}
F_{\Al}(q) = - \frac{\pi}{\Delta_l} \frac{\ud F(q)}{\ud \ln q}
\sum_k F_l^{(0)}(q) \clq_\Al^{(0)} , \qquad l \geq 1,
\end{equation}
where $F(q)$ is the zero-order monopole of the distribution
function, Eq.~(\ref{eq:sn1}). Here, we have left the harmonic
index $(k)$ implicit on the scalar harmonic functions and their
coefficients. The momentum-dependent prefactor is chosen so that
\emph{for massless particles} the $F_l^{(0)}$ are independent of
$q$ (see Eq.~(\ref{eq:sn2}) below) and
\begin{eqnarray}
q_a &=& \rho \sum_k F_1^{(0)} \clq^{(0)}_a , \qquad (m=0) , \\
\pi_{ab} &=& \rho \sum_k F_2^{(0)} \clq^{(0)}_{ab}, \qquad (m=0).
\label{eq:sn1a}
\end{eqnarray}
In the massless case, $\sum_k F_l^{(0)} \clq^{(0)}_{\Al}$ are
proportional to the multipoles of the neutrino temperature
anisotropy. For the gradient of the monopole we use
\begin{equation}
\clv_a(q) = \frac{k}{4} \frac{\ud F(q)}{\ud \ln q} \sum_k
F_0^{(0)}(q) \clq_a^{(0)} ,
\end{equation}
so that, on integrating over $q$,
\begin{equation}
\Delta_a  = \frac{a}{\rho} \D_a \rho = - \sum_k k F_0^{(0)}
\clq^{(0)}_a , \qquad (m=0) ,
\end{equation}
and hence $\sum_k F_0^{(0)}\clq^{(0)}$ is essentially $\delta \rho
/ \rho$ on hypersurfaces orthogonal to $u^a$ (up to a constant).

After they decouple, the neutrino multipoles satisfy
Eqs.~(\ref{qpoles}) and (\ref{propv}) with vanishing collision
terms. Expanding in harmonics, we find~\cite{2002PhRvD..66b3531L}
\begin{eqnarray}
\dot{F}_l^{(0)}(q) &+& \frac{k}{a}\frac{q}{\epsilon}\left\{
\frac{l+1}{2l+1} \left[1-\left((l+1)^2
-1\right)\frac{K}{k^2}\right] F_{l+1}^{(0)}(q)
- \frac{l}{2l+1} F_{l-1}^{(0)}(q)\right\} \nonumber \\
&&\mbox{} + 4 \delta_{l0} \dot{h} +
\delta_{l1}\frac{4}{3}\frac{k}{a}\left(\frac{q}{\epsilon} h +
\frac{\epsilon}{q}A\right) + \delta_{l2} \frac{8}{15} \frac{k}{a}
\sigma = 0 , \label{eq:sn2}
\end{eqnarray}
where the kinematic quantities are
\begin{equation}
h_a = - \sum_k kh \clq^{(0)}_a ,\qquad A_a = \sum_k \frac{k}{a} A
\clq^{(0)}_a , \quad \sigma_{ab} = \sum_k \frac{k}{a} \sigma
\clq^{(0)}_{ab} . \label{eq:hAsig}
\end{equation}
When most of the neutrinos are relativistic, we can set
$q=\epsilon$ in Eq.~(\ref{eq:sn2}) and we recover massless
dynamics. On sub-horizon scales, neutrinos are unable to cluster
and free-streaming excites multipoles higher than $l=1$. At late
times, for those species that are non-relativistic, free-streaming
effectively turns off for $k/a \ll H \epsilon/q$ which defines the
neutrino Jeans length $\lambda_{\text{J}} \sim v_*/H$ where $v_*$
is the typical thermal velocity. When free-streaming is not
operating, the $F_l^{(0)}(q)$ are constant for $l>2$. During
matter domination, the comoving Jeans length \emph{falls} (as
$a^{-1/2}$) so that neutrinos start to cluster again at late times
for modes that are sub-Hubble at the non-relativistic transition
once their thermal velocities have redshifted sufficiently. For
modes that are inside the Jeans length today, the clustered
baryons and CDM have never felt the gravity of clustered
neutrinos. All other things being equal, increasing the neutrino
mass increases the expansion rate by a constant factor when
neutrinos are very non-relativistic, and so structure formation is
slowed down on small scales and the matter power spectrum is
suppressed by a constant factor proportional to the summed
neutrino masses~\cite{1998PhRvL..80.5255H}. There is no such
effect on large scales since the enhanced expansion rate is
mitigated by neutrino clustering. A comprehensive review of
neutrino scalar perturbations is given in
Ref.~\cite{2006PhR...429..307L}.

We have focussed on the effects of massive neutrinos on matter
clustering. However, if in the future it were possible to detect
the cosmic neutrino background directly, its anisotropies would
open up a rich new source of cosmological information. The power
spectra of the anisotropies can easily be computed from the
solution of Eq.~(\ref{eq:sn2}) using the methods described for the
CMB in \S~\ref{ssCMB}; see Ref.~\cite{2007JCAP...01...14M} for
further details.

\subsubsection{Tensor perturbations}\label{sssNEUTRINO_T}

For tensor perturbations, we expand the $F_\Al(q)$ as
\begin{equation}
F_{\Al}(q) = - \frac{\pi}{\Delta_l} \frac{\ud F(q)}{\ud \ln q}
\sum_k F_l^{(\pm 2)}(q) \clq_\Al^{(\pm 2)} , \qquad l \geq 2,
\end{equation}
where the sum over modes $\sum_k$ includes the two parity states
labelled with a superscript $(\pm 2)$. Expanding
Eq.~(\ref{qpoles}) in harmonics gives~\cite{2002PhRvD..66b3531L}
\begin{eqnarray}
\dot{F}_l^{(\pm 2)}(q) &+& \frac{k}{a}\frac{q}{\epsilon}\left\{
\frac{(l+3)(l-1)}{(2l+1)(l+1)}\left[1-\left((l+1)^2 -3\right)
\frac{K}{k^2}\right] F_{l+1}^{(\pm 2)}(q) - \frac{l}{2l+1}
F_{l-1}^{(\pm 2)}(q)\right\}
\nonumber \\
&&\mbox{} \phantom{xxxxxxxxxxx} + \delta_{l2} \frac{8}{15}
\frac{k}{a} \sigma^{(\pm 2)} = 0 , \label{eq:tn1}
\end{eqnarray}
where the shear $\sigma_{ab} = \sum_k (k/a) \sigma^{(\pm 2)}
\clq_{ab}^{(\pm 2)}$. The physics of Eq.~(\ref{eq:tn1}) is
straightforward: the anisotropic expansion due to the shear of
gravitational waves continually sources quadrupole anisotropy;
advection moves this power up to higher $l$ with $\Delta l= 1$
taking the typical thermal crossing time across the wavelength of
the perturbation. For perturbations outside the neutrino Jeans
scale, advection is ineffective within an expansion time and the
$l>2$ multipoles become approximately constant.

An important application of Eq.~(\ref{eq:tn1}) is the damping of
gravitational waves by neutrino
free-streaming~\cite{2004PhRvD..69b3503W,1972ApJ...176..323S}. The
evolution of cosmological gravitational waves was considered in
\S~\ref{ssGWPs} in terms of the shear. However, to make contact
with the metric-based literature, it is convenient to work with
the transverse, trace-free metric perturbation $\clh_{ab} = a^2
\clh_{ij} (\ud x^i)_a (\ud x^j)_b$, where
\begin{equation}
\ud s^2 = a^2 \left[-\ud \eta^2 + (\gamma_{ij} + \clh_{ij}) \ud
x^i \ud x^j \right]
\end{equation}
and $\gamma_{ij}$ the background (conformal) spatial metric. Note
that $\clh_{ij}$ is gauge-invariant~\cite{1980PhRvD..22.1882B}. In
terms of the metric, the shear and the electric part of the Weyl
tensor are (e.g.\ Ref.~\cite{1990CQGra...7.1169S})
\begin{eqnarray}
\sigma_{ab} &=& \frac{1}{2} \dot{\clh}_{ab} \, \\
E_{ab} &=& -\frac{1}{4}\frac{1}{a^2}\clh_{ab}''+ \frac{K}{2a^2}
\clh_{ab} - \frac{1}{4} \D^2 \clh_{ab} ,
\end{eqnarray}
where primes denote the action of $a u^a \nabla_a$. The equation
of motion for $\clh_{ab}$ follows from the shear propagation
equation; we find
\begin{equation}
\clh_{ab}'' + 2 \frac{a'}{a}\clh_{ab}' - \left(a^2 \D^2 - 2
K\right)\clh_{ab} = 2a^2 \pi_{ab} . \label{eq:tn2a}
\end{equation}
After neutrino decoupling, but during radiation domination, the
neutrino anisotropic stress is dynamically important for
sub-Hubble scales. Before decoupling, scattering keeps the
neutrinos isotropic, while after matter-radiation equality, the
role of the neutrino stress is suppressed by the ratio $f_\nu$ of
neutrino energy density to the total energy density. During
radiation domination, $f_\nu \approx 0.405$ is constant and the
harmonic expansion of $\clh_{ab}$ evolves with conformal time as
\begin{equation}
\clh^{(\pm 2)\prime\prime} + \frac{2}{\eta} \clh^{(\pm 2)\prime} +
(k^2 + 2K) \clh^{(\pm 2)} = \frac{6 f_\nu}{\eta^2} F_2^{(\pm 2)}
\label{eq:tn2}
\end{equation}
for massless neutrinos. (We have used the tensor-mode version of
Eq.~(\ref{eq:sn1a}) to relate $F_2^{(\pm 2)}$ to the neutrino
anisotropic stress.) In the absence of anisotropic stress, the
regular solution of Eq.~(\ref{eq:tn2}) is $\clh^{(\pm 2)} \propto
j_0(k\eta)$ in a flat universe. The metric perturbation is thus
constant outside the horizon but then decays as $a$ after horizon
crossing.

Quite generally, Eqs.~(\ref{eq:tn1}) and (\ref{eq:tn2a}) can be
recast as an integro-differential using the following integral
solution for the $F_l^{(\pm 2)}$ in the massless limit in a flat
universe,
\begin{equation}
F_l^{(\pm 2)} = -4 l(l-1) \int^\eta \ud \eta' k \sigma^{(\pm
2)}(\eta') \frac{j_l(k \Delta \eta)}{(k\Delta \eta)^2} ,
\end{equation}
where $\Delta \eta \equiv \eta - \eta'$.\footnote{Unsurprisingly,
the integral solution for massless neutrinos is the same as for
the CMB in the absence of scattering; see \S~\ref{sssCMB_S}.} The
lower limit of integration should be the end of neutrino
decoupling, but it is harmless to approximate this as $\eta=0$.
Combining the quadrupole solution with the harmonic expansion of
Eq.~(\ref{eq:tn2}) for $K=0$, gives~\cite{2004PhRvD..69b3503W}
\begin{equation}
\clh^{(\pm 2)\prime\prime} + \frac{2a'}{a} \clh^{(\pm 2)\prime} +
k^2 \clh^{(\pm 2)} = -24 f_\nu(\eta) \left(\frac{a'}{a}\right)^2
\int_0^\eta \ud \eta' \, \clh^{(\pm 2)\prime}(\eta')
\frac{j_2(k\Delta \eta)}{(k\Delta \eta)^2} , \label{eq:tn3}
\end{equation}
where we have used $\clh^{(\pm 2)\prime} = 2 k \sigma^{(\pm 2)}$.
Modes that enter the horizon well before matter-radiation
equality, but well after neutrino decoupling, have their amplitude
damped by a factor $\approx 0.803$ with essentially no phase
shift, i.e.\ $\clh^{(\pm 2)}(\eta)\approx 0.803 \clh^{(\pm 2)}(0)
j_0(k\eta)$ during radiation
domination~\cite{2004PhRvD..69b3503W}. However, the observational
consequences of such modes is limited since their wavelengths are
too short to affect the CMB and too long for direct detection with
laser interferometers. CMB polarization is dominated by modes
entering the horizon around recombination and the universe is not
fully matter-dominated at this epoch. The asymptotic results of
Ref.~\cite{2004PhRvD..69b3503W} are not applicable in this limit
and a full numerical solution of Eq.~(\ref{eq:tn3}) is required
for accurate results. Both the phase shift and damping of the
amplitude turn out to be important and their interplay can both
enhance or reduce the polarization power by a few percent
depending on scale.\footnote{For further discussion of this point,
see the unpublished notes by Antony Lewis that accompany the CAMB
code; they are available as
http://cosmologist.info/notes/CAMB.ps.gz.}

\subsection{Cosmic microwave background}\label{ssCMB}

Since the detection of anisotropies in the temperature of CMB
radiation by COBE~\cite{1992ApJ...396L...1S}, the CMB has played a
major role in establishing quantitative constraints on the
cosmological model. The small $O(10^{-5})$ amplitude of these
fluctuations means they are well described by linear perturbation
theory and the physics of the CMB is thus very well understood. For
detailed recent reviews, see
Refs.~\cite{2002ARA&A..40..171H,2003AnPhy.303..203H,%
2005LNP...653...71C,2008CL}

The prediction of angular variations in the temperature of the
radiation, due to the propagation of photons through an
inhomogeneous universe~\cite{1967ApJ...147...73S}, followed shortly
after the (definitive) discovery of the CMB in 1965 by Penzias \&
Wilson~\cite{1965ApJ...142..419P}. Shortly after, polarization in
the CMB was predicted in models with anisotropy in the expansion
around the time of recombination~\cite{1968ApJ...153L...1R}. The
detailed physics of CMB fluctuations in almost-FLRW models was
essentially understood by the early 1970s~\cite{1968ApJ...151..459S,%
1968ApJ...153....1P,1968ZhETF..55..278Z,1970ApJ...162..815P,%
1970Ap&SS...7....3S}for models with only baryonic matter;
cold-dark-matter models were considered a decade
later~\cite{1982ApJ...263L...1P,1984ApJ...285L..45B}. Further
important developments included the effect of spatial
curvature~\cite{1983ApJ...273....2W},
polarization~\cite{1983MNRAS.202.1169K,1984ApJ...285L..45B} and
gravitational waves~\cite{1969MNRAS.144..255D,1985SvA....29..607P}.
All of these works used the standard metric-based approach to
cosmological perturbation theory, but the physics of the CMB has
also been studied extensively in the 1+3-covariant
approach~\cite{1999ApJ...513....1C,2000CQGra..17..871C,%
2000AnPhy.282..285G,1995ApJ...443....1S,1998PhRvD..58b3001C,%
2000PhRvD..62d3004C,2004PhRvD..70d3518L,1995PhRvD..51.1525M,%
1995PhRvD..51.5942M,1997ApJ...476..435S,1997CQGra..14.3391D,%
1999PhRvD..59h3506M,%
2000AnPhy.282..321G,2000GReGr..32.1059C,2000ApJ...538..473L,%
2004PhRvD..70d3011L}. This brings to the CMB the benefits described
in \S~\ref{sIRCs}, and in particular: (i) clarity in the definition
of the variables employed; (ii) covariant and gauge-invariant
perturbation theory around a variety of background models; (iii)
provision of a sound basis for studying non-linear effects; and (iv)
freedom to employ any coordinate system or tetrad.

In this subsection our main focus is on the linear theory of CMB
anisotropies, but we present most of the basic 1+3-covariant
framework in a non-perturbative manner. The approach therefore
provides a convenient starting point for non-linear treatments of
CMB fluctuations~\cite{1999PhRvD..59h3506M}. There are several
non-linear effects that are expected to affect the CMB at an
important level on scales below a few arcminutes. These include
gravitational
lensing~\cite{1987A&A...184....1B,1989MNRAS.239..195C,%
1990MNRAS.243..353L,1996ApJ...463....1S} (see, also,
Ref.~\cite{2006PhR...429....1L} for a comprehensive review), various
scattering effects either during~\cite{1998ApJ...508..435G} or after
the universe reionized~\cite{1972CoASP...4..173S,%
1986ApJ...306L..51O,1987ApJ...322..597V} (see e.g.\
Ref.~\cite{1999PhR...310...97B} for a review) and gravitational
redshifting effects~\cite{1968Natur.217..511R}. For polarization,
the gravitational lensing effect may even be the dominant
contribution to the $B$-mode
component~\cite{1997PhRvL..78.2058K,1997PhRvL..78.2054S} on all
angular scales~\cite{1998PhRvD..58b3003Z}. A complete non-linear
computation of the CMB anisotropies in almost-FLRW models is still
lacking, but there has been some progress made recently at
second-order in perturbation
theory~\cite{2006JCAP...06..024B,2007JCAP...01...19B}. (However,
note that non-perturbative effects are known to be important for
gravitational lensing of the CMB~\cite{2005PhRvD..71j3010C}).

\subsubsection{CMB observables}\label{sssPOL}

We are interested in both the total intensity and polarization
properties of the CMB. Therefore, we describe the CMB photons by a
one-particle distribution function that is tensor-valued: $f_{b
c}(x^a,p^a)$~\cite{1982qed..book.....L}. It is a Hermitian tensor
defined so that the expected number of photons contained in a
proper phase-space element $\ud^3 \vx \ud^3 \vp$, and with
polarization state $\epsilon^a$ is $\epsilon^a{}^* f_{a b}
\epsilon^b \ud^3 \vx \ud^3 \vp$. The complex polarization 4-vector
$\epsilon^a$ is orthogonal to the photon momentum, $\epsilon^a p_a
= 0$ (adopting the Lorentz gauge), and is normalised as
$\epsilon_a^* \epsilon^a = 1$. The distribution function is also
defined to be orthogonal to $p^a$ so $f_{ab} p^a = 0$. We can make
a 1+3-covariant decomposition of the photon 4-momentum as
\begin{equation}
p^a = E(u^a + e^a) ,
\end{equation}
where, now, the magnitude of the momentum $\lambda$ equals the
energy $E$ since photons are massless. For a photon in a pure
polarization state $\epsilon^a$, the direction of the electric
field relative to $u^a$ is $s^a{}_b \epsilon^b$ where, recall,
$s_{ab} \equiv h_{ab} - e_a e_b$ is the screen projection tensor.

The (Lorentz-gauge) polarization 4-vector is only unique up to
constant multiples of $p^a$, reflecting the remaining gauge
freedom, but the \emph{observed} polarization vector $s^a{}_b
\epsilon^b$ is unique. As the residual gauge freedom also affects
the distribution function $f_{ab}$, it is sometimes convenient to
work directly with the (screen-)projected polarization tensor,
\begin{equation}
P_{ab} \propto E^3 s_a{}^c s_b{}^d f_{cd} \label{eq:obspol}
\end{equation}
which governs the observable properties of the radiation field
from the perspective of $u^a$. It is unaffected by the residual
electromagnetic gauge freedom. The factor $E^3$ is included in the
definition of $P_{ab}$ so that it relates simply to the observed
Stokes brightness parameters for the radiation field. Decomposing
$P_{ab}$ into its irreducible components,
\begin{equation}
P_{ab}(E,e^d) = \frac{1}{2}I(E,e^d) s_{ab} + \clp_{ab}(E,e^d)
+\frac{1}{2} i V(E,e^d) \varepsilon_{abc}e^c, \label{eq:5}
\end{equation}
defines the total intensity brightness, $I$, the circular
polarization, $V$, and the linear polarization tensor $\clp_{ab}$
which is PSTF and transverse to $e^a$. The projected polarization
tensor can, alternatively, be interpreted in terms of classical
electromagnetic fields: for quasi-monochromatic radiation with
electric field $\Re [E^a(t) \exp(-i \omega t)]$, where $\omega$ is
the angular frequency and the complex representative $E^a$ varies
little over a wave period,
\begin{equation}
P^{ab} \propto \langle E^a E^{b*} \rangle .
\end{equation}
Here, the angle brackets denote time averaging. The linear
polarization is often described in terms of Stokes brightness
parameters $Q$ and $U$ (e.g.\ Ref.~\cite{1960ratr.book.....C})
which, operationally, measure the difference in intensity between
radiation transmitted by a pair of orthogonal polarizers (for
$Q$), and the same but after a right-handed rotation of the
polarizers by 45 degrees about the propagation direction $e^a$
(for $U$). If we introduce a pair of orthogonal polarization
vectors $(e_1)^a$ and $(e_2)^a$, which are perpendicular to $u^a$
and $e^a$, i.e.\ $s_b^a (e_i)^b = (e_i)^a$ for $i=1$ and $2$, and
are oriented so that $\{ u^a, (e_1)^a, (e_2)^a, e^a \}$ form a
right-handed orthonormal tetrad, we have
\begin{equation}
\clp_{ab}(e_i)^a (e_j)^b = \frac{1}{2} \left( \begin{array}{cc}
Q & U \\
U  & - Q \end{array} \right) . \label{eq:4}
\end{equation}
The invariant $2\clp^{ab} \clp_{ab} = Q^2+U^2$ is the magnitude
(squared) of the linear polarization.

Since $I(E,e^c)$ and $V(E,e^c)$ are scalar functions on the sphere
$e^a e_a = 1$ at a point in spacetime, their local angular
dependence can be handled by an expansion in PSTF tensor-valued
multipoles, as in Eq.~(\ref{eq:multipoles}):
\begin{eqnarray}
I(E,e^c) & = & \sum_{l=0}^\infty I_{A_l}(E) e^{A_l} , \label{eq:8}\\
V(E,e^c) & = & \sum_{l=0}^\infty V_{A_l}(E) e^{A_l} . \label{eq:9}
\end{eqnarray}
For $\clp_{ab}$, we use the fact that any STF tensor on the sphere
can be written in terms of angular derivatives of two scalar
potentials, $P_E$ and $P_B$, as (e.g.\
Ref.~\cite{1997PhRvL..78.2058K})
\begin{equation}
\clp_{ab} = {}^{(2)}\nabla_{\langle a} {}^{(2)}\nabla_{b\rangle}
P_E + \epsilon^c{}_{\langle a}{}^{(2)}\nabla_{b\rangle}
{}^{(2)}\nabla_c P_B ,
\end{equation}
where ${}^{(2)}\nabla_a$ and $\epsilon_{ab} = \varepsilon_{abc}
e^c$ are the covariant derivative and alternating tensor on the
two-sphere. The scalar fields $P_E$ and $P_B$ are even and odd
under parity respectively, and define the  {electric} and
 {magnetic} parts of the linear polarization. Expanding $P_E$
and $P_B$ in PSTF multipoles in the usual way, and evaluating the
angular derivatives, we can
write~\cite{2000PhRvD..62d3004C,1980RvMP...52..299T}
\begin{equation}
\clp_{ab}(E,e^c) = \sum_{l=2}^\infty[\cle_{ab C_{l-2}}(E)
e^{C_{l-2}}]^{\TT} - \sum_{l=2}^\infty [e_{d_1}\varepsilon^{d_1
d_2}{}_{(a}\clb_{b)d_2 C_{l-2}}(E) e^{C_{l-2}}]^{\TT}.
\label{eq:11}
\end{equation}
Here, TT denotes the transverse (to $e^a$), trace-free part: for a
general second-rank tensor $F_{ab}$
\begin{equation}
[F_{ab}]^{\TT} = s_a^c s_b^d F_{cd} - \frac{1}{2} s_{ab} s^{cd}
F_{cd} . \label{eq:7}
\end{equation}
Equation~(\ref{eq:11}) can be inverted to determine the PSTF
tensors $\cle_\Al$ and $\clb_\Al$ as
\begin{eqnarray}
\cle_{A_l}(E) & = & M_l{}^2 \Delta_l{}^{-1} \int \d\Omega\,
e_{\langle A_{l-2}} \clp_{a_{l-1} a_l \rangle} (E,e^c) ,
\label{eq:12} \\
\clb_{A_l}(E) & = & M_l{}^2 \Delta_l{}^{-1} \int \d\Omega\, e_b
\epsilon^{bd}{}_{\langle a_l} e_{A_{l-2}} \clp_{a_{l-1}\rangle d}
(E,e^c), \label{eq:13}
\end{eqnarray}
where $M_l \equiv \sqrt{2l(l-1)/[(l+1)(l+2)]}$. The multipole
expansion in Eq.~(\ref{eq:11}) is the coordinate-free version of
the tensor spherical harmonic expansion introduced to the analysis
of CMB polarization in Ref.~\cite{1997PhRvL..78.2058K}. An
alternative expansion, whereby $Q \pm i U$ is expanded in
spin-weighted spherical harmonics (e.g.\
Ref.~\cite{1967JMP..8..2155G}), is also commonly
employed~\cite{1997PhRvL..78.2054S}. The expansion~(\ref{eq:11})
is also equivalent to that introduced in the 1970s by Dautcourt \&
Rose~\cite{1978AN....299...13D}.

In observational cosmology, the CMB anisotropy is usually
expressed in terms of thermodynamic equivalent temperature, i.e.\
the distribution function $I(E,e^a)/E^3$ divided by $\partial
f_{\text{Pl}}/\partial T_{\text{CMB}}$, where
$f_{\text{Pl}}(E/T_{\text{CMB}})$ is the Planck function at the
average CMB temperature.\footnote{$T_{\text{CMB}} = 2.725\,
\text{K}$~\cite{2002ApJ...581..817F}.} Ignoring spectral
distortions, the linear-theory CMB \emph{anisotropy} and
polarization brightness are independent of energy when expressed
as thermodynamic temperatures. This is because the \emph{linear}
perturbations in $f_{ab}$ inherit the spectral dependence of
$\epsilon \partial f_{\text{Pl}} / \partial
\epsilon$.\footnote{The spectral dependence follows from, for
example, the form of the source terms in Eq.~(\ref{qpoles}) for
massless particles, and from the energy dependence of the
linearised scattering term in Eq.~(\ref{eq:35b}) below.} For the
CMB, it follows that we can integrate over energy without loss of
information so we define bolometric multipoles
\begin{equation}
I_{A_l} = \Delta_l \int_0^\infty \ud E\, I_{A_l}(E) \label{eq:14}
\end{equation}
for $l \geq 0$. The normalisation is chosen so that the three
lowest multipoles give the radiation energy and momentum densities
and anisotropic stress respectively (c.f.\
Eqs.~(\ref{eq:rhodef})--(\ref{eq:pidef})):
\begin{equation}
I=\rho, \qquad I_a = q_a, \qquad I_{ab} = \pi_{ab}^{(\gamma)}.
\label{eq:15}
\end{equation}
We define $\cle_{A_l}$, $\clb_{A_l}$ and $V_{A_l}$ similarly. The
fractional anisotropy in the CMB temperature, $\delta_T(e^a)$, is
then related to the $I_{\Al}$ by
\begin{eqnarray}
\delta_T(e^c) &=& \frac{\pi}{I} \int_0^\infty \ud E\, I'(E,e^a) \\
              &=& \frac{\pi}{I} \sum_{l=1}^\infty \Delta_l{}^{-1}
I_{A_l} e^{A_l} \label{eq:17}
\end{eqnarray}
to first order, where $I'(E,e^a)$ is the brightness anisotropy.

For theories, such as single-field inflation, that predict initial
perturbations that are very close to being Gaussian distributed
(see Ref.~\cite{2004PhR...402..103B} for a recent review), the CMB
fluctuations should also be Gaussian distributed where linear
theory applies. If we further assume that the statistical
properties of the fluctuations are invariant under the isometries
of the background cosmological model (i.e.\ translations and
rotations for FLRW), the CMB power spectra fully characterise the
statistics of the CMB anisotropies and polarization.\footnote{We
shall not be concerned here with spectral distortions in the CMB,
which are an important probe of the energetics of the universe.
For a recent review of this topic, and future prospects, see e.g.\
Ref.~\cite{2003MNRAS.342..543B}.} The temperature power spectrum
is defined in terms of the $I_{\Al}$
by~\cite{2000AnPhy.282..285G}:
\begin{equation}
\left(\frac{\pi}{I}\right)^2 \langle I_{A_l} I^{B_{l'}}\rangle =
\Delta_l C_l^{T}\delta_l^{l'} h_{\langle A_l \rangle}^{\langle B_l
\rangle}. \label{eq:16}
\end{equation}
where $h_{\langle A_l \rangle}^{\langle B_l \rangle}\equiv
h_{\langle a_1}^{\langle b_1}\dots h_{a_l \rangle}^{b_l \rangle}$.
The angle brackets denote a statistical average over the ensemble
of fluctuations. Equation~(\ref{eq:16}) is entirely equivalent to
the usual definition of the anisotropy power spectrum in terms of
the variance of $a_{lm}$ with $\delta_T(e) = \sum_{l > 0} a_{lm}
Y_{lm}(e)$. The temperature correlation function evaluates to
\begin{equation}
\langle \delta_T(e^c) \delta_T(e^{\prime c}) \rangle =
\sum_{l=1}^\infty \frac{(2l+1)}{4\pi} C_l^{T} P_l(\cos\theta),
\label{eq:18}
\end{equation}
where $\theta$ is the angle between the directions $e$ and $e'$,
and $P_l$ is a Legendre polynomial. In deriving Eq.~(\ref{eq:18})
we used the result $e^{\langle A_l \rangle} e'_{\langle A_l
\rangle} = (2l+1)\Delta_l P_l(\cos\theta)/(4\pi)$.

The power spectra for the polarization are defined similarly, but,
since we choose our conventions for the power spectra to conform
with Ref.~\cite{1997PhRvL..78.2054S}, it it is necessary to
include an additional factor of $M_l/\sqrt{2}$ for each factor of
the polarization. For example, for $E$-modes
\begin{equation}
\left(\frac{\pi}{I}\right)^2 \langle \cle_{A_l}
\cle^{B_{l'}}\rangle = \frac{l(l-1)}{(l+1)(l+2)}\Delta_l C_l^{E}
\delta_l^{l'} h_{\langle A_l \rangle}^{\langle B_l \rangle}.
\label{eq:19}
\end{equation}
Unfortunately, the definitions of the polarization power spectra
given in Ref.~\cite{1997PhRvL..78.2054S} differ from those in
Ref.~\cite{1997PhRvL..78.2058K} by factors of $\sqrt{2}$; see
Ref.~\cite{1997PhRvD..55.7368K} for details. For a
parity-symmetric ensemble, $B$-mode polarization is uncorrelated
with the $E$-mode and the temperature anisotropies. The
correlation functions of linear polarization are most simply
expressed in terms of the Stokes parameters. For two propagation
directions $e$ and $e'$, define Stokes parameter $\bar{Q}$,
$\bar{U}$ and $\bar{Q}'$ and $\bar{U}'$ using the direction of the
tangent to the spherical geodesic connecting $e$ and $e'$ to
define the $(e_1)^a$ basis vectors at the two points. Then the
non-zero correlation functions assuming parity symmetry
are~\cite{1997PhRvD..55.7368K,1999IJMPD...8...61N,2004MNRAS.350..914C}
\begin{eqnarray}
\langle \bar{Q}\bar{Q}' \rangle &=& \frac{1}{2} \sum_l
\frac{2l+1}{4\pi}[C_l^E (d^l_{2\,2} + d^l_{2\,-2})(\theta) + C_l^B
(d^l_{2\,2} - d^l_{2\,-2})(\theta)]
\nonumber \\
\langle \bar{U} \bar{U}' \rangle &=& \frac{1}{2} \sum_l
\frac{2l+1}{4\pi}[C_l^E (d^l_{2\,2} - d^l_{2\,-2})(\theta) + C_l^B
(d^l_{2\,2} + d^l_{2\,-2})(\theta)] , \label{pol:eq11}
\end{eqnarray}
where $d^l_{mn}$ are the reduced Wigner functions and the Stokes
parameters are expressed in thermodynamic temperature, i.e.\
$Q(e^a) = (\pi /I)\int \ud E \, Q(E,e^a)$. The correlation
properties of the linear polarization tensor $\clp_{ab}(e^c)
\equiv (\pi/I) \int \d E\, \clp_{ab}(E,e^c)$
are easily expressed in terms of these results. For example,\footnote{%
This result corrects Eq.~(2.20) of
Ref.~\cite{2000PhRvD..62d3004C}.}
\begin{equation}
2 \langle \clp_{ab}(e^c) \clp^{ab}(e^{\prime c}) \rangle =
\frac{1}{2}(1+\cos^2\theta) \langle \bar{Q}\bar{Q}' \rangle +
\cos\theta \langle \bar{U} \bar{U}' \rangle . \label{eq:20}
\end{equation}
The geometric factors $(1+\cos^2\theta)$ and $\cos\theta$ appear
due to the contractions between the $(e_1)^a$ basis vectors at the
two points. These would not be present if, instead, we
parallel-transported $\clp_{ab}(e^c)$ on the 2-sphere to the point
$e'$ and performed the contraction there.

\subsubsection{Transformation properties under change of frame}
\label{sssTRANS}

The phase-space volume element $\ud^3 \vx \ud^3 \vp$ is
Lorentz-invariant and so $\epsilon^a{}^* f_{ab} \epsilon^b$ is a
Lorentz scalar for any polarization 4-vector $\epsilon^a$. It
follows that $f_{ab}$ is properly covariant under Lorentz
transformations. To see the implications of this for the observed
polarization, consider a new velocity field
$\tilde{u}^a=\gamma(u^a+v^a)$, where $v^a$ is the projected relative
velocity in the $u^a$ frame and $\gamma$ is the associated Lorentz
factor. For a given photon with 4-momentum $p^a$, the energy and
propagation directions in the $\tilde{u}_a$ frame are given by the
Doppler and aberration formulae:
\begin{eqnarray}
\tilde{E} & = & \gamma E (1-e^a v_a), \label{eq:21} \\
\tilde{e}^a & = & [\gamma(1-e^b v_b)]^{-1}(u^a +
e^a)-\gamma(u^a+v^a). \label{eq:22}
\end{eqnarray}
Note that $\tilde{e}^a$ is a projected vector relative to
$\tilde{u}_a$. The screen projection tensor for a given null
direction transforms to
\begin{equation}
\tilde{s}_{ab} = s_{ab} + \frac{2\gamma}{\tilde{E}}
p_{(a}s_{b)c}v^c + \frac{\gamma^2}{\tilde{E}^2}p_a p_b s_{cd} v^c
v^d . \label{eq:23}
\end{equation}
For any vector orthogonal to $p^a$, for example the 4-polarization
$\epsilon^a$,
\begin{equation}
\tilde{s}^a{}_c s^c{}_b \epsilon^b = \tilde{s}^a{}_b \epsilon^b
\qquad (p^a \epsilon_a = 0) .
\end{equation}
>From this it follows that the screen-projected direction of
polarization transforms as
\begin{equation}
s^a{}_b \epsilon^b \mapsto \tilde{s}^a{}_b \epsilon^b =
\tilde{s}^a{}_c (s^c{}_b \epsilon^b) ,
\end{equation}
and the observed polarization tensor by~\cite{2000PhRvD..62d3004C}
\begin{equation}
\tilde{E}^{-3}\tilde{P}_{ab}(\tilde{E},\tilde{e}^c)=E^{-3}\tilde{s}_a^{d_1}
\tilde{s}_b^{d_2} P_{d_1 d_2}(E,e^c). \label{eq:24}
\end{equation}
Under this transformation law the intensity, circular polarization
and linear polarization do not mix. Moreover, $I(E,e^c)/E^3$ and
$V(E,e^c)/E^3$ are frame-invariant for a given null direction, and
the transformation law for $\clp_{ab}(E,e^c)$ follows that for
$P_{ab}(e^c)$. The degree of linear polarization
$[2\clp_{ab}(E,e^c)\clp^{ab}(E,e^c)]^{1/2}/I(E,e^c)$ is invariant
under changes of frame. The transformation law for $P_{ab}(E,e^c)$
ensures that the tetrad components of $P_{ab}(E,e^c)/E^3$, and
hence the Stokes parameters divided by $E^3$, are invariant if the
polarization basis vectors are transformed as
\begin{equation}
(\tilde{e}_i)^a = \tilde{s}^a_b (e_i)^b, \label{eq:25}
\end{equation}
for $i=1,2$. An alternative way of viewing Eq.~(\ref{eq:25}) is to
note that, in terms of components on an orthonormal tetrad in the
$u^a$ frame and its Lorentz-boosted counterpart in the
$\tilde{u}^a$ frame, the basis vector $\ve_1$ at direction $\ve$
is parallel-propagated along the geodesic connecting $\ve$ and
$\tilde{\ve}$ on the 2-sphere to obtain
$\tilde{\ve}_1$~\cite{2002PhRvD..65j3001C}.

Under changes of frame, multipoles with different $l$ mix because
of Doppler and beaming effects. Using the invariance of
$I(E,e^a)/E^3$, the multipoles $I_\Al(E)$ and the
energy-integrated $I_\Al$ transform as
\begin{eqnarray}
\tilde{I}_\Al(\tilde{E}) &=& \Delta_l^{-1} \sum_{l'} \int \ud
\Omega \, \gamma(1-e^b v_b) I_{B_{l'}}[\gamma^{-1}\tilde{E}/(1-
e^c v_c)] e^{B_{l'}} \tilde{e}_{\langle \Al \rangle} , \\
\tilde{I}_\Al &=& \sum_{l'} \Delta_{l'}^{-1} I_{B_{l'}} \int \ud
\Omega \, [\gamma(1-e^b v_b)]^2 e^{B_{l'}} \tilde{e}_{\langle \Al
\rangle} ,
\end{eqnarray}
where the PSTF $\tilde{e}_{\langle \Al \rangle}$ is with respect
to $\tilde{u}^a$. The transformation of the energy-dependent
multipoles is non-local in energy and therefore a little messy;
expansions in terms of $v^a$ can be found in
Ref.~\cite{2002PhRvD..65j3001C}. The transformation of the
energy-integrated multipoles are somewhat simpler: to first order
in the relative velocity~\cite{2000PhRvD..62d3004C},
\begin{equation}
\tilde{I}_{A_l} = \tilde{h}^{\langle B_l \rangle}_{\langle A_l
\rangle} I_{B_l} + (l-2) v^b I_{b A_l} - \frac{l(l+3)}{(2l+1)}
v_{\langle a_l}I_{A_{l-1}\rangle}. \label{eq:26}
\end{equation}
An equivalent result holds for $V_{A_l}$. The linear polarization
is more complicated since $E$ and $B$ mix. To first-order in
$v^a$,
\begin{eqnarray}
\tilde{\cle}_{A_l} &=& \tilde{h}^{\langle B_l \rangle}_{\langle
A_l \rangle} \cle_{B_l} - \frac{l(l+3)}{(2l+1)} v_{\langle
a_l}\cle_{A_{l-1}\rangle}
+ \frac{(l-2)(l-1)(l+3)}{(l+1)^2} v^b \cle_{b A_l} \nonumber \\
&&\mbox{} +\frac{6}{(l+1)} v_b\varepsilon^{bc}{}_{\langle
a_l}\clb_{A_{l-1}\rangle c},
\label{eq:27}\\
\tilde{\clb}_{A_l} &=& \tilde{h}^{\langle B_l \rangle}_{\langle
A_l \rangle} \clb_{B_l} - \frac{l(l+3)}{(2l+1)} v_{\langle
a_l}\clb_{A_{l-1}\rangle}
+ \frac{(l-2)(l-1)(l+3)}{(l+1)^2} v^b \clb_{b A_l} \nonumber \\
&&\mbox{} -\frac{6}{(l+1)} v_b\varepsilon^{bc}{}_{\langle
a_l}\cle_{A_{l-1}\rangle c}. \label{eq:28}
\end{eqnarray}
In an almost-FLRW model the polarization is a first-order
quantity, as are physically-defined relative velocities. It
follows that the $E$ and $B$-mode multipoles are frame-invariant
in linear theory.

\subsubsection{Radiative transfer}
\label{sssTRANSFER}

The phase-space volume element $\ud^3 \vx \ud^3 \vp$ is conserved
along a photon path and the polarization 4-vector $\epsilon^a$ is
parallel-transported. It follows that $f_{ab}$ is also
parallel-propagated in phase space if there are no collisions (see
e.g.\ Ref.~\cite{1989CQGra...6.1171B} for a rigorous discussion).
The observed polarization tensor is given by
Eq.~(\ref{eq:obspol}); multiplying by $E^3$ and taking the
derivative in phase space gives~\cite{2000PhRvD..62d3004C}
\begin{equation}
\cll [E^{-3}P_{ab}(E,e^c)]= 0, \label{eq:29a}
\end{equation}
where the Liouville operator $\cll$ acts on transverse tensors
$A_{ab}=[A_{ab} ]^{\TT}$, like $P_{ab}$, as
\begin{equation}
\cll[A_{ab}(E,e^c)] \equiv s_a^{d_1} s_b^{d_2} p^e \nabla_e A_{d_1
d_2} (E,e^c) . \label{eq:30}
\end{equation}
Physically, this means the observed polarization tensor is
propagated as parallel as its projection properties allow. If we
now include collisions, we obtain the exact Boltzmann equation
\begin{equation}
\cll [E^{-3}P_{ab}(E,e^c)]=K_{ab}(E,e^c), \label{eq:29}
\end{equation}
where $K_{ab}(E,e^c)$ is the fully projected (with $s_{ab}$) form
of the invariant collision tensor. If we change frame to
$\tilde{u}^a \equiv \gamma (u^a + v^a)$, the projected collision
tensor transforms like $P_{ab} / E^3$, i.e.\
\begin{equation}
\tilde{K}_{ab}(\tilde{E},\tilde{e}^c)=\tilde{s}_a^{d_1}
\tilde{s}_b^{d_2} K_{d_1 d_2}(E,e^c). \label{eq:34}
\end{equation}
This result is useful since the scattering tensor is often
simplest to evaluate in some preferred frame, picked out by the
physics of the scattering process. The scattering tensor in a
general frame then follows from the transformation~(\ref{eq:34}).

The Liouville operator $\cll$ preserves the irreducible
decomposition of the polarization tensor [Eq.~(\ref{eq:5})], so
that
\begin{eqnarray}
\cll[E^{-3}P_{ab}(E,e^c)] &=& \frac{1}{2} \frac{\d}{\d\lambda}
[E^{-3}I(E,e^c)]s_{ab} + \cll[E^{-3}\clp_{ab}(E,e^c)] \nonumber \\
&&\mbox{}+\frac{1}{2} i
\frac{\d}{\d\lambda}[E^{-3}V(E,e^c)]\varepsilon_{abd}e^d.
\label{eq:31}
\end{eqnarray}
If we propagate the (transverse) polarization basis vectors
$(e_i)^a$ according to
\begin{equation}
s_{ab} p^c \nabla_c (e_i)^b =0, \label{eq:32}
\end{equation}
then all four Stokes brightness parameters (divided by $E^3$) are
constant along the photon path when there are no collisions.

For the CMB, the dominant collisional process over the epochs of
interest for the formation of anisotropies and polarization
(around recombination and reionization) is Compton scattering. To
an excellent approximation we can ignore the effects of Pauli
blocking, induced scattering, and electron recoil in the rest
frame of the scattering electron, so that the scattering may be
approximated by classical Thomson scattering in the electron rest
frame with no change in the photon energy. Furthermore, we can
neglect the small velocity dispersion of the electrons due to
their finite temperature and so consider scattering off a cold gas
of electrons with proper number density $\tilde{n}_e$ in the
electron rest frame.\footnote{An important exception of relevance
for secondary anisotropies in the CMB is Compton scattering in hot
intra-cluster gas -- the Sunyaev-Zel'dovich
effect~\cite{1972CoASP...4..173S} (see
Ref.~\cite{1999PhR...310...97B} for a review) -- where the
electron temperatures can be $\sim O(10) \,\text{keV}$. The
low-frequency decrement in total intensity of the CMB in the
direction of clusters is proportional to the electron temperature.
Finite temperature effects for polarization are considered in
Refs.~\cite{2000MNRAS.312..159C,2000ApJ...533..588I,2004astro.ph.12095P}
and can give up to $\sim O(10)\%$ corrections to the results for
cold intra-cluster gas at the spectral peak of the signal.}
Denoting the rest frame by $\tilde{u}^a$, the \emph{exact}
projected collision tensor in the Thomson limit in that frame
is~\cite{2000PhRvD..62d3004C}\footnote{%
This corrects two sign errors in the right-hand side of Eq.~(3.7)
of Ref.~\cite{2000PhRvD..62d3004C}. There, the term involving
$\tilde{\cle}_{ab}$ in the first line and $\tilde{I}_{ab}$ in the
second have the wrong sign. } (see also
Ref.~\cite{2004astro.ph.12094P} for Klein-Nishina corrections)
\begin{eqnarray}
\tilde{E}^2 \tilde{K}_{ab}(\tilde{E},\tilde{e}^c) &=&
\tilde{n}_{{\text{e}}} \sigt \biggl\{\frac{1}{2} \tilde{s}_{ab}
\left[-\tilde{I}(\tilde{E},\tilde{e}^c) + \tilde{I}(\tilde{E}) +
{\frac{1}{10}} \tilde{I}_{d_1 d_2}(\tilde{E})\tilde{e}^{d_1}
\tilde{e}^{d_2} - {\frac{3}{5}} \tilde{\cle}_{d_1 d_2}(\tilde{E})
\tilde{e}^{d_1}
\tilde{e}^{d_2}\right] \nonumber \\
&&\mbox{}+ \left[- \tilde{\clp}_{ab}(\tilde{E},\tilde{e}^c) -
{\frac{1}{10}} [\tilde{I}_{ab}(\tilde{E})]^{\TT}
+ {\frac{3}{5}} [\tilde{\cle}_{ab}(\tilde{E})]^{\TT} \right] \nonumber \\
&&\mbox{}+ {\frac{1}{2}}i \tilde{\varepsilon}_{abd_1}
\tilde{e}^{d_1} \left[- \tilde{V}(\tilde{E},\tilde{e}^c) +
{\frac{1}{2}} \tilde{V}_{d_2} (\tilde{E})
\tilde{e}^{d_2}\right]\biggr\}, \label{eq:35}
\end{eqnarray}
where $\sigt$ is the Thomson cross section. This expression for
the scattering tensor follows from inserting the multipole
decomposition of the polarization tensor into the Kernel for
Thomson in-scattering (e.g. Ref.~\cite{1960ratr.book.....C}), and
integrating over scattering directions. An expression equivalent
to Eq.~(\ref{eq:35}) appears to have been first derived in
Ref.~\cite{1978AN....299...13D}.

Some general observations follow from
Eq.~(\ref{eq:35})~\cite{2000PhRvD..62d3004C,1997PhRvD..56..596H,%
1978AN....299...13D}. It is written in irreducible form with the first set of terms on the right affecting the total intensity, the second set the linear polarization and the third the circular polarization. In each case, scattering out of the phase-space element is described by $-\tilde{n}_{\text{e}} \sigt P_{ab}$. For $I$, in-scattering couples to the monopole and quadrupole in total intensity, and to the $E$-mode quadrupole. Comparison with Eqs.~(\ref{eq:rhodot}) and~(\ref{eq:qdot}) show that there is no change in energy density in the electron rest frame due to Thomson scattering, but there is momentum exchange if the radiation has a dipole moment. Linear polarization is generated by in-scattering of the quadrupoles in total intensity and $E$-mode polarization. Comparison with Eq.~(\ref{eq:11}) shows that in the electron rest-frame, the polarization is \emph{generated} purely as an $E$-mode quadrupole. The transformations~(\ref{eq:27}) and~(\ref{eq:28}) show that this is generally not true in some other frame. Finally, we see that circular polarization is fully decoupled from total intensity and linear polarization. In consequence, in any frame the circular polarization will remain exactly zero if it is initially.

The scattering tensor~(\ref{eq:35}) is first-order in small
quantities about an FLRW background. Transforming to a general
frame $u^a$, and keeping only first-order terms, we find
\begin{eqnarray}
E^2 K_{ab}(E,e^c) &=& n_{\text{e}}\sigt\biggl\{\frac{1}{2}
s_{ab}\left[-I(E,e^c) + I(E) - e^c v_c E^4
\frac{\partial}{\partial E}
\left(\frac{I(E)}{E^3}\right) \right. \nonumber \\
&&\mbox{} \phantom{n_{\text{e}}\sigt\biggl\{\frac{1}{2} s_{ab}++}
\left. +\frac{1}{10} I_{d_1 d_2}(E) e^{d_1}e^{d^2}
-\frac{3}{5}\cle_{d_1 d_2}(E) e^{d_1} e^{d_2}\right] \nonumber \\
&&\mbox{} \phantom{n_{\text{e}}\sigt\biggl\{\frac{1}{2} s_{ab}++}
+ \left[-\clp_{ab}(E,e^c)-\frac{1}{10}[I_{ab}(E,e^c)]^{\text{TT}}
+ \frac{3}{5}[\cle_{ab}(E,e^c)]^{\text{TT}}\right] \nonumber \\
&&\mbox{} \phantom{n_{\text{e}}\sigt\biggl\{\frac{1}{2} s_{ab}++}
+\frac{1}{2} i
\varepsilon_{abd_1}e^{d_1}\left[-V(E,e^c)+\frac{1}{2}
V_{d_2}(E)e^{d_2}\right] \biggr\} , \label{eq:35b}
\end{eqnarray}
where $n_e$ is the electron density relative to $u^a$.

The Boltzmann equation~(\ref{eq:29}) can be written in multipole
form by expressing $P_{ab}(E,e^c)$ as a multipole expansion using
Eqs.~(\ref{eq:8}), (\ref{eq:9}), and (\ref{eq:11}), and
decomposing the resulting equation into multipoles. This leads to
four sets of multipole hierarchies for $I_{A_l}(E)$,
$\cle_{A_l}(E)$, $\clb_{A_l}(E)$, and $V_{A_l}(E)$. In linear
theory, those for $I_{A_l}(E)$ and $\cle_{A_l}(E)$ are coupled by
Thomson scattering, and, generally, those for $\cle_{A_l}(E)$ and
$\clb_{A_l}(E)$ are coupled by advection (see below). The exact
equations for total intensity and circular polarization follow
Eq.~(\ref{FAl}) but with, for example, $F_\Al(\lambda)$ replaced
by $I_\Al(E)/E^3$, $E=\lambda$ and the collision multipoles
\begin{eqnarray}
E^2 C_\Al[f] &=& n_{\text{e}}\sigt \left[ -I_\Al(E) + I(E)
\delta_{l0} - v_{a_1} E^4 \frac{\partial}{\partial E}
\left(\frac{I(E)}{E^3}\right) \delta_{l1} \right. \nonumber \\
&&\mbox{} \phantom{n_{\text{e}}\sigt ++} \left.
+\left(\frac{1}{10} I_{a_1 a_2}(E) -\frac{3}{5}\cle_{a_1
a_2}(E)\right) \delta_{l2} + \dots \right].
\end{eqnarray}
The multipole equations for $\cle_{A_l}(E)$ and $\clb_{A_l}(E)$
are more involved than those for the intensity or circular
polarization and only the linearised equations (about FLRW) have
been calculated to date~\cite{2000PhRvD..62d3004C}. (The $l=2$
equation was given in an orthonormal tetrad in
Ref.~\cite{1978AN....299...13D} under the assumption that the
higher multipoles vanish.) In linear form, for the $E$-mode
polarization
\begin{eqnarray}
\dot{\cle}_{A_l}(E) &-& \frac{1}{3}\Theta
E^4\frac{\partial}{\partial E} [E^{-3}\cle_{A_l}(E)] + \D_{\langle
a_{l}}\cle_{A_{l-1}\rangle} (E) + \frac{(l+3)(l-1)}{(l+1)(2l+3)}
\D^b \cle_{b A_l}(E)
\nonumber \\
&-& \frac{2}{(l+1)} \curl \clb_{A_l}(E) = -n_{\text{e}} \sigt
\cle_{A_l}(E) - \frac{1}{10}n_{\text{e}} \sigt[I_{a_1 a_2}(E) - 6
\cle_{a_1 a_2}(E)] \delta_{l2} , \label{eq:43}
\end{eqnarray}
and for the $B$-mode
\begin{eqnarray}
\dot{\clb}_{A_l}(E) &-& \frac{1}{3}\Theta
E^4\frac{\partial}{\partial E} [E^{-3}\clb_{A_l}(E)] + \D_{\langle
a_l}\clb_{A_{l-1}\rangle} (E) + \frac{(l+3)(l-1)}{(l+1)(2l+3)}
\D^b \clb_{b A_l}(E)
\nonumber \\
&+& \frac{2}{(l+1)} \curl \cle_{A_l}(E) = -n_{\text{e}} \sigt
\clb_{A_l}(E). \label{eq:44}
\end{eqnarray}
The $E$ and $B$-mode multipoles are coupled by the curl terms
where, recall, the curl of a PSTF tensor is defined by
Eq.~(\ref{eq:curl}). In a general almost-FLRW cosmology, $B$-mode
polarization is generated only by advection of the $E$-mode. This
does not happen if the perturbations about FLRW are curl-free, as
is the case for scalar perturbations. We thus have the important
result that linear scalar perturbations do not generate $B$-mode
polarization~\cite{1997PhRvL..78.2058K,1997PhRvL..78.2054S}; see
also \S~\ref{sssCMB_S}--\ref{sssCMB_V}.

The equations for the energy-integrated multipoles follow from
integrating Eqs.~(\ref{FAl}), (\ref{eq:43}) and~(\ref{eq:44}) over
energy. In linear form,
\begin{eqnarray}
\dot{\cle}_{A_l} &+&\frac{4}{3}\Theta \cle_{A_l} +
\frac{(l+3)(l-1)}{(l+1)^2} \D^b \cle_{b A_l} + \frac{l}{(2l+1)}
\D_{\langle a_l}\cle_{A_{l-1}\rangle}
\nonumber \\
&-& \frac{2}{(l+1)} \curl \clb_{A_l} = - n_{\text{e}} \sigt \left[
\cle_{A_l} + \left(\frac{1}{10} I_{a_1 a_2} - \frac{3}{5}
\cle_{a_1 a_2}\right)\delta_{l2}\right] , \label{eq:47}\\
\dot{\clb}_{A_l} &+&\frac{4}{3}\Theta \clb_{A_l} +
\frac{(l+3)(l-1)}{(l+1)^2} \D^b \clb_{b A_l} + \frac{l}{(2l+1)}
\D_{\langle a_l}\clb_{A_{l-1}\rangle}
\nonumber \\
&+& \frac{2}{(l+1)} \curl \cle_{A_l} = - n_{\text{e}} \sigt
\clb_{A_l}. \label{eq:48}
\end{eqnarray}
For the circular polarization,
\begin{equation}
\dot{V}_{A_l} + \frac{4}{3}\Theta V_{A_l} + \D^b V_{b A_l} +
\frac{l}{(2l+1)}\D_{\langle a_l} V_{A_{l-1} \rangle}  =
-n_{\text{e}} \sigt \left(V_{A_l} -\frac{1}{2} V_{a_1} \delta_{l1}
\right), \label{eq:49}
\end{equation}
and for the total intensity
\begin{eqnarray}
\dot{I}_{A_l} &+& \frac{4}{3}\Theta I_{A_l} + \D^b I_{b A_l} +
\frac{l}{(2l+1)} \D_{\langle a_l} I_{A_{l-1}\rangle} + \frac{4}{3}
I A_{a_1}\delta_{l1} + \frac{8}{15} I \sigma_{a_1 a_2}
\delta_{l2} \nonumber \\
&=& -n_{\text{e}} \sigt \left[ I_{A_l} - I \delta_{l0} -
\frac{4}{3} I v_{a_1} \delta_{l1} - \left(\frac{1}{10}I_{a_1 a_2}
- \frac{3}{5}\cle_{a_1 a_2} \right) \delta_{l2} \right] .
\label{eq:50}
\end{eqnarray}
The monopole moment does not vanish in a homogeneous background so
we use its projected gradient to characterise the perturbation in
the radiation energy density. Defining $\Delta_a \equiv a \D_a I /
I$ for the radiation, the projected gradient of the $l=0$ moment
of Eq.~(\ref{eq:50}) gives
\begin{equation}
\dot{\Delta}_a + \frac{a}{I}\D_a \D^b I_b + 4 \dot{h}_a =0,
\label{eq:Deltadot}
\end{equation}
where, to linear order, $3\dot{h}_a = a(3 H A_a + \D_a \Theta)$
from Eq.~(\ref{eq:hdot}). The above equation also follows from
integrating Eq.~(\ref{propv}) with $\lambda^3 \, \ud \lambda$ and
noting that the linear Thomson collision term has no monopole.

Equations~(\ref{eq:47})--(\ref{eq:Deltadot}) provide a complete
description of the linear evolution of the CMB anisotropies and
polarization in general almost-FLRW models. In particular, they
are valid for all types of perturbation since no harmonic
expansion has been made. We also see that the highest rank of the
source terms is two so that only scalar, vector and tensor modes
can be excited.  In the following subsections we give integral
solutions of the multipole equations for scalar, vector and tensor
modes and briefly discuss the physics of each. We shall not
consider circular polarization any further since it is not
generated by Thomson scattering.

\subsubsection{Scalar perturbations}
\label{sssCMB_S}

Following the discussion in \S~\ref{sssLIN_KIN}, we expand the
PSTF multipoles in the harmonic tensors $\clq^{(0)}_{\Al}$ defined
in Eq.~(\ref{eq:52}):
\begin{eqnarray}
I_{A_l} &=& I \sum_k \left( \prod_{n=0}^l
\kappa_n^{(0)}\right)^{-1} I_l^{(0)} \clq_{A_l}^{(0)},
\qquad l \geq 1,\label{eq:53}\\
\cle_{A_l} &=& I \sum_k \left(\prod_{n=0}^l
\kappa_n^{(0)}\right)^{-1}
 \cle_l^{(0)} \clq_{A_l}^{(0)},
\qquad l \geq 2 , \label{eq:54}
\end{eqnarray}
where, for later convenience, we have introduced
\begin{equation}
\kappa_l^{(m)} \equiv [1-(l^2-1-m)K/k^2]^{1/2} , \qquad l \geq m ,
\label{eq:55}
\end{equation}
with $\kappa_0^{(0)}=1$. It follows from Eq.~(\ref{eq:scalcurl})
that $\curl \cle_\Al = 0$ and so the $B$-mode of polarization is
not excited by advection (or, as always, Thomson scattering) for
scalar perturbations so we need not include it. For the projected
gradient of the radiation energy density $I$, we expand as
\begin{equation}
\Delta_a \equiv \frac{a\D_a I}{I} = - \sum_k k I_0^{(0)} \clq_a .
\end{equation}
In scalar harmonic form, the linearised multipole equations now
become
\begin{eqnarray}
\dot{I}^{(0)}_l &+&
\frac{k}{a}\left[\frac{(l+1)}{(2l+1)}\kappa_{l+1}^{(0)}
I^{(0)}_{l+1} - \frac{l}{(2l+1)}\kappa_l^{(0)} I^{(0)}_{l-1}
\right] + 4 \dot{h} \delta_{l0} + \frac{4}{3}\frac{k}{a} A
\delta_{l1}
+ \frac{8}{15}\frac{k}{a} \kappa_2^{(0)} \sigma \delta_{l2} \nonumber \\
&&\mbox{} = -n_{\text{e}} \sigt\left[I^{(0)}_l -  I^{(0)}_0
\delta_{l0} - \frac{4}{3} v \delta_{l1} - \left(\frac{1}{10}
I^{(0)}_2 - \frac{3}{5} \cle^{(0)}_2\right) \delta_{l2} \right]
\label{eq:59}
\end{eqnarray}
for $l\geq 0$ and, for $E$-mode polarization,
\begin{equation}
\dot{\cle}^{(0)}_l +
\frac{k}{a}\left[\frac{(l+3)(l-1)}{(2l+1)(l+1)}
\kappa_{l+1}^{(0)}\cle^{(0)}_{l+1} -
\frac{l}{(2l+1)}\kappa_l^{(0)} \cle^{(0)}_{l-1} \right] =
-n_{\text{e}} \sigt\left[\cle^{(0)}_l + \left(\frac{1}{10}
I^{(0)}_2 - \frac{3}{5} \cle^{(0)}_2 \right) \delta_{l2} \right] .
\label{eq:escal}
\end{equation}
In deriving these, we have used Eqs.~(\ref{eq:52})
and~(\ref{eq:58}). The kinematic quantities that enter
Eq.~(\ref{eq:59}) have been mode-expanded following
Eq.~(\ref{eq:hAsig}) with $v_a = \sum v \clq^{(0)}_a$ for the
relative velocity of the electron-baryon plasma. These multipole
equations hold for a general FLRW model and are fully equivalent
to those obtained in Ref.~\cite{1998PhRvD..57.3290H} using the
total-angular momentum method.

Note that in closed models, the $\clq^{(m)}_\Al$ necessarily
vanish globally for $l \geq \nu$, so the same is true of the
$I_\Al$ and $\cle_\Al$ from a given harmonic mode. Power thus
streams up the hierarchy as far as the $l= \nu -1$ multipole, but
is then reflected back down. This is enforced in
Eqs.~(\ref{eq:59}) and (\ref{eq:escal}) by $\kappa^{(m)}_\nu = 0$.
That there is a maximum multipole, hence minimum angular scale,
that can arise from projection of a given harmonic mode is due to
the focusing of geodesics in closed FLRW models.

Early computer codes to compute the CMB anisotropy integrated a
carefully truncated version of the multipole equations directly. A
major advance was made in Ref.~\cite{1996ApJ...469..437S} where
the Boltzmann hierarchy was formally integrated thus allowing a
very efficient solution for the CMB anisotropy. This procedure was
implemented in the CMBFAST code\footnote{http://www.cmbfast.org}
and, later, in parallelised derivatives such as
CAMB~\cite{2000ApJ...538..473L}\footnote{http://camb.info/}.

The integral solution for the total intensity for general spatial
curvature is~\cite{2000PhRvD..62d3004C,1998PhRvD..57.3290H,%
1998ApJ...494..491Z}
\begin{eqnarray}
I^{(0)}_l &=& 4 \int^{t_R}\ud t\,
e^{-\tau}\Bigg\{\left[-\frac{k}{a} \sigma_k + \frac{3}{16}
n_{\text{e}} \sigt (\kappa_2^{(0)})^{-1} \left( I^{(0)}_2 - 6
\cle^{(0)}_2\right)\right] \left[\frac{1}{3} \Phi_l^\nu(x) +
\frac{1}{(\nu^2+1)} \frac{\d^2}{\d x^2}
\Phi_l^\nu(x)\right] \nonumber \\
&&\mbox{}
 - \left(\frac{k}{a}A - n_{\text{e}} \sigt v \right)\frac{1}{\sqrt{\nu^2+1}}
\frac{\d}{\d x}\Phi_l^\nu(x) -
\left[\dot{h}-\frac{1}{4}n_{\text{e}}\sigt I^{(0)}_l \right]
\Phi_l^\nu(x) \Bigg\}. \label{eq:66}
\end{eqnarray}
Here, $\tau \equiv \int n_{\text{e}} \sigt \ud t$ is the optical
depth back along the line of sight and, recall, $x=\sqrt{|K|}
\chi$ with $\chi$ the comoving radial distance (or, equivalently,
conformal look-back time) along the line of sight and
$\Phi_l^\nu(x)$ are the ultra-spherical Bessel functions with
$\nu^2 = (k^2 + K)/|K|$ for scalar perturbations. The geometric
factors $\Phi_l^\nu/3 + (\nu^2+1)^{-1}\d^2 \Phi_l^\nu/\d x^2$ and
$(\nu^2+1)^{-1/2} \d \Phi_l^\nu/\d x$ arise from the projections
of $\clq_{ab}^{(0)}e^a e^b$ and $\clq^{(0)}_a e^a$ respectively,
at $x$ back along the line of sight. Source terms of these forms
enter the Boltzmann equation through the shear and the quadrupole
dependence of Thomson scattering, for $\clq_{ab}^{(0)}e^a e^b$,
and the acceleration and baryon velocity, for $\clq^{(0)}_a e^a$.
Their angular projections follow from Eqs.~(\ref{eq:clqrecurs})
and~(\ref{eq:clqmproj}).

The integral solution for the $E$-mode polarization
is~\cite{2000PhRvD..62d3004C,1998PhRvD..57.3290H,1998ApJ...494..491Z}
\begin{equation}
\cle^{(0)}_l = - \frac{l(l-1)}{(\nu^2+1)} \int^{t_R} \d t\,
n_{\text{e}} \sigt e^{-\tau} (\kappa_2^{(0)})^{-1}
\left(\frac{3}{4}I^{(0)}_2 - \frac{9}{2} \cle^{(0)}_2\right)
\frac{\Phi_l^\nu(x)}{\sinh^2 \! x}. \label{eq:67}
\end{equation}
The geometric term $\Phi^\nu_l(x) / \sinh^2 x$ now arises from the
projection of source terms that go like
$[\clq^{(0)}_{ab}]^{\text{TT}}$; see the form of the
linear-polarization source terms in the scattering
tensor~(\ref{eq:35b}). Equations~(\ref{eq:66}) and (\ref{eq:67})
are valid in an open universe. For closed models one should
replace the hyperbolic functions by their trigonometric
counterparts, and $\nu^2+n$ by $\nu^2-n$ where $n$ is an integer.

Since we are working in linear theory, the coefficients
$I^{(0)}_l$ and $\cle^{(0)}_l$ will depend linearly on the
primordial perturbation $\phi_{(k)}$. Introducing transfer
functions, $T_l^T(k)$ and $T_l^E(k)$, we can write
\begin{equation}
I^{(0)}_l = T_l^T(k) \phi_{(k)}, \qquad \cle^{(0)}_l =
\frac{M_l}{\sqrt{2}} T_l^E(k) \phi_{(k)}, \label{eq:69}
\end{equation}
where the normalisation for polarization is to account for the
additional $l$-dependent factors in Eq.~(\ref{eq:19}). Note that
the symmetry of the background model ensures that transfer
functions depend only on the magnitude of the wavenumber $k$. The
choice of $\phi_{(k)}$ is one of convention. For the adiabatic,
growing-mode initial conditions that follow from single-field
inflation, the convenient choice is the (constant) curvature
perturbation $\clr_{(k)}$ on comoving hypersurfaces. For models
with isocurvature fluctuations, the relative entropy gradient is
appropriate. More generally, in models with mixed initial
conditions having $N$ degrees of freedom per harmonic mode, the
transfer functions generalise to $N$ functions per $l$ and $k$.

We shall express the CMB power spectra in terms of the power
spectrum of the $\phi_{(k)}$ and, since we discuss higher-rank
perturbations later, we shall sketch the derivation for rank-$m$
perturbations in which case the $\phi_{(k)}^{(\pm m)}$ are the
coefficients of the expansion of the PSTF tensor $\phi_{A_m}$ in
terms of the $\clq^{(\pm m)}_{A_m}$.\footnote{The derivation here
follows that given by Antony Lewis in the unpublished CAMB notes
available as http://cosmologist.info/notes/CAMB.ps.gz.} It is
always possible to choose the $\clq^{(m)}_{A_m}$ such that
\begin{equation}
\langle \phi^{(\pm m)}_{(k)} \phi^{(\pm m)}_{(k)}{}^* \rangle =
f_\phi(k) \delta_{kk'},
\end{equation}
with no correlations between the opposite parity modes, for
perturbations that are statistically isotropic, homogeneous and
parity-invariant; see Ref.~\cite{1997PhRvD..55.4596G} for further
details, and, e.g.\ Ref.~\cite{2000PhRvD..62d3004C} for specific
constructions for scalar and tensor perturbations. Here, the
symbolic delta-function enforces $\sum_{k'} A_{(k')}^{(\pm m)}
\delta_{kk'} = A_{(k)}^{(\pm m)}$, and $f_\phi(k)$ depends only on
the magnitude of the wavenumber $k$. The power spectrum of
$\phi_{A_m}$ is proportional to $f_\phi(k)$ and here we
\emph{define} it so that
\begin{eqnarray}
\langle \phi_{A_m} \phi^{A_m} \rangle &=& \sum_k f_\phi(k)
\clq^{(\pm m)}_{A_m}
(\clq^{(\pm m) A_m})^* \nonumber \\
&=& \sum_k f_\phi(k) \frac{1}{V} \int \ud V \, \clq^{(\pm
m)}_{A_m}
(\clq^{(\pm m) A_m})^* \nonumber \\
&\equiv & \int \frac{\nu \ud \nu}{(\nu^2+1)} \clp_\phi(k) .
\label{eq:clpower1}
\end{eqnarray}
Here, we have used statistical homogeneity to replace the
correlator by an integral over a spatial volume $V$ that we shall
let tend to all space. Note that for scalar modes the integration
measure $\nu \ud \nu / (\nu^2+1) =\ud \ln k$, but this is not the
case for higher rank perturbations. For non-flat models there is
no universal convention for the definition of $\clp_\phi(k)$ for
$m >0$ but, of course, any ambiguity is removed when a definite
physical model for the generation of fluctuations is considered.
For example, for $m=2$ and using the (gauge-invariant) metric
perturbation $\clh_{ab}$ for $\phi_{ab}$,
\begin{equation}
\clp_s(k) \propto \frac{(\nu^2+4)}{\nu^2} \tanh (\pi \nu /2)
\end{equation}
for the minimal scale-invariant open inflation model of
Ref.~\cite{1997PhRvD..55.7461B}. Forming the power spectra from
Eqs.~(\ref{eq:16}) and~(\ref{eq:19}), we now have
\begin{eqnarray}
\frac{(2l+1)\Delta_l}{\pi^2} C_l^{XY} &=& \sum_k (\prod_{n=m}^l
\kappa_n^{(m)})^{-2} T^X_l(k) T^Y_l(k) f_\phi(k) \clq^{(\pm
m)}_{\Al} (\clq^{(\pm m)\Al})^* \nonumber \\ &=& \sum_k \biggl[
(\prod_{n=m}^l \kappa_n^{(m)})^{-2}
T^X_l(k) T^Y_l(k) f_\phi(k) \nonumber \\
&&\mbox{} \times \left(\frac{-a}{k}\right)^{l-m} \frac{1}{V} \int
\ud V \, \D_{A_{l-m}} \clq^{(\pm m)}_{a_{l-m+1}\ldots a_l}
(\clq^{(\pm m) \Al})^* \biggr] \nonumber \\
&=& \sum_k \biggl[ (\prod_{n=m}^l \kappa_n^{(m)})^{-2}
T^X_l(k) T^Y_l(k) f_\phi(k) \nonumber \\
&&\mbox{} \times \left(\frac{a}{k}\right)^{l-m} \frac{1}{V} \int
\ud V \, \clq^{(\pm m)}_{a_{l-m+1}\ldots a_l}
\D_{A_{l-m}} (\clq^{(\pm m)\Al})^* \biggr] \nonumber \\
&=& \frac{2^{l-m}(l+m)!(l-m)!}{(2l)!} \sum_k \biggl[
(\kappa_m^{(m)})^{-2}
T^X_l(k) T^Y_l(k) f_\phi(k) \nonumber \\
&& \mbox{} \times \frac{1}{V} \int \ud V \, \clq_{A_m}^{(\pm m)}
(\clq^{(\pm m)A_m})^* \biggr], \label{eq:cl1}
\end{eqnarray}
where we have integrated by parts and made repeated use of the
divergence relations for the $\clq^{(m)}_{\Al}$, e.g.\
Eq.~(\ref{eq:58}). Here, $XY$ is equal to $TT$, $TE$, $EE$ and
$BB$ only, by parity. If we now compare Eqs.~(\ref{eq:clpower1})
and~(\ref{eq:cl1}), we see that
\begin{equation}
C_l^{XY} = \frac{\pi}{4} \frac{(l+m)!(l-m)!}{2^m (l!)^2} \int
\frac{\nu \ud \nu}{(\nu^2+1)} (\kappa_m^{(m)})^{-2} T^X_l(k)
T^Y_l(k) \clp_\phi(k) , \label{eq:clpower}
\end{equation}
where, recall, $\kappa_0^{(0)} = 1$. In closed models, we replace
$\nu^2+1$ by $\nu^2-1$ and the integral becomes a discrete sum
over integer $\nu$.

The simple physics of CMB anisotropies for scalar perturbations is
more apparent if we integrate the shear and acceleration terms in
Eq.~(\ref{eq:66}) by parts. We then find
\begin{multline}
\frac{1}{4}I^{(0)}_l - \left(\frac{a}{k}\dot{\sigma} + A\right)
\delta_{l0} - \frac{1}{3} \sigma \delta_{l1} = \int^{t_R} \ud t\,
e^{-\tau}\Bigg\{n_{\text{e}} \sigt \left[\frac{1}{4}I^{(0)}_0 -
\left(\frac{a}{k}\dot{\sigma} + A\right)\right]
 \Phi^\nu_l(x) \\
+ n_{\text{e}} \sigt (v-\sigma) \frac{1}{\sqrt{\nu^2+1}}
\frac{\d}{\d x}\Phi_l^\nu(x) + \frac{3}{16} n_{\text{e}} \sigt
(\kappa_2^{(0)})^{-1} \left(I^{(0)}_2 - 6 \cle^{(0)}_2\right)
\left[\frac{1}{3} \Phi_l^\nu(x) + \frac{1}{(\nu^2+1)}
\frac{\d^2}{\d x^2}
\Phi_l^\nu(x)\right] \\
+ 2 \dot{\Phi} \Phi^\nu_l(x) \Bigg\} , \label{eq:66b}
\end{multline}
where $\Phi$ is the scalar potential for the Weyl tensor,
\begin{equation}
E_{ab} = \sum_k \left(\frac{k}{a}\right)^2 \Phi \clq^{(0)}_{ab} ,
\end{equation}
which plays the role of the (conformally-invariant) Newtonian
potential. We have written Eq.~(\ref{eq:66b}) in such a form that
the combination of terms on the left-hand side (which are
evaluated at $R$) are independent of the choice of $u^a$, as are
each set on the right-hand side. In deriving Eq.~(\ref{eq:66b}) we
have used the linearised $\dot{\sigma}_{ab}$, $\dot{E}_{ab}$
equations and the shear constraint of \S~\ref{sssKs}. As we can
evaluate Eq.~(\ref{eq:66b}) in any frame, it is convenient to
choose the  {conformal Newtonian gauge} in which the velocity
field $u^a$ has zero shear and vorticity. In this frame, the
linearised $\dot{\sigma}_{ab}$ equation becomes a constraint that
determines the acceleration:
\begin{equation}
\D_{\langle a}A_{b\rangle} = E_{ab} - \frac{1}{2} \pi_{ab} .
\end{equation}
In harmonic form, this is $A = - \Phi + \rho a^2 \pi / 2$, where
$\pi_{ab} = \sum_k \rho \pi \clq^{(0)}_{ab}$. In terms of the
usual conformal Newtonian metric (e.g.\
Ref.~\cite{1995ApJ...455....7M})
\begin{equation}
\ud s^2 = a^2 [-(1+2\Psi_N)\ud \eta^2 + (1+2\Phi_N)\gamma_{ij} \ud
x^i \ud x^j] ,
\end{equation}
where $\gamma_{ij}$ is the conformal spatial metric in the FLRW
background, $A_a = \D_a \Psi_N$ or, in harmonics, $A = - \Psi_N$.
Note also that the conformally-invariant Weyl potential $2 \Phi =
-\Phi_N + \Psi_N$ and $\Phi_N=-\Psi_N$ in the absence of anisotropic stress. If we further approximate to the case of sharp
last scattering, which is valid for $k^{-1} \gg 50\, \text{Mpc}$,
and ignore reionization and the anisotropic nature of Thomson
scattering, the temperature anisotropy in the Newtonian gauge
reduces to
\begin{eqnarray}
\left[\frac{1}{4}I^{(0)}_l + \Psi_N \delta_{l0}\right]_R &=&
\left(\frac{I^{(0)}_0}{4} +\Psi_N\right) \Phi^\nu_l(x_*) +v_N
\left. \frac{1}{\sqrt{\nu^2+1}}
\frac{\d}{\d x}\Phi_l^\nu(x)\right|_{x_*} \nonumber \\
&&\mbox{} + 2 \int_{t_*}^{t_R}
 \dot{\Phi} \Phi^\nu_l(x) \, \ud t ,
\label{eq:SW}
\end{eqnarray}
where, e.g.\ $t_*$ is the time at last scattering. The temperature
anisotropy can now be interpreted in terms of projections of three
terms at last scattering: (i) the intrinsic temperature variations
$I_0^{(0)}/4$ (recall, $I^{(0)}_0$ is related to the projected
gradient of the energy density of the CMB); (ii) the Newtonian
potential $\Psi_N$, which appears because of gravitational
redshifting; and (iii) Doppler shifts of the form $e_a v^a_N$,
where $v^a_N$ is the baryon velocity relative to the zero-shear
$u^a$. The integrated (Sachs-Wolfe) term in Eq.~(\ref{eq:SW})
contributes when the Weyl potential evolves in time, such as when
dark energy starts to dominate the expansion dynamics at low
redshift, and arises because of the net blueshift accrued as a
photon traverses a decaying potential well.

On comoving scales $\sim 30\,\text{Mpc}$ or greater, photon
diffusion due to the finite mean-free path to Thomson scattering
can be ignored. In this limit, the dynamics of the source terms in
Eq.~(\ref{eq:SW}) can be reduced to that of a driven
oscillator~\cite{1995ApJ...444..489H}. To see this, note from
Eq.~(\ref{eq:50}) that in the limit of tight-coupling, $I_a = 4 I
v_a / 3$ and that $I_{\Al} = \cle_{A_l} = 0$ for $l \geq 2$. The
CMB is therefore isotropic in the baryon rest-frame and the
linearised momentum evolution for the combined photon-baryon fluid
gives
\begin{equation}
\dot{v}_a + \frac{H R}{(1+R)}v_a +
\frac{1}{4(1+R)}\frac{\Delta_a}{a} + A_a = 0 , \label{eq:osc1}
\end{equation}
where we have ignored baryon pressure. Here, $R \equiv 3
\rho^{(b)}/(4 \rho^{(\gamma)})$, which scales as $a$, and
$\Delta_a$ is the fractional comoving density gradient for the
CMB. The evolution of $\Delta_a$ was given in
Eq.~(\ref{eq:Deltadot}) which now becomes
\begin{equation}
\dot{\Delta}_a + 4 \dot{h}_a + \frac{4}{3} a \D_a \D^b v_b = 0 .
\label{eq:osc2}
\end{equation}
Switching to conformal time and combining Eqs.~(\ref{eq:osc1})
and~(\ref{eq:osc2}), we find
\begin{equation}
\Delta_a^{\prime \prime} + \frac{\clh R}{(1+R)} \Delta_a^{\prime}
- \frac{1}{3(1+R)} a^2 \D_a \D^b \Delta_b = -4 h_a^{\prime \prime}
- \frac{4 \clh R}{(1+R)} h_a^{\prime} + \frac{4}{3} a^3 \D_a \D^b
A_b ,
\end{equation}
where $\clh = a H$ is the conformal Hubble parameter. This
equation is valid in any frame; it describes a driven oscillator
whose free oscillations are at frequency $k c_s$, where the sound
speed $c_s^2 = 1/[3(1+R)]$, and are damped by the expansion of the
universe acting on the baryon velocity. If we now specialise to
the Newtonian frame, we can express the driving terms on the right
in terms of $\Phi_N$ and $\Psi_N$. We use Eq.~(\ref{eq:hdot}) and
the zero-shear $\dot{E}_{ab}$ equation and shear constraint to
find
\begin{equation}
a\D_{\langle a} \dot{h}_{b \rangle} = (a^2 \D_{\langle a}
\D_{b\rangle} \Phi_N)^{\cdot} ,
\end{equation}
or $\dot{h} = \dot{\Phi}_N$ in harmonics. Finally, we recover
the standard harmonic form of the oscillator equation in the
Newtonian frame:
\begin{equation}
\Delta'' + \frac{\clh R}{(1+R)} \Delta ' + \frac{k^2}{3(1+R)}
\Delta = -4 \Phi_N'' - \frac{4\clh R}{(1+R)} \Phi_N' - \frac{4}{3}
k^2 \Psi_N . \label{eq:osc3}
\end{equation}
For adiabatic initial conditions, the cosine solution is excited
and \emph{all} modes with $k \int^{\eta_*} c_s \, \ud \eta = n
\pi$ are at extrema of their oscillation at last scattering. This
gives a series of acoustic oscillations in the temperature power
spectrum~\cite{1969Ap&SS...4..301Z}, of which the first five have
now been observed by a combination of sub-orbital experiments and
the WMAP satellite~\cite{2008arXiv0803.0593N}. An example of the
CMB power spectra in a $\Lambda$CDM model is shown in
Fig.~\ref{fig:cls}, computed with the CAMB
code~\cite{2000ApJ...538..473L}. The acoustic peaks are a rich
source of cosmological information. Their relative heights depends
on the baryon density (i.e.\ $R$) and matter density since these
affect the midpoint of the acoustic oscillation and the efficacy
of the gravitational driving in
Eq.~(\ref{eq:osc3})~\cite{1995ApJ...444..489H}. The angular
position of the peaks depends on the type of initial condition and
the angular diameter distance to last scattering. Moreover, the
general morphology of the spectra is related to the distribution
of primordial power with scale, i.e.\ the power spectrum
$\clp_\phi(k)$. For current constraints from the CMB see
Ref.~\cite{2008arXiv0803.0547K}.

\begin{figure}[t!]
\begin{center}
\includegraphics[height=12cm,angle=-90]{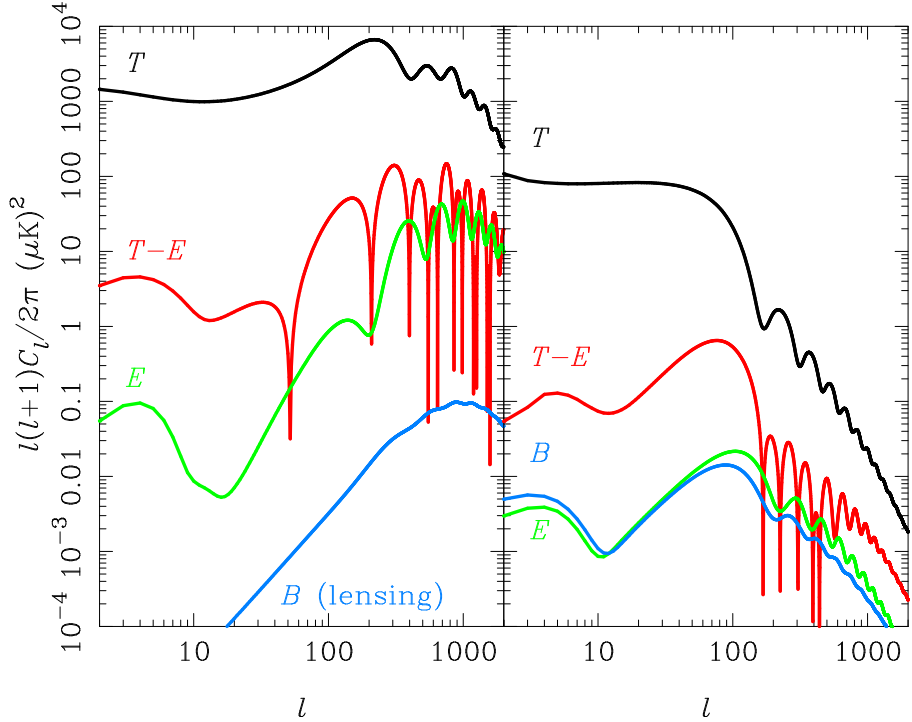}
\end{center}
\caption{Power spectra produced by adiabatic scalar perturbations
(left) and tensor perturbations (right) for a tensor-to-scalar
ratio $r=0.20$ and optical depth to reionization of $0.08$. The
power spectrum of the $B$-modes produced by gravitational lensing
of the scalar $E$-mode polarization is also shown on the left.
\label{fig:cls}}
\end{figure}

On smaller scales photon diffusion becomes important. The
breakdown of tight-coupling has two important effects on the CMB.
First, the acoustic oscillations are exponentially damped and this
gives the rapid decline of $C_l^T$ at high $l$ apparent in
Fig.~\ref{fig:cls}. Second, anisotropies can start to grow in the
CMB intensity and this produces linear polarization on Thomson
scattering. To first-order in the ratio of the mean-free time to
the expansion time or the wavelength of the perturbation, the
polarization is an $E$-mode quadrupole:
\begin{equation}
\cle_{ab} \approx \frac{8}{45} \frac{I}{n_{\text{e}}\sigt}\left(
\sigma_{ab} + D_{\langle a} v_{b \rangle}\right) ,
\end{equation}
which follows quite generally from Eqs.~(\ref{eq:47})
and~(\ref{eq:50}). For scalar perturbations, the polarization thus
traces the projected derivative of the baryon velocity relative to
the Newtonian frame. The peaks in the $C_l^E$ spectrum thus occur
at the minima of $C_l^T$ as the baryon velocity oscillates $\pi/2$
out of phase with $\Delta$. This behaviour can be seen in
Fig.~\ref{fig:cls}. The large-angle polarization from
recombination is necessarily small by causality, but a large-angle
signal is generated by re-scattering at
reionization~\cite{1997PhRvD..55.1822Z,2007ApJS..170..335P}.

\subsubsection{Vector perturbations}
\label{sssCMB_V}

Vector modes describe vortical motions of the cosmic fluids. They
have received considerably less attention than scalar and tensor
modes mainly because they are not excited during inflation.
Furthermore, due to conservation of angular momentum, the
vorticity of radiation decays as $1/a$ and matter as $1/a^2$ so
that
vector modes are generally singular.\footnote{%
It is possible to construct solutions with compensating singular
vortices in the neutrinos and radiation that leave the
perturbations to the spacetime geometry non-singular (e.g.\
Ref.~\cite{2004PhRvD..70d3518L}), but these are rather contrived.}
Vector modes are important in models with sources, such as
magnetic fields (see Ref.~\cite{2007PhR...449..131B} for a recent
review) or topological defects~\cite{1998PhRvD..58b3506T}.

The CMB anisotropies from vector modes were first studied
comprehensively in Ref.~\cite{1986ApJ...308..546A}; the full
kinetic theory treatment was developed in the
total-angular-momentum method in
Refs.~\cite{1997PhRvD..56..596H,1998PhRvD..57.3290H}. Here we
shall extend the $1+3$-covariant treatment of
Ref.~\cite{2004PhRvD..70d3518L} to general spatial curvature.

We expand the radiation anisotropies and polarization in the PSTF
derivatives of the vector harmonics as
\begin{eqnarray}
I_{A_l} &=& I \sum_k \left( \prod_{n=1}^l
\kappa_n^{(1)}\right)^{-1}
I_l^{(\pm 1)} \clq_{A_l}^{(\pm 1)} , \qquad l \geq 1,\label{eq:vec74}\\
\cle_{A_l} &=& I \sum_k \left(\prod_{n=1}^l
\kappa_n^{(1)}\right)^{-1}
 \cle_l^{(\pm 1)} \clq_{A_l}^{(\pm 1)}, \qquad l \geq 1 ,\label{eq:vec75} \\
\clb_{A_l} &=& I \sum_k \left(\prod_{n=1}^l
\kappa_n^{(1)}\right)^{-1}
 \clb_l^{(\pm 1)} \clq_{A_l}^{(\pm 1)}, \qquad l \geq 1 ,\label{eq:vec76}
\end{eqnarray}
where the sum over harmonics, $\sum_k$, includes a sum over the
two parity states of vector harmonics labelled $\pm 1$. For
non-scalar perturbations, $B$-mode polarization is produced from
$E$-modes by advection so we must now include $\clb_{\Al}$.
Substituting in Eq.~(\ref{eq:50}), we find
\begin{eqnarray}
\dot{I}^{(\pm 1)}_l &+& \frac{k}{a}\left[\frac{l(l+2)}{(l+1)(2l+1)}
\kappa_{l+1}^{(1)} I^{(\pm 1)}_{l+1} - \frac{l}{(2l+1)}
\kappa_l^{(1)} I^{(\pm 1)}_{l-1} \right] \nonumber\\ &+& \frac{4}{3}
\frac{k}{a} \kappa_1^{(1)} A^{(\pm 1)} \delta_{l1}+ \frac{8}{15}
\frac{k}{a} \kappa_1^{(1)} \kappa_2^{(1)} \sigma^{(\pm 1)}
\delta_{l2} \nonumber
\\=&-&n_{\text{e}} \sigt \left[I^{(\pm 1)}_{l} -\frac{4}{3}
\kappa_1^{(1)} v^{(\pm 1)}\delta_{l1} - \left(\frac{1}{10} I^{(\pm
1)}_2 - \frac{3}{5} \cle^{(\pm 1)}_2\right) \delta_{l2} \right] ,
\raisetag{-\baselineskip} \label{eq:vec79}
\end{eqnarray}
for $l \geq 1$. Here, $v^{(\pm 1)}$ and $\sigma^{(\pm 1)}$ are the
harmonic expansions of the baryon peculiar velocity and shear
respectively:
\begin{equation}
v_a = \sum_k v^{(\pm 1)} \clq_a^{(\pm 1)} , \qquad \sigma_{ab} =
\sum_k \frac{k}{a} \sigma^{(\pm 1)} \clq_{ab}^{(\pm 1)} .
\end{equation}
The $l=1$ moment $I_a$ is generated from the projected gradient of
the photon energy density by advection. For vector perturbations,
$\Delta_a$ is non-vanishing in a general gauge. Indeed, the
linearised identity
\begin{equation}
\curl \D_a \rho = - 2 \dot{\rho} \omega_a
\end{equation}
relates the curl of $\Delta_a$ for any species to the vorticity
and the time evolution of that species. Of course, if the frame is
chosen to be hypersurface orthogonal, $\omega_a = 0$ and $\Delta_a
= 0$ for vector perturbations. However, in general we have a
non-zero $\Delta_a$ with
\begin{equation}
\sqrt{1+\frac{2K}{k^2}} \Delta^{(\pm 1)} = -\frac{8}{3}
\frac{a}{k} \Theta \omega^{(\mp 1)} , \label{eq:vec1}
\end{equation}
where $\Delta_a = - \sum_k k \Delta^{(\pm 1)} \clq_a^{(\pm 1)}$
and $\omega_a = \sum_k (k/a) \omega^{(\pm 1)} \clq_a^{(\pm 1)}$.
Note that since $\Delta_a$ and $\omega_a$ are related by a curl,
Eq.~(\ref{eq:vec1}) links the coefficients of vector harmonics of
opposite parity. In Eq.~(\ref{eq:vec79}), it should be understood
that the quantity $I_0^{(\pm 1)} \equiv \Delta^{(\pm 1)}$; it can
be considered a source term since it is algebraic in the
vorticity.

The integral solution for the vector-mode anisotropies in a
general almost-FLRW model is~\cite{1998PhRvD..57.3290H}
\begin{eqnarray}
I^{(\pm 1)}_l &=& \frac{4l}{\sqrt{\nu^2+2}} \int^{t_R} \d t\,
e^{-\tau} \Bigg\{ \left[ \frac{k}{a}\left(\frac{1}{4}\Delta^{(\pm
1)}-A^{(\pm 1)} \right) + n_\text{e} \sigt v^{(\pm 1)} \right]
\frac{\Phi^\nu_l(x)}{\sinh x} \nonumber \\
&-& \left[\frac{k}{a} \sigma^{(\pm 1)}
 + \frac{3}{16} n_\text{e} \sigt
\left(\kappa_1^{(1)}\kappa_2^{(1)}\right)^{-1}\left(I_2^{(\pm 1)}
- 6 \cle_2^{(\pm 1)}\right)\right] \frac{1}{\sqrt{\nu^2+2}}
\frac{\ud}{\ud x} \left(\frac{\Phi^\nu_l(x)}{\sinh  x}
\right)\Bigg\} . \label{eq:vec84}
\end{eqnarray}
This is valid for $l \geq 1$ in an open universe; the equivalent
result for $K > 0$ follows from the usual replacements of $\nu^2 +
n$ with $\nu^2 - n$ and hyperbolic functions by their
trigonometric counterparts. The geometric terms
$\Phi^\nu_l(x)/\sinh x$ arise from the projection of sources going
like $e^a \clq^{(\pm 1)}_a$ along the line of sight [see
Eq.~(\ref{eq:clqmproj})]. The terms involving the derivative of
this function arise from shear and scattering sources going like
$e^a e^b \clq_{ab}^{(\pm 1)}$, as follows from
Eq.~(\ref{eq:clqrecurs}). If we integrate the shear term by parts
in Eq.~(\ref{eq:vec84}) we find
\begin{eqnarray}
\frac{1}{4}I^{(\pm 1)}_l - \frac{1}{3} \kappa_1^{(1)} \sigma^{(\pm
1)} \delta_{l1} &=& \frac{l}{\sqrt{\nu^2+2}} \int^{t_R} \d t\,
e^{-\tau} \Bigg\{ \left[ \frac{k}{a}\left(\frac{1}{4}\Delta^{(\pm
1)}-A^{(\pm 1)}\right) -\dot{\sigma}^{(\pm
1)}\right]\frac{\Phi^\nu_l(x)}{\sinh x} \nonumber \\&&+ n_\text{e}
\sigt \left(v^{(\pm 1)}-\sigma^{(\pm 1)}\right)
\frac{\Phi^\nu_l(x)}{\sinh  x} + \frac{3}{16} n_\text{e} \sigt
\left(\kappa_1^{(1)}\kappa_2^{(1)}\right)^{-1} \nonumber
\\&&\times\left(I_2^{(\pm 1)} -
6 \cle_2^{(\pm 1)}\right) \frac{1}{\sqrt{\nu^2+2}} \frac{\ud}{\ud x}
\left(\frac{\Phi^\nu_l(x)}{\sinh x} \right)\Bigg\} .
\label{eq:vec84b}
\end{eqnarray}
Each term in this expression can be shown to frame-invariant. We
see that the anisotropy is sourced at last scattering by the
baryon velocity in a zero-shear frame (i.e.\ by the Doppler
effect) and by anisotropic Thomson scattering. There is also an
integrated effect which involves
\begin{equation}
 \frac{k}{a}\left(\frac{1}{4}\Delta^{(\pm 1)}-A^{(\pm 1)}\right)
-\dot{\sigma}^{(\pm 1)} = -\frac{2}{\sqrt{1+2K/k^2}} \dot{H}^{(\mp
1)} , \label{eq:vec2}
\end{equation}
where the vector-mode contribution to the magnetic part of the
Weyl tensor is
\begin{equation}
H_{ab} = \sum_k \left(\frac{k}{a}\right)^2 H^{(\pm 1)}
\clq_{ab}^{(\pm 1)} ,
\end{equation}
and we have used Eq.~(\ref{eq:vec1}), the vorticity propagation
equation and the magnetic Weyl equation. The integrated
contribution to the vector-mode temperature anisotropies is
similar to the scalar-mode ISW effect, but involves the
gravito-magnetic part of the Weyl curvature rather than
gravito-electic. Equation~(\ref{eq:vec84b}) extends the analysis
of Ref.~\cite{2004PhRvD..70d3518L} to non-flat models; the
velocity source term and the integrated contribution are
consistent with the gauge-invariant analysis of
Ref.~\cite{1986ApJ...308..546A} though the (small) anisotropic
scattering term is not included there.

The polarization multipole equations for vector modes are
\begin{eqnarray}
\dot{\cle}^{(\pm 1)}_l &+&
\frac{k}{a}\left[\frac{(l-1)l(l+2)(l+3)}{(l+1)^3 (2l+1)}
\kappa_{l+1}^{(1)} \cle^{(\pm 1)}_{l+1} - \frac{l}{(2l+1)}
\kappa_l^{(1)} \cle^{(\pm 1)}_{l-1} \right] - \frac{2}{l(l+1)}
\frac{k}{a} \sqrt{1+\frac{2K}{k^2}}
\clb^{(\mp 1)}_l \nonumber \\
&=& -n_{\text{e}} \sigt\left[\cle^{(\pm 1)}_l + \left(\frac{1}{10}
I^{(\pm 1)}_2 - \frac{3}{5}
\cle^{(\pm 1)}_2\right) \delta_{l2} \right] , \label{eq:vec82} \\
\dot{\clb}^{(\pm 1)}_l &+&
\frac{k}{a}\left[\frac{(l-1)l(l+2)(l+3)}{(l+1)^3 (2l+1)}
\kappa_{l+1}^{(1)} \clb^{(\pm 1)}_{l+1} - \frac{l}{(2l+1)}
\kappa_l^{(1)} \clb^{(\pm 1)}_{l-1} \right] + \frac{2}{l(l+1)}
\frac{k}{a} \sqrt{1+\frac{2K}{k^2}}
\cle^{(\mp 1)}_l \nonumber \\
&=& -n_{\text{e}} \clb^{(\pm 1)}_l . \label{eq:vec83}
\end{eqnarray}
Note here how the $\text{curl}$ couplings between $E$ and
$B$-modes in Eqs.~(\ref{eq:47}) and~(\ref{eq:48}) lead to a
coupling between the $\cle^{(\pm 1)}_l$ and $\clb^{(\mp 1)}_l$. If
we choose the primordial perturbation variable $\phi_{(k)}^{(\pm
1)}$ so that the radiation anisotropies $I_l^{(\pm 1)}$ are linear
in them (for example, $\phi_{(k)}^{(\pm 1)} = \sigma_{(k)}^{(\pm
1)} + v_{(k)}^{(\pm 1)}$), $\clb_l^{(\pm 1)}$ will be linear in
the primordial perturbation of the opposite parity,
$\phi_{(k)}^{(\mp 1)}$. It follows that, if the primordial
fluctuations are parity-symmetric in the mean, there will be no
correlations between $\clb_{\Al}$ and either of $\cle_{\Al}$ or
$I_{\Al}$.

The integral solutions for the polarization
are~\cite{1998PhRvD..57.3290H}
\begin{eqnarray}
\cle^{(\pm 1)}_l &=& -\frac{3l(l-1)}{4(l+1)}\frac{1}{(\nu^2+2)}
\int^{t_R} \d t\, n_{\text{e}} \sigt e^{-\tau} (\kappa_1^{(1)}
\kappa_2^{(1)}){}^{-1}
\Bigg[ \left(I_2^{(\pm 1)} - 6 \cle_2^{(\pm 1)}\right) \nonumber \\
&&\mbox{} \phantom{xxxxxxxxxxxxxxxxxxxx} \times \left(
\frac{1}{\sinh x} \frac{\ud}{\ud x}\Phi_l^\nu(x) + \frac{\cosh x}{%
\sinh^3 x} \Phi_l^\nu(x)\right) \Bigg] ,
\label{eq:vec95} \\
\clb^{(\pm 1)}_l &=& \frac{3l(l-1)}{4(l+1)}\frac{\nu}{(\nu^2+2)}
\int^{t_R} \d t\, n_{\text{e}} \sigt e^{-\tau} (\kappa_1^{(1)}
\kappa_2^{(1)}){}^{-1} \left(I_2^{(\mp 1)} - 6 \cle_2^{(\mp
1)}\right) \frac{\Phi_l^\nu(x)}{\sinh x} , \label{eq:vec96}
\end{eqnarray}
in an open universe. Examples of CMB power spectra from vector
modes are given in Fig.~2 of Ref.~\cite{2004PhRvD..70d3518L}.

\subsubsection{Tensor perturbations}
\label{sssCMB_T}

The imprint of tensor perturbations, or gravitational waves, is
implicit in the original work of Sachs \&
Wolfe~\cite{1967ApJ...147...73S}, although the first detailed
calculations for temperature were reported in
Ref.~\cite{1969MNRAS.144..255D} and for polarization in
Ref.~\cite{1985SvA....29..607P}. Other important milestones
include constraints on the gravitational wave background from the
large-angle temperature anisotropy from
COBE~\cite{1992PhRvL..69..869K,1992MNRAS.258P..57L,1992PhLB..291..391L},
the introduction of the $E$-$B$ decomposition (which was already
implicit in the early work of Dautcourt \&
Rose~\cite{1978AN....299...13D}) and the realisation that $B$-mode
polarization is a particularly sensitive probe of tensor
modes~\cite{1997PhRvL..78.2058K,1997PhRvL..78.2054S}, and
constraints from the WMAP data~\cite{2008arXiv0803.0547K}. The
effect of tensor modes on the CMB from the 1+3-covariant
perspective is discussed in
Refs.~\cite{2000CQGra..17..871C,2000PhRvD..62d3004C}.

For tensor perturbations we expand the temperature and
polarization multipoles in the $\clq^{(\pm 2)}_{\Al}$:
\begin{eqnarray}
I_{A_l} &=& I \sum_k \left( \prod_{n=2}^l
\kappa_n^{(2)}\right)^{-1}
I_l^{(\pm 2)} \clq_{A_l}^{(\pm 2)} , \qquad l \geq 2,\label{eq:74}\\
\cle_{A_l} &=& I \sum_k \left(\prod_{n=2}^l
\kappa_n^{(2)}\right)^{-1}
 \cle_l^{(\pm 2)} \clq_{A_l}^{(\pm 2)}, \qquad l \geq 2 ,\label{eq:75} \\
\clb_{A_l} &=& I \sum_k \left(\prod_{n=2}^l
\kappa_n^{(2)}\right)^{-1}
 \clb_l^{(\pm 2)} \clq_{A_l}^{(\pm 2)}, \qquad l \geq 2 ,\label{eq:76}
\end{eqnarray}
where implicit in the sum over harmonics, $\sum_k$, is a sum over
the parity states labelled $\pm 2$. As for vector modes, we must
now include the $B$-mode polarization. The linearised multipole
hierarchy for the temperature anisotropies, Eq.~(\ref{eq:50}),
becomes
\begin{eqnarray}
\dot{I}^{(\pm 2)}_l &+&
\frac{k}{a}\left[\frac{(l+3)(l-1)}{(l+1)(2l+1)} \kappa_{l+1}^{(2)}
I^{(\pm 2)}_{l+1} - \frac{l}{(2l+1)} \kappa_l^{(2)} I^{(\pm
2)}_{l-1} \right] + \frac{8}{15} \frac{k}{a} \kappa_2^{(2)}
\sigma^{(\pm 2)}
\delta_{l2} \nonumber \\
&=& - n_{\text{e}} \sigt \left[I^{(\pm 2)}_{l} -
 \left(\frac{1}{10} I^{(\pm 2)}_2 - \frac{3}{5}
\cle^{(\pm 2)}_2\right) \delta_{l2} \right] , \label{eq:79}
\end{eqnarray}
for $l \geq 2$, where $\sigma_{ab} = \sum_k (k/a) \sigma^{(\pm 2)}
\clq_{ab}^{(\pm 2)}$. For the polarization, we have
\begin{eqnarray}
\dot{\cle}^{(\pm 2)}_l &+& \frac{k}{a}\left[\frac{(l+3)^2(l-1)^2}{%
(2l+1)(l+1)^3} \kappa_{l+1}^{(2)} \cle^{(\pm 2)}_{l+1} -
\frac{l}{(2l+1)} \kappa_l^{(2)} \cle^{(\pm 2)}_{l-1} \right] -
\frac{4}{l(l+1)} \frac{k}{a} \sqrt{1+\frac{3K}{k^2}}
\clb^{(\mp 2)}_l \nonumber \\
&=& -n_{\text{e}} \sigt\left[\cle^{(\pm 2)}_l + \left(\frac{1}{10}
I^{(\pm 2)}_2 - \frac{3}{5}
\cle^{(\pm 2)}_2\right) \delta_{l2} \right] , \label{eq:82} \\
\dot{\clb}^{(\pm 2)}_l &+& \frac{k}{a}\left[\frac{(l+3)^2(l-1)^2}{%
(2l+1)(l+1)^3} \kappa_{l+1}^{(2)} \clb^{(\pm 2)}_{l+1} -
\frac{l}{(2l+1)} \kappa_l^{(2)} \clb^{(\pm 2)}_{l-1} \right] +
\frac{4}{l(l+1)} \frac{k}{a} \sqrt{1+\frac{3K}{k^2}}
\cle^{(\mp 2)}_l \nonumber \\
&=& -n_{\text{e}} \clb^{(\pm 2)}_l . \label{eq:83}
\end{eqnarray}
Note, again, how the $\text{curl}$ couplings between $E$ and
$B$-modes in Eqs.~(\ref{eq:47}) and~(\ref{eq:48}) lead to a
coupling between the $\cle^{(\pm 2)}_l$ and $\clb^{(\mp 2)}_l$ so
there will be no correlations between $\clb_{\Al}$ and either of
$\cle_{\Al}$ or $I_{\Al}$ for fluctuations that are
parity-invariant in the mean.

The integral solution for the tensor contribution to the intensity
anisotropy in a general almost-FLRW model
is~\cite{2000PhRvD..62d3004C,1998PhRvD..57.3290H}
\begin{equation}
I^{(\pm 2)}_l =
\frac{4l(l-1)}{[(\nu^2+1)(\nu^2+3)]^{1/2}}\int^{t_R} \d t\,
e^{-\tau} \left(-\frac{k}{a}\sigma^{(\pm 2)} +
\frac{3}{16}n_{\text{e}}\sigt (\kappa_2^{(2)}){}^{-1} (I_2^{(\pm
2)} - 6 \cle^{(\pm 2)})\right) \frac{\Phi^\nu_l(x)}{\sinh^2\! x},
\label{eq:84}
\end{equation}
in an open universe. The sources for the tensor-mode Boltzmann
equation are of the form $\clq^{(\pm 2)}_{ab} e^a e^b$ and the
projection of these at $x$ along the line of sight gives rise to
the geometric factor $l(l-1)\Phi^\nu_l(x)/\sinh^2\! x$; see
Eq.~(\ref{eq:clqmproj}). The solution for the polarization is
\begin{eqnarray}
\cle^{(\pm 2)}_l &=& \frac{-3M_l{}^2}{8[(\nu^2+1)(\nu^2+3)]^{1/2}}
\int^{t_R} \d t\, n_{\text{e}} \sigt e^{-\tau}
(\kappa_2^{(2)}){}^{-1} \left(I_2^{(\pm 2)} - 6 \cle_2^{(\pm
2)}\right) \phi_l^\nu(x),
\label{eq:95} \\
\clb^{(\pm 2)}_l &=& \frac{-3M_l{}^2}{8[(\nu^2+1)(\nu^2+3)]^{1/2}}
\int^{t_R} \d t\, n_{\text{e}} \sigt e^{-\tau}
(\kappa_2^{(2)}){}^{-1} \left(I_2^{(\mp 2)} - 6 \cle_2^{(\mp
2)}\right) \psi_l^\nu(x), \label{eq:96}
\end{eqnarray}
where the geometric terms $\phi_l^\nu(x)$ and $\psi^\nu_l(x)$ are
\begin{eqnarray}
\phi^\nu_l(x) &=& \frac{\d^2}{\d x^2}\Phi^\nu_l(x) + 4\coth\! x
\frac{\d}{\d x}\Phi^\nu_l(x) - (\nu^2-1-2\coth^2 x) \Phi^\nu_l(x),
\label{eq:97} \\
\psi^\nu_l(x) &=& -2\nu \left[ \frac{\d}{\d x}\Phi^\nu_l(x) +
2\coth\! x \Phi^\nu_l(x) \right] \label{eq:98}
\end{eqnarray}
in an open universe. The results for $K > 0$ follow from the usual
replacements.

We relate the radiation multipoles to the gauge-invariant
primordial metric perturbation $\clh_{ab}$ of
\S~\ref{sssNEUTRINO_T} via transfer functions
\begin{equation}
I^{(\pm 2)}_l = T_l^T(k) \clh_{(k)}^{(\pm 2)}, \qquad \cle^{(\pm
2)}_l = \frac{M_l}{\sqrt{2}} T_l^E(k) \clh_{(k)}^{(\pm 2)}, \qquad
\clb^{(\pm 2)}_l = \frac{M_l}{\sqrt{2}} T_l^B(k) \clh_{(k)}^{(\mp
2)} ,
\end{equation}
where is should be noted that $\clb_l^{(\pm 2)}$ are linear in
$\clh_{(k)}^{(\mp 2)}$. The primordial tensor power spectrum is
defined so that
\begin{equation}
\langle \clh_{ab} \clh^{ab} \rangle = \int \frac{\nu \ud
\nu}{(\nu^2+1)} \clp_\clh(k) ,
\end{equation}
and the non-vanishing CMB power spectra from tensor modes are then
given by Eq.~(\ref{eq:clpower}) with $m=2$.

Examples of CMB power spectra from tensor modes are shown in the
right-hand panel of Fig.~\ref{fig:cls}. CMB anisotropies are
produced by the anisotropic expansion of gravitational waves along
the line of sight after last-scattering. Since gravitational waves
damp away once they become sub-Hubble, with amplitude falling as
$1/a$, the CMB spectra fall away rapidly above $l \sim 100$
(corresponding to the horizon size at recombination).
Gravitational waves produce roughly equal power in $E$- and
$B$-mode polarization, but with a somewhat sharper projection
between linear scales $k$ and angular scales $l$ for $E$-modes
than $B$~\cite{1997PhRvD..56..596H}. As for scalar perturbations,
reionization produces additional large-angle polarization.

The ratio of primordial tensor modes to curvature perturbations is
set to $r=0.20$ in Fig.~\ref{fig:cls}, corresponding to the
current upper limit on tensor modes from a combination of WMAP temperature
and $E$-mode polarization data, and distance measures from the baryon
acoustic oscillations in large-scale structure data and
supernovae~\cite{2008arXiv0803.0547K}.  The
constraint comes mainly from the (large-angle) temperature
anisotropies. Simple inflation models naturally produce a
background of gravitational waves with an almost scale-invariant
power spectrum~\cite{1979JETPL..30..682S}
\begin{equation}
\clp_\clh(k) = \frac{16}{\pi}
\left(\frac{H}{m_{\text{Pl}}}\right)^2 ,
\end{equation}
where $H$ is the Hubble parameter during inflation and
$m_{\text{Pl}} = 1.22 \times 10^{19} \, \text{GeV}/c^2$ is the
Planck mass. The current upper limit on tensor modes implies $H <
1.2\times 10^{14}\, \text{GeV}$ and so an  {energy scale of
inflation} $E_{\text{inf}} < 2.2 \times 10^{16}\, \text{GeV}$. The
$B$-mode of polarization is yet to be observed and the constraints
$B$-modes place on gravitational waves are not yet competitive
with the temperature anisotropies. However, because of cosmic
variance from the dominant scalar perturbations, the constraint on
tensor modes from the temperature anisotropies will not improve
much as instruments get more sensitive but that from $B$-modes
will. Ratios $r \sim 10^{-2}$ are targets for a new generation of
sensitive $B$-mode surveys from ground and balloon-borne
platforms. Ultimately, the detection of tensor modes via the
$B$-mode polarization may only be limited by our ability to
control instrumental systematic effects
(e.g.~\cite{2003PhRvD..67d3004H,2007MNRAS.tmp..292O}) and clean
out foreground contaminants~\cite{2005MNRAS.360..935T} and
secondary processes such as gravitational
lensing~\cite{2002PhRvD..65b3003H,2002PhRvD..65b3505L}.

\subsubsection{Cosmic microwave background in other cosmological models}
\label{sssCMB_NONFLRW}

Anisotropies in the CMB in other cosmological models have been
considered by a number of authors. A very general approach to
constrain departures from FLRW symmetries with the CMB was
pioneered by Ref.~\cite{1995PhRvD..51.1525M}, subsequently
improved in Ref.~\cite{1995PhRvD..51.5942M}, and applied to COBE data in Ref.~\cite{1997ApJ...476..435S} (with an erratum in Ref.~\cite{1999ApJ...522..559S}). The idea was to use observational constraints on the $F_\Al$ plus the 1+3-covariant dynamical equations to constrain the geometry in a model-independent way. Given that we can only measure the anisotropies here and now, assumptions about the size of temporal and spatial derivatives are required to extract useful constraints. For the large-angle anisotropies, the above papers assumed the expansion-normalised derivative of the multipoles were bounded by the multipoles themselves. This is indeed the case for almost FLRW models, but examples are known where the assumption is violated and an isotropic CMB observed now can still be accompanied by large spacetime anisotropy~\cite{1999ApJ...522L...1N}.

The CMB has also been investigated extensively in models with a
subset of FLRW symmetry. Spatially-homogeneous but
(globally-)anisotropic models were considered in the pioneering
work of Collins \& Hawking~\cite{1973MNRAS.162..307C} and its
subsequent extensions (e.g.\ Ref.~\cite{1985MNRAS.213..917B}).
Bianchi models that can be considered small perturbations of FLRW
are particularly well studied and the anisotropies have a rich
phenomenology due to geodesic focusing and spiralling. Constraints
on global rotation and shear in these early papers were superseded
by analysis based on the COBE
data~\cite{1996PhRvL..77.2883B,1997PhRvD..55.1901K} for Bianchi
$\text{VII}_h$ models. More recently, a curious correlation has
been found between the anisotropy template in this model and the
WMAP data~\cite{2005ApJ...629L...1J}. The result is
statistically-significant, and removal of the correlated pattern
can explain a number of anomalous features in the WMAP data.
However, the cosmological parameters required are at odds with
those needed to explain the CMB anisotropy on smaller scales and,
moreover, the predicted polarization anisotropy has now been
computed and it exceeds the WMAP polarization observations
on large angular scales~\cite{2007MNRAS.380.1387P}.

Spherically-symmetric models have also received considerable
attention. They provide a useful analytic model of the secondary
anisotropies due to non-linear gravitational effects (such as the
imprint of a forming cluster of galaxies)~\cite{1968Natur.217..511R,1992ApJ...388..225P} and a
simple way of relaxing the Copernican assumption.

\section{Beyond the linear regime}\label{sBLR}
Linear perturbation theory is a good approximation only at the initial stages of gravitational collapse, when the density contrast is well below unity. Most of the observed structures in the universe, however, have density contrasts well in excess of unity. The density within a cluster of galaxies, for example, is between $10^2$ and $10^3$ times greater than the average density of the universe, while that of a galaxy is about $10^5$ times larger. To understand the evolution of these objects we need to go beyond the limits of the linear regime.

\subsection{Nonlinear peculiar kinematics}\label{ssNPKs}
In an unperturbed, idealised Friedmann universe, comoving particles have velocities that follow Hubble's law. When perturbations are present, however, the Hubble flow is distorted and matter acquires `peculiar' velocities. The dipole anisotropy of the CMB seems to suggest that our Local Group is moving with respect the smooth Hubble flow, which defines the frame where the CMB dipole vanishes, at a speed of approximately 600~km/sec (e.g.~see~\cite{1993sfu..book.....P}). Such, rather large, velocity perturbations can have important implications for any nonlinear structure formation scenario. After recombination and on scales well inside the Hubble length, one can use the Newtonian theory to study peculiar velocities. As we are successively probing significant fractions of the Hubble radius, however, the need for a relativistic treatment increases.

\subsubsection{1+3 peculiar-velocity
decomposition}\label{sss1+3PVD}
When studying peculiar motions one needs to define the associated velocities relative to a preferred reference frame. The latter is not comoving with the fluid, since there are no peculiar velocities relative to the matter frame by construction. Following~\cite{1998CQGra..15.3545V,2001CQGra..18.5115E,%
2002PhRvD..66l4015E}, we choose our reference velocity field ($u_a$) to be both irrotational and shear-free. Then, we assume the presence of matter moving with 4-velocity
\begin{equation}
\tilde{u}_a=\gamma(u_a+v_a)\,,  \label{tua}
\end{equation}
where $v_a$ is the peculiar velocity of the fluid (with $u_av^a=0$) and $\gamma=(1-v^2)^{-1/2}$ is the Lorentz-boost factor (see also \S~\ref{sss4vFs}). The $u_av^a=0$ condition guarantees that $\tilde{u}_a$ is also timelike (i.e.~that $\tilde{u}_a\tilde{u}^a=-1$), irrespective of the value of the $\gamma$-factor. Here, however, we will be dealing with non-relativistic peculiar motions, which means that $v^2\ll1$ and consequently that $\gamma\simeq1$. The instantaneous rest space of the `tilded' observer is defined by means of the tensor
\begin{equation}
\tilde{h}_{ab}= g_{ab}+ \tilde{u}_a\tilde{u}_b\,,  \label{thab}
\end{equation}
which projects orthogonal to $\tilde{u}_a$.\footnote{The peculiar velocity field is orthogonal to $u_a$ by construction but does not lie in the rest frame of the `tilded' observers, even for non-relativistic peculiar velocities. Indeed, following (\ref{tua}) and (\ref{thab}), we find that $\tilde{u}_av^a=\gamma v^2\neq0$ and that $\tilde{h}_a{}^bv_b=v_a+\gamma v^2\tilde{u}_a\neq v_a$.} When the $u_a$-frame has $\sigma_{ab}=0=\omega_{ab}$, it corresponds to Bardeen's quasi-Newtonian gauge (see~\cite{1980PhRvD..22.1882B}) and is related to the comoving (Lagrangian) reference system via the transformation laws given in Appendix~\ref{AssTU4VB}.

The peculiar kinematics are covariantly determined by the irreducible variables of the motion, obtained by decomposing the gradient of the $v_a$-field. To simplify the equations we choose as our time direction the one along $\tilde{u}_a$. Then, $\tilde{\rm D}_a=\tilde{h}_a{}^b\nabla_b$ defines the associated orthogonally projected covariant derivative operator and the projected gradient of the peculiar velocity splits as
\begin{equation}
\tilde{\rm D}_bv_a= {1\over3}\,\hat{\Theta}\tilde{h}_{ab}+ \hat{\sigma}_{ab}+ \hat{\omega}_{ab}\,,  \label{Dbva}
\end{equation}
with $\hat{\Theta}=\tilde{\rm D}_av^a$, $\hat{\sigma}_{ab}=\tilde{\rm D}_{\langle b}v_{a\rangle}$ and $\hat{\omega}_{ab}=\tilde{\rm D}_{[b}v_{a]}$. In analogy with ${\rm D}_bu_a$ (see \S~\ref{sssKs}), the tensor $\tilde{\rm D}_bv_a$ describes the relative motion of neighbouring peculiar flow lines, while $\hat{\Theta}$, $\hat{\sigma}_{ab}$ and $\hat{\omega}_{ab}$ represent the volume expansion (or contraction), the shear and the vorticity of the peculiar motion respectively. Thus, a region that has `decoupled' from the background expansion and is collapsing has $\hat{\Theta}<0$.

\subsubsection{Nonlinear peculiar motions}\label{sssNLPMs}
In cosmological studies, it helps to identify the preferred 4-velocity field, that is the reference $u_a$-frame of the previous section, with the one setting the CMB dipole to zero. Then, an observer moving with the matter (i.e.~with the $\tilde{u}_a$-frame of \S~\ref{sss1+3PVD}) monitors its motion using the nonlinear evolution equation of the peculiar velocity field
\begin{equation}
\dot{v}_{\langle a\rangle}= \tilde{A}_a- A_{\langle a\rangle}- {1\over3}\,(\tilde{\Theta}- \hat{\Theta})v_{\langle a\rangle}- (\tilde{\sigma}_{ab}-\hat{\sigma}_{ab})v^b- (\tilde{\omega}_{ab}-\hat{\omega}_{ab})v^b\,,  \label{vadot}
\end{equation}
where $\dot{v}_{\langle a\rangle}=\tilde{h}_a{}^b\dot{v}_b$ and $v_{\langle a\rangle}=\tilde{h}_a{}^bv_b$, and that of its projected gradients. The latter, which is obtained by applying the Ricci identity to $v_a$ (see~\cite{2002PhRvD..66l4015E} for the technical details), reads
\begin{eqnarray}
\tilde{h}_b{}^d\tilde{h}_a{}^c\left(\tilde{\rm D}_dv_c\right)^{\cdot}&=& \tilde{\rm D}_b\dot{v}_a- {1\over9}\,\tilde{\Theta}\hat{\Theta}\tilde{h}_{ab}- {1\over3}\,\left[\tilde{\Theta}(\hat{\sigma}_{ab}+\hat{\omega}_{ab}) +\hat{\Theta}(\tilde{\sigma}_{ab}+\tilde{\omega}_{ab})\right] \nonumber\\ &&-\hat{\sigma}_{ca}\tilde{\sigma}^c{}_b- \hat{\sigma}_{ca}\tilde{\omega}^c{}_b+ \hat{\omega}_{ca}\tilde{\sigma}^c{}_b+ \hat{\omega}_{ca}\tilde{\omega}^c{}_b+ \dot{v}_{\langle a\rangle}\tilde{A}_b- {1\over3}\,\tilde{\Theta}\tilde{A}_av_{\langle b\rangle} \nonumber\\ &&-\tilde{A}_a(\tilde{\sigma}^c{}_b +\tilde{\omega}^c{}_b)v_c+ \tilde{A}_a\tilde{\rm D}_bv^2- \tilde{h}_a{}^c\tilde{h}_b{}^dR_{csdq}v^s\tilde{u}^q\,, \label{Dbvadot}
\end{eqnarray}
where $\tilde{A}_a=\dot{\tilde{u}}_a=\tilde{u}^b\nabla_b\tilde{u}_a$ is the 4-acceleration of the `tilded' observer. Note the Riemann-curvature term in the right-hand side of the above, which couples the geometry of the spacetime with both $\tilde{u}_a$ and $v_a$. As we will see next, the latter coupling has profound implications for the kinematics of peculiar motions.

The trace, the symmetric trace-free component and the skew part of the above lead to the propagation equations of $\hat{\Theta}$, $\hat{\sigma}_{ab}$ and $\hat{\omega}_{ab}$ respectively~\cite{2002PhRvD..66l4015E}. Thus, the time evolution of the volume scalar is monitored by
\begin{eqnarray}
\dot{\hat{\Theta}}&=& -{1\over3}\,\tilde{\Theta}\hat{\Theta}- \tilde{\sigma}_{ab}\hat{\sigma}^{ab}+ \tilde{\omega}_{ab}\hat{\omega}^{ab}+ \tilde{\rm D}_a\dot{v}^a+ \tilde{A}_a\dot{v}^a+ \tilde{q}_av^a+ \left[{1\over2}\,(\tilde{\rho}+3\tilde{p})-\Lambda\right]v^2 \nonumber\\ &&-\left(\tilde{\sigma}_{ab}-\tilde{\omega}_{ab} +{1\over3}\,\tilde{\Theta}\tilde{h}_{ab}\right)\tilde{A}^av^b+ \tilde{A}^a\tilde{\rm D}_av^2\,,  \label{pRay}
\end{eqnarray}
which is the Raychaudhuri analogue of the peculiar flow and shares several close analogies with its standard counterpart (compare to Eq.~(\ref{Ray}) in \S~\ref{sssKs}).\footnote{We remind the reader that in this section overdots denote time-derivatives along $\tilde{u}_a$ and angled brackets projection orthogonal to $\tilde{u}_a$. Also, `tilded' quantities are measured in the matter ($\tilde{u}_a$) frame.} There are differences, however, and the main one is in the role played by the matter fields. According to the seventh term in the right-hand side of (\ref{pRay}), the total gravitational mass of the system tends to increase the average separation between two neighbouring peculiar flow lines. A positive cosmological constant, on the other hand, brings the aforementioned  flow lines closer together.

Taking the symmetric trace-free and the antisymmetric parts of (\ref{Dbvadot}) leads to the respective propagation formulae of the shear and the vorticity. Following~\cite{2002PhRvD..66l4015E}, we have
\begin{eqnarray}
\tilde{h}_{\langle b}{}^d\tilde{h}_{a\rangle}{}^c \dot{\hat{\sigma}}_{cd}&=& -{1\over3}\,\left(\tilde{\Theta}\hat{\sigma}_{ab} +\hat{\Theta}\tilde{\sigma}_{ab}\right)- \hat{\sigma}_{c\langle a}\tilde{\sigma}^c{}_{b\rangle}- \hat{\sigma}_{c\langle a}\tilde{\omega}^c{}_{b\rangle}+ \hat{\omega}_{c\langle a}\tilde{\sigma}^c{}_{b\rangle}+ \hat{\omega}_{c\langle a}\tilde{\omega}^c{}_{b\rangle}+ \tilde{\rm D}_{\langle b}\dot{v}_{a\rangle} \nonumber\\ &&+ \tilde{A}_{\langle a}\tilde{h}_{b\rangle}{}^c\dot{v}_c- {1\over3}\,\tilde{\Theta}\tilde{A}_{\langle a}\tilde{h}_{b\rangle}{}^cv_c- \tilde{A}_{\langle a} (\tilde{\sigma}^c{}_{b\rangle}+\tilde{\omega}^c{}_{b\rangle})v_c+ \tilde{A}_{\langle a}\tilde{\rm D}_{b\rangle}v^2+ v^2\tilde{E}_{ab} \nonumber\\ &&-\tilde{\varepsilon}_{cd\langle a}v^c\tilde{H}^d{}_{b\rangle}- {1\over2}\,v^2\tilde{\pi}_{ab}- {1\over2}\,\tilde{q}_{\langle a}\tilde{h}_{b\rangle}{}^cv_c\,, \label{psigmadot}
\end{eqnarray}
for the `peculiar shear', and
\begin{eqnarray}
\tilde{h}_{[b}{}^d\tilde{h}_{a]}{}^c\dot{\hat{\omega}}_{cd}&=& -{1\over3}\,\left(\tilde{\Theta}\hat{\omega}_{ab} +\hat{\Theta}\tilde{\omega}_{ab}\right)- \hat{\sigma}_{c[a}\tilde{\sigma}^c{}_{b]}- \hat{\sigma}_{c[a}\tilde{\omega}^c{}_{b]}+ \hat{\omega}_{c[a}\tilde{\sigma}^c{}_{b]}+ \hat{\omega}_{c[a}\tilde{\omega}^c{}_{b]}+ \tilde{\rm D}_{[b}\dot{v}_{a]} \nonumber\\ &&- \tilde{A}_{[a}\tilde{h}_{b]}{}^c\dot{v}_c- {1\over3}\,\tilde{\Theta}\tilde{A}_{[a}\tilde{h}_{b]}{}^cv_c- \tilde{A}_{[a}(\tilde{\sigma}^c{}_{b]}+\tilde{\omega}^c{}_{b]})v_c+ \tilde{A}_{[a}\tilde{\rm D}_{b]}v^2\nonumber\\ &&-\tilde{\varepsilon}_{cd[a}v^c\tilde{H}^d{}_{b]}- {1\over2}\,\tilde{q}_{[a}\tilde{h}_{b]}{}^cv_c\,, \label{pomegadot}
\end{eqnarray}
for the peculiar vorticity. When compared to their conventional analogues (see Eqs.~(\ref{sigmadot}) and (\ref{omegadot}) in \S~\ref{sssKs}), these expressions show a high degree of complexity. Among others, the coupling between spacetime geometry and peculiar velocity seen in (\ref{Dbvadot}), has led to a extra (magnetic) Weyl effects on both the shear and the vorticity.

\subsubsection{The case of dust}\label{sssCD}
To this point, the peculiar velocity field is arbitrary and $v_a$ has not been associated with any particular matter source. The system (\ref{vadot})-(\ref{pomegadot}) is completely general and $\gamma\simeq1$ is the only restriction imposed so far.

Consider now an almost-FLRW spacetime with a family of observers having a non-relativistic peculiar velocity relative to the comoving frame, which is irrotational and shear-free. This defines the covariant analogues of the quasi-Newtonian cosmologies (see also \S~\ref{sss1+3PVD} above). In what follows we will identify the quasi Newtonian frame with the reference 4-velocity field $u_a$ and align $\tilde{u}_a$ with a non-relativistic (dust) component; a picture that reflects the post-recombination universe. In that case $\tilde{p}=0=\tilde{\pi}_{ab}=\tilde{A}_a$ and, given that $\tilde{q}_a=0$ in the matter frame, expressions (\ref{vadot}) and (\ref{pRay})-(\ref{pomegadot}) reduce to
\begin{eqnarray}
\dot{v}_a&=& -A_{\langle a\rangle}- {1\over3}\,(\tilde{\Theta}- \hat{\Theta})v_{\langle a\rangle}- (\tilde{\sigma}_{ab}-\hat{\sigma}_{ab})v^b- (\tilde{\omega}_{ab}-\hat{\omega}_{ab})v^b\,,  \label{dvadot}\\
\dot{\hat{\Theta}}&=& -{1\over3}\,\tilde{\Theta}\hat{\Theta}- \tilde{\sigma}_{ab}\hat{\sigma}^{ab}+ \tilde{\omega}_{ab}\hat{\omega}^{ab}+ \tilde{\rm D}_a\dot{v}^a+ \left({1\over2}\,\tilde{\rho}-\Lambda\right)v^2\,,  \label{pdRay}\\ \nonumber\\ \dot{\hat{\sigma}}_{ab}&=& -{1\over3}\,\left(\tilde{\Theta}\hat{\sigma}_{ab} +\hat{\Theta}\tilde{\sigma}_{ab}\right)- \hat{\sigma}_{c\langle a}\tilde{\sigma}^c{}_{b\rangle}- \hat{\sigma}_{c\langle a}\tilde{\omega}^c{}_{b\rangle}+ \hat{\omega}_{c\langle a}\tilde{\sigma}^c{}_{b\rangle}+ \hat{\omega}_{c\langle a}\tilde{\omega}^c{}_{b\rangle} \nonumber\\ &&+\tilde{\rm D}_{\langle b}\dot{v}_{a\rangle}+ v^2\tilde{E}_{ab}- \tilde{\varepsilon}_{cd\langle a}v^c\tilde{H}^d{}_{b\rangle}  \label{pdsigmadot}
\end{eqnarray}
and
\begin{eqnarray}
\dot{\hat{\omega}}_{ab}&=& -{1\over3}\,\left(\tilde{\Theta}\hat{\omega}_{ab} +\hat{\Theta}\tilde{\omega}_{ab}\right)- \hat{\sigma}_{c[a}\tilde{\sigma}^c{}_{b]}- \hat{\sigma}_{c[a}\tilde{\omega}^c{}_{b]}+ \hat{\omega}_{c[a}\tilde{\sigma}^c{}_{b]}+ \hat{\omega}_{c[a}\tilde{\omega}^c{}_{b]} \nonumber\\ &&+\tilde{\rm D}_{[b}\dot{v}_{a]}- \tilde{\varepsilon}_{cd[a}v^c\tilde{H}^d{}_{b]}\,,  \label{pdomegadot}
\end{eqnarray}
respectively. Following (\ref{pdsigmadot}), the effect of electric (the tidal) part of the Weyl field on the shear anisotropy of the peculiar motion is of higher perturbative order, compared to that of its magnetic counterpart. This will make a difference when taking the second-order limit of the above named expressions around a quasi-Newtonian background.

\subsection{The mildly nonlinear regime}
Linear perturbation theory provides an adequate description of the early stages of galactic collapse when the distortions from homogeneity and isotropy are still relatively small. As the perturbations grow stronger, however, the linear approximation brakes down and one needs to incorporate nonlinear effects. During the transition from the linear to the fully nonlinear regime, a period which one might call the ``mildly nonlinear era'', one can monitor the perturbed quantities by means of the second-order equations instead of the fully nonlinear ones.

\subsubsection{Second order evolution}\label{sssSOE}
Assuming an Einstein-de Sitter background, the linear relations between the quasi-Newtonian reference system and the matter frame are $\tilde{\Theta}=\Theta+\hat{\Theta}$, $\tilde{\sigma}_{ab}=\hat{\sigma}_{ab}$ and $\tilde{\omega}_{ab}=\hat{\omega}_{ab}$. Also to first order, $\tilde{\rho}=\rho$, $\tilde{E}_{ab}=E_{ab}$ and $\tilde{H}_{ab}=0$, while $A_a$ can be shown to derive from a potential~\cite{2001CQGra..18.5115E}. Then, Eq.~(\ref{dvadot}) reduces to the linear expression
\begin{equation}
\dot{v}_a= -{1\over3}\,\Theta v_a- \tilde{\rm D}_a\phi\,,  \label{ldvadot}
\end{equation}
where $\phi$ is the peculiar gravitational potential. Following~\cite{2001CQGra..18.5115E}, the above accepts a single growing mode with $v_a\propto a^{1/2}$ and has the same linear evolution in both the quasi-Newtonian and the matter frames. Given that, it helps to describe the peculiar motions by means of the rescaled velocity field $V_a=a^{-1/2}v_a$, which remains constant to leading order both in the reference and the matter frame (see~\cite{2001CQGra..18.5115E,2002PhRvD..66l4015E} and also \S~\ref{sssRVs} below).

The mildly nonlinear regime monitors the early proto-galactic collapse of a dust cloud that has decoupled from the background expansion. On these grounds we may set $\Theta\ll\hat{\Theta}$. Then, on a quasi-Newtonian Einstein-de Sitter background, the nonlinear system (\ref{dvadot})-(\ref{pdomegadot}) leads to the second-order set~\cite{2002PhRvD..66l4015E}
\begin{eqnarray}
\dot{v}_a&=& -A_{\langle a\rangle}- {1\over3}\,\Theta v_{\langle a\rangle}\,,  \label{d2vadot}\\
\dot{\hat{\Theta}}&=& -{1\over3}\,\hat{\Theta}^2- 2\hat{\sigma}^2+ 2\hat{\omega}^2+ \tilde{\rm D}_a\dot{v}^a+ {1\over2}\,\rho v^2\,,  \label{pd2Ray}\\ \nonumber\\ \dot{\hat{\sigma}}_{ab}&=& -{2\over3}\,\hat{\Theta}\hat{\sigma}_{ab}- \hat{\sigma}_{c\langle a} \hat{\sigma}^c{}_{b\rangle}+ \hat{\omega}_{c\langle a} \hat{\omega}^c{}_{b\rangle}+ \tilde{\rm D}_{\langle b} \dot{v}_{a\rangle}\,,  \label{pd2sigmadot}\\ \nonumber\\
\dot{\hat{\omega}}_{ab}&=& -{2\over3}\,\hat{\Theta}\hat{\omega}_{ab}- 2\hat{\sigma}_{c[a} \hat{\omega}^c{}_{b]}+ \tilde{\rm D}_{[b}\dot{v}_{a]}\,. \label{pd2omegadot}
\end{eqnarray}
Hence, at this perturbative level, both the electric and the magnetic Weyl tensors have decoupled and the peculiar kinematics evolve unaffected by the long-range gravitational field. The absence of the electric Weyl component in (\ref{pd2sigmadot}), in particular, is a consequence of the two-frame approach. Next, we will see the implications of this absence for the asymptotic final shape of a collapsing overdensity.

\subsubsection{Rescaling the variables}\label{sssRVs}
For our purposes, the key linear result is that the peculiar velocity evolves as $v_a\propto a^{1/2}$. This then means that, to leading order, the rescaled velocity field $V_a=a^{-1/2}v_a$ remains constant (i.e.~$\dot{V}_a=0$) and that $V_a$ comes from a potential and it is therefore non-rotating, namely that $\tilde{\rm D}_{[b}V_{b]}=0$ (see~\cite{2001CQGra..18.5115E,%
2002PhRvD..66l4015E} for details). Hence, one can always assume that the linear peculiar velocity of the dust component is both acceleration-free and irrotational. To fully exploit this linear result we rescale our kinematical variables according to $\hat{\Theta}=a^{1/2}\bar{\Theta}$, $\hat{\sigma}_{ab}=a^{1/2}\bar{\sigma}_{ab}$ and $\hat{\omega}_{ab}=a^{1/2}\bar{\omega}_{ab}$, with $\bar{\Theta}=\tilde{\rm D}_aV^a$, $\bar{\sigma}_{ab}=\tilde{\rm D}_{\langle b}V_{a\rangle}$ and $\bar{\omega}_{ab}=\tilde{\rm D}_{[b}V_{a]}$. Then, the system (\ref{d2vadot})-(\ref{pd2omegadot}) also rescales to~\cite{2002PhRvD..66l4015E}
\begin{eqnarray}
\dot{V}_a&=& -a^{-1/2}A_{\langle a\rangle}- {1\over2}\,\Theta V_{\langle a\rangle}\,,  \label{d2Vadot}\\
\dot{\bar{\Theta}}&=& -{1\over3}\,a^{1/2}\bar{\Theta}^2- 2a^{1/2}\bar{\sigma}^2+ 2a^{1/2}\bar{\omega}^2+ \tilde{\rm D}_a \dot{V}^a+ {1\over2}\,a^{1/2}\rho V^2\,,  \label{pd2Rayr}\\ \nonumber\\ \dot{\bar{\sigma}}_{ab}&=& -{2\over3}\,a^{1/2}\bar{\Theta}\bar{\sigma}_{ab}- a^{1/2}\bar{\sigma}_{c\langle a}\bar{\sigma}^c{}_{b\rangle}+ a^{1/2}\bar{\omega}_{c\langle a}\bar{\omega}^c{}_{b\rangle}+ \tilde{\rm D}_{\langle b}\dot{V}_{a\rangle}\,,  \label{pd2sigmadotr}\\ \nonumber\\
\dot{\bar{\omega}}_{ab}&=& -{2\over3}\,a^{1/2}\bar{\Theta}\bar{\omega}_{ab}- 2a^{1/2}\bar{\sigma}_{c[a}\bar{\omega}^c{}_{b]}+ \tilde{\rm D}_{[b} \dot{V}_{a]}\,. \label{pd2omegadotr}
\end{eqnarray}

\subsection{The relativistic Zeldovich approximation}
Most analytical models of nonlinear structure formation are based on the assumption of spherically symmetry~\cite{1972ApJ...176....1G} -- see also~\cite{1993sfu..book.....P}. The spherical collapse scenario became popular because of its simplicity, but in reality it stops short from explaining key features of the observed universe. Galactic collapse does not seem to proceed isotropically. Galaxy surveys show complicated triaxial structures, which require non-spherical analysis if they were to be explained.

\subsubsection{The Zeldovich ansatz}
The Zeldovich approximation is not restricted to spherical symmetry and addresses the mildly nonlinear collapse of protogalactic clouds. It applies to scales well within the Hubble radius, as they decouple from the background expansion and `turn around'. The approximation works by extrapolating into the nonlinear regime the exact linear result of the acceleration-free and irrotational motion of the dust component. This considerably simplifies the equations and allows for analytical solutions.

When the Zeldovich ansatz $\dot{V}_a=0=\bar{\omega}_{ab}$ is applied to the second-order set (\ref{d2Vadot})-(\ref{pd2omegadot}), the motion of the collapsing pressure-free matter is determined by the reduced pair
\begin{eqnarray}
\dot{\bar{\Theta}}&=& -{1\over3}\,a^{1/2}\bar{\Theta}^2- 2a^{1/2}\bar{\sigma}^2+ {1\over2}\,a^{1/2}\rho V^2\,,  \label{ZRay1}\\ \nonumber\\ \dot{\bar{\sigma}}_{ab}&=& -{2\over3}\,a^{1/2}\bar{\Theta}\bar{\sigma}_{ab}- a^{1/2}\bar{\sigma}_{c\langle a}\bar{\sigma}^c{}_{b\rangle}\,.  \label{Zsigmadot1}
\end{eqnarray}
Note the matter term in the right-hand side of (\ref{ZRay1}). Given that $V_a$ is constant and that $\rho\propto a^{-3}$ for dust, the impact of the background matter upon $\bar{\Theta}$ decays away. This means that the collapse is increasingly dominated by the kinematics and that gravity becomes progressively less important. The situation is closely analogous to that seen in studies of silent universes (e.g.~see~\cite{1995ApJ...445..958B}), or during the Kasner regime of the Bianchi~I cosmologies (see \S~\ref{sssB1Cs}).

Referring the reader to~\cite{2002PhRvD..66l4015E} for the details of the relativistic analysis and also to~\cite{1996ASPC...94...31B} for the original (Newtonian) covariant treatment of the Zeldovich approximation, we introduce the new `time' variable $\tau$ (constructed so that $\dot{\tau}=-a^{1/2}\bar{\Theta}$). The minus sign compensates for the fact that we are dealing with a collapsing region (i.e.~$\bar{\Theta}<0$) and guarantees that $\dot{\tau}>0$ always. Note that $\tau\rightarrow\infty$ as we approach the singularity, where $\bar{\Theta}\rightarrow-\infty$. Then, the system (\ref{ZRay1}) and (\ref{Zsigmadot1}), the former without the matter term, transforms into the set
\begin{eqnarray}
\bar{\Theta}^{\prime}&=& {1\over3}\,\bar{\Theta}+ 2\bar{\Theta}^{-1}\bar{\sigma}^2\,,  \label{ZRay2}\\ \nonumber\\ \bar{\sigma}^{\prime}_{ab}&=& {2\over3}\,\bar{\sigma}_{ab}+ \bar{\Theta}^{-1}\bar{\sigma}_{c\langle a}\bar{\sigma}^c{}_{b\rangle}\,,  \label{Zsigmadot2}
\end{eqnarray}
where primes indicating differentiation with respect to $\tau$. Our last step is to assume the shear eigenframe, where $\bar{\sigma}_{ab}={\rm diag}(\bar{\sigma}_1,\bar{\sigma}_2,\bar{\sigma}_3)$ with $\bar{\sigma}_3=-(\bar{\sigma}_1+\bar{\sigma}_2)$. Relative to this frame Eqs.~(\ref{ZRay2}) and (\ref{Zsigmadot2}) lead to
\begin{eqnarray}
\bar{\Theta}^{\prime}&=& {1\over3}\,\bar{\Theta}+ 2\bar{\Theta}^{-1}\left(\bar{\sigma}_1^2+\bar{\sigma}_2^2 +\bar{\sigma}_1\bar{\sigma}_2\right)\,,  \label{ZRay3}\\ \nonumber\\ \bar{\sigma}^{\prime}_1&=& {2\over3}\,\bar{\sigma}_1+ {1\over3}\,\bar{\Theta}^{-1}\bar{\sigma}_1^2- {2\over3}\,\bar{\Theta}^{-1}\left(\bar{\sigma}_1 +\bar{\sigma}_2\right)\bar{\sigma}_2\,,  \label{Zsigma1dot}\\ \nonumber\\ \bar{\sigma}^{\prime}_2&=& {2\over3}\,\bar{\sigma}_2+ {1\over3}\,\bar{\Theta}^{-1}\bar{\sigma}_2^2- {2\over3}\,\bar{\Theta}^{-1}\left(\bar{\sigma}_1 +\bar{\sigma}_2\right)\bar{\sigma}_1\,.  \label{Zsigma2dot}
\end{eqnarray}
with the behaviour of $\bar{\sigma}_3$ determined from that of $\bar{\sigma}_1$ and $\bar{\sigma}_2$~\cite{2002PhRvD..66l4015E}. The above second-order system provides a fully covariant formulation of the Zeldovich approximation, which governs the small-scale evolution of pressure-free matter, as the latter decouples from the background expansion and starts to turn around and collapse.

\subsubsection{The Zeldovich pancakes}
The question is whether or not the relativistic analysis also predicts that one-dimensional pancakes are the final configurations of any generic overdensity. Given the qualitative nature of the question, one can employ a dynamical-systems approach to look for the answer. To begin with, consider the dimensionless variables (see~\cite{2002PhRvD..66l4015E} and also~\cite{1996ASPC...94...31B})
\begin{equation}
\Sigma_+= {3\over2}\,\bar{\Theta}^{-1} \left(\bar{\sigma}_1+\bar{\sigma}_2\right) \hspace{10mm} {\rm and} \hspace{10mm} \Sigma_-= {\sqrt{3}\over2}\,\bar{\Theta}^{-1} \left(\bar{\sigma}_1-\bar{\sigma}_2\right)\,,  \label{Sigmas}
\end{equation}
which measure the anisotropy of the collapse. Clearly, when both $\Sigma_+$ and $\Sigma_-$ vanish, we are dealing with a spherically symmetric collapse. On introducing $\Sigma_{\pm}$, Eq.~(\ref{ZRay3}) transforms into
\begin{equation}
\bar{\Theta}^{\prime}= {1\over3}\,\bar{\Theta}+ {2\over3}\,\bar{\Theta}\left(\Sigma_+^2+\Sigma_-^2\right)\,,  \label{ZRay4}
\end{equation}
while expressions (\ref{Zsigma1dot}), (\ref{Zsigma2dot}) become
\begin{equation}
\Sigma_+^{\prime}= {1\over3}\,\left[1-\Sigma_+ -2\left(\Sigma_+^2+\Sigma_-^2\right)\right]\Sigma_++ {1\over3}\,\Sigma_-^2  \label{ZSigma1dot}
\end{equation}
and
\begin{equation}
\Sigma_-^{\prime}= {1\over3}\,\left[1+2\Sigma_+ -2\left(\Sigma_+^2+\Sigma_-^2\right)\right]\Sigma_-\,, \label{ZSigma2dot}
\end{equation}
respectively. Thus, the evolution of $\Sigma_+$ and $\sigma_-$ has decoupled from that of $\bar{\Theta}$ and the shape of the collapsing overdensity is monitored by the subsystem (\ref{ZSigma1dot}), (\ref{ZSigma2dot}). Technically speaking, the problem has been reduced to the study of the planar dynamical system depicted in Fig.~\ref{Zp}. Physically, this dimensional reduction means that the shape of the collapsing dust cloud does not depend on the collapse timescale.

\begin{figure*}
\begin{center}
\vspace{-25mm}\includegraphics[height=5.5in]{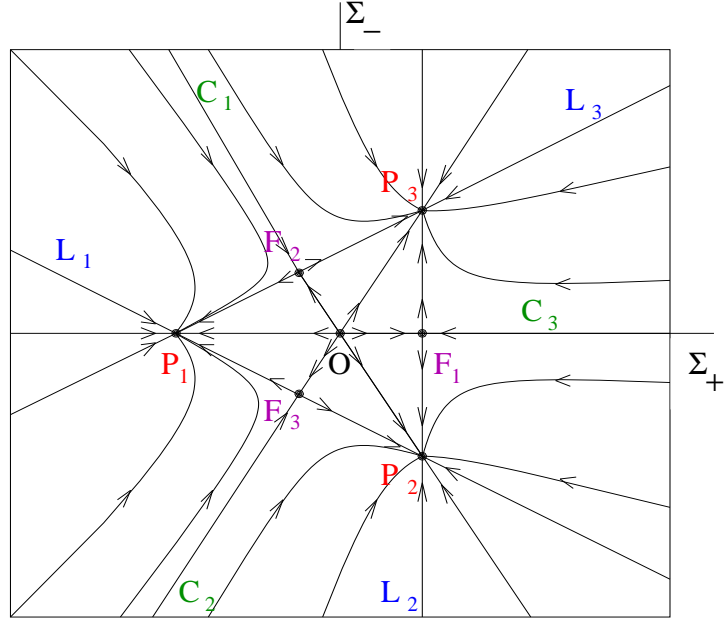}\quad
\end{center}
\vspace{-25mm}\caption{Phase plane with $\Sigma_+\equiv X$ and $\Sigma_-\equiv Y$. The lines $L_1$, $L_2$ and $L_3$ that form the central triangle correspond to $\sigma_i=-\bar{\Theta}/3$ ($i=1,2,3$), with the three pancakes located at $P_1$, $P_2$ and $P_3$ where these lines intersect. The points $F_1$, $F_2$, $F_3$ represent filamentary solutions and spindle-like singularities, while $O$ corresponds to spherically symmetric, isotropic collapse (see~\cite{2002PhRvD..66l4015E} and also~\cite{1996ASPC...94...31B}).}\label{Zp}
\end{figure*}

Referring to~\cite{2002PhRvD..66l4015E} for the technical details, we point out that the vertices $P_1$, $P_2$ and $P_3$ of the triangle seen in Fig.~\ref{Zp} are stationary points of the system (\ref{ZSigma1dot}), (\ref{ZSigma2dot}) and also act as attractors. Each one of the three vertices corresponds to an one-dimensional pancake solution, which is stationary along two directions and collapses along the third shear eigendirection. Generic solutions are asymptotic to one of these pancakes (one for each eigendirection). The bisecting lines $C_1$, $C_2$ and $C_3$ intersect to the stationary point $O$ that represents shear-free spherically symmetric collapse. Finally, where the bisecting lines intersect the triangle, we have the stationary points $F_1$, $F_2$ and $F_3$ that correspond to exact filamentary solutions. The pancakes are stable nodes, the filaments are saddle points and the spherically symmetric collapse is a unstable node. In other words, once the collapse sets in, the pancakes are the natural attractors for a generic overdensity. This result is in disagreement with dynamical studies of silent universes, which argue for spindle-like rather than pancake singularities (see~\cite{1995ApJ...445..958B} and also~\cite{1994ApJ...431..486B}). The reason for this difference appears to lie in the role of the tidal field. Silent universes allow for a nonzero electric Weyl component, but set its magnetic counterpart to zero. In the presence of this ``truncated'' tidal field, the collapsing overdesnity appears to evolve towards a Kasner-type singularity where pancakes are a set of measure zero. The analysis of peculiar velocities in Einstein-de Sitter models, however, has shown that the long-range gravitational field has negligible input at second order. As a result, the relativistic equations have been reduced to the Newtonian ones and pancakes have been reinstated as the natural attractors of generic protogalactic collapse, this time within the realm of general relativity. This also appears in agreement with numerical simulations which also seem to favor pancake formation over all other types of singularities~\cite{1995PhRvL..75....7S}.

\subsection{Averaging and backreaction}
The increasing complexity of the equations, as one moves beyond the linear regime, means that analytical studies of nonlinear structure formation are always bound within the limits of certain approximations. So, a common feature in every (local) nonlinear analytical work is that some effects are sidestepped in favour of others. The spherical collapse scenario and the Zeldovich anzatz are probably the two best known examples. It therefore seems likely that only via a consistent averaging method one can achieve a tractable, all inclusive (and nonlocal) nonlinear study.

\subsubsection{Spatial averaging}
The question of whether an inhomogeneous spacetime behaves on average like a homogeneous solution of Einstein's equations is a long-standing issue, directly related to the problem of averaging general-relativistic spacetimes~\cite{1963SvA.....6..699S,%
1984grg..conf..215E}. The reason is the generic nonlinearity of general relativity and the difficulty of establishing a unique and unambiguous way of averaging the spacetime metric, without throwing away crucial information during the process. This ongoing ambiguity has lead a number of cosmologists to argue that the averaging problem may be crucial in our understanding of the recent expansion of the visible universe. Thus, when it was recently suggested that structure formation may be responsible for the acceleration of the expansion, averaging techniques were employed to study the dark energy problem in terms of kinematic backreaction effects from spatial inhomogeneities~\cite{2004JCAP...02..003R,%
2005PhRvD..71f3537B,2005hep.th....3117K,2005PhRvD..72b3517G,%
2005PhRvL..95o1102C,2005gr.qc.....7057N,2006NJPh....8..322K,%
2005astro.ph.10523M,2006astro.ph..1699M,2006AIPC..861..987M,%
2006CQGra..23.6955P}. Although the backreaction idea has been criticised, primarily on the basis of currents observations~\cite{2005PhRvD..71j3521F,2005PhRvD..72h3501H,%
2005JCAP...09..009G,2007JCAP...01....7A,2006CQGra..23..235I,%
2006CQGra..23..817B,2006PhRvD..74b3506V,2008JCAP...03..023P}, it is attractive in principle because it can solve the coincidence problem without appealing to a cosmological constant, speculative quintessence fields, or nonlinear corrections to Einstein's gravity. For recent reviews on the backreaction question the reader is referred to~\cite{2008GReGr..40..467B,2008arXiv0801.2692R}.

General relativity has been applied to a range of averaging scales, depending on the scale of the physical system under study. Cosmology applies to the largest astrophysical scales, which typically extend over a significant fraction of the Hubble length (see~\cite{2005PhLA..347...38E} for a discussion and further references). Nevertheless, the major issue of defining a suitable averaging process remains open. The literature contains more than one averaging methods that have been applied to cosmology~\cite{1968PhRv..166.1263I,1997BASI...25..401Z,%
1997A&A...320....1B,2000GReGr..32..105B,2001GReGr..33.1381B}. Here we will follow the Buchert approach, which builds on the Newtonian theory (where spatial averaging is relatively simple) before extending to general relativity by confining to scalar variables~\cite{1997A&A...320....1B,2000GReGr..32..105B}. These averaged scalars become the effective dynamical sources that an observer will expect to measure. In the averaged equations one isolates an explicit source term, which is commonly referred to as the {\em backreaction term}. The latter quantifies the deviation from a given `fitting model' that usually coincides with the Friedmann spacetime. From our point of view, the Buchert scheme has an additional advantage because it applies naturally to the 1+3 covariant formulae used here.

Following~\cite{1997A&A...320....1B,2000GReGr..32..105B}, the spatial averaging of an arbitrary scalar field $\phi=\phi(x^a)$, over a simply connected domain $D$ is a covariant operation, defined by the averaging operator
\begin{equation}
\langle\phi\rangle_D= V_D^{-1}\int_D\phi\mathcal{H}{\rm d}^3x^a\,,  \label{<phi>}
\end{equation}
where the angled brackets indicate spatial averaging, $V_D$ is the volume of the domain in question and $\mathcal{H}=\sqrt{{\rm det}(h_{ab})}$. Since the volume is time-dependent in principle, the average of the time derivative of a locally defined scalar generally differs form the time derivative of the average. This non-commutativity between spatial averaging and temporal evolution can be formulated in a simple mathematical rule. For a rest-mass preserving domain the commutation between the volume-averaging and the time-averaging operators leads to
\begin{equation}
\langle\phi\rangle_D^{\cdot}- \langle\dot{\phi}\rangle_D= \langle\Theta\phi\rangle_D- \langle\Theta\rangle_D\langle\phi\rangle_D\,,  \label{avrule}
\end{equation}
with $\Theta$ representing the local expansion rate. The right-hand side of the above generally introduces source terms in the evolution equations which can be interpreted as the backreaction effects of the averaging process. Such backreaction terms have recently been considered as a possible solution to the dark energy question.

\subsubsection{The averaged equations}
The first step is to write the average equations in the form of a chosen family of local equations, with any deviation between the two sets treated as a backreaction effect. Here, we will write the averaged formulae in a Friedmann-type form. To be precise, assuming irrotational dust moving along worldlines tangent to the 4-velocity $u_a$-field and applying rule (\ref{avrule}) to the continuity, the Friedmann and the Raychaudhuri equations we arrive at the system~\footnote{The absence of rotation is necessary to guarantee the existence of flow-orthogonal hypersurfaces.}
\begin{eqnarray}
&&\langle\rho\rangle^{\cdot}+ \langle\Theta\rangle\langle\rho\rangle= 0\,,  \label{avce}\\ \nonumber\\ &&\langle\Theta\rangle^{\cdot}+ {1\over3}\,\langle\Theta\rangle^2+ {1\over2}\,\langle\rho\rangle- \Lambda= {2\over3}\,\langle\left(\Theta-\langle\Theta\rangle\right)^2\rangle- 2\langle\sigma^2\rangle\,,  \label{avRay}\\ \nonumber\\ &&{1\over2}\,\langle\mathcal{R}\rangle- \langle\rho\rangle+ {1\over3}\,\langle\Theta\rangle^2- \Lambda= -{1\over3}\,\langle\left(\Theta-\langle\Theta\rangle\right)^2\rangle+ \langle\sigma^2\rangle\,,  \label{avFried1}
\end{eqnarray}
where the right-hand sides of (\ref{avRay}) and (\ref{avFried1}) involve the identity $\langle(\phi-\langle\phi\rangle)^2\rangle= \langle\phi^2\rangle-\langle\phi\rangle^2$. The above provides a (non-closed) set of three effective Einstein equations for the spatially averaged scalars of non-rotating inhomogeneous universes containing pressure-free matter~\cite{2004JCAP...02..003R,%
2006CQGra..23.1823R}. Within this environment, Eqs.~(\ref{avRay}), (\ref{avFried1}) are exact and there is no need to assume that the inhomogeneity and the anisotropy are small perturbations. Note that, despite the non-commutativity between spatial averaging and temporal evolution, the averaged continuity equation has retained the form of its local counterpart. This means that $\langle\rho\rangle\propto a_D^{-3}$, where $a_D$ is the average scale factor smoothed over the domain $D$ and defined so that $\dot{a}_D/a_D=\langle\Theta\rangle/3$. However, the averaging process has led to extra terms in the right-hand sides of both (\ref{avRay}) and (\ref{avFried1}), collectively given by the domain-dependent scalar
\begin{equation}
\mathcal{Q}_D= {2\over3}\,\langle\left(\Theta-\langle\Theta\rangle\right)^2\rangle- 2\langle\sigma^2\rangle= -{2\over3}\left(\langle\Theta\rangle^2-\langle\Theta^2\rangle\right)- 2\langle\sigma^2\rangle\,.  \label{cQ}
\end{equation}
We interpret this quantity, which in principle can be either positive or negative, as the kinematic backreaction of spatial averaging upon pressure-free FRW models, since in its absence the averaged formulae recover the form of their local counterparts.\footnote{The fitting model does not always need to be the FRW spacetime. For example, in~\cite{2007CQGra..24.1023B} the Buchert averaged formulae were written in Bianchi type-$I$ form to study analogous backreaction effects on anisotropic cosmological models.} We also note that Eqs.~(\ref{avce})-(\ref{avFried1}) may be recast into the standard form of a spatially flat LFRW universe, provided the (effective) density and pressure are given by $\rho_{eff}=\langle\rho\rangle- (\mathcal{Q}_D+\langle\mathcal{R}\rangle)/2$ and $p_{eff}= -\mathcal{Q}_D/2+\langle\mathcal{R}\rangle/6$ respectively~\cite{2001GReGr..33.1381B}. Thus, given an effective equation of state, the set (\ref{avce})-(\ref{avFried1}) is solvable and provides the scalar characteristics of the inhomogeneous universe on a given spatial scale.

\subsection{Backreaction and accelerated expansion}
The most striking feature of the $\Lambda{\rm CDM}$ model is that 95\% of the matter in the universe today is in some unknown form. In fact, the bulk of the matter (approximately 70\% of it) is in the form of dark energy, triggering the accelerated expansion. Nevertheless, we have not actually measured the cosmic acceleration nor detected the dark-energy component of the universe. What we have observed, are cosmological parameters sensitive to the expansion history of the universe. The results show that the Enstein-de Sitter model does not fit the current data (see~\cite{2007arXiv0709.3102K} for an up-to-date discussion).

\subsubsection{Conditions for acceleration}
Observations also seem to indicate that the expansion of the universe started accelerating relatively recently; roughly at the time galaxy formation was moving from its linear phase into the nonlinear regime. This coincidence has led a number of cosmologists to suggest that cosmic acceleration could be the direct consequence of structure formation and of the fact that, at least on certain scales, the universe does not obey the FLRW symmetries. These ideas have lead to what are now known as `kinematic backreaction' scenarios.

Following (\ref{avRay}) and in the absence of a cosmological constant, the backreaction effects will cause a given domain of the averaged universe to expand at an accelerated pace (i.e.~with $\ddot{a}_D>0$) provided that
\begin{equation}
\mathcal{Q}_D> {1\over2}\,\langle\rho\rangle\,.  \label{aacon1}
\end{equation}
This means that $\mathcal{Q}_D$ needs to be positive, which implies that shear fluctuations must be superseded by those in the volume expansion (see (\ref{cQ})). In the opposite case, $\mathcal{Q}_D$ will be negative and the domain will decelerate further instead of accelerating. Condition (\ref{aacon1}) also implies that, if backreaction is to work, $\mathcal{Q}_D$ should decay slower than the average density.

To examine the above condition further we introduce the dimensionless, domain-dependent parameters $\Omega_{\rho}=3\langle\rho\rangle/\langle\Theta\rangle^2$, $\Omega_{\mathcal{R}}=-3\langle\mathcal{R}\rangle/ 2\langle\Theta\rangle^2$ and $\Omega_{\mathcal{Q}}= -3\mathcal{Q}_D/2\langle\Theta\rangle^2$ and then recast expression (\ref{avFried1}) into
\begin{equation}
\Omega_{\rho}+ \Omega_{\mathcal{R}}+ \Omega_{\mathcal{Q}}= 1\,.  \label{avFried2}
\end{equation}
At the same time, condition (\ref{aacon1}) takes the form
\begin{equation}
\Omega_{\mathcal{Q}}< -{1\over4}\,\Omega_{\rho}\,,  \label{aacon2}
\end{equation}
ensuring that $\mathcal{Q}_D$ is positive as required. Substituting this result into Eq.~(\ref{avFried2}), the latter leads to the constraint
\begin{equation}
{3\over4}\,\Omega_{\rho}+ \Omega_{\mathcal{R}}> 1\,,  \label{aacon3}
\end{equation}
between the matter and the 3-curvature contribution to the effective $\Omega$-parameter of the averaged universe. Note that $\Omega_{\rho}$ is always positive, by construction, though $\Omega_{\mathcal{Q}}$ can take negative values as well.

It is conceivable that one can arrive to the same qualitative result, namely that the aforementioned backreaction effects can lead to accelerated expansion, through a perturbative approach as well. In~\cite{2005hep.th....3117K}, contributions from super-Hubble perturbations to the scale-factor evolution of a spatially flat FLRW background were claimed capable of accelerating the expansion rate of the inhomogeneous model, provided the initial perturbation had the right sign (see also~\cite{2005PhRvD..72h3501H} for the relevant counter-arguments).

\subsubsection{Attractive aspects and caveats}
The observed universe is full of nonlinear structures that, at least on certain scales, can seriously distort its perceived homogeneity and isotropy. As yet, the implications of these nonlinearities have not been fully appreciated, since the mathematical complexity of the problem has discouraged a systematic and in depth study. In this respect, spatial averaging may provide an alternative approach. An advantage of the Buchert averaging scheme is that it encodes the nonlinear effects mentioned above within scalars that are relatively simple to analyse (see~\cite{2008GReGr..40..467B} for recent review). It is not surprising therefore that the method has been applied to study the backreaction idea by several authors. Different averaging schemes, primarily that of Zalaletdinov~\cite{1997BASI...25..401Z}, have also been employed with similar qualitative results~\cite{2005PhRvL..95o1102C,%
2007arXiv0704.1734C,2007PhRvD..76d4006P}. The common factor is that backreaction can substantially change the behaviour of the exact model and therefore potentially explain the accelerated expansion of our universe without appealing to exotic matter fields, or introducing corrections to standard gravity. Moreover, since the effects are triggered by the onset of structure formation, they can offer a natural answer to the coincidence problem. All these make the backreaction idea attractive in principle.

However, current observations support a low density universe with $\Omega_{\rho}$ considerably less than unity. If accelerated expansion is to be triggered solely by backreaction effects, condition (\ref{aacon3}) is only satisfied in models with a considerable amount of negative (recall that $\langle\mathcal{R}\rangle= -2\langle\Theta\rangle^2\Omega_{\mathcal{R}}/3$) spatial curvature~\cite{2006CQGra..23.1823R,2005CQGra..22L.113B}. In other words, condition (\ref{aacon3}) seems to contradict the widespread belief that the 3-curvature of our universe is very small.

Even if we assume that the aforementioned backreaction effects are strong enough to supersede the `background' kinematics, nothing guarantees that they will lead to accelerated expansion. The sign of the backreaction term, which will decide which way the effects will go to, has yet to be decided. In fact, because of this sign-ambiguity, the same kinematic backreaction term, which is now proposed as a conventional solution to the dark-energy problem, was earlier suggested as an effective dark-matter source~\cite{1996ASPC...94..349B}.

We may not have clear information on the sign of $\mathcal{Q}_D$, but we can monitor its dynamical evolution by means of a consistency/integrability condition. In particular, taking the time derivative of (\ref{avFried1}) and then using (\ref{avce}) and (\ref{avRay}) one arrives at
\begin{equation}
\dot{\mathcal{Q}}_D+ 2\langle\Theta\rangle\mathcal{Q}_D= -\langle\mathcal{R}\rangle^{\cdot}- {2\over3}\,\langle\Theta\rangle \langle\mathcal{R}\rangle\,.  \label{cQdot}
\end{equation}
The above means that if $\langle\mathcal{R}\rangle\propto a_D^{-2}$, namely if the averaged spatial curvature behaves like its FRW counterpart, the backreaction term will scale as $\mathcal{Q}_D\propto a_D^{-6}$ mimicking a `stiff' dilatonic fluid. Clearly, the same evolution law holds when $\langle\mathcal{R}\rangle=0$ as well.\footnote{As expected, when $\mathcal{Q}_D$ vanishes, $\langle\mathcal{R}\rangle\propto a_D^{-2}$ and the averaged model behaves like its standard LFRW counterpart (see Eqs.~(\ref{avce})-(\ref{avFried1})). Generally speaking, however, the reduced system still describes inhomogeneous non-equilibrium states.} In either case, the density term in (\ref{avRay}) will quickly dominate the backreaction effects, even when condition (\ref{aacon3}) is initially satisfied. Recall that $\langle\rho\rangle\propto a_D^{-3}$, just like in the standard Einstein-de Sitter model. Therefore, a change in the scale-factor dependance of the average 3-Ricci scalar seems necessary if the backreaction idea is to work~\cite{2005CQGra..22L.113B}. Overall, if condition (\ref{aacon1}) holds, a typical Hubble volume would not correspond to a perturbative state near the LFRW model~\cite{2005CQGra..22L.113B}.

\section{Summary and outlook}\label{sSO}
Cosmology is now firmly a data-driven science. The principal drivers over the past fifteen years have been large surveys of galaxy
redshifts and hence their three-dimensional clustering, precision
measurements of the CMB temperature anisotropies and, recently,
polarization, and measurements of the magnitude-redshift relation of distant supernovae. Together, these have revolutionised our understanding of the constituents, geometry and initial conditions
for the observable universe.

General relativity, or its Newtonian approximation on sub-Hubble
scales, appears to provide a satisfactory description of the
dynamics of the universe and its fluctuations. This is a
considerable triumph for the theory since it extends the range of
scales over which it has passed observational tests by many orders
of magnitude. Bold theoretical predictions of the simplest models,
in which initially small, adiabatic and Gaussian fluctuations evolve passively under gravity and hydrodynamics on an FRW background, have been impressively verified. The observed acoustic peaks in the CMB
and galaxy clustering power spectrum are a particularly noteworthy
example. Our aim here has been to provide a comprehensive review of
the dynamics and perturbations of cosmological models that are based in general relativity, and to describe how a simple class of these
models provides an excellent and remarkably efficient description of current data. By uniformly working with 1+3-covariant methods
throughout, we hope to have presented a unified treatment of a range of topics in contemporary cosmology.

However, general relativity can only explain the observed structure
and evolution of the universe by supplementing the known particles
of the standard model with additional components that have, as yet,
not yielded to direct detection. Cold dark matter is invoked to
reconcile the rotation curves of galaxies with their distribution of luminous matter and the observed clustering of galaxies with the
small amplitude of CMB fluctuations. One or more potential-dominated scalar fields are invoked in the early universe to drive inflation
and hence remove some of the fine-tuning issues that beset the
standard hot big-bang scenario. A further violation of the strong
energy condition is required at late times to explain the observed
accelerated expansion. This requires that a further component dubbed dark energy -- either a woefully small contribution from the vacuum, or an additional, unclustered dynamical component -- be introduced.
Alternatives to CDM, a fundamental inflaton field and dark energy
are all being pursued actively. For example, in attempts to realise
inflation in string/M theory, the role of the inflaton can be played by scalar fields describing the geometry of the compactified
dimensions (see Ref.~\cite{2007hep.th....2059K} for a recent
review). Infra-red modifications to general relativity can lead to
late-time acceleration without dark energy, but do not address the
problem of why the vacuum does not gravitate~\cite{2004PhRvD..70d3528C}. Introducing additional scalar, vector and tensor degrees of freedom (not all of which are
dynamical) yields a relativistic theory of gravity~\cite{2004PhRvD..70h3509B} which reduces to the
phenomenology of Milgrom's modified Newtonian dynamics~\cite{1983ApJ...270..365M} on galactic scales and may
therefore offer an alternative to CDM. Future astronomical
observations should be able to discriminate between some of these
alternatives. For example, a detection of CMB $B$-mode polarization
may signal trouble for many string-theory scenarios of
inflation~\cite{2007PhRvD..75l3508B}; gravitational lensing
observations may provide further support for a dominant,
non-interacting, cold component (i.e.\ CDM) in galaxy clusters and
large-scale structures~\cite{2006ApJ...648L.109C,%
2007ApJ...661..728J}; and accurately mapping out the expansion rate of the universe to high redshift with future supernovae surveys and baryon acoustic oscillations, and the growth rate of structure with tomographic lensing surveys, may distinguish between modifications of gravity or a physical dark energy component~\cite{2006ApJ...648..797B}.

{\bf Acknowledgements:} CT wishes to thank John Barrow and George Ellis for many helpful discussions. AC acknowledges a Royal Society University Research Fellowship and gratitude to his long-term collaborators in this area, particularly Antony Lewis. The work of RM is supported by STFC and he thanks George Ellis and Malcolm MacCallum for useful discussions.

\newpage

\appendix

\section{Appendix}\label{sA}
\subsection{1+3 Covariant decomposition}\label{AssCD}
The skew part of a projected rank-2 tensor is spatially dual to the projected vector $S_a=\varepsilon_{abc}S^{[bc]}/2$, and any projected second-rank tensor has the irreducible covariant
decomposition
\begin{equation}
S_{ab}= {1\over3}\,Sh_{ab}+ \varepsilon_{abc}S^c+ S_{\langle
ab\rangle}\,,
\end{equation}
where $S=S_{cd}h^{cd}$ is the spatial trace and $S_{\langle
ab\rangle}=S_{(ab)}-Sh_{ab}/3$ is the projected symmetruc and
trace-free (PSTF) part of $S_{ab}$. In the 1+3 covariant formalism,
all quantities are either scalars, projected vectors or PSTF
tensors.

The projected derivative operator, $\D_a=h_a{}^b\nabla_b$, further
splits irreducibly into a 1+3 covariant spatial divergence
\begin{equation}
\D^aV_a\,, \hspace{15mm} \D^bS_{ab}\,,
\end{equation}
a spatial curl
\begin{equation}
\curl V_a= \varepsilon_{abc}\D^bV^c\,, \hspace{15mm} \curl S_{ab}=
\varepsilon_{cd(a}\D^cS_{b)}{}^d\,,
\end{equation}
and a 1+3 covariant spatial distortion
\begin{equation}
\D_{\langle a}V_{b\rangle}= \D_{(a}V_{b)}-
{1\over3}\,\D_cV^c\,h_{ab}\,, \hspace{15mm} \D_{\langle a}
S_{bc\rangle}= \D_{(a}S_{bc)}- {2\over5}\,h_{(ab}\D^d S_{c)d}\,.
\end{equation}
Note that, as a result of the commutation laws between projected
covariant derivatives (see next section), the ${\rm div}\,{\rm
curl}$ operation is {\em not} in general zero.

The covariant irreducible decompositions of the derivatives of
scalars, vectors and rank-2 tensors are given in exact (nonlinear)
form by
\begin{eqnarray}
\nabla_a\psi&=& -\dot{\psi}u_a+ \D_a\psi\,,  \label{a1}\\
\nabla_bV_a&=& -u_b\left(\dot{V}_{\langle a\rangle}+
A_cV^cu_a\right)+ u_a\left[{1\over3}\,\Theta V_b+\sigma_{bc}V^c+
\varepsilon_{bde}\omega^dV^e\right] \nonumber\\ &&+{1\over3}\,\D_c
V^c\,h_{ab}-{1\over2}\,\varepsilon_{abc}\curl V^c+\D_{\langle a}
{V}_{b\rangle}\,, \label{a2}
\end{eqnarray}
and
\begin{eqnarray}
\nabla_cS_{ab}&=& -u_c\left(\dot{S}_{\langle
ab\rangle}+2u_{(a}S_{b)d}A^d\right)+ 2u_{(a}\left[{1\over3}\,\Theta
S_{b)c}+ S_{b)}{}^d\left(\sigma_{cd}
-\varepsilon_{cde}\omega^e\right)\right] \nonumber\\
&&+{3\over5}\,\D^dS_{\langle a|d|}h_{b\rangle c}
-{2\over3}\,\varepsilon_{dc(a}S_{b)}{}^d+\D_{\langle
a}{S}_{bc\rangle}\,,  \label{a3}
\end{eqnarray}
respectively. The algebraic correction terms in equations (\ref{a2}) and (\ref{a3}) arise from the relative motion of comoving observers, as encoded in the kinematic quantities.

\subsection{Transformations under a 4-velocity
boost}\label{AssTU4VB}
Consider an observer moving with 4-velocity $\tilde{u}_a$ relative
to the $u_a$-frame. The two motions are related by the
transformation
\begin{equation}
\tilde{u}_a = \gamma(u_a+v_a)\,, \label{B25}
\end{equation}
where $\gamma=(1-v^2)^{-1/2}$ is the Lorentz-boost factor and $v_a$
is the `peculiar' velocity (with $v_au^a=0$). Note that for
non-relativistic peculiar motions $v\ll1$ and $\gamma\simeq1$. Also,
\begin{eqnarray}
\tilde{h}_{ab}&=& h_{ab}+ \gamma^2\left(v^2u_au_b
+2u_{(a}v_{b)}+v_av_b\right)\,,  \label{ab1} \\
\tilde{\varepsilon}_{abc}&=& \gamma\varepsilon_{abc}+
\gamma\left(2u_{[a}\varepsilon_{b]cd}
+u_c\varepsilon_{abd}\right)v^d\,,  \label{ab2}
\end{eqnarray}
are the relations between the fundamental algebraic tensors of the
two frames.

The irreducible kinematic quantities, as measured in the
$\tilde{u}_a$-frame, are defined by means of the decomposition (see
Eq.~(\ref{Nbua}) in \S~\ref{sssKs})
\begin{equation}
\nabla_b\tilde{u}_a=
{1\over3}\,\tilde{\Theta}\tilde{h}_{ab}+\tilde{\sigma}_{ab}
+\tilde{\varepsilon}_{abc}\tilde{\omega}^c-\tilde{A}_a\tilde{u}_b\,.
\label{tNbua}
\end{equation}
Using the above, the relation $\nabla_a\gamma=
\gamma^3v^b\nabla_av_b$ and expression (\ref{a2}) from Appendix~\ref{AssCD}, we arrive at the following kinematic transformation laws:
\begin{eqnarray}
\tilde{\Theta}&=& \gamma\Theta+ \gamma\left(\D_av^a+A^av_a\right)+
\gamma^3W\,,  \label{ab3}\\ \tilde{A}_a&=& \gamma^2A_a+
\gamma^2\left[\dot{v}_{\langle a\rangle}+{1\over3}\,\Theta v_a
+\sigma_{ab}v^b-\varepsilon_{abc}\omega^bv^c+
\left({1\over3}\,\Theta v^2+A^bv_b
+\sigma_{bc}v^bv^c\right)u_a\right. \nonumber\\ &&\left.
+{1\over3}\,(\D_bv^b)v_a+{1\over2}\,\varepsilon_{abc}v^b\curl v^c
+v^b\D_{\langle b}v_{a\rangle}\right]+\gamma^4W(u_a+v_a)\,,
\label{ab4}\\ \tilde{\omega}_a&=& \gamma^2\left[\left(1-
{1\over2}\,v^2\right)\omega_a-{1\over2}\,\curl v_a
+{1\over2}\,v_b\left(2\omega^b-\curl v^b \right)u_a+
{1\over2}\,v_b\omega^bv_a \right. \nonumber\\ &&\left.
+{1\over2}\,\varepsilon_{abc}A^bv^c
+{1\over2}\,\varepsilon_{abc}\dot{v}^bv^ca
+{1\over2}\,\varepsilon_{abc}\sigma^b{}_dv^cv^d\right]\,,
\label{ab5}\\ \tilde{\sigma}_{ab}&=& \gamma\sigma_{ab}+
\gamma\left(1+\gamma^2\right)u_{(a}\sigma_{b)c}v^c+
\gamma^2A_{(a}\left[v_{b)}+v^2u_{b)}\right] \nonumber\\
&&+\gamma\D_{\langle a}v_{b\rangle}- {1\over3}\,h_{ab}
\left[A_cv^c+\gamma^2\left(W-\dot{v}_cv^c\right)\right] \nonumber\\
&&+\gamma^3u_au_b\left[\sigma_{cd}v^cv^d+{2\over3}\,v^2A_cv^c
-v^cv^d\D_{\langle c}v_{d\rangle}
+\left(\gamma^4-{1\over3}\,v^2\gamma^2 -1\right)W\right]\nonumber\\
&&+\gamma^3u_{(a}v_{b)}\left[A_cv^c+\sigma_{cd}v^cv^d-\dot{v}_cv^c+
2\gamma^2\left(\gamma^2-{1\over3}\,\right)W\right]\nonumber\\
&&+{1\over3}\,\gamma^3v_av_b\left[\D_cv^c-A_cv^c
+\gamma^2\left(3\gamma^2-1\right)W\right]+\gamma^3v_{\langle a}
\dot{v}_{b\rangle }+v^2\gamma^3u_{(a}\dot{v}_{\langle b\rangle )} \nonumber\\
&&+\gamma^3v_{(a}\sigma_{b)c}v^c-\gamma^3\omega^bv^c\varepsilon_{bc(a}
\left(v_{b)}+v^2u_{b)}\right)+ 2\gamma^3v^c\D_{\langle c}
v_{(a\rangle }\left(v_{b)}+u_{b)}\right)\,,  \label{ab6}
\end{eqnarray}
with
\begin{equation}
W\equiv \dot{v}_cv^c+ {1\over3}\,v^2\D_cv^c+ v^cv^d\D_{\langle c}
v_{d\rangle}\,. \label{W}
\end{equation}

Similarly, one can decompose the energy-momentum tensor of the
matter with respect to the $\tilde{u}_a$-frame (see expression
(\ref{Tab1}) in \S~\ref{sssMFs}). Then,
\begin{equation}
T_{ab}=\tilde{\rho}\tilde{u}_a\tilde{u}_b+
\tilde{p}\,\tilde{h}_{ab}+ 2\tilde{q}_{(a}\tilde{u}_{b)}+
\tilde{\pi}_{ab}\,,  \label{tilTab}
\end{equation}
and the transformed dynamic quantities are given by
\begin{eqnarray}
\tilde{\rho}&=& \rho+ \gamma^2\left[v^2(\rho+p)-2q_av^a
+\pi_{ab}v^av^b\right]\,, \label{ab7}\\ \tilde{p}&=& p+
{1\over3}\,\gamma^2\left[v^2(\rho+p)-2q_av^a
+\pi_{ab}v^av^b\right]\,, \label{ab8}\\ \tilde{q}_a&=& \gamma q_a-
\gamma\pi_{ab}v^b- \gamma^3\left[(\rho+p)-2q_bv^b
+\pi_{bc}v^bv^c\right]v_a \nonumber\\ &&-\gamma^3\left[v^2(\rho+p)-
(1+v^2)q_bv^b+\pi_{bc}v^bv^c\right]u_a\,, \label{ab9}\\
\tilde{\pi}_{ab}&=& \pi_{ab}+ 2\gamma^2v^c\pi_{c(a}\left\{u_{b)}
+v_{b)}\right\}-2v^2\gamma^2q_{(a}u_{b)}-2\gamma^2q_{\langle a}
v_{b\rangle} \nonumber\\ &&-{1\over3}\,\gamma^2\left[v^2(\rho+p)
+\pi_{cd}v^cv^d\right]h_{ab} \nonumber\\
&&+{1\over3}\,\gamma^4\left[2v^4(\rho+p)-4v^2q_cv^c
+(3-v^2)\pi_{cd}v^cv^d\right]u_au_b \nonumber\\
&&+{2\over3}\,\gamma^4\left[2v^2(\rho+p)-(1+3v^2)q_cv^c+2
\pi_{cd}v^cv^d\right]u_{(a}v_{b)} \nonumber\\
&&+{1\over3}\,\gamma^4\left[(3-v^2)(\rho+p)-4q_cv^c+2
\pi_{cd}v^cv^d\right]v_av_b\,. \label{ab10}
\end{eqnarray}

Finally, relative to the $\tilde{u}_a$-frame, the
Gravito-electric/magnetic field decomposes according to (see
Eq.~(\ref{rmweyl}) in \S~\ref{sssWC}),
\begin{equation}
C_{ab}{}{}^{cd}= 4\left(\tilde{u}_{[a}\tilde{u}^{[c}
+\tilde{h}_{[a}{}^{[c}\right)\tilde{E}_{b]}{}^{d]}+
2\tilde{\varepsilon}_{abe}\tilde{u}^{[c}\tilde{H}^{d]e}+
2\tilde{u}_{[a}\tilde{H}_{b]e}\tilde{\varepsilon}^{cde}\,,
\label{tWeyl}
\end{equation}
where $\tilde{E}_{ab}$ and $\tilde{H}_{ab}$ are respectively the
Lorentz-boosted electric and magnetic components of the free
gravitational field. Then,
\begin{eqnarray}
\tilde{E}_{ab}&=& \gamma^2\left\{(1+v^2)
E_{ab}+v^c\left[2\varepsilon_{cd(a}H_{b)}{}^d+
2E_{c(a}u_{b)}\right.\right. \nonumber\\
&&\left.\left.+(u_au_b+h_{ab})E_{cd}v^d-2E_{c(a}v_{b)}
+2u_{(a}\varepsilon_{b)cd}H^{de}v_e\right]\right\} \,, \label{ab11}\\
\tilde{H}_{ab} &=& \gamma^2\left\{(1+v^2)
H_{ab}+v^c\left[-2\varepsilon_{cd(a}E_{b)}{}^d+
2H_{c(a}u_{b)}\right.\right. \nonumber\\
&&\left.\left.+(u_au_b+h_{ab})H_{cd}v^d-2H_{c(a}v_{b)}
-2u_{(a}\varepsilon_{b)cd}E^{de}v_e\right]\right\}\,. \label{ab12}
\end{eqnarray}

The transformation laws of the electric and magnetic Weyl components, may be compared to those of their Maxwell counterparts, namely to
\begin{eqnarray}
\tilde{E}_a&=& \gamma\left(E_a+\varepsilon_{abc}v^bB^c
+v^bE_bu_a\right)\,, \label{tEa}\\ \tilde{B}_a&=& \gamma\left(B_a
-\varepsilon_{abc}v^bE^c+v^bB_bu_a\right)\,, \label{tBa}
\end{eqnarray}
where (see decomposition (\ref{Fab}) in \S~\ref{sssEMFs})
\begin{equation}
F_{ab}= 2u_{[a}E_{b]}+\varepsilon_{abc}B^c=
2\tilde{u}_{[a}\tilde{E}_{b]}+\tilde{\varepsilon}_{abc}\tilde{B}^c\,.
\label{tFab}
\end{equation}

Note that all the transformations are given explicitly in terms of
irreducible quantities (i.e.~irreducible in the original
$u_a$-frame).

\subsection{Covariant commutation laws}\label{AssCCLs}
According to definition (\ref{deriv}a), the orthogonally projected
covariant derivative operator satisfies the condition
$\mathrm{D}_{a}h_{bc}=0$. This means that we can use $h_{ab}$ to
raise and lower indices in equations acted upon by this operator.
Following Frobenius' theorem, however, rotating spaces do not
possess integrable 3-D submanifolds
(e.g.~see~\cite{1984gere.book.....W,2004rtmb.book.....P}).
Therefore, the $\mathrm{D}_{a}$-operator cannot be used as a
standard 3-D derivative in such spaces and it does not always
satisfy the usual commutation laws (see below and
also~\cite{1990PhRvD..42.1035E}).

When acting on a scalar quantity the orthogonally projected
covariant derivative operators commute according to
\begin{equation}
\mathrm{D}_{[a}\mathrm{D}_{b]}f= -\omega_{ab}\dot{f}\,.
\label{A1}
\end{equation}
The above is a purely relativistic result and underlines the
different behaviour of rotating spacetimes within Einstein's theory. Similarly, the commutation law for the orthogonally projected
derivatives of spacelike vectors reads
\begin{equation}
\mathrm{D}_{[a}\mathrm{D}_{b]}v_{c}= -\omega_{ab}\dot{v}_{\langle
c\rangle}+ {\frac{1}{2}}\,\mathcal{R}_{dcba}v^{d}\,.  \label{A2}
\end{equation}
where $v_{a}u^{a}=0$ and $\mathcal{R}_{abcd}$ represents the Riemann
tensor of the observer's local rest-space. Finally, when dealing
with orthogonally projected tensors, we have
\begin{equation}
\mathrm{D}_{[a}\mathrm{D}_{b]}S_{cd}=
-\omega_{ab}h_{c}{}^{e}h_{d}{}^{f}\dot{S}_{ef}+
{\frac{1}{2}}\,(\mathcal{R}_{ecba}S^{e}{}_{d}
+\mathcal{R}_{edba}S_{c}{}^{e})\,,  \label{A3}
\end{equation}
with $S_{ab}u^{a}=0=S_{ab}u^{b}$. Note that in the absence of
rotation, $\mathcal{R}_{abcd}$ is the Riemann tensor of the
(integrable) 3-D hypersurfaces orthogonal to the $u_{a}$-congruence
For details on the definition, the symmetries and the key equations
involving $\mathcal{R}_{abcd}$, the reader is referred to
\S~\ref{sssSC}. We also note that the above equations are fully
nonlinear and hold at all perturbative levels.

In general relativity, time derivatives do not generally commute
with their spacelike counterparts. For scalars, in particular, we have
\begin{equation}
\mathrm{D}_{a}\dot{f}-
h_{a}{}^{b}(\mathrm{D}_{b}f)^{\displaystyle\cdot}= -\dot{f}A_{a}+
{\frac{1}{3}}\,\Theta\mathrm{D}_{a}f+
\mathrm{D}_{b}f\left(\sigma^{b}{}_{a}+\omega^{b}{}_{a}\right)\,,
\label{A4}
\end{equation}
at all perturbative levels. Assuming an FLRW background, we find
that the orthogonally projected gradient and the time derivative of
the first-order vector $v_{a}$ commute as
\begin{equation}
a\mathrm{D}_{a}\dot{v}_{b}=
\left(a\mathrm{D}_{a}v_{b}\right)^{\displaystyle\cdot}\,,
\label{A5}
\end{equation}
to linear order. Similarly, when dealing with first-order spacelike
tensors, we have the following linear commutation law
\begin{equation}
a\mathrm{D}_{a}\dot{S}_{bc}=
\left(a\mathrm{D}_{a}S_{bc}\right)^{\displaystyle\cdot}\,.
\label{A6}
\end{equation}

\subsection{Scalar, vector and tensor modes}\label{AssSVTMs}
In the coordinate-based approach, perturbations are decomposed from
the start into scalar, vector and tensor modes, using appropriate
harmonics. The covariant approach does not depend on a priori
splitting into harmonic modes and it is independent of any
Fourier-type decomposition. Instead, all the perturbative quantities are described as spatial vectors $V_a=V_{\langle a\rangle}$ or as
spatial, symmetric and trace-free (PSTF) rank-2 tensors
$S_{ab}=S_{\langle ab\rangle}$ (higher-rank PSTF tensors are needed
in kinetic theory -- see \S~\ref{sKINETIC}).

The scalar modes are characterised by the fact that all vectors and
tensors are generated by scalar potentials. For instance,
\begin{equation}
V_a=\D_a V \hspace{15mm} {\rm and} \hspace{15mm} S_{ab}=\D_{\langle
a}\D_{b\rangle}S\,, \label{smodes}
\end{equation}
for some $V, S$. This implies that $\curl V_a=0=\curl S_{ab}$.

For vector modes, all vectors are transverse (solenoidal) and
proportional to $\omega_a$. Also, all tensors are generated by
transverse vector potentials. Thus,
\begin{equation}
\D^aV_a=0 \hspace{15mm} {\rm and} \hspace{15mm} S_{ab}= \D_{\langle
a}S_{b\rangle}\,, \label{vmodes}
\end{equation}
where $\D^aS_a=0$. Vector modes are nonzero if and only if the
vorticity is nonzero.

Tensor modes are characterised by the vanishing of all vectors and
by the transverse traceless nature of all tensors. In other words,
\begin{equation}
V_a=0 \hspace{15mm} {\rm and} \hspace{15mm} \D^bS_{ab}=0\,.
\label{tmodes}
\end{equation}
This way no perturbative scalars or vectors can be formed.

We can expand these modes in harmonic basis functions (Fourier modes in the case $K=0$). For example, for scalar modes, the harmonics are time-independent eigenfunctions that satisfy the scalar
Laplace-Beltrami equation. In other words, $\dot{\clq}^{(k)}=0$ and
\begin{equation}
{\rm D}^2_a\clq^{(k)}=-\left({k\over a}\right)^2\clq^{(k)}\,,
\label{sLB}
\end{equation}
where $k$ is the eigenvalue of the associated harmonic mode and ${\rm D}^2={\rm D}^a{\rm D}_a$. The latter takes continuous values when $K=0,-1$ and discrete ones for $K=+1$. In particular, $k=\nu\geq0$ when the 3-space has Euclidean geometry and $k^2=\nu^2+1\geq0$ for hyperbolic spatial sections, with $\nu$ representing the comoving wavenumber of the mode in all cases. Supercurvature modes have $\lambda=a/k>a$ and in open FLRW models correspond to $0\leq k^2<1$. Those with $k^2>1$, on the other hand, span scales smaller than the curvature radius and are therefore termed subcurvature. Clearly, the $k^2=1$ threshold indicates the curvature scale, with $\lambda=\lambda_K=a$ (see also \S~\ref{sssFLRWCs}). Note that, although they are often ignored (e.g.~see~\cite{1967RvMP...39..862H}), supercurvature modes are necessary if we want perturbations with correlations lengths bigger than the curvature radius~\cite{1995PhRvD..52.3338L}. Finally, when the 3-curvature is positive, $k^2=\nu(\nu+2)$ and $\nu=1,2,\ldots$.

\section{Notation}\label{BsN}
\begin{itemize}
\item \textbf{Spacetime Geometry}\newline Line element: $\mathrm{d}s^{2}=g_{ab}\mathrm{d}x^{a}\mathrm{d}x^{b}= -\mathrm{d}\tau^{2}$, with $c=1$.\newline 4-velocity: $u^{a}=\mathrm{d}x^{a}/\mathrm{d}\tau$, 3-D projection tensor: $h_{ab}=g_{ab}+u_{a}u_{b}$.\newline 4-D permutation tensor: $\eta_{abcd}$, 3-D permutation tensor: $\varepsilon_{abc}=\eta_{abcd}u^{d}$.\newline Covariant derivative: $\nabla_{b}T_{a}=\partial T_{a}/\partial x^{b}-\Gamma_{ab}^{c}T_{c}$.\newline Time derivative: $\dot{T}_{a}=u^{b}\nabla_{b}T_{a}$, 3-D covariant derivative: $\mathrm{D}_{b}T_{a}=h_{b}{}^{d}h_{a}{}^{c}\nabla_{d}T_{c}$.\newline Riemann tensor: $R_{abcd}$, Ricci tensor: $R_{ab}=R^{c}{}_{acb}$, Ricci scalar: $R=R^{a}{}_{a}$.\newline 3-Riemann tensor: $\mathcal{R}_{abcd}$, 3-Ricci tensor: $\mathcal{R}_{ab}=\mathcal{R}^{c}{}_{acb}$, 3-Ricci scalar: $\mathcal{R}=\mathcal{R}^{a}{}_{a}$.\newline 3-curvature index: $K=0,\,\pm1$, with $\mathcal{R}=6K/a^{2}$ (in FLRW models).\newline Weyl Tensor: $C_{abcd}$, electric Weyl: $E_{ab}=C_{acbd}u^{c}u^{d}$, magnetic Weyl: $H_{ab}=\varepsilon_{a}{}^{cd}C_{cdbe}u^{e}/2$.

\item \textbf{Kinematics}\newline Expansion scalar: $\Theta=\nabla^{a}u_{a}=\mathrm{D}^{a}{}u_{a}$, scale factor: $a$, with $\dot{a}/a=\Theta /3$. \newline Conformal time: $\eta$, with $\dot{\eta}=1/a$. \newline Vorticity tensor: $\omega_{ab}=\mathrm{D}_{[b}u_{a]}$, vorticity vector: $\omega_{a}=\varepsilon_{abc}\omega^{bc}/2$. \newline Shear tensor: $\sigma_{ab}=\mathrm{D}_{\langle b} u_{b\rangle}= \mathrm{D}_{(b}u_{b)}- (\mathrm{D}^{c}u_{c})h_{ab}/3$, 4-acceleration: $A_{a}=u^{b}\nabla_{b}u_{a} $. \newline Hubble parameter: $H=\dot{a}/a$, deceleration parameter: $q=-\ddot{a}a/\dot{a}^2$.

\item \textbf{Matter Fields}\newline Field equations: $R_{ab}-(R/2)g_{ab}=T_{ab}$, with $\kappa=8\pi G=1$.\newline Matter energy-momentum tensor: $T_{ab}=\rho u_{a}u_{b}+ph_{ab}+ 2u_{(a}q_{b)}+\pi_{ab}$.\newline Matter density: $\rho=T_{ab}u^{a}u^{b}$, isotropic pressure: $p=T_{ab}h^{ab}/3$.\newline Barotropic index: $w=p/\rho$, adiabatic sound speed: $c^{2}_{\mathrm{s}}=\dot{p}/\dot{\rho}$.\newline Energy flux: $q_{a}=h_{a}{}^{b}T_{bc}u^{c}$, anisotropic pressure: $\pi_{ab}=T_{\langle ab\rangle}= T_{(ab)}-(T/3)h_{ab}$. \newline Particle flux vector: $N_a=nu_a+\cln_a$. \newline Particle number density: $n=-N_au^a$, particle drift: $\cln_a=h_a{}^bN_b$. \newline Entropy flux: $S_a=Su_a$ (in equilibrium). \newline Entropy density: $S=-S_au^a$, specific entropy: $s=S/n$.

\item \textbf{Electromagnetism}\newline Electromagnetic tensor: $F_{ab}$, magnetic field: $B_{a}=\varepsilon_{abc}F^{bc}/2$, electric field: $E_{a}=F_{ab}u^{b}$.\newline Energy density: $(B^2+E^2)/2$, isotropic pressure: $(B^2+E^2)/6$. \newline Poynting vector: $\clp_a=\varepsilon_{abc}E^bB^c$, anisotropic pressure: $\Pi_{ab}=-E_{\langle a} E_{b\rangle}-B_{\langle  a}B_{b\rangle }$. \newline Alfv\'{e}n speed: $c_{\mathrm{a}}^{2}=B^{2}/(\rho+p+B^{2})$. \newline Electric 4-current: $J_{a}$, electric 3-current: $\mathcal{J}_{a}=J_{\langle a\rangle}=h_{a}{}^{b}J_{b}$. \newline Charge density: $\mu=-J_{a}u^{a}$, electrical conductivity: $\varsigma$.

\item \textbf{Minimally coupled scalar fields}\newline Scalar field: $\varphi$, with $\dot{\varphi}=-\nabla_a\nabla^a\varphi>0$ and ${\rm D}_a\varphi=0$. \newline Potential: $V(\varphi)$, with $\nabla_a\nabla^a\varphi-V^{\prime}(\varphi)=0$. \newline Energy density: $\rho^{(\varphi)}= \dot{\varphi}^2/2+V(\varphi)$, pressure: $p^{(\varphi)}=\dot{\varphi}^2/2-V(\varphi)$.

\item \textbf{Perturbations}\newline Matter density gradients: $\Delta_{a}=(a/\rho)\mathrm{D}_{a}\rho$, with $\Delta_{ab}=a\mathrm{D}_{b}\Delta_{a}$ and $\Delta=\Delta^{a}{}_{a}$.\newline Matter vortices: $\mathcal{W}_{ab}=\Delta_{[ab]}$, with $\mathcal{W}_{a}=\varepsilon_{abc}\mathcal{W}^{bc}/2$.\newline Volume expansion gradients: $\mathcal{Z}_{a}=a\mathrm{D}_{a}\Theta$, with $\mathcal{Z}_{ab}=a\mathrm{D}_{b}\mathcal{Z}_{a}$ and $\mathcal{Z}=\mathcal{Z}^{a}{}_{a}$.\newline Magnetic density gradients: $\mathcal{B}_{a}=(a/B^{2})\mathrm{D}_{b}B^{2}$ and $\mathcal{B}_{ab}=a\mathrm{D}_{b}\mathcal{B}_{a}$, with $\mathcal{B}=\mathcal{B}^{a}{}_{a} $.\newline Effective entropy perturbations: $\mathcal{E}_{a}$, $\mathcal{S}_{a}^{(ij)}$, with $\mathcal{S}_{a}^{(ij)}= -\mathcal{S}_{a}^{(ji)}$. \newline Peculiar velocity: $v_a$, with $v_au^a=0$.
\end{itemize}

\bibliography{covariant_review_chap1235,covariant_review_chap46}
\bibliographystyle{h-elsevier}

\end{document}